\documentclass[epjH]{svjour}
\usepackage{epsfig}
\usepackage{graphicx}
\usepackage{hyperref}
\usepackage {amsmath}
\usepackage {amssymb}
\usepackage {bm}  
\usepackage{latexsym}
\usepackage[usenames,dvipsnames]{color}
\def\lsim{\raise0.3ex\hbox{$<$\kern-0.75em\raise-1.1ex\hbox{$\sim$}}}
\def\gsim{\raise0.3ex\hbox{$>$\kern-0.75em\raise-1.1ex\hbox{$\sim$}}}

\def\mean#1{\left<#1\right>}
\def\Journal#1#2#3#4{{#4}. {\it #1} {\bf #2}: #3}
\def\IJMPA{{Int. J. Mod. Phys. A}}
\def\IJMPE{{Int. J. Mod. Phys. E}}
\def\EPJC{{Eur. Phys. J. C}}

\def\NCA{Nuovo Cimento\ }

\def\NIMA{{Nucl. Instrum. Methods A}}
\def\NPA{{Nucl. Phys. A}}
\def\NPB{{Nucl. Phys. B}}
\def\PLB{{Phys. Lett. B}}
\def\PLC{Phys. Repts.\ }
\def\PL{Phys. Lett.\ }
\def\PRL{Phys. Rev. Lett.\ }
\def\PRD{{Phys. Rev. D}}
\def\PRC{{Phys. Rev. C}}
\def\PR{Phys. Rev.}

\def\RMP{Rev. Mod. Phys.\ }
\def\RPP{Rep. Prog. Phys.\ }
\def\ZPC{{Z. Phys. C}}
\def\ARNPS{{Ann. Rev. Nucl. Part. Sci.\ }} 
\def\RMP{Rev. Mod. Phys.\ }

\def\QGP{{\color{Red} Q}{\color{Blue} G}{\color{Green} P}} 
\def\QCD{{\color{Red} Q}{\color{Green} C}{\color{Blue} D}} 
\begin{document}
\title{How hadron collider experiments contributed to the development of \QCD:\\from hard-scattering to the perfect liquid}
\author{M.~J.~Tannenbaum\thanks{\email{mjt@bnl.gov}}}
%
\institute{Physics Department, Brookhaven National Laboratory Upton, NY 11973-5000 USA \ }
\abstract{A revolution in elementary particle physics occurred during the period from the ICHEP1968 to the ICHEP1982 with the advent of the parton model from discoveries in Deeply Inelastic electron-proton Scattering at SLAC, neutrino experiments, hard-scattering observed in p$+$p collisions at the CERN ISR, the development of \QCD,  the discovery of the J/$\Psi$ at BNL and SLAC and the clear observation of high transverse momentum jets at the CERN SPS $\bar{p}+p$ collider. These and other discoveries in this period led to the acceptance of \QCD\ as the theory of the strong interactions. The desire to understand nuclear physics at high density such as in neutron stars led to the application of \QCD\ to this problem and to the prediction of a Quark-Gluon Plasma (\QGP ) in nuclei at high energy density and temperatures. This  eventually led to the construction of the Relativistic Heavy Ion Collider (RHIC) at BNL to observe superdense nuclear matter in the laboratory. This article discusses how experimental methods and results which confirmed \QCD\ at the first hadron collider, the CERN ISR, played an important role in experiments at the first heavy ion collider, RHIC, leading to the discovery of the \QGP\ as a perfect liquid as well as discoveries at RHIC and the LHC which continue to the present day.}
\maketitle
\section{Introduction}

The beginning of the revolution in understanding the strong interactions of elementary particles took place in the period from the 14th International Conference on High Energy Physics in Vienna, Austria in 1968 (ICHEP1968) to the 16th ICHEP at Fermilab \& Chicago in 1972. 
\subsection{ICHEP1968}
Ironically it was a result from Deeply Inelastic Scattering (DIS) that started it. This was presented at the end of Pief Panofsky's rapporteur talk in the Electromagnetic Interactions - Experimental session at ICHEP1968~\cite{Pief-Vienna}. The preliminary results from the SLAC-MIT electron+proton scattering experiment showed two very striking features, the first in the structure function and the second in the ratio of the inelastic to elastic scattering cross sections.

\begin{figure}[!ht]
\begin{center}
\hspace*{-0.1in}\includegraphics[width=0.70\linewidth]{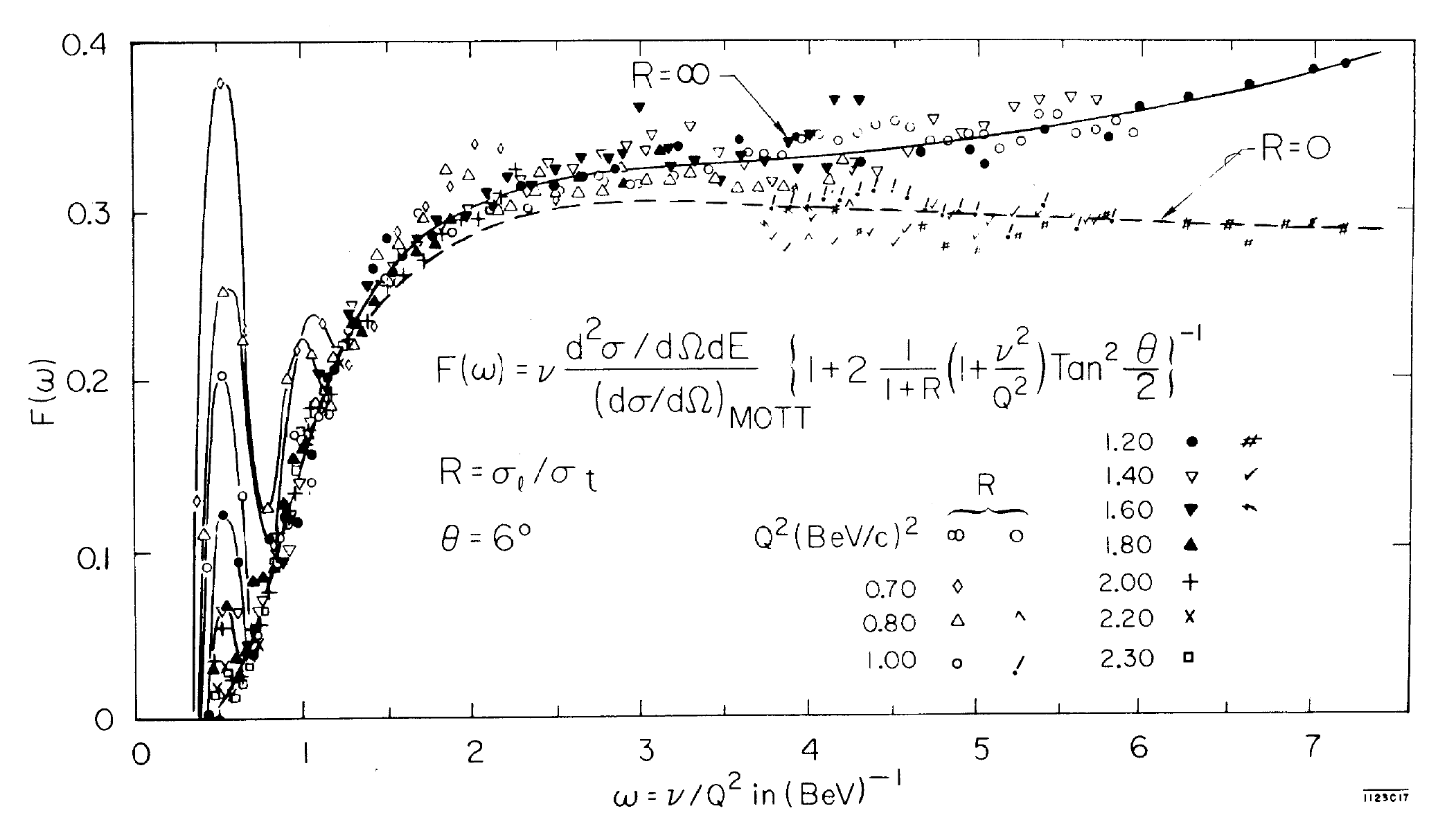} 
\hspace*{-0.1in}\raisebox{0.0pc}{\includegraphics[width=0.33\linewidth]{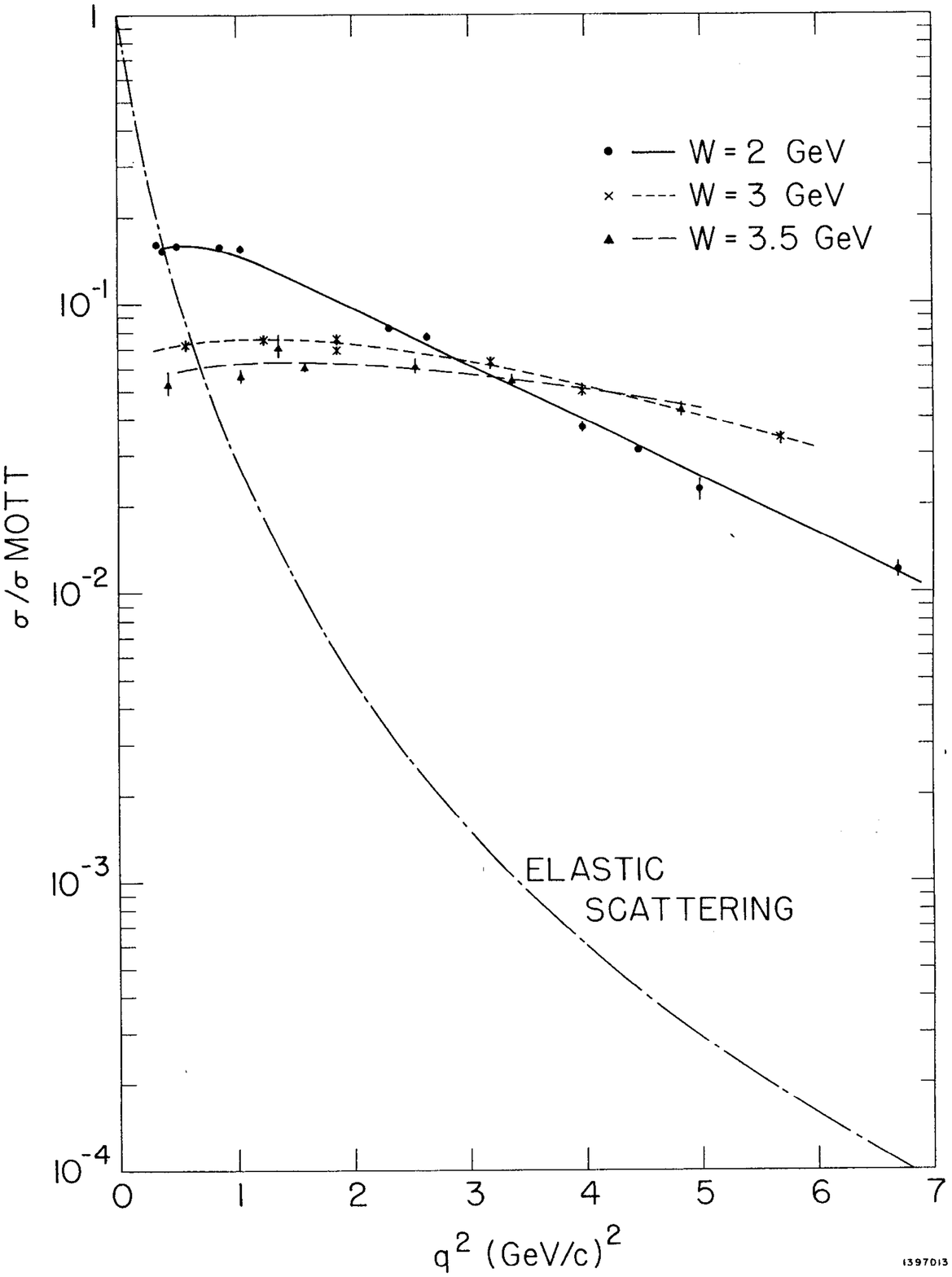}}
\end{center}
\caption[]
{a) (left) $F({\nu}/{Q^2})$ as a function of $\omega=\nu/Q^2$ with values of $Q^2$ indicated for $\theta= 6^o$~\cite{Pief-Vienna} b) (right) ($d^2\sigma/d\Omega dE')/\sigma_{\rm Mott}$) in GeV$^{-1}$ vs $q^2$ for 3 values of $W$ the invariant mass of the recoil hadron state, indicated, compared to ($d\sigma/d\Omega)_{\rm elastic}/\sigma_{\rm Mott}$) where $\sigma_{\rm Mott}=(4\pi \alpha^2\cos^2\theta/2)/{Q^4}$, calculated for $10^\circ$~\cite{DIS}.    
\label{fig:Bjscaling} }
\end{figure}

The inelastic scattering cross section for the case when only the outgoing electron is detected is given by the formula~\cite{DrellWalecka64}
\begin{equation}
\frac{d^2 \sigma}{dQ^2 d\nu}=\frac{4\pi\alpha^2}{Q^4} \left[ W_2(Q^2,\nu)\left( 1-\frac{\nu}{E}-\frac{Q^2}{4E^2}\right) + 2 W_1(Q^2,\nu) \frac{Q^2}{4E^2} \right] 
\label{eq:MJTNIMA}
\end{equation}
where $W_2(Q^2,\nu)$ and $W_1(Q^2,\nu)$ are structure functions which each depend on the 4-momentum transfer squared, $Q^2$, and the energy loss $\nu=E-E'$ of the incoming electron with energy $E$ and outgoing energy $E'$.    
The amazing result was that although one might expect many plots of $W_1$ and $W_2$ as functions of various values of $Q^2$ and $\nu$, they all collapse to one curve 
as suggested by Bjorken~\cite{BjPR179}  
\begin{equation} F_2(Q^2,\nu)\equiv\nu W_2(Q^2,\nu)=F_2(\frac{Q^2}{\nu}) \label{eq:F2}\end{equation}
and thus known as Bjorken Scaling (Fig.~\ref{fig:Bjscaling}a). 
The other important point made by Panofsky (Fig.~\ref{fig:Bjscaling}b)~\cite{DIS} was that the inelastic  cross sections in the continuum ``are very large and decrease much more slowly with $Q^2$ than the cross-sections for elastic-scattering and the specific resonant states.''

All this led Bjorken (inspired by Feynman)~\cite{BjPaschosPR185} to the concept of a proton composed of fundamental pointlike constituents (partons) from which the electrons scatter elastically, incoherently  as viewed in a frame in which the proton has infinite momentum. The free partons each have longitudinal momentum a fraction $x$ of the proton longitudinal momentum $\mathbf{P}$ where the probability of the parton to have the momentum $x\mathbf{P}$ is given by $F_2(x)$ where $x=Q^2/(2M\nu)$ and $M$ is the mass of the proton. As an experimentalist, I had done muon-proton elastic scattering in the more usual frame where the proton is at rest in a liquid hydrogen target. The proton recoils with kinetic energy $\nu$ from an elastic scattering by the $\mu$ (or $e$) with 4-momentum-transfer$^2$ $Q^2$ where $\nu=Q^2/(2M)$ so that it was easy for me to understand that DIS of an $e$ from a proton is just quasi-elastic scattering from a parton at rest with a fracton $x$ of the mass of the proton which would become a fraction of the momentum in any moving frame. 

\subsection{ICHEP1972}
\begin{figure}[!ht]
\begin{center}
\hspace*{-0.0pc}\includegraphics[width=0.45\linewidth]{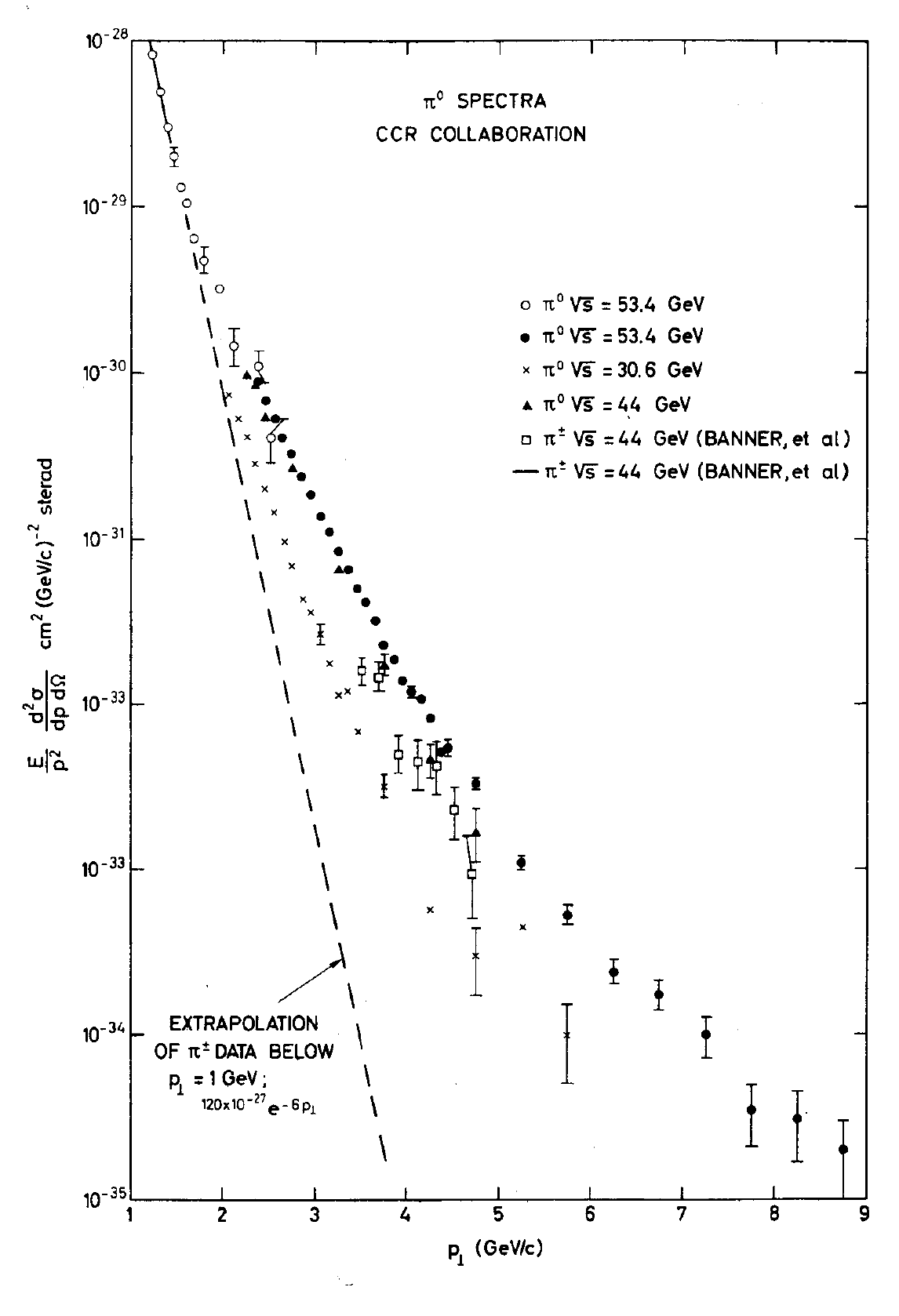} 
\hspace*{0.5pc}\raisebox{0.0pc}{\includegraphics[angle=1.2,width=0.50\linewidth]{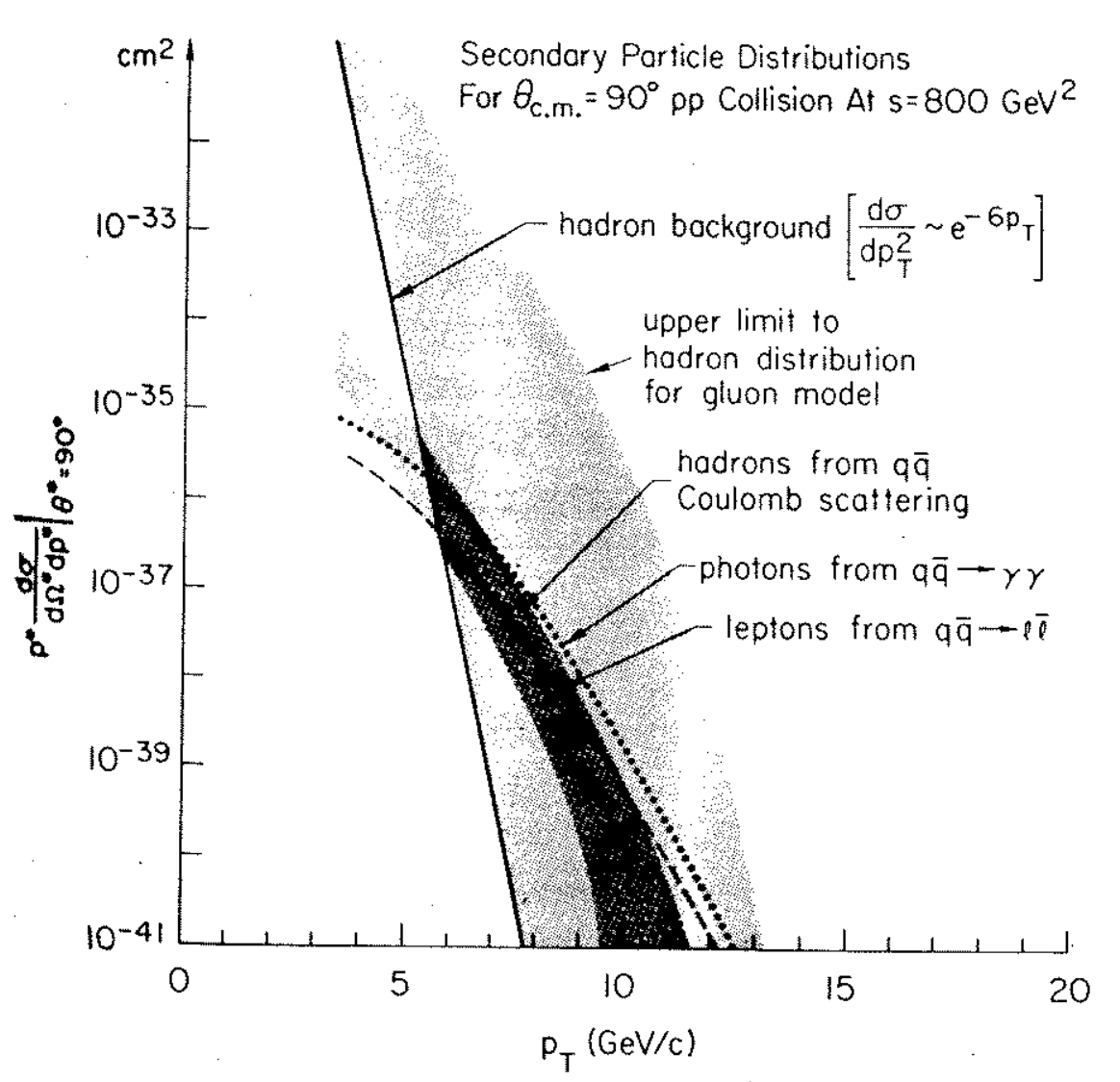}}  
\end{center}
\caption[]
{a) (left) $p_T$ spectra of invariant cross sections $Ed^3\sigma/dp^3$ of $\pi^0$ and $\pi^+$ for several $\sqrt{s}$ at the CERN-ISR~\cite{CoolICHEP72}. b)(right) Predicted $p_T$ spectrum of hadrons from Coulomb scattering of partons in p$+$p collisions at $\sqrt{s}=28.3$ GeV compared to the $e^{-6p_T}$ ~\cite{Cocconi}formula background~\cite{BBK}.    
\label{fig:CoolBBK} }
\end{figure}

Frankly many physicists did not accept the idea that the proton was composed of a gigantic number of massless constituents (partons)~\cite{FriedmanNobel}. In fact we used to call Manny Paschos, who worked at Brookhaven National Laboratory (BNL) in this period, ``Manny Parton''. Everything changed for most people at the 16th ICHEP in 1972 at Fermilab and Chicago where three spectacular new results were presented.
\begin{enumerate}
\item[1.] Don Perkins in his conclusions~\cite{PerkinsICHEP72} from new measurements of neutrino scattering cross sections and nucleon structure functions from the CERN Gargamelle collaboration emphasized: i) that both $\nu$ and $\bar{\nu}$ cross sections in the range 2-10 GeV are linear with energy, in accord with Bjorken scaling; \ldots iv) In terms of constituent models, the fractional charged (Gell-Mann/Zweig) quark model~\cite{GellMannQ,ZweigA} is the \underbar{only} one that fits both the neutrino and electron data ($\int F_2^{\nu N}(x)\, dx/\int F_2^{e N}(x)\, dx=3.4\pm 0.7\approx 18/5$); v) The fractional nucleon 4-momentum carried by gluons is 50\% and vi) by antiquark constituents is only $\sim 10$\% of that carried by quarks and antiquarks together.
\item[2.] The birth of \QCD\ where Fritzsch and Gell-Mann~\cite{QCDbirth} proposed that the three distinct colors assigned to the 3 $s$-quarks in the \mbox{$\Omega{^-}(sss)$} \cite{Gell-Mann2013} to allow them to be in the same state, avoiding simple Fermi statistics by obeying ``para-Fermi statistics of rank 3''~\cite{Greenberg1964}, could also apply to gluons ``which could form a color-octet of neutral vector fields obeying the Yang-Mills equations''. 
\item[3.] The discovery of enhanced particle production at large transverse momentum $p_T$ in p$+$p collisions at the CERN-ISR~\cite{CoolICHEP72} with a power-law shape, which depended on the c.m. energy of the collision, $\sqrt{s}$, quite distinct from the exponential $e^{-6p_T}$~\cite{Cocconi} spectrum at low $p_T<1$ GeV/c which depended minimally if at all on $\sqrt{s}$ (Fig.~\ref{fig:CoolBBK}a). This was proof that the partons of DIS interacted strongly with each other rather than simply scatter electromagnetically which must occur in a p$+$p collision since the partons are charged (Fig.~\ref{fig:CoolBBK}b)~\cite{BBK}. 

\end{enumerate}

This last result changed the focus of many experimenters from the small angle, low transverse momentum region where most particles were produced, typical of fixed-target experiments in beams of high energy particles from an accelerator, to the region of large transverse momentum production, which at the CERN-ISR, the first hadron collider, was perpendicular to the axis of the colliding beams, a perfect location for the study of ``high $p_T$ physics''. 

The first proposals of experiments to the ISR in early 1969 had been dominantly in the forward direction so that the ISR committee decided to divide up the angular regions for different experiments at several interaction points to three regions~\cite{Russo}: small angles (up to 150 mrad); medium angles (100-300 mrad) and large angles (300 mrad to $\pi/2$ rad). The ISR did make discoveries and obtain important results on low $p_T$ ``soft'' physics~\cite{Jacob1984} so one may ask why some of the experiments wanted to run at the largest angles, where few if any particles were expected. The answer is that they were looking for $W^{\pm}$-bosons, the proposed quanta of the weak interactions~\cite{LYW60}.  

\section{Weak Bosons from strong interactions? 1960-1970}
The completion of the two highest energy ($\sim 30$ GeV) proton accelerators, the PS at CERN (November 1959) and the AGS at BNL (August 1960) opened the possibility of studying weak interactions at high energy rather than only by radioactive decay. The idea of the left-handed parity violating $W^{\pm}$ bosons as the quanta of the weak interactions~\cite{LYW60} led to proposals for neutrino beams and experiments at the two laboratories~\cite{Pontecorvo59} \cite{MS60} to measure neutrino scattering and to discover the $W$ which would prevent the otherwise point-like neutrino interaction cross sections from increasing monotonically with energy until they violated unitarity~\cite{LeeCERN61}. Although a Nobel Prize winning discovery, the muon neutrino~\cite{Danby62}, was made with the first accelerator neutrino experiment, the $W$ was not observed in the early neutrino experiments. Relatively low limits were set for the mass of the $W^{\pm}$ ($>2$ GeV)~\cite{CERNnoW64}\cite{AGS65} because of the low energy neutrino beams with small neutrino cross sections and it was realized that p$+$p collisions might be more favorable for producing the $W$, e.g p$+$p$\rightarrow d+W^+$~\cite{BernsteinPR129} \cite{Nearing1963} \cite{GoodPiccioni64}. This last reference is interesting for three reasons:
\begin{enumerate}
\item[1.] There is a comment from A. (Antonino) Zichichi~\cite{Nino64} on studies at CERN to observe the $W^{\pm}$ in p$+$p collisions, which is so clear that it deserves a verbatim quote:``{\it We would observe the $\mu$'s from $W$-decays. By measuring the angular and momentum distribution at large angles of $K$ and $\pi$'s, we can predict the corresponding $\mu$-spectrum. We then see if the $\mu$'s found at large angles agree with or exceed the expected numbers.}'' The $W^{\pm}$ would be visible above the background as a Jacobian peak at lepton transverse momentum $p_T^\mu=M_W/2$. 
\item[2.] The proposed experiment was never done because the original calculation did not include the $p$ or $d$ form factors and thus the original cross section calculation needed be reduced by a factor of $\approx 3000$~\cite{Nearing64}.  
\item[3.] This is exactly the way that the $W$ was discovered at the CERN $S\bar{p}pS$ collider 19 years later~\cite{UA1W,UA2W} as best shown in Ref.~\cite{UA2ZPC30}.
\end{enumerate}

\subsection{First p$+$p experiments searching for the $W$ 1965--1970}
Experiments overcame the low cross section for large angle muons from the reaction $p+p\rightarrow W$+anything by interacting the entire extracted proton beam in a well-shielded dense target where only muons could penetrate the shielding (beam dump experiments); but no signal was observed~\cite{BNLnoW65} \cite{ZGS65}. This brought up the question of how to know how many $W$ bosons should have been produced. Chilton, Saperstein and Shrauner~\cite{ChiltonPR148}  
emphasized the need to know the time-like form factor of the proton to calculate the cross section for producing $W$'s and used the Conserved Vector Current (CVC) theory to estimate the $W$ production cross section by relating it to di-muon production in the reaction $p+p\rightarrow \mu^+ + \mu^-$+anything. Yamaguchi~\cite{YamaguchiNC43} then proposed that the time-like form factor could be found by measuring the number of $e^+ + e^-$ or $\mu^+ + \mu^-$ ``massive virtual photons'' of the same invariant mass as the $W$ in the p$+$p reaction but warned that the individual leptons from these electromagnetically produced pairs or from real photons from the decay $\pi^0\rightarrow \gamma+\gamma$ from the huge number of $\pi^0$ produced  might mask the leptons from the $W^{\pm}$. 

These developments led Leon Lederman to propose in August 1967 a more elegant beam dump di-muon experiment at the BNL-AGS (AGS 420) with a thick Uranium target and muon momentum measurement by range in a series of iron and concrete absorbers to search for $W$ bosons, measure the time-like form factor and possibly look for `massive' vector meson production. I had proposed a di-muon experiment in May 1967 (AGS 412) modifying an existing Harvard photoproduction experiment (AGS 310)  so as to search for $\mu^+ \mu^-$ pairs produced in a Pb target by an incident muon beam---muon tridents. With three muons in the final state, two of them identical, we could test whether muons obeyed Fermi statistics or para-statistics and whether there was a force between muons other than electromagnetic. 

Leon and I had been in contact about $\mu^+ \mu^-$ pairs via mail over the summer but I decided to stick with my trident experiment which did in fact measure that the muon was a 
Fermion~\cite{Trident-PRL}. My story then loops around to Panofsky's talk at the ICHEP1968~\cite{MJTW}; but the outcome of Leon's di-muon experiment turned out to be one of the pillars of the parton model and just missed a truly revolutionary discovery at the BNL-AGS in 1974, the $J/\Psi$~\cite{Ting1974}, that cemented the parton model and belief in real quarks, which will be discussed in its chronological order below. 
\subsection{Drell-Yan leads to better $W$ searches and other discoveries} 
\begin{figure}[!th]
\begin{center}
\hspace*{-0.0pc}\includegraphics[width=0.50\linewidth]{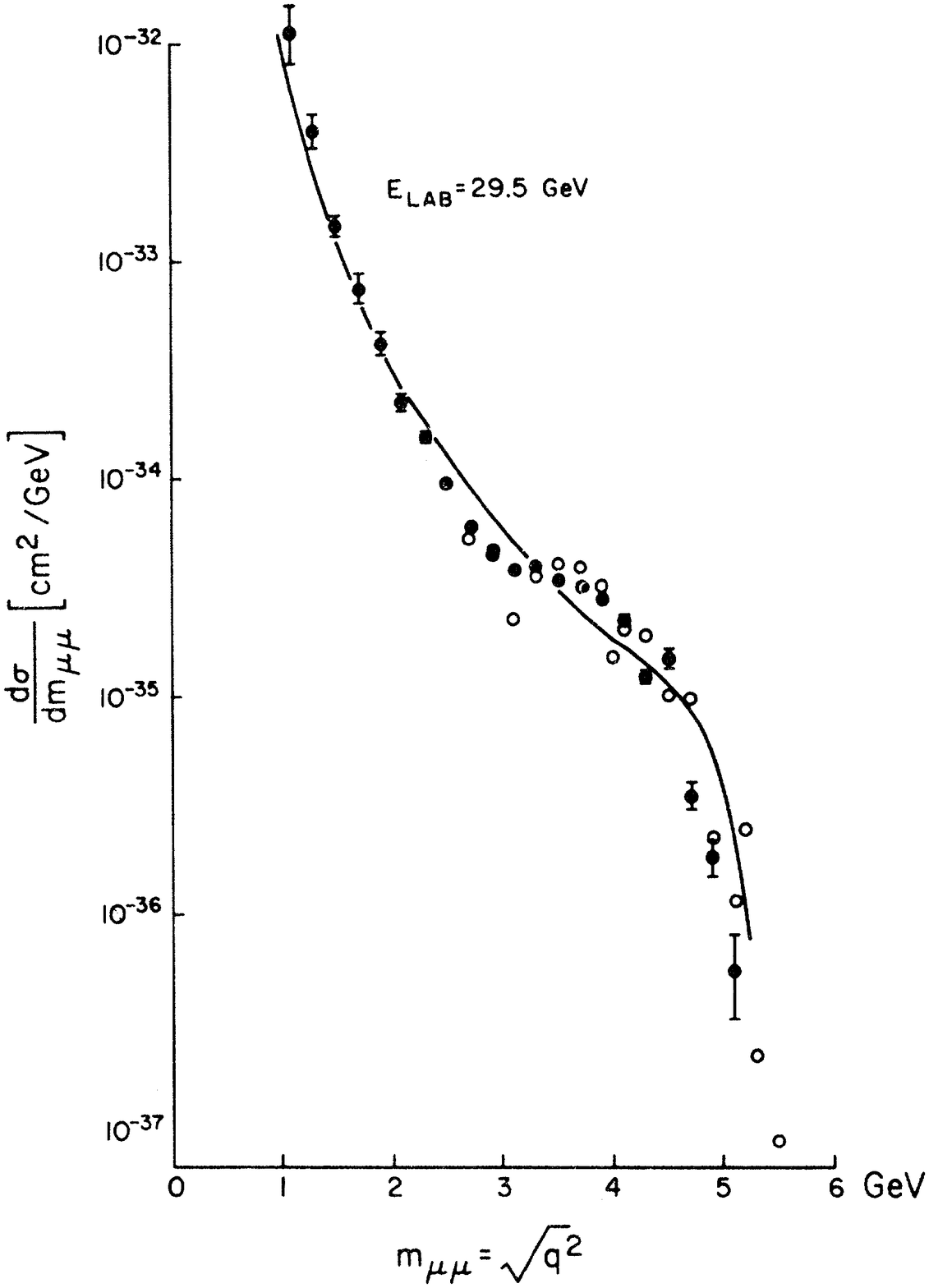} 
\hspace*{-1.05pc}\raisebox{1.0pc}{\includegraphics[angle=-0.5,width=0.46\linewidth]{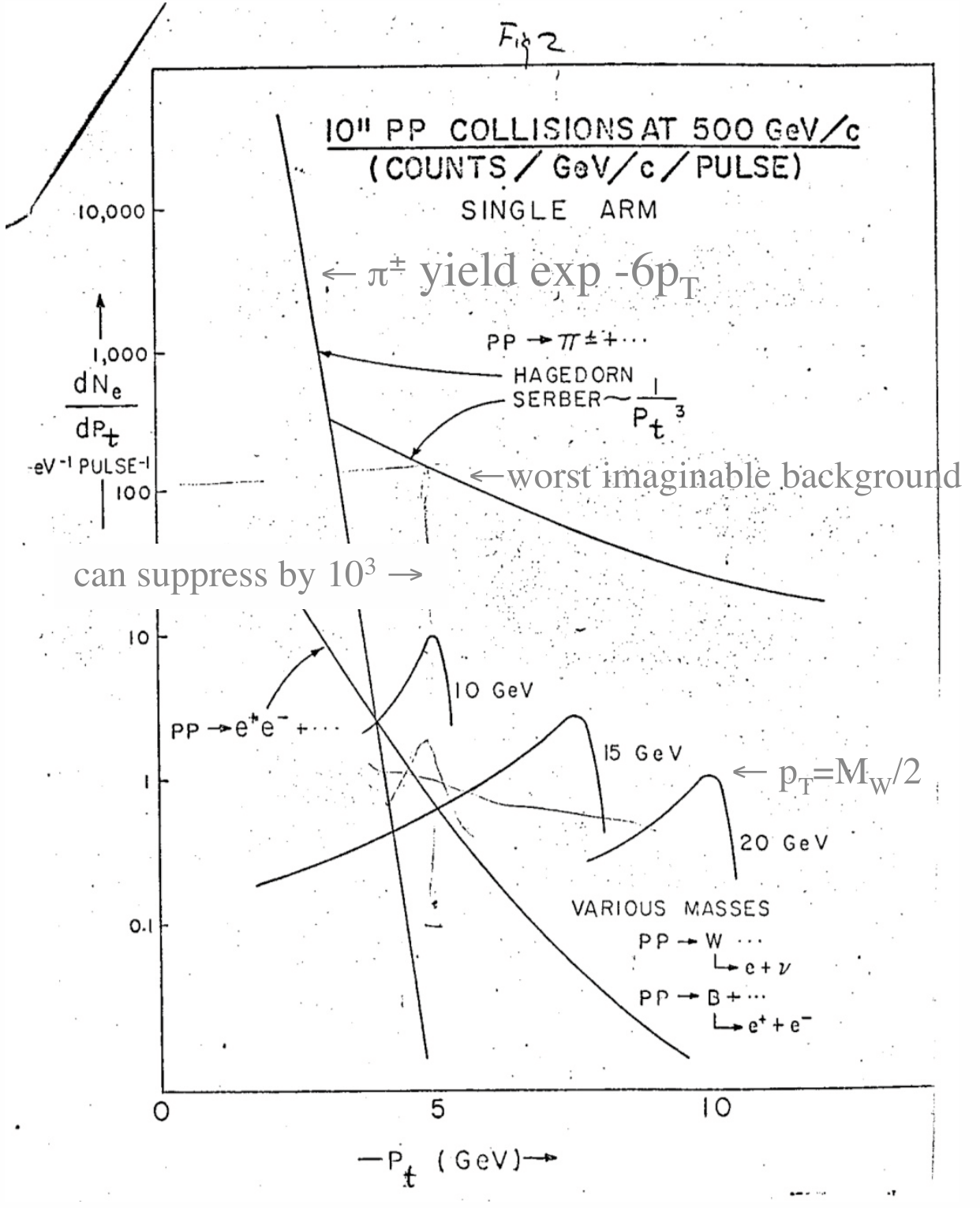}}  
\end{center}
\caption[]
{a) (left) Di-muon invariant mass spectrum $d\sigma/dm_{\mu\mu}$~\cite{LMLPRL25} with the totally incorrect theoretical prediction~\cite{GuidoPRL26} b)(right) the $p_T$ distribution of p$+$p$\rightarrow \mu^+ +\mu^- + X$. The cross section was calculated with Bjorken scaling from (a) with the $W$ cross section as $\sigma_W=0.05\ m_W^2\  d\sigma/dm_{\mu\mu}$ from CVC~\cite{ChiltonPR148}. The `Zichichi signatures' for possible $W$ masses are indicated~\cite{E70prop}Addendum\label{fig:LML}. }
\end{figure}

Having seen Lederman's preliminary results at an American Physical Society meeting, Sid Drell and Tung-Mow Yan at SLAC~\cite{DrellYanPRL25} used the parton model and Bjorken scaling in $m^2/s$ to calculate large mass  timelike virtual-photon i.e. lepton pair production in hadron-hadron `inelastic collisions' by parton-antiparton annihilation, or in today's terminology $\bar{q}+q\rightarrow \mu^+ + \mu^-$. There was also another calculation described in Ref.~\cite{DrellYanPRL25} which ``employs light-cone commutators and Regge theory, yielding a relative cross section in good agreement with the shape of the experimental distribution.''~\cite{GuidoPRL26} (Fig.~\ref{fig:LML}a). 

Lederman was very excited about this agreement with the measured $d\sigma/dm_{\mu\mu}$ distribution~\cite{LMLPRL25}. This was because with the use of the CVC theory~\cite{ChiltonPR148} and Bjorken scaling he could calculate the $W$ cross section at any $\sqrt{s}$ and thus predict the sensitivity of his di-muon proposal (E70,  which I initially joined~\cite{E70prop}) at the newly opening Fermilab (Fig.~\ref{fig:LML}b), as well as the di-electron proposal at the CERN ISR~\cite{CCRprops} which became the experiment that discovered the unexpectedly large high $p_T$ $\pi^0$ spectrum announced at ICHEP1972~\cite{CoolICHEP72}. 
	
The high $p_T$ discovery represented totally new physics, strongly interacting partons, which excited me very much. I resigned from all my Fermilab proposals, and being on the faculty of the Rockefeller University since 1971 I was able to move to Geneva to be with the Rockefeller group there and continue the work of the CERN-Columbia-Rockefeller (CCR) series of experiments which helped lead to the confirmation of \QCD\  as the theory of the strong interactions. 

Before returning to the aftermath of the high $p_T$ discovery, a comment from the di-muon measurement paper~\cite{LMLPRL25} is worth noting. Resonances were one of the original purposes of the experiment stated in the AGS 420 proposal, and were mentioned in the di-muon publication~\cite{LMLPRL25}: ``As seen in both the mass spectrum and the resultant cross section $d\sigma/dm$, there is no forcing evidence of any resonant structure'' although ``Indeed in the mass region near 3.5 GeV/c$^2$, the observed spectrum may be reproduced by a composite of a resonance and a steeper continuum.'' A proposal was actually made in September 1970 (AGS 549) by some younger members of the Lederman group to follow up the resonance possibility; but it never really got off the ground for various reasons.\footnote[1]{See Ref.~\cite{RATCUP} for further information about this issue and other issues raised but not discussed in full detail in this article.} As noted above the situation was finally cleared up in November 1974; but there were also many exciting developments in high $p_T$ physics and \QCD\ from 1971 to 1974.

\section{1971-1973 \QCD, Asymptotic Freedom, $\mathbf{x_T}$ scaling and more.}
Theoretical physicsists were very busy in this period inventing the theory of asymptotic freedom and \QCD\  
``from a Yang-Mills gauge model based on colored quarks and color octet gluons''~\cite{FGLPLB47},\cite{HDPPRL30}, \cite{GrossWilczekPRL30},  \cite{tHooft99} A quick summary of asymptotic freedom in \QCD\  is that the coupling constant obeys the equation~\cite{AltarelliParisiNPB126}:
\begin{equation}
\alpha_s(Q^2)=\frac{12\pi}{(11N_c-2n_f)\ln(Q^2/\Lambda^2)}=\frac{12\pi}{25\ln(Q^2/\Lambda^2)}
\label{eq:alphanf}
\end{equation} 
for $N_c=3$ colors and, as will turn out, $n_f=4$ flavors of quarks. The asymptotic freedom is the decrease  
of $\alpha_s(Q^2)$ as $Q^2$ increases. 
However, at this point in time asymptotic freedom had minimal if any impact on experimentalists at hadron colliders who were much more interested in the scaling rules and models directly related to the high $p_T$ measurements.  

\subsection{First scaling law 1971}
In addition to calculating the cross sections for Coulomb scattering of partons in p$+$p collisions, BBK~\cite{BBK} presented a scaling law for the Coulomb scattering which must exist for charged partons: 
\begin{equation}
E \frac{d^3\sigma}{dp^3}=\frac{4\pi\alpha^2}{p_T^4} {\cal F}
(\frac{-\hat{u}}{\hat{s}},\frac{-\hat{t}}{\hat{s}})
\label{eq:bbkscaling}
\end{equation}
where $\alpha=e^2/\hbar c$ is the fine structure constant.  
The two factors are a $1/p_T^4$ term, 
characteristic of single photon exchange, and a form factor ${\cal F}$ that scales, i.e. is only a function of the ratios of the Mandelstam variables, $\hat{s}$, $\hat{t}$ and 
$\hat{u}$, the constituent-scattering invariants for scattering of a parton with 4-momentum $p_1$, on another parton with 4-momentum $p_2$ to give an outgoing parton pair with $p_3$ and $p_4$, i.e. 
$p_1+p_2\rightarrow p_3 +p_4$. \vspace*{-1.0pc}

\subsubsection{Quick kinematics review}
      Relativistic kinematics are important for understanding parton-parton scattering and eventually \QCD\ so I give a brief description here with the Minkowski convention~\footnotemark[1]. The { four-vector} 
momentum $p$ of a particle with 3-vector momentum ${\bf P}$, energy $E$, in a 
particular rest frame, and invariant mass $m$, in units where the speed of light $c$ is taken as unity, is denoted such that the 4-th (time) component $\equiv i E$ :
\begin{equation}
p=({\bf P}, i E)\quad\mbox{or}\quad p=(P_x,P_y,P_z,i E)\quad \mbox{and}\quad p^2=p\cdot p={\bf P\cdot P} - E^2=-m^2
\quad .\label{eq:4vector}\end{equation}
The squared modulus of the 4-vector, $p^2$, is invariant under a Lorentz transformation. 

The Lorentz Invariant quantities to describe two-two scattering kinematics are:      
    \begin{eqnarray}
    -\hat{s}&=&(p_1 + p_2)^2=(p_3 +p_4)^2 \cr
    -\hat{t}&=&(p_1 - p_3)^2=(p_4 - p_2)^2 \cr
    -\hat{u}&=&(p_1 - p_4)^2=(p_3 - p_2)^2 \cr 
    \hat{s}+\hat{t}+\hat{u}&=& m^2_1 + m^2_2 + m^2_3 + m^2_4\qquad . \label{eq:stu} 
    \end{eqnarray}
where $\hat{s}$ is the parton-parton center of mass energy squared, $Q^2=-\hat{t}$ is the 4-momentum-transfer-squared in the parton-parton elastic scattering. See appendix~\ref{app:pscat} for a full 
discussion of parton-parton scattering kinematics.
\setcounter{footnote}{1}
\subsection{A better scaling rule, $\mathbf{x_T}$ scaling, 1972}
\label{sec:BBG}
Inspired by ``The recent measurements at CERN ISR of single-particle inclusive scattering at 90$^\circ$ and large transverse momentum'' which ``revealed several dramatic features of pion inclusive reactions'', Blankenbecler, Brodsky and Gunion (BBG) proposed a new general scaling formula, $x_T$ scaling ~\cite{BBG1972}: 
\begin{equation}
E \frac{d^3\sigma}{dp^3}=\frac{1}{p_T^{n_{\rm eff}}} F(x_T) \qquad \mbox{where} 
\qquad x_T\equiv \frac{2 p_T}{\sqrt{s}} \quad .
\label{eq:BBG}
\end{equation} 
The power ${n_{\rm eff}}$ is determined by the force-law 
between constituents, the quantum-exchange governing the reaction. Thus, ${n_{\rm eff}}=4$ for QED or ``Vector Gluon Exchange''. However, BBG also included a new model of their own, the ``constituent interchange model (CIM)'' which predicted that the ``proton beam at the ISR behaved like a meson beam of much lower energy'' so that high $p_T$ pions were produced by ``quark-meson scattering by the exchange of a quark'' which gave ${n_{\rm eff}}=8$ in their model ``which does give an excellent account of the data''~\cite{CCRPLB46}. 
\subsection{The 3 ISR experiments publish, 1973}
Three ISR experiments published observation of enhanced high $p_T$ production at 90$^\circ$ in the c.m. system in 1973. The first publication, from the British-Scandinavian ISR Collaboration~\cite{AlperPLB44} measured non-identified charged hadrons with $1.5\leq p_T\leq 4.5$ GeV/c at $\sqrt{s}=44$ and 53 GeV, with the comment ``no strong energy dependence is observed for these transverse momenta''. The second publication by the Saclay-Strasbourg group~\cite{BannerPLB44} used a Cerenkov counter to identify $\pi^{\pm}$ with $3.5\leq p_T\leq 5$ GeV/c and noted an $e^{-2.4 p_T}$ dependence for the invariant cross section in this range ``in sharp contrast with the'' $e^{-6 p_T}$ shape of the spectrum below 1 GeV/c. More importantly, they also measured $\pi^0$ by conversion of the $\pi^0\rightarrow \gamma+\gamma$ decay photons in the ISR vacuum pipe but found no single $e^{\pm}$ with $p_T >3$ GeV/c so set a limit for single electron production.   

\begin{figure}[!th]
\begin{center}
\includegraphics[width=0.46\linewidth]{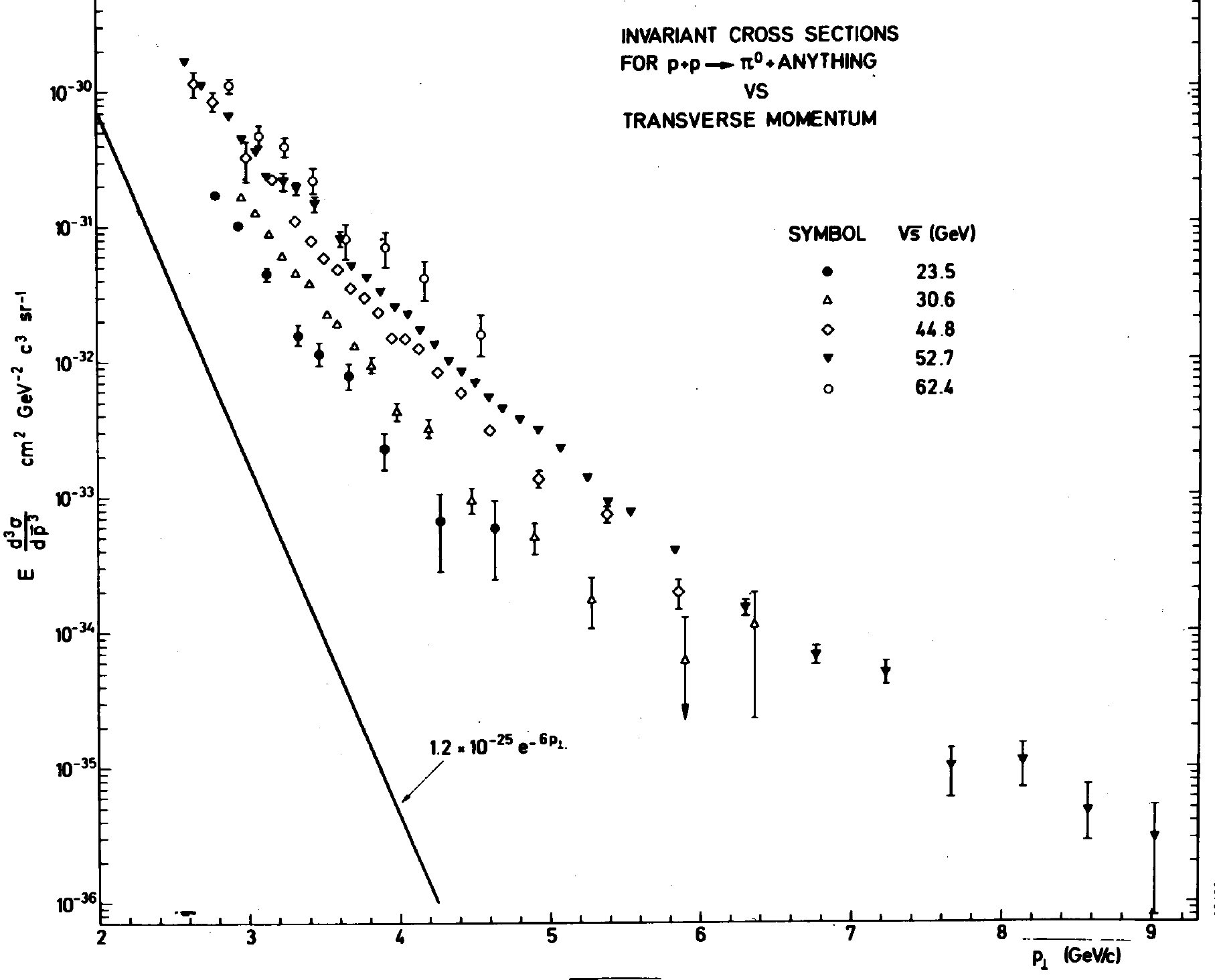}
\includegraphics[width=0.53\linewidth]{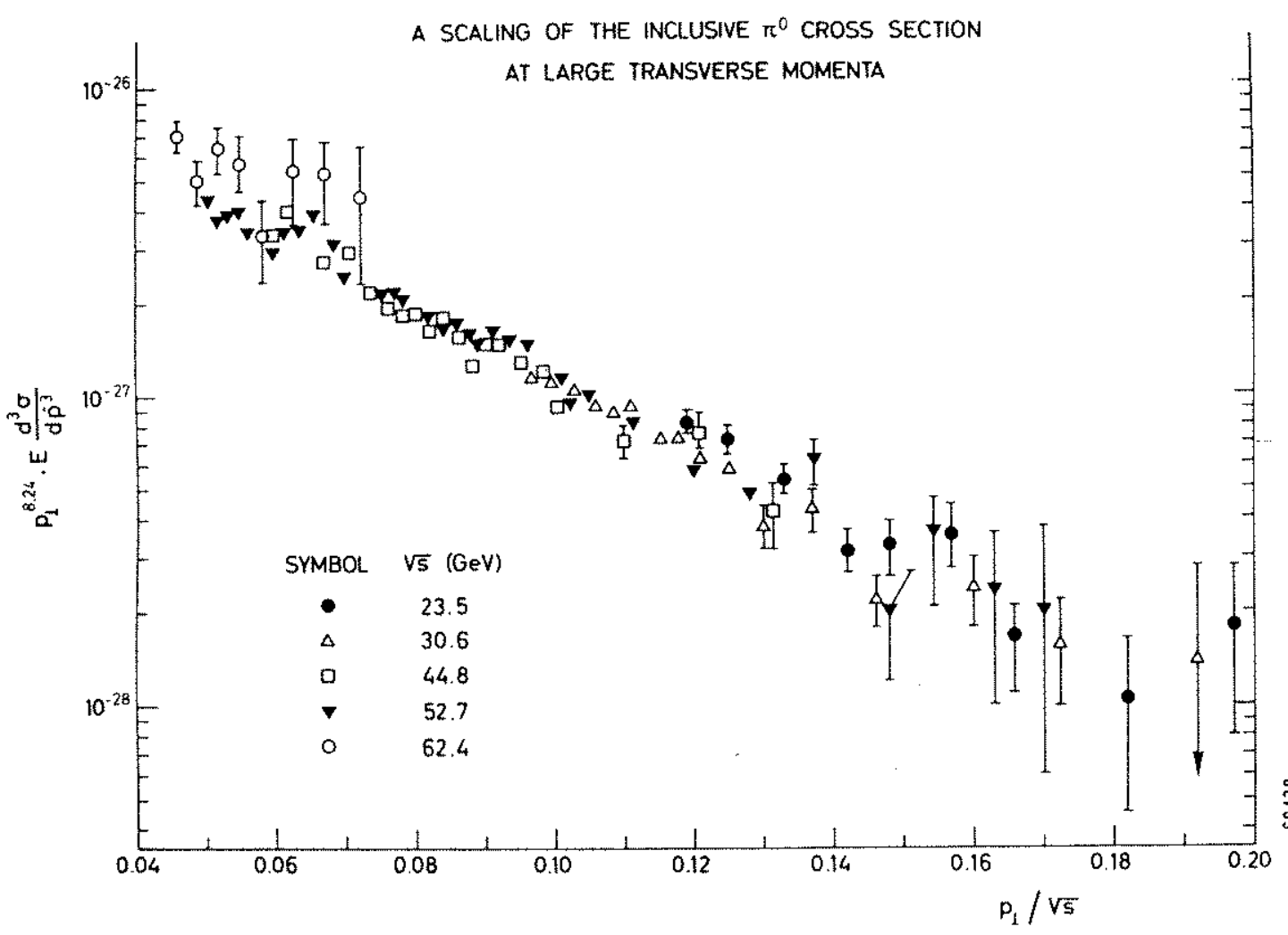}
\caption[] {\cite{CCRPLB46} a) (left) The invariant cross section $E d^3\sigma/dp^3$ for p$+$p$\rightarrow \pi^0 +X$ vs.  $p_\perp$ for 5 c.m. energies. b) (right) \label{fig:CCRpi0} The function $F(x_T)=p_{\perp}^{n_{\rm eff}} E d^3\sigma/dp^3$ deduced from the measurements in (a) using the best fit value,  $n=8.24$. The errors are statistical only.}
\end{center}
\end{figure} 

The best and most influential measurement was the final publication of the preliminary data (Fig.~\ref{fig:CoolBBK}a) shown at ICHEP1972 by the CERN Columbia Rockefeller (CCR) group (Fig.~\ref{fig:CCRpi0}a)~\cite{CCRPLB46}. The $\pi^0$ were detected in two identical Pb Glass electromagnetic calorimeters  preceded by tracking chambers located at 90$^\circ$ on both sides of the colliding proton beams. The $\pi^0$ spectrum was measured in the range $2.4\leq p_T\leq 9$ GeV/c at 5 c.m. energies $\sqrt{s}=23.5,30.6,44.8, 52.7$ and 62.4 GeV.  A  very large $\sqrt{s}$ dependence was observed for the cross sections, which were all ``orders of magnitude'' higher than an extrapolation of the ``well known exponential'', $e^{-6 p_\perp}$. 
The break in the cross-section due to parton-parton electromagnetic scattering anticipated by BBK (Fig.~\ref{fig:CoolBBK}b)~\cite{BBK} was observed but did not agree with magnitude or the form of their Coulomb prediction, $E d^3\sigma/dp^3\propto p_\perp^{-4} F(x_T)$. However, as noted above, the data for all $\sqrt{s}$ could be well fit by the BBG $x_T$ scaling form (Eq.~\ref{eq:BBG}) with the value $n_{\rm eff}=8.24\pm 0.05$(stat)$\pm 0.70$(sys) with $F(x_T)=A \exp (-bx_T)$, in agreement with the CIM prediction~\cite{BBG1972}, as illustrated by a plot of the data in the form $p_\perp^{8.24} \times E d^3\sigma/dp^3$ (Fig.~\ref{fig:CCRpi0}b).
\section{1974-1976 a charming period.} 
\subsection{The discovery of charm particles by direct $\mathbf{e^{+/-}}$ production at the CERN ISR} 
\label{sec:CCRS1}   
This discovery~\cite{CCRSPLB53} is not generally credited to the ISR, nor even mentioned in many articles about charm physics~\cite{Tavernier} or the ISR~\cite{Darriulat1} because charm quarks ($c$) were only a theoretical idea~\cite{BJShelly},\cite{GIM} at that time. 
The discovery came about because two of the experiments, CCR~\cite{CCRPLB46} and Saclay-Strasbourg \cite{BannerPLB44}, whose searches for $W$ bosons by detecting single electrons at large $p_T$ were overwhelmed by the high $p_T$ pion discovery, decided to combine. 
 
The combined experiment, the CERN, Columbia, Rockefeller, Saclay (CCRS) collaboration \cite{CCRS1973}, had  two spectrometer arms located around 90$^\circ$ on opposite sides of ISR intersection region I1 with vastly improved $e^{\pm}$, $e^+ + e^-$, $\pi^{\pm}$ and and $\pi^0$ detection as well as outstanding  features to eliminate the background in this difficult measurement:
 
\begin{enumerate}
\item[1.] $\geq 10^5$ charged hadron rejection from electron identification in gas Cerenkov counters combined with matching the momentum and energy of an electron candidate in the magnetic spectrometer and the PbGlass Electromagnetic Calorimeter; 
\item[2.] precision measurement of $\pi^0\rightarrow \gamma +\gamma$ and $\eta\rightarrow \gamma+\gamma$ decays, the predominant background source;\footnote{This was better in `arm2' with the CCR PbGlass than in `arm1' with the Saclay 3$X_o$ shower counters.} 
\item[3.] a minimum of material in the aperture to avoid external conversions of photons;
\item[4.] zero magnetic field on the axis to avoid de-correlating conversion pairs; 
\item[5.] rejection of conversions in the vacuum pipe (and small opening angle internal conversions) by requiring single-particle ionization in a hodoscope ($H$) of scintillation counters close to the vacuum pipe, preceded by a thin track chamber to avoid conversions in the hodoscope; 
\item[6.] precision background determination in the direct-single-$e^{\pm}$ signal channel by adding external converter---to distinguish direct single-$e^{\pm}$ from $e^{\pm}$ from photon conversion.
\end{enumerate}

Preliminary results on ``the observation of leptons at large $p_T$'' from 6 experiments were presented at the ICHEP in London in July 1974: one for $e^{\pm}$ by CCRS~\cite{CCRSLondon74}, \cite{CCRSPLB53}; one for both $e^{\pm}$ and $\mu^{\pm}$ channels by Lederman's E70~\cite{Appel74lepton} at Fermilab; and four for $\mu^{\pm}$: one by Cronin and collaborators, E100 at Fermilab~\cite{Boymond74}; one from Serpukhov~\cite{Lebedev1973} that had previously claimed that the observed ratio $\mu/\pi=2.5\times 10^{-5}$ at $\sqrt{s}=12$ GeV was from ``electromagnetic production of muon pairs'' (later improved~\cite{AbramovPLB64}); and two others that were upper limits on $W^{\pm}\rightarrow\mu^{\pm} +X$ production. CCRS and the Fermilab experiments all found that the ratio $e,\mu/\pi\approx 10^{-4}$. There were no clear ideas about the source of these direct leptons at this time, as well summarized by Lederman at ICHEP1974~\cite{LMLLondon74}.
   
Following these results, many experiments that were not originally designed for these measurements, which are extremely difficult, also attempted to look for direct single leptons. This only added confusion to the situation because they mostly found results that were incorrect. A beautiful summary of all these experiments and their difficulty was again presented by Leon Lederman at the 1975 Lepton Photon Symposium~\cite{LMLSLAC75} \cite{LMLPLC76} paraphrased succintly as:
\begin{enumerate}
\item[$\mu^{\pm}$:] Muon detection requires an understanding of the ÒtrivialÓ but devastating background of pion and kaon decay.
\item[$e^{\pm}$:] Electron detection suffers from problems of conversions of photons (principally from $\pi^0$ decay)in any residual material, from Dalitz pairs, from electron decay of long-lived sources (K, $\Lambda$, $\Sigma$, ...)
\end{enumerate}
Here I point out that Leon left out $\eta\rightarrow \gamma +\mu^+ +\mu^-$ (Dalitz) decay as a background for muons. In fact, in 1977 at the Lepton Photon Symposium, Jim Cronin~\cite{Cronin1977} made a strong claim in his talk from which I quote the abstract: ``A review of direct production of leptons and photons in hadron-hadron collisions is presented. Production of lepton pairs with large mass is well accounted for by the Drell-Yan process. The origin of direct single leptons is principally due to the production of lepton pairs. A dominant source of lepton pairs is at low effective mass, $m<600$ MeV.'' 
I strenuously disagreed with him for the CCRS result in a comment~\cite{MJT1977comment}. Here is why.
\subsection{The CCRS measurement} 
\begin{figure}[!ht]
\begin{center}
\begin{tabular}{cc}
\hspace*{-0.1in}\raisebox{0pc}{\includegraphics[width=0.51\linewidth]{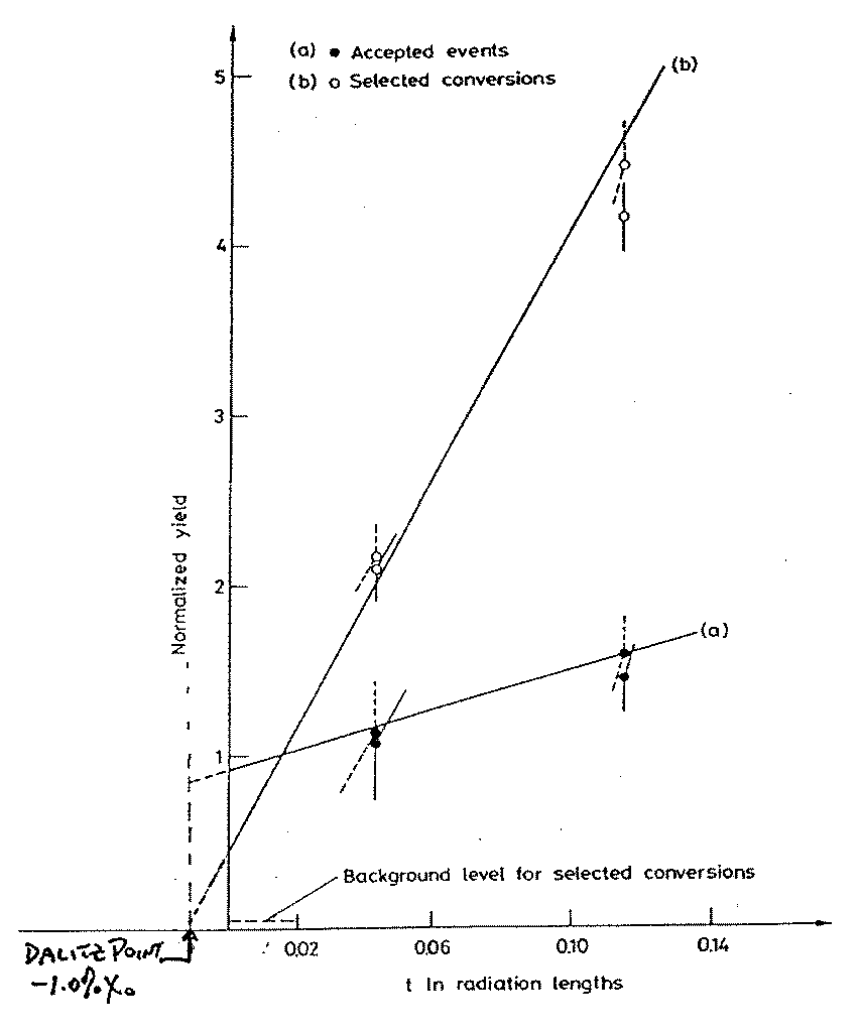}} &
\hspace*{-0.15in}\raisebox{0pc}{\includegraphics[width=0.47\linewidth]{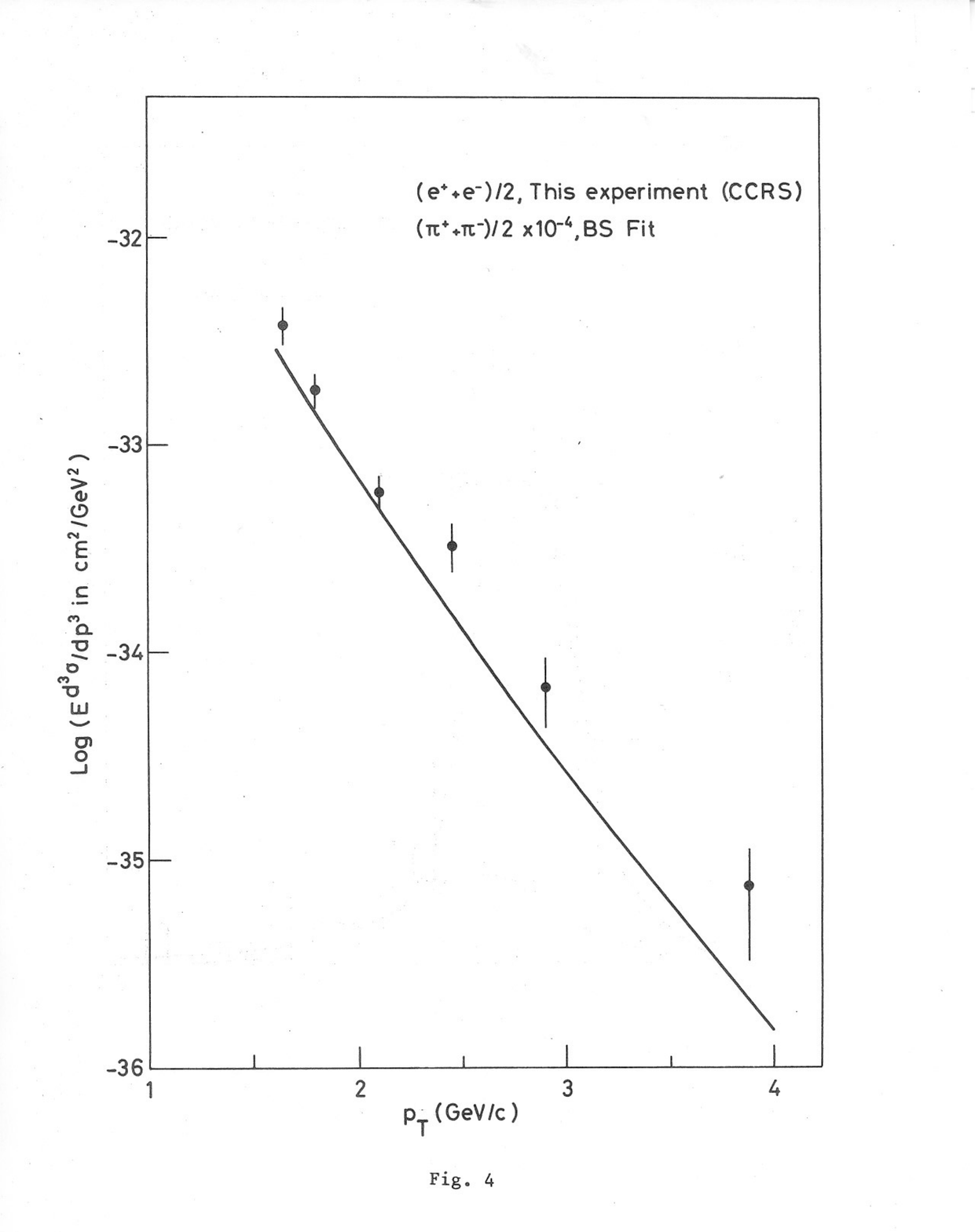}}
\end{tabular}
\end{center}
\caption[]
{a) (left) Yield of inclusive $e^\pm$ vs. total external material $t$ in radiation lengths for accepted prompt $e^\pm$ events ($\bullet$)  and selected conversions ($\circ$)~\cite{CCRSNPB113}.  b) (right) Charged averaged cross section of ($e^+ +e^-)/2$ at $\sqrt{s}=52.7$ GeV as a function of $p_T$ ($\bullet$)~\cite{CCRSPLB53}. Solid line is fit to British-Scandinavian ($\pi^+ +\pi^-)/2$ cross section multiplied by $10^{-4}$~\cite{BSLondon74}.
\label{fig:CCRS746} }
\end{figure}

We made an extraordinary effort to measure all the possible backgrounds, the primary one being external conversions of decay photons or internal conversions (Dalitz pairs) from the decay $\pi^0\rightarrow \gamma +e^+ +e^-$ and $\eta\rightarrow \gamma+e^+ +e^-$. We did this by adding material of a few percent of a radiation length $X_o$ in addition to the 0.016 $X_o$ thickness of the stainless steel corrugated vacuum pipe (Fig.~\ref{fig:CCRS746}a)~\cite{CCRSNPB113}. Measurements were made for the accepted single $e^{\pm}$ events and also for selected conversions by requiring double (or more) ionization in the $H$ hodoscope. The data are normalized to the normal running conditions with only the ISR vacuum pipe of 0.016 $X_o$. 
\subsubsection{Dalitz pairs--Internal conversion}
\label{sec:convertermethod} 
  The converter method to determine the external and internal conversions, $\gamma\rightarrow e^+ + e^-$ from $\pi^0\rightarrow \gamma +\gamma$ and $\eta\rightarrow \gamma+\gamma$ decay is straightforward to understand with a few equations.
  The total Dalitz (internal conversion) probability is~\cite{KrollWada}: 
\begin{equation}
\delta=\frac{\Gamma_{ee\gamma}}{\Gamma_{\gamma\gamma}}=\frac{2\alpha}{3\pi}\left(\ln \frac{M^2}{m_e^2}-\frac{7}{2}\right)=\begin{array}{l} 1.19\% \mbox{\ for\ } \pi^0\\1.62\% \mbox{\ for\ } \eta \end{array}\qquad ,
\label{eq:Dalitztotal}
\end{equation} 
where $M$ is the mass of the $\pi^0$ or $\eta$. With additional external converter present
the probability of internal and external conversion per $\gamma$ is 
\begin{equation}
 {e^{-}|_{\gamma} \over \gamma} = {e^{+}|_{\gamma} \over \gamma} 
={ \delta_{2} \over 2} + {t \over { {9\over 7} X_0} } \equiv \delta_{eff} 
\label{eq:efromgamma}
\end{equation}  
where: $t/X_o$, is the total external converter thickness in radiation lengths;  $\delta_2 /2$ is the Dalitz (internal conversion) branching ratio per photon 
= 0.6\% for $\pi^0\rightarrow \gamma \gamma$, 0.8\% for $\eta\rightarrow \gamma \gamma$; and the factor 9/7 
comes from the ratio of conversion length to radiation length. 
For $\pi^0$ and $\eta$ decays, the yield extrapolates to zero at the ``Dalitz 
point''  $-{9\over 14}\delta_2 \sim -1\%$---actually $-0.8$\% for $\pi^0$ and 
$-1.0$\% for $\eta$---so the method has the added advantage that it depends 
very little on the $\eta/\pi^0$ ratio. In Fig.~\ref{fig:CCRS746}a, the extrapolation 
for the $e^\pm$ from selected conversions ($\circ$) extrapolates nicely to zero at the Dalitz point, indicating a photonic source, while the accepted prompt $e^\pm$ events ($\bullet$)
show only a small decrease from the normal thickness used 
for data-collecting ($t/X_o=1.6$\%) to the Dalitz point: the value at the Dalitz point is the normalized yield of the prompt single $e^\pm$ with the photonic background subtracted. 

The same method doesn't work at all for the single $\mu^{\pm}$ which have the `devastating background' from meson decay. Furthermore the  $\eta\rightarrow \gamma+\mu^+ +\mu^-$ decay which would fit Cronin's description of the source of his single muons is not given correctly by Eq.~\ref{eq:Dalitztotal} when $m_\mu$ is substituted for $m_e$ because it is valid only to leading order in $m^2/M^2$~\cite{JarkskogNPB1}. The correct answer is $\delta_{\eta\mu\mu\gamma}=(0.08\pm0.01)\%$~\cite{LandsbergPLC} \cite{Djhelyadin80}. Furthermore, the converter method can not be used because a radiation length for muons is $({m_\mu/m_e})^2=42727$ times as large as $X_o$, the radiation length for $e^{\pm}$ and $\gamma$. In fact, what I used to say in the 1970's is still valid today "all those who try to measure direct leptons become the world's experts on $\eta$ Dalitz decay''~\cite{JanePLB59}.
\subsubsection{A final obstacle, then publication.}
The last possible background source was the decay of the vector meson $\phi\rightarrow e^+ e^-$ with the cross section $\sigma_{\phi}=1.1\sigma_{\pi^0}$. Such copious $\phi$ production should be easy to detect via the principal decay mode $K^+ K^-$, so we removed the Cerenkov counter from the trigger in `arm2' and made a short run looking for charged hadron $h^+ +h^-$ pairs with $p_T>1.6$ GeV/c. We didn't find any and set a limit $\sigma_{\phi}=\leq 0.4\sigma_{\pi^0}$ to 90\% confidence which implied that the possible $\phi$ contribution was less than 0.25 to 0.5 the observed single $e^{\pm}$ signal. The result was that the direct single $e^{\pm}$ cross section in the range $1.6<p_T<4.7$ GeV/c at $\sqrt{s}=52.7$ GeV is $(1.2\pm 0.2)\times 10^{-4}$ that of the $\pi^{\pm}$ cross section (Fig.~\ref{fig:CCRS746}b)~\cite{CCRSPLB53}. 
We submitted the publication on 28 October 1974 but of course we had spent the time since the ICHEP in July 1974 trying to find the $e^{\pm}$ that was balancing the $p_T$ of the observed $e^{\pm}$ spectrum in the opposite spectrometer `arm1', but didn't find any thing. However, we did find out the answer two weeks later with the `November revolution' in particle physics caused by the discovery of a new ``heavy particle $J$ with  mass $m=3.1$ GeV/c$^2$ and width approximately zero''~\cite{Ting1974} in p+Be collisions at the BNL-AGS (Fig.~\ref{fig:Tingetal}a) simultaneously with the same particle except called $\Psi$ in $e^+ +e^- \rightarrow$ hadrons in the storage ring (SPEAR) at SLAC~\cite{Richter1974}. 
\subsection{Charmonia---the November Revolution that made everybody believe in partons}
\subsubsection{First: new proposals, new results} 
\label{sec:ISRproposals}
CERN went into a frenzy over the $J/\Psi$ discovery~\cite{Russoa} as did the individual experiments. Within another 2 weeks, just in time for a ``second narrow resonance'' the $\Psi'$ with $m=3.7$ GeV/c$^2$ to be discovered at SLAC~\cite{Abrams1974}, two proposals for new $e^+ e^-$ measurements from members of CCRS were submitted to the ISRC. Also in this interval CCRS found the reason why we didn't find anything balancing our single $e^{\pm}$ in the opposite spectrometer: a computer bug in `arm1' that rejected  tracks in the entire $H$ hodoscope rather than just near the spaces between counters. In fact in the CCOR proposal~\cite{CCOR74} there were 7 $e^+ e^-$ events shown of which 4 clustered near $m_{ee}=3$ GeV/c$^2$. The issue of why there were two proposals from CCRS members is also interesting. 

In May 1973, the CCR part of CCRS had made a proposal~\cite{CCOR73}\footnote{Joined by Oxford in January 1974 (CERN/ISRC/73-13 Add. 2).} for a totally new detector with much larger solid angle than either the original CCR or CCRS experiments, which was approved in March 1974. It consisted of a solenoid magnet with a thin aluminum-stabilized superconducting coil, only 1.0 $X_o$ thick, outside of which were two PbGlass EMcalorimeters, opposite in azimuth, each covering a polar angle $\theta=\pm 30^\circ$ around $90^\circ$ and azimuthal angle $\phi=\pm 20^\circ$ around the median plane. Tracking and momentum measurement of charged particles was accomplished by a set of cylindrical drift chambers inside the solenoid which covered the full azimuth in a polar angle range $\theta=\pm 37^\circ$ around $90^\circ$.  Ironically, the day that the $J/\Psi$ discovery was announced, November 11, 1974, was the same day that removal of the CCRS experiment from the ISR intersection region I1 began in order to start building the CCOR detector. 

Immediately after the $J/\Psi$ announcement, the CCRS group together with Pierre Darriulat's group started discussing a ``mini'' experiment to look for the single $e^{\pm}$ and $e^+ e^-$ pairs; but quickly couldn't agree. Thus, CCOR~\cite{CCOR74} and the new Darriulat+Saclay group~\cite{CERNSaclay74} quickly submitted new proposals to the ISRC. The CCOR proposal~\cite{CCOR74} was basically the CCR experiment with Cherenkov counters added for $e^{\pm}$ identification and triggering, so was orthogonal to the approved CCOR experiment~\cite{CCOR73} and was wisely rejected by the ISRC. The CERN-Saclay proposal, which was basically CCRS with larger acceptance and PbGlass EMcalorimeters in both arms, was approved on December 11 and eventually became the CERN-Saclay-Zurich (CSZ) experiment with a small muon spectrometer added perpendicular to the $e^{\pm}$ spectrometers. 
\begin{figure}[t]
\centering
{\footnotesize a)}\includegraphics[width=0.35\linewidth]{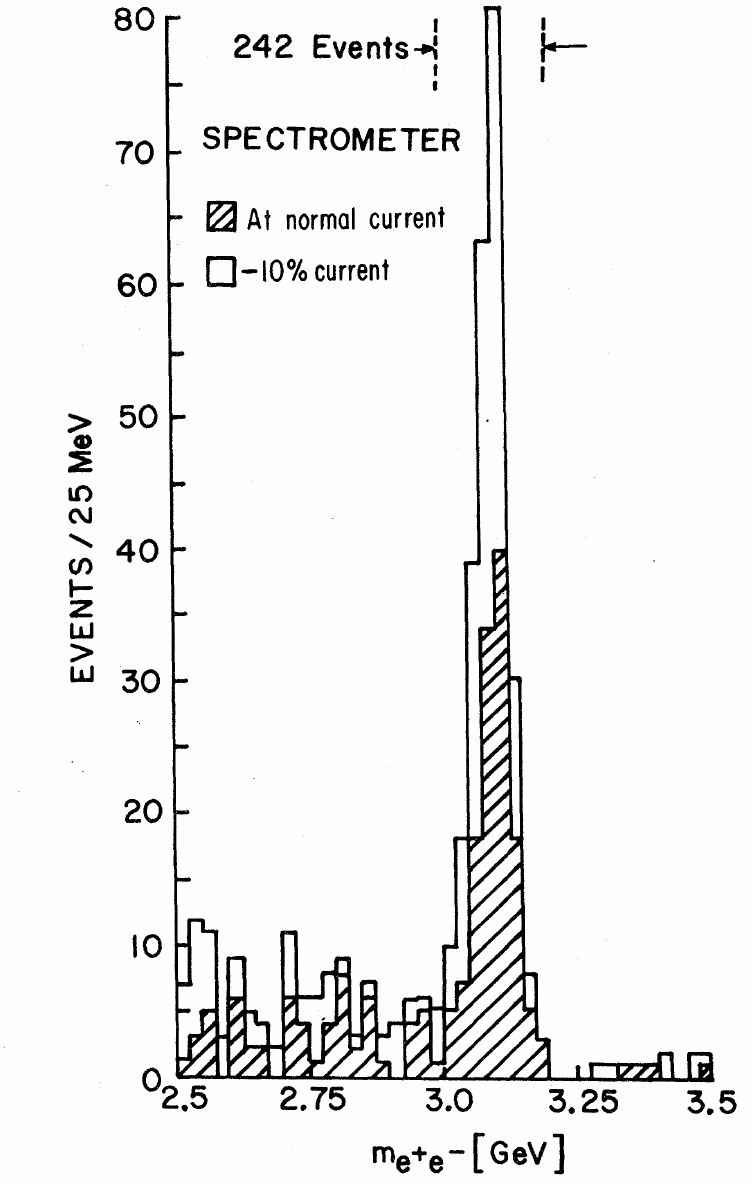}\hspace{0.5pc}
\raisebox{0.1pc}{\begin{minipage}[b]{0.50\linewidth}
\hspace*{2pc}{\footnotesize b)}\includegraphics[angle=-0.9,width=0.8\linewidth]{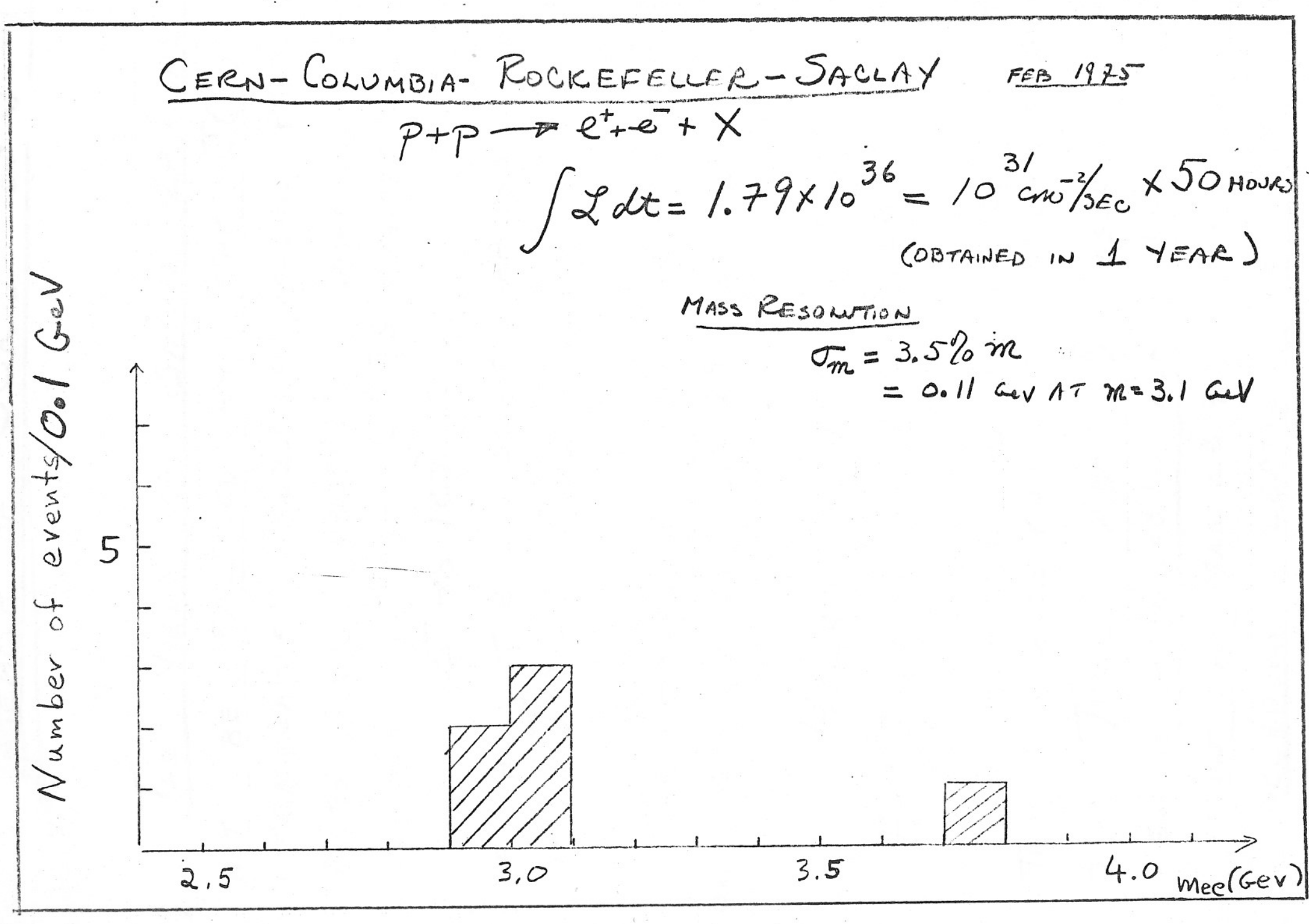}\vspace*{1.0pc}
\hspace*{0.5pc}{\footnotesize c)}\includegraphics[width=0.98\linewidth]{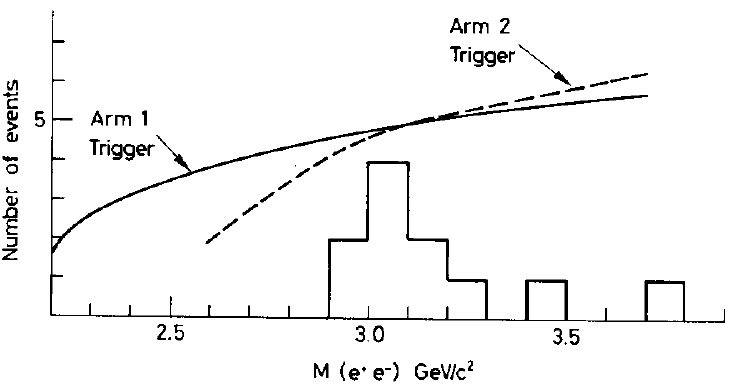}
\end{minipage}}
\caption[] {a) AGS $J$ discovery~\cite{Ting1974}: the mass remains the same with two different spectrometer settings. 
b) CCRS plot shown by Amaldi. c) First $J/\Psi$ at CERN~\cite{CCRSPLB56}.
\label{fig:Tingetal}}\vspace*{-1.0pc}
\end{figure}
\subsubsection{Then more work, discussions and publications}   
Work continued on looking for more $e^+ e^-$ events in CCRS eventually ending in 11 total events. Meanwhile the rest of the CERN physicists were not idle and a series of Discussion Meetings was held in the CERN auditorium, the most memorable for me being the one on February 21, 1975: ``The new particles: what are CERN experimentalists doing?'' We in CCRS had the only $J/\Psi$ data and I had collected the events we had found up to that time and submitted a plot to the ISR coordinator, Ugo Amaldi, who gave the ISR presentation. I was sitting in the front row. Carlo Rubbia (my boss when I was a CERN post-doc in 1965-66) was seated in the second row, one seat to my left. When Ugo showed the plot (Fig.~\ref{fig:Tingetal}b), Carlo got very excited and pointed at the event at 3.7 GeV/c$^2$ and asked loudly repeatedly ``What is that event at 3.7, what is that event at 3.7?'' We couldn't really claim that it was the $\Psi'$, so I turned to Carlo and said, ``Oh Carlo, you are the expert in one-event experiments.'' It was amazing: laughter filled the auditorium, moving up the rows in a wave until everybody in the audience (except Carlo) was laughing---very memorable! The CCRS result was submitted for publication on 22 April 1975~\cite{CCRSPLB56} (Fig.~\ref{fig:Tingetal}c). 
\subsubsection{OK so we found the $J/\Psi$ (a bit late): was it the source of the single $e^{\pm}$?}
\begin{figure}[h]
    \centering
\raisebox{0.0pc}{\includegraphics[angle=+0.4,width=0.29\linewidth]{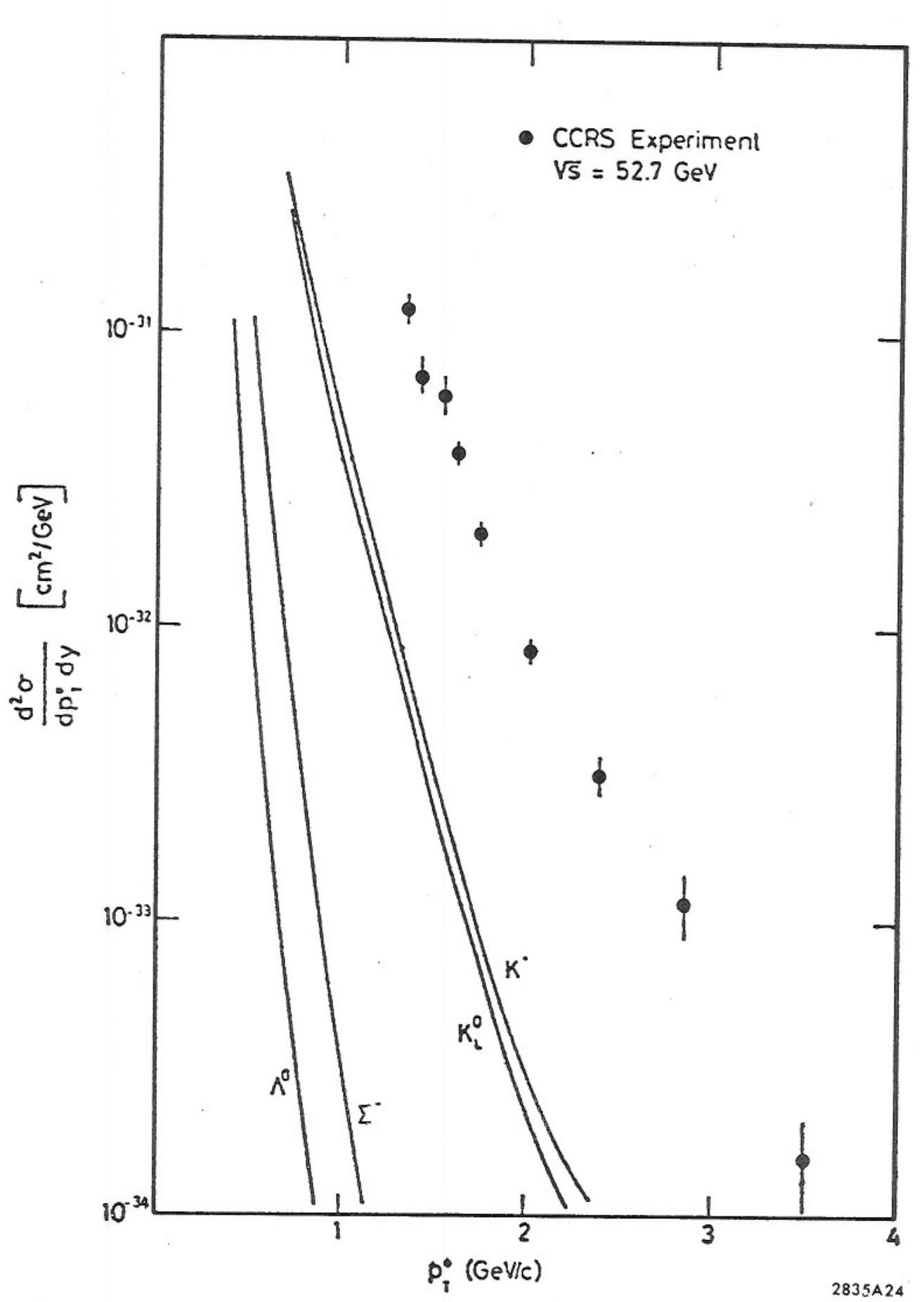}} 
\raisebox{-0.5pc}{\includegraphics[width=0.35\linewidth]{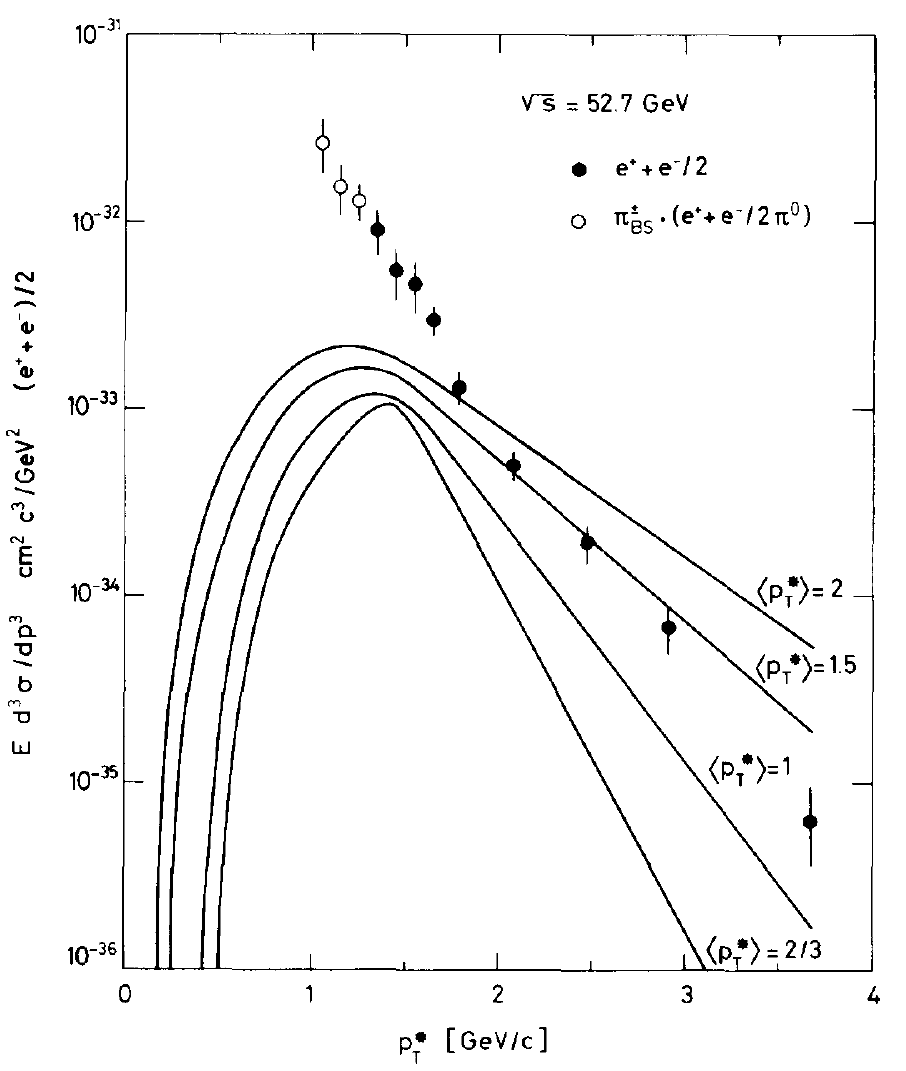}}
\raisebox{-0.25pc}{\includegraphics[angle=-1.1,width=0.35\linewidth]{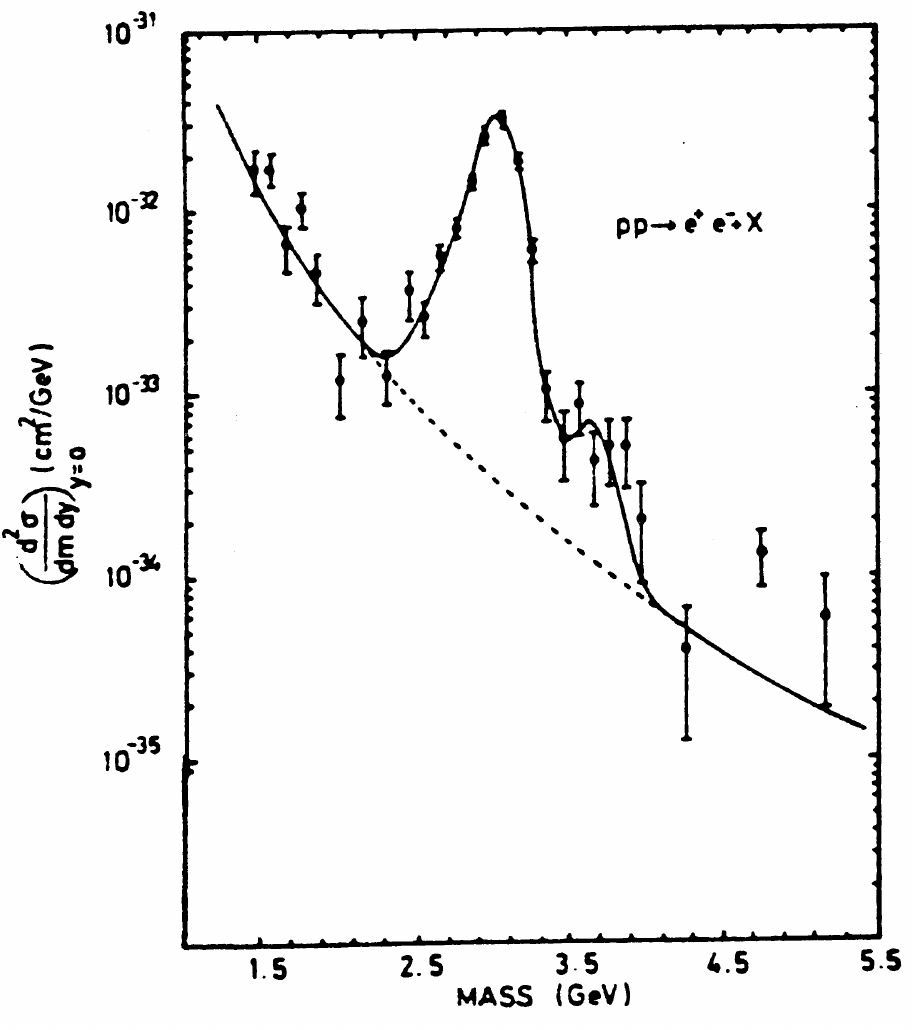}}
\caption[]{CCRS direct single $e^{\pm}$ $p_T$ spectrum at $\sqrt{s}=52.7$ GeV with background from a)(left) charged hadron decays~\cite{LMLSLAC75} and b)(center) calculated $e^{\pm}$ spectrum from $J/\Psi$ decay with several values of $\mean{p_T}$~\cite{CCRSNPB113}. c)(right) Best measurement of the $J/\Psi$ $d^2\sigma/dm_{ee}dy|_{y=0}$ at the CERN ISR~\cite{ClarkNPB142}.
\label{fig:bkge}} 
\end{figure}

    In addition to the observation of the `new particle' there was an important issue to address: was the $J/\Psi$ the source of the single $e^{\pm}$? The answer was NO as shown in the final CCRS paper on ``Electrons at the ISR'' (Fig.~\ref{fig:bkge}b)~\cite{CCRSNPB113} and also shown for charged hadron decays in Lederman's talk~\cite{LMLSLAC75} (Fig.~\ref{fig:bkge}a).

    Another important issue addressed by CCRS~\cite{CCRSPLB56} was the measurement of the cross section for the reaction:
\begin{equation} 
      p+p\rightarrow J(3.1) + X \rightarrow e^+ + e^- +X \label{eq:Jxsec}
\end{equation} 
with result
\begin{equation} 
B_{ee}\times \frac{d\sigma}{dy} (p+p\rightarrow J+ X)=(7.5\pm 2.5)\times 10^{-33} {\rm cm}^2
\label{eq:Jcross}
\end{equation} 
averaged over the rapidity acceptance $|y|\leq 0.30$. This result at $\sqrt{s}\approx 50$ GeV is two-orders of magnitude larger than the value for the BNL discovery~\cite{Ting1974} at $\sqrt{s}\approx 7.5$ GeV. The reason that CCRS~\cite{CCRSPLB56} was able to measure the cross section was that thanks to Cerenkov counters together with the EMcalorimeter that allowed a trigger on $e^{\pm}$ in the range $1.0\leq p_T\leq 4.7$ GeV/c, the cross section could be measured for $J/\psi\rightarrow e^+ e^-$ for all $p_{T_{ee}}>0$. This was the first of three such measurements at the CERN ISR~\cite{CCRSPLB56,ClarkNPB142,WillisPLB91J} with similar $e^{\pm}$ triggers and not repeated at a hadron collider until RHIC~\cite{PHENIXPRL92} and eventually CDF at the Fermilab Tevatron~\cite{CDFPRD71}. The best measurement of the $J/\Psi$ at the ISR~\cite{ClarkNPB142} is shown in Fig.~\ref{fig:bkge}c.

\subsubsection{What are the $J/\Psi$ and $\Psi'$?} 
In less than a month, the $J/\Psi$ was understood to be a narrow bound state of the `heavy' $c\bar{c}$ charm quarks~\cite{GIM,AlvaroShelly} and was given the name charmonium~\cite{AppelquistPolitzer}. The quarks were proposed to be bound with a simple potential (the Cornell potential~\cite{Cornell1,Cornell2}) that incorporated both the Coulomb and confining forces:
\begin{equation}
V(r)=-\frac{4}{3}\frac{\alpha_s}{r} + \sigma r
\label{eq:CornellVr}
\end{equation}
where $\alpha_s$ is the \QCD\ coupling constant (Eq.~\ref{eq:alphanf}) and $\sigma$ represents the string-tension in a string-like confining potential. Also predictions were made for heavier quarks $Q$ with stronger binding of $Q\bar{Q}$ leading to `a rich assortment of rather narrow resonances'~\cite{Cornell2}.
\subsubsection{The Upsilon; and the ISR recognition problem}
It is worth noting that in early 1976 a measurement of $p+ Be \rightarrow e^+ +e^-$ + {\it anything} using a 400 GeV proton beam at Fermilab~\cite{HomPRL36} claimed that a ``statistically significant clustering near 6 GeV suggests the existence of a narrow resonance''. This was made by the Lederman group  who suggested that ``the name $\Upsilon$ (Upsilon) be given either to the resonance at 6 GeV if confirmed or to the onset of high-mass dilepton physics.'' This later became known as the oopsLeon when, in a measurement of high mass $\mu^+ +\mu^-$ pairs which allowed a higher data taking rate made possible by the filtering of most hadrons, the same group~\cite{HomPRL37} observed that ``the dimuon mass spectrum provides no evidence for fine structure above 5 GeV''. 

    \begin{figure}[!h]
    \begin{center}
\raisebox{0.0pc}{\includegraphics[angle=+1.0,width=0.32\linewidth]{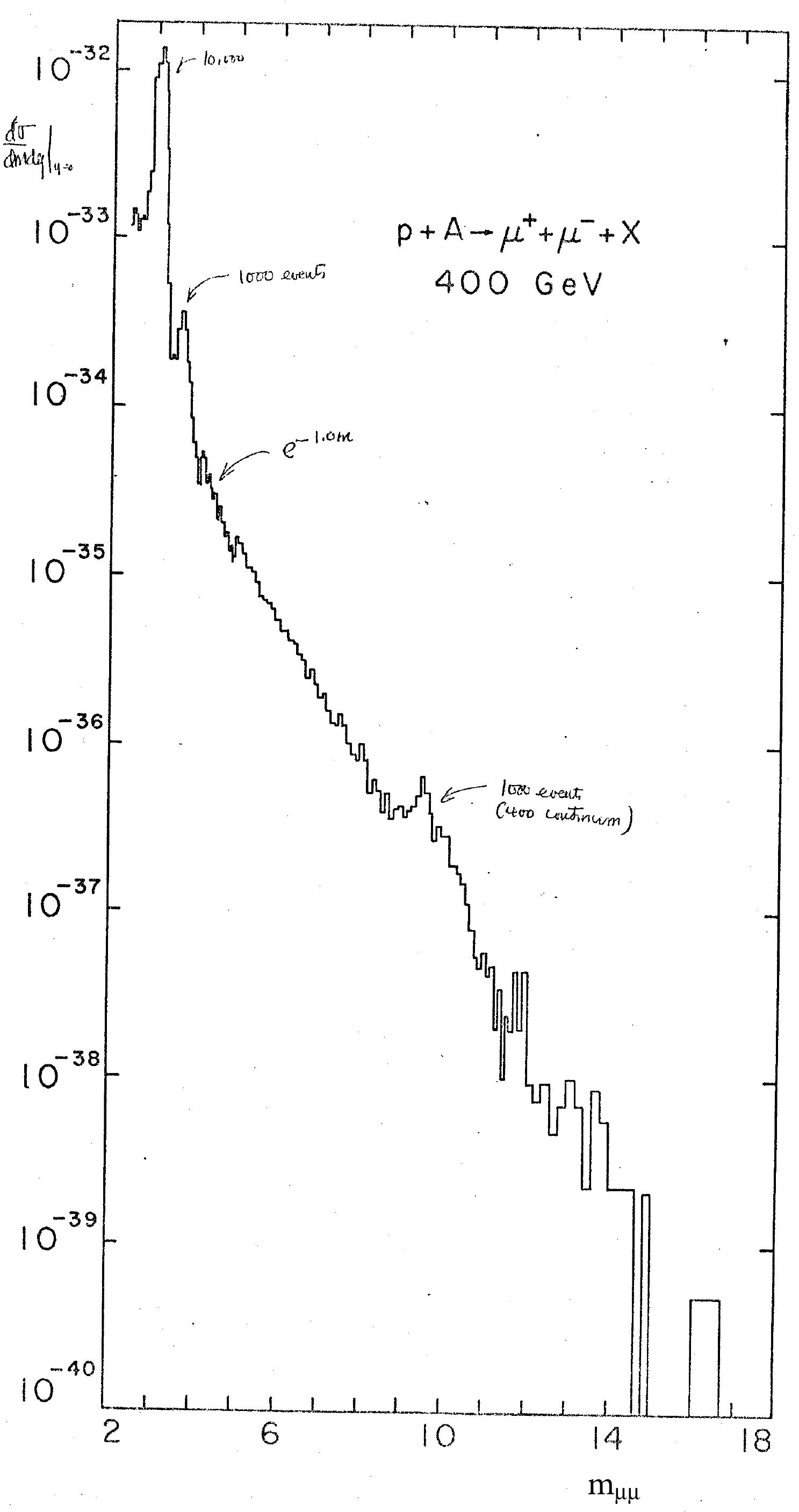}} \hspace*{1.0pc}
\raisebox{-0.23pc}{\includegraphics[angle=+0.0,width=0.58\linewidth]{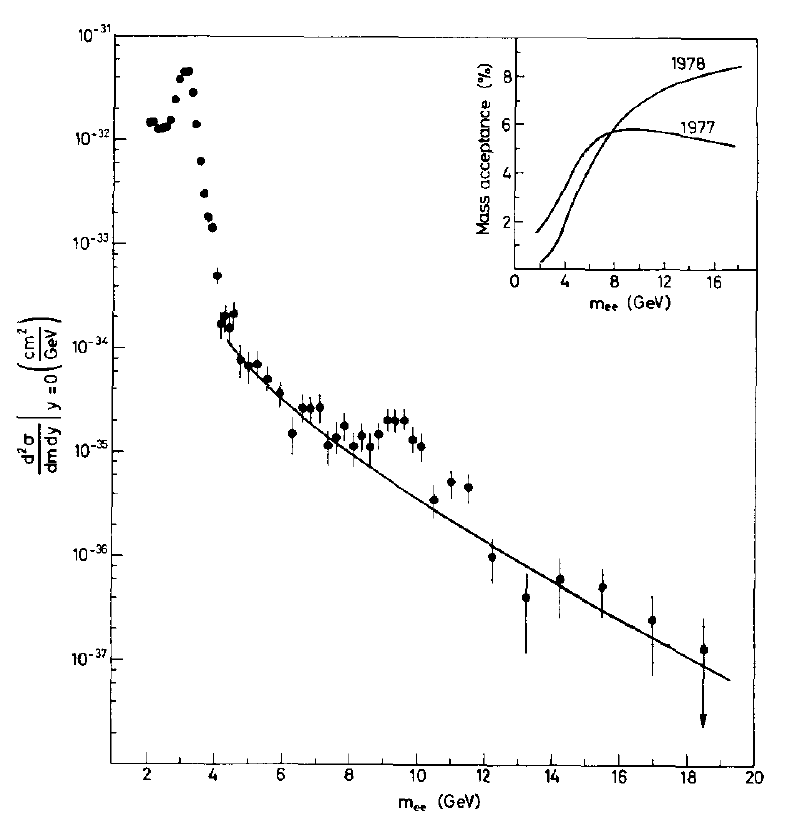}}%
\end{center}
\caption[]{a) (left) CFS dimuon spectrum~\cite{HerbPRL39} with $\Psi$, $\Psi'$, Drell-Yan exponential and $\Upsilon$ indicated. b) (right) ABCSY $d^2\sigma/dm_{ee} dy|_{y=0}$ averaged for $\sqrt{s}=53$ and 63 GeV~\cite{ABCSYDY}\label{fig:ups}} 
\end{figure} 

The situation was again reversed following a major redesign of the dimuon experiment, when the real $\Upsilon \rightarrow \mu^+ +\mu^-$, a narrow resonance at 9.5 GeV, was discovered~\cite{HerbPRL39} (Fig.~\ref{fig:ups}a). This was a narrow bound state of heavier quarks as predicted~\cite{Cornell2} and turned out to be the discovery of the bottom quark $b$, a member of a third family of quarks much heavier than the first two families. 

This discovery cemented a negative view of the ISR by many CERN physicists~\cite{Russo} for missing the discoveries of the $c$ quark (actually quarkonia) and now the $b$ quark in dilepton production by lower energy but much larger luminosity conventional fixed target accelerators. They were drawn instead to the new 400 GeV Super proton Synchrotron (SpS) which started physics operation at the end of 1976 and which ironically made its greatest contributions by being turned into a proton-antiproton collider in mid 1981.   Unfortunately this negative pervasive attitude totally missed the most important physics contributions of the ISR which were in hadron physics and which made major contributions to the development of \QCD. 

Measurements of the Upsilons were also made at the ISR~\cite{CCORUps},\cite{BCSYwrong}, \cite{WillisPLB91J} but the best measurement (Fig.~\ref{fig:ups}b)~\cite{ABCSYDY} was more concerned about Drell-Yan production.\footnote{Drell-Yan production, or $\bar{q}+q\rightarrow \mu^+ + \mu^- +X$ is predominantly an electromagnetic process which measures the $\bar{q}$ composition of a proton. It is mainly a parton model reaction. \QCD\ enters only in the next to leading order.}  However, the Upsilon measurement \cite{WillisPLB91J} did correct an anomalously large value of the cross section claimed in their previous publication~\cite{BCSYwrong}. 
\subsection{The direct single $\mathbf{e^\pm}$ are from the semi-leptonic decay of charm particles.} 
\begin{figure}[h]
\begin{center}
\includegraphics[width=1.0\linewidth]{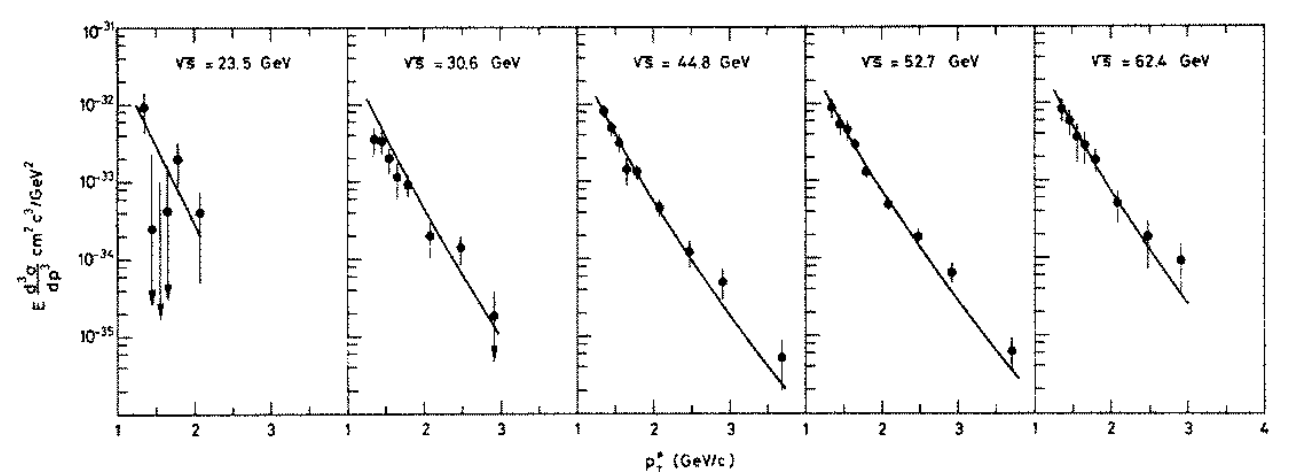}
\end{center}
\caption[]{Invariant cross sections of $(e^+ + e^-)/2$ (points) at mid-rapidity for 5 values of $\sqrt{s}$ in p$+$p collisions at the CERN-ISR~\cite{CCRSNPB113}. Lines represent a fit of the $(\pi^+ +\pi^-)/2$ British-Scandinavian data~\cite{BS2}, multiplied by $10^{-4}$}
\label{fig:eroots}
\end{figure}

\begin{figure}[h]
\begin{center}
\includegraphics[width=0.66\linewidth]{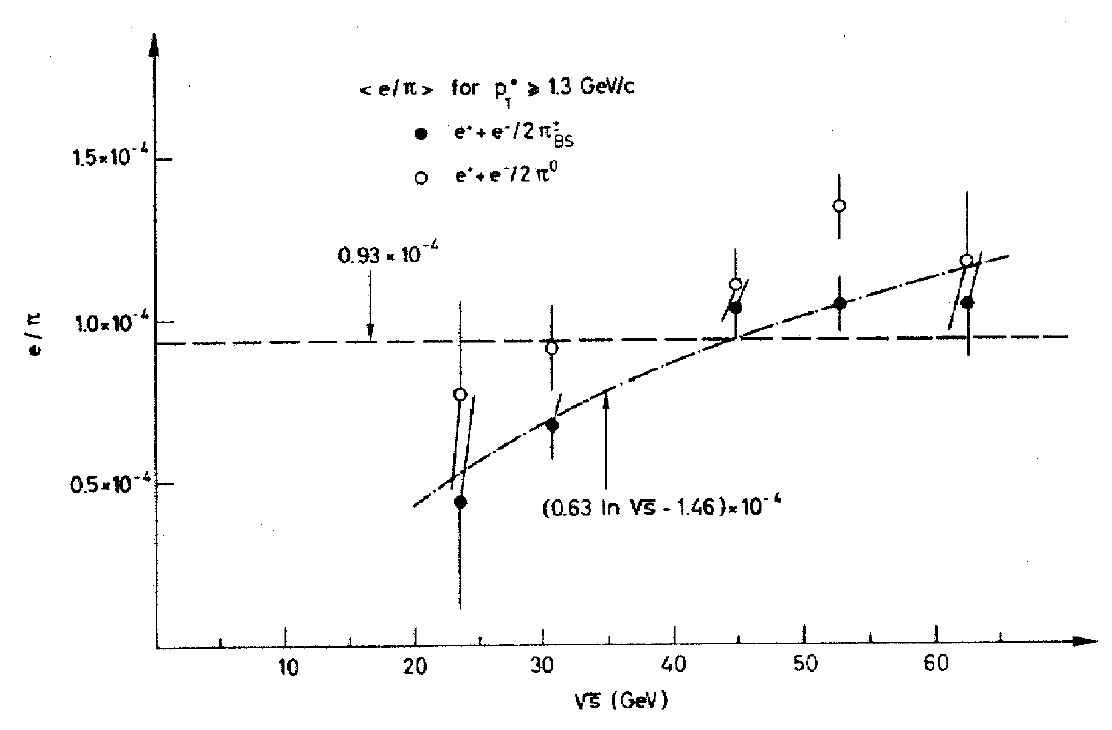}
\end{center}\vspace*{-0.15in}
\caption[]{The ratios $(e^+ + e^-)/(\pi^+ +\pi^-)$ and $(e^+ + e^-)/2\pi^0$ for $p_T\geq 1.3$ GeV/c as a function of 
c.m. energy $\sqrt{s}$ from CCRS~\cite{CCRSNPB113}. The $\pi^0$ data were obtained from the measured spectra of selected conversions and the $\pi^{\pm}$ from the British-Scandinavan data~\cite{BS2} (solid points). The two curves shown are fits to the solid points.   \label{fig:charm2}}
\end{figure}

With the $J/\Psi$ issue out of the way, the comprehensive CCRS single $e^{\pm}$ measurements~\cite{CCRSNPB113} remained a very interesting challenge (Fig.~\ref{fig:eroots}).
The direct-single $e^{\pm}$ spectra in the range $1.3<p_T<3.2$ GeV/c were roughly constant at a level of $10^{-4}$ of charged pion production at all 5 c.m. energies measured.  However, in detail, the ratio of $e/\pi$ increased systematically by a factor of $\sim$ 1.8 from $\sqrt{s}=$ 30 to 60 GeV\cite{CCRSNPB113} (Fig.~\ref{fig:charm2}).    
\subsubsection{The theorists rise to the occasion.}
Although discovered in 1974, it was not until 1976 that the direct single $e^{\pm}$ were shown to be from  the semi-leptonic decay of charm particles. Of course without the intervening $J/\Psi$ charmonium discovery, it is unlikely that charm particles would have been considered. In fact, the first attempt of explaining the CCRS measurements was by Farrar and Frautschi~\cite{Glennys76} who proposed that the direct-single-$e^{\pm}$ were due to the internal conversion of direct photons with a ratio $\gamma/\pi^0\sim$10-20\%. However CCRS was able to cleanly detect (and reject) both external and internal conversions, since there was zero magnetic field on the axis, so was able to set limits excluding this explanation to less than 6.6\% of the observed signal~\cite{CCRSNPB113}.

\begin{figure}[h]
   \begin{center}
\raisebox{0.0pc}{\includegraphics[width=0.53\linewidth]{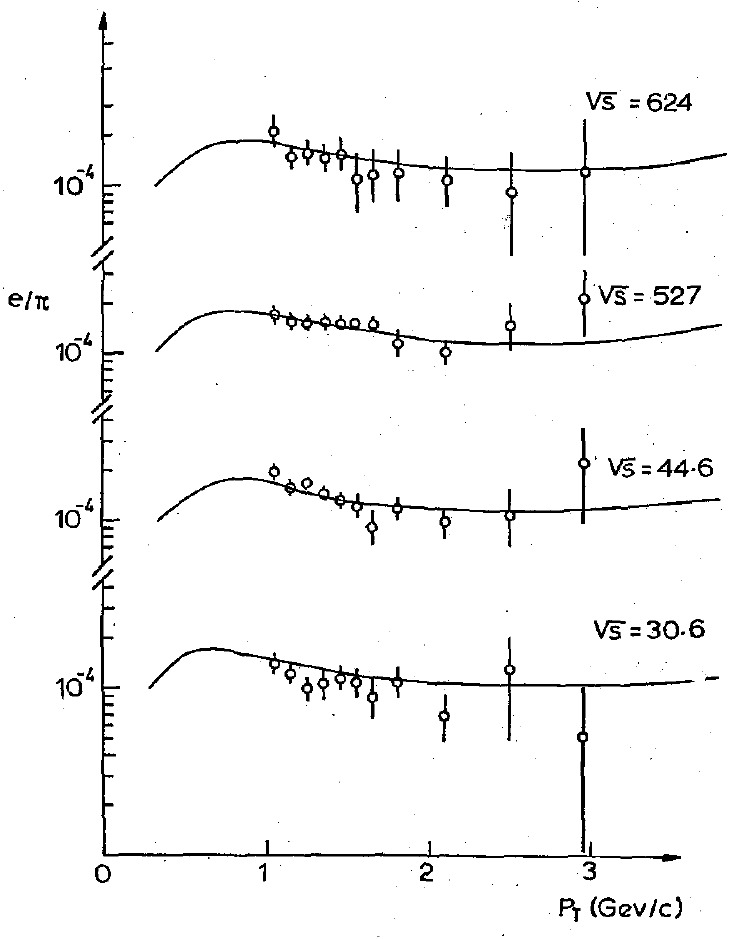}}\hspace*{1.0pc}
\raisebox{-0.5pc}{\includegraphics[width=0.45\linewidth]{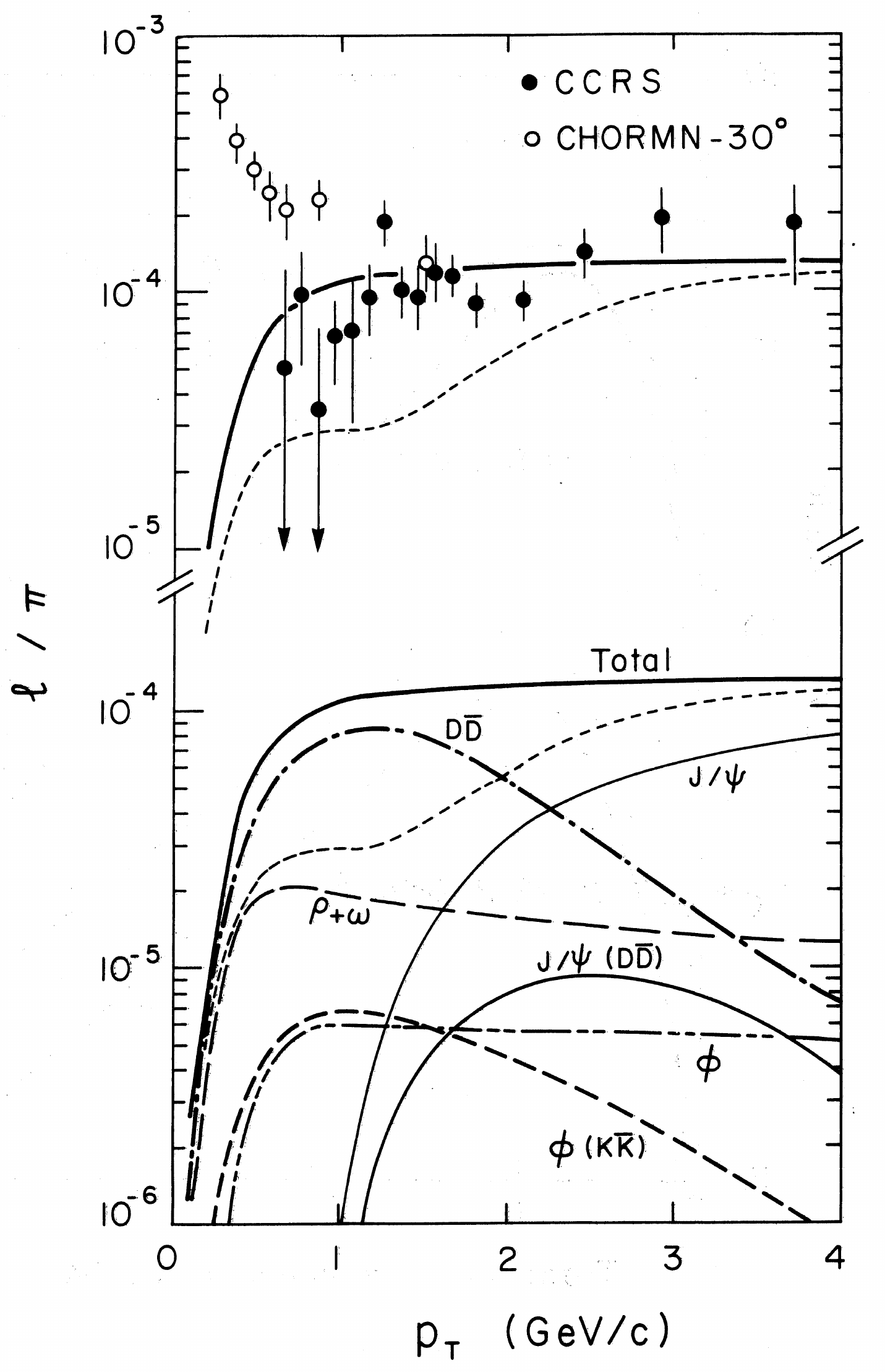}}

\end{center}
\caption[]{a)(left) CCRS $e/\pi$ in two spectrometer arms at two values of $\sqrt{s}$~\cite{CCRS75} with prediction, see Ref.~\cite{HLLSPLB} for details. b) (right) Bourquin-Gaillard~\cite{BG76} calculation of a cocktail of decays to the $e/\pi$ ratio.
\label{fig:HLLSBourqGaill}}
\end{figure}

 	The first correct explanation of the CCRS direct-single-$e^{\pm}$ (prompt leptons) was given by Hinchliffe and Llewellyn-Smith~\cite{HLLSPLB} as due to semi-leptonic decay of charm particles. The predictions from charm decay~\cite{HLLSPLB} are in excellent agreement with the CCRS data~\cite{CCRS75}  submitted to the SLAC conference as shown in Fig.~\ref{fig:HLLSBourqGaill}a.
A similar explanation was published by Maurice Bourquin and Jean-Marc Gaillard~\cite{BG76}, two experimentalists, who compared the measured $e/\pi$ ratios to a cocktail of all known leptonic decays including {\em ``Possible contributions from the conjectured charm meson...''} (Fig.~\ref{fig:HLLSBourqGaill}b). It is interesting to note the usage of ``conjectured'' because it seems that this was the notation for $D^+$ and $D^0$ charm mesons from the first SLAC paper which failed to find open charm~\cite{BoyarskiPRL35} a year before it was actually discovered at SLAC in August 1976~\cite{OpenCharm}.

Something important to note about the charm calculations~\cite{HLLSNPB}, \cite{BG76} and clearly shown in Fig.~\ref{fig:HLLSBourqGaill}b is that neither calculation could fit the data points 
for $p_T<1$ GeV/c at $30^\circ$ from the CHORMN experiment~\cite{chormn} at the ISR. 
The important issue from a historical perspective is that many experiments not designed for the purpose wanted to get into the prompt-lepton act, which resulted in questionable (i.e. incorrect) results, and this is only one example. See Ref.~\cite{BG76} for more details. This led to the unfortunate  situation that later papers about charm ignored the many disagreeing ISR measurements of the direct-single-$e^{\pm}$ from charm, which were essentially ridiculed, or in the case of the original CCRS discovery~\cite{CCRSNPB113}  downplayed in review articles~\cite{FabjanPLC}; or ignored~\cite{Tavernier} because it was published before either charm or the $J/\Psi$ were discovered so there is no reference to the word ``charm'' in the publication. 

Later experiments at the CERN-ISR using the upgraded Split Field Magnet were designed for a program of heavy flavor searches. The CCRS direct $e^{\pm}$ measurement at $\sqrt{s}=62.4$ GeV was confirmed by a well designed single $e^\pm$ experiment performed by the Zichichi group~\cite{Basile81}. Also the first direct evidence of charm hadron production in hadron interactions was the observation at the CERN-ISR of the charm $D^+$ meason via its $D^+\rightarrow K^+ \pi^+ \pi^-$ decay by the CCHK experiment~\cite{CCHKPLB81}.
\subsubsection{Why do I dwell on charm at the CERN-ISR?}
One reason is the lack of recognition in the charm literature as noted. However the main reason is that the measurement of charm by direct single $e^{\pm}$ in p$+$p and Au$+$Au collisions~\cite{ppg066} 30 years later at the Relativistic Heavy Ion Collider (RHIC) at Brookhaven National Laboratory (BNL) was one of the principal observations that led to the discovery of the Quark Gluon Plasma and the conclusion that it was a perfect liquid with a viscosity to entropy density ratio ``intriguingly close to the conjectured quantum lower bound'' $1/4\pi$~\cite{QuantumBound}. 
\section{\QCD\ meets high $\mathbf{p_T}$ 1977-1979}
\label{sec:QCD1}
It wasn't until 1977 that \QCD\  was applied to hard-scattering with the first theoretical calculations of the basic equations for the elementary \QCD\ subprocesses at leading order~\cite{CombridgePLB70} \cite{CutlerPRD17} \cite{FritzschPLB69}. However the first application of ``asymptotic freedom (AF)'' to ``wide-angle hadronic collisions'' was in 1975 in a publication~\cite{CahalanPRD11} with an incorrect conclusion;  but with an improved $x_T$ scaling formula, that experimenters liked, which included the evolution effects of ``AF'' and is still in use at present. 
\subsection{Improved $x_T$ scaling.}
The incorrect conclusion of Ref.~\cite{CahalanPRD11} was that ``single vector gluon exchange contributes insignificantly to wide angle hadronic collisions''; but, in my humble opinion, they may have been overly influenced by at least one author of the original (BBG) $x_T$ scaling~\cite{BBG1972}. This is based on an acknowledgement in Ref.~\cite{CahalanPRD11}:``Two of us (J.~K. and L.~S.) also thank S.~Brodsky for emphasizing to us repeatedly that the present data on wide-angle hadron scattering show no evidence for vector exchange''.

Nevertheless the new $x_T$ scaling formula, which introduced the ``effective index'' $n_{\rm eff}(x_T, \sqrt{s})$ to account for `scale breaking' (i.e. \QCD\ evolution):
\begin{equation}
E \frac{d^3\sigma}{dp^3}={1 \over {p_T^{{n_{\rm eff}(x_T,\sqrt{s})}} }  } 
F(x_T)={1\over {\sqrt{s}^{{\,n_{\rm eff}(x_T,\sqrt{s})}} } } 
\: G(x_T) \quad {\rm where}\quad x_T\equiv\frac{2p_T}{\sqrt{s}}
\label{eq:nxt}
\end{equation}
 was very useful in understanding the next round of high $p_T$ measurements. 
\subsection{Second generation high $\mathbf{p_T}$ measurements}
\begin{figure}[!t] 
\centering
\includegraphics[width=0.39\linewidth]{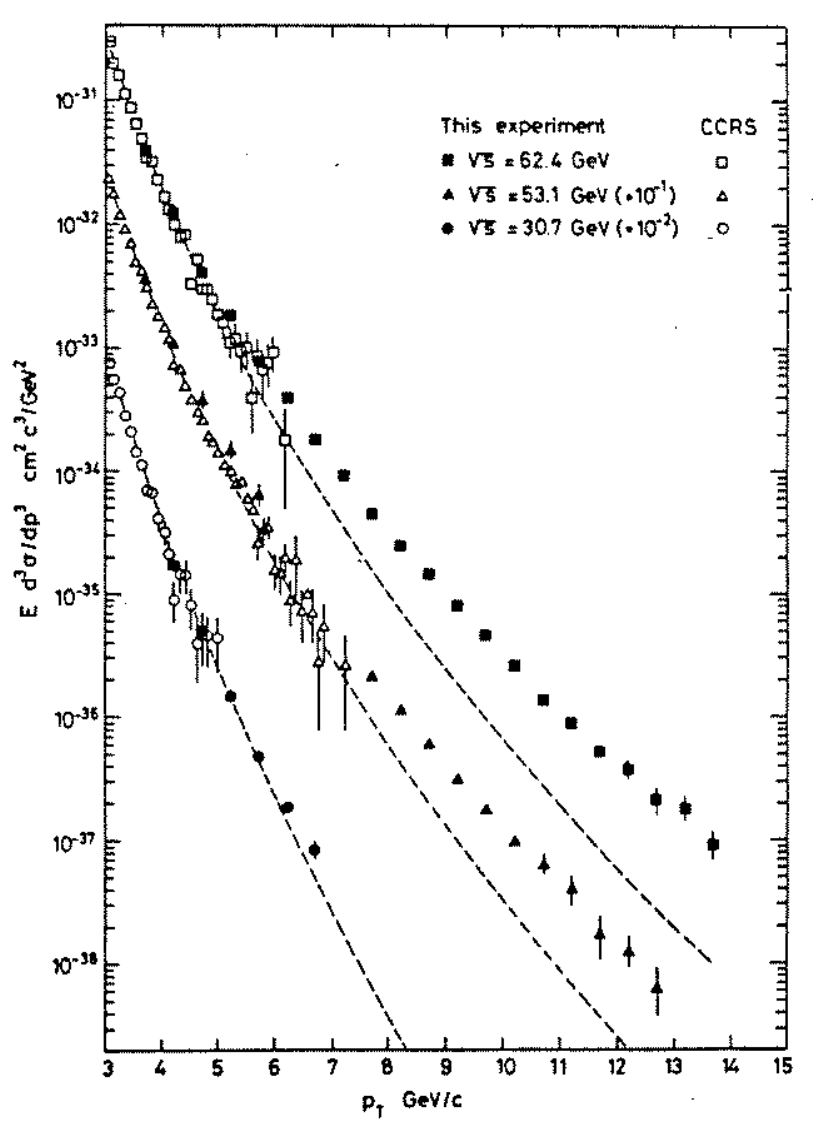}
\caption[] {Invariant $\pi^0$ cross sections $E d^3\sigma/dp^3$ vs $p_T$ for $\sqrt{s}$ indicated~\cite{CCORPLB79}. Open points and dashed lines are data and fit are from a previous measurement~\cite{CCRSNPB106} \label{fig:ccorpi0}}
\end{figure}

The second generation of high $p_T$ experiments at the CERN-ISR straightened out the $n=8$ issue of the first round of experiments (recall Fig.~\ref{fig:CCRpi0}) and made good use of Eq.~\ref{eq:nxt}.
The CCOR experiment presented their first results (Fig.~\ref{fig:ccorpi0})~\cite{CCORPLB79} together with data and a fit from a previous CCRS measurement~\cite{CCRSNPB106} that had an order of magnitude more data than the original CCR and Saclay-Strasbourg measurements. Thanks to further increased integrated luminosity and acceptance, the CCOR data~\cite{CCORPLB79} extended another 4 orders of magnitude in cross section to $p_T\approx 14$ GeV/c. The CCRS fit, $A e^{-bx_T}/p_T^n$, to their data at all three $\sqrt{s}$ with $n=8.6$ and $b=12.5$ (shown as dashed lines on Fig.~\ref{fig:ccorpi0}) works in the CCRS $p_T$ range but fails badly for $p_T>7$ GeV/c and remains well below the CCOR data out to $p_T\approx 14$ GeV/c. 
\begin{figure}[!h]
\centering
{\begin{minipage}[b]{0.405\linewidth}
\hspace*{0pc}{\footnotesize a)}\includegraphics[width=0.9\linewidth]{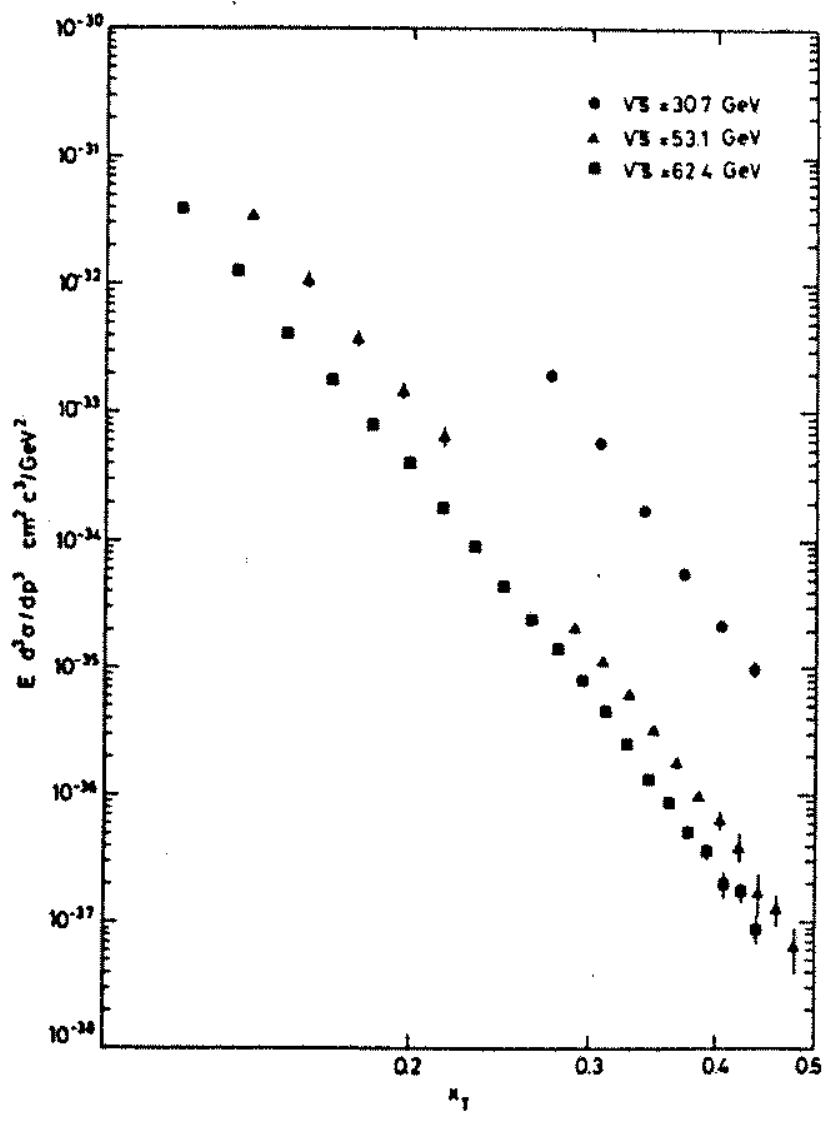}\vspace*{0.5pc}
\hspace*{0.5pc}{\footnotesize b)}\includegraphics[width=0.9\linewidth]{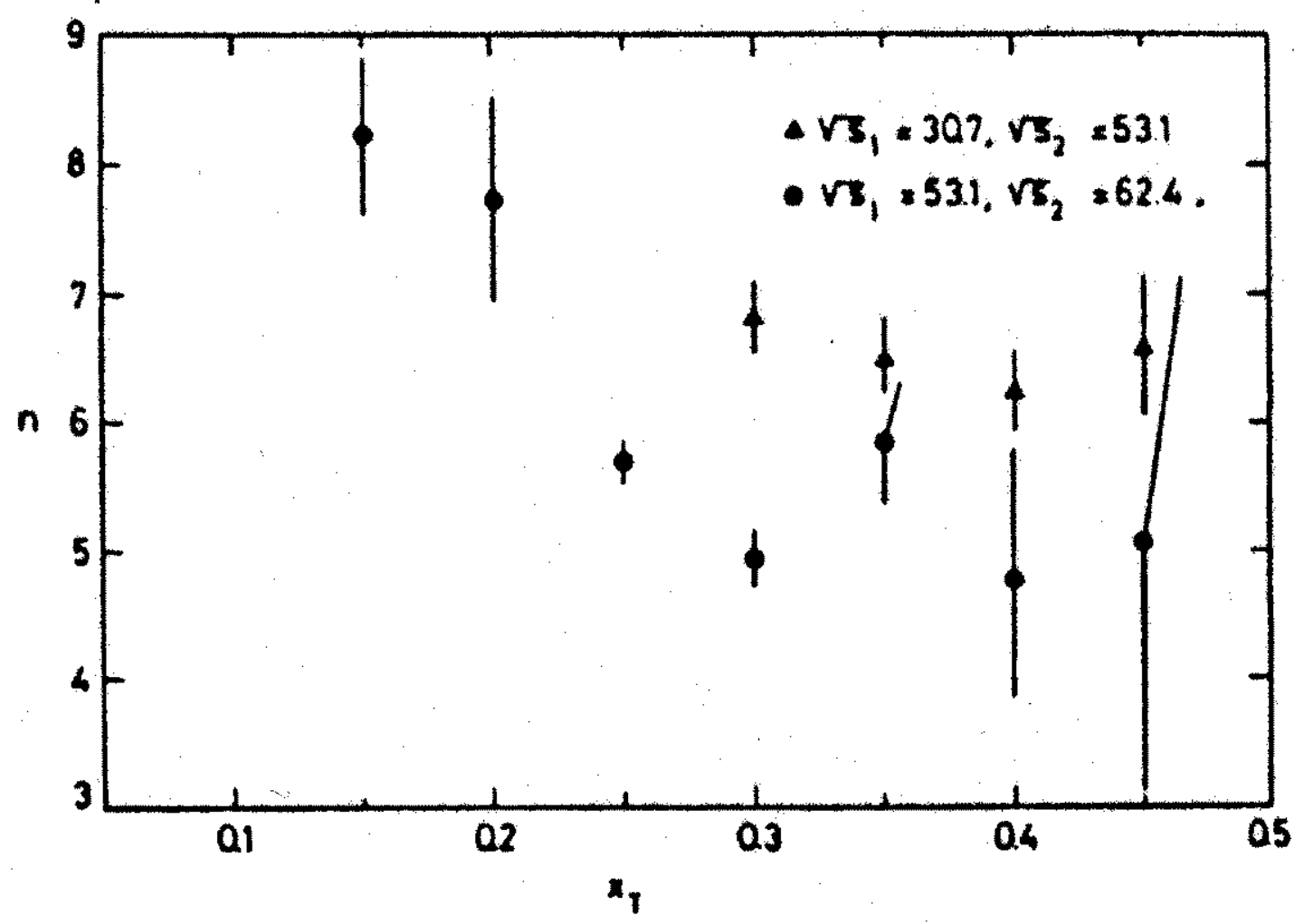}
\end{minipage}}
{\begin{minipage}[b]{0.486\linewidth}
\hspace*{0pc}{\footnotesize c)}\includegraphics[width=0.90\linewidth,height=1.0\linewidth]{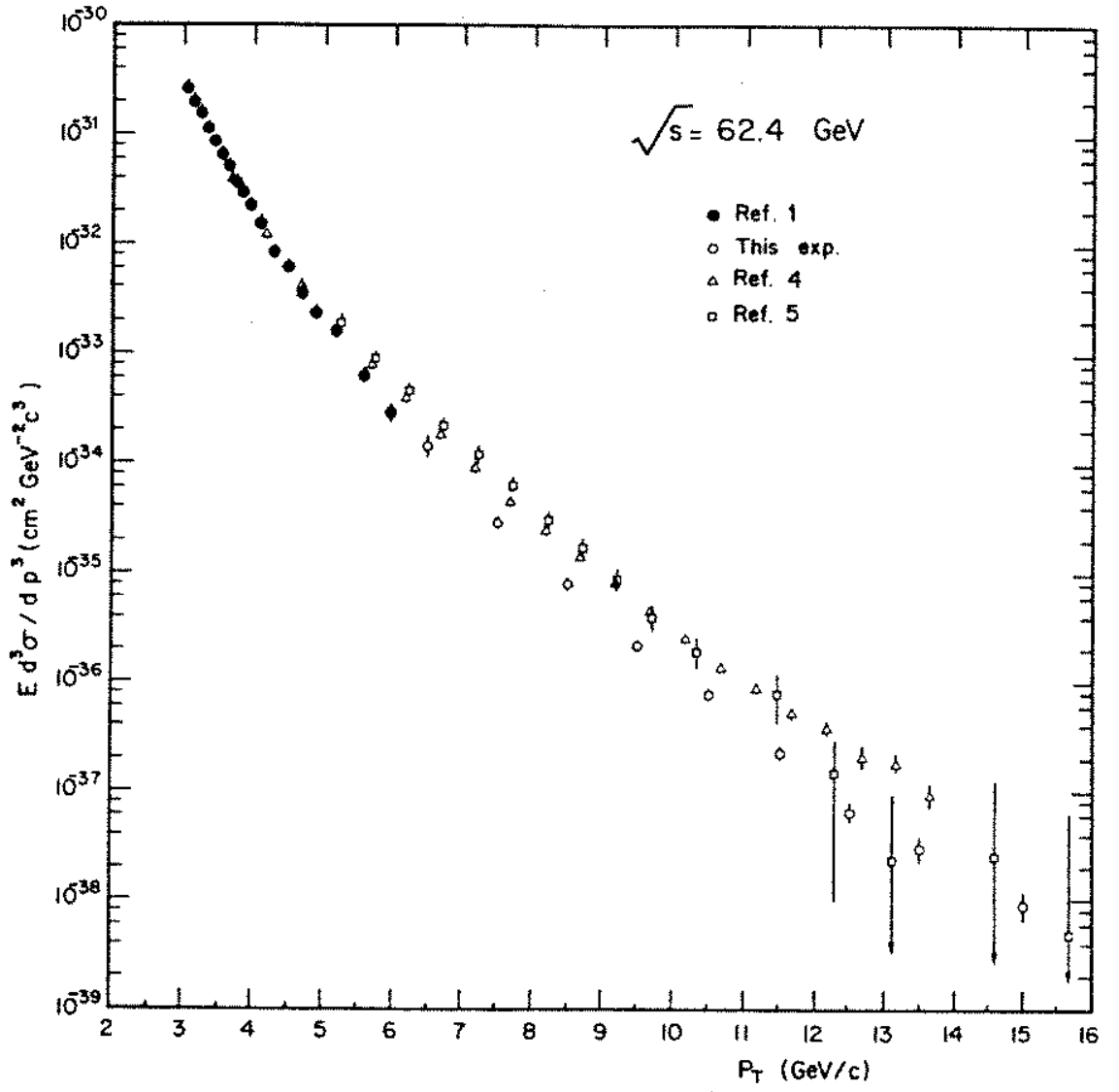}\vspace*{0.5pc}
\hspace*{0.5pc}{\footnotesize d)}\includegraphics[width=0.90\linewidth]{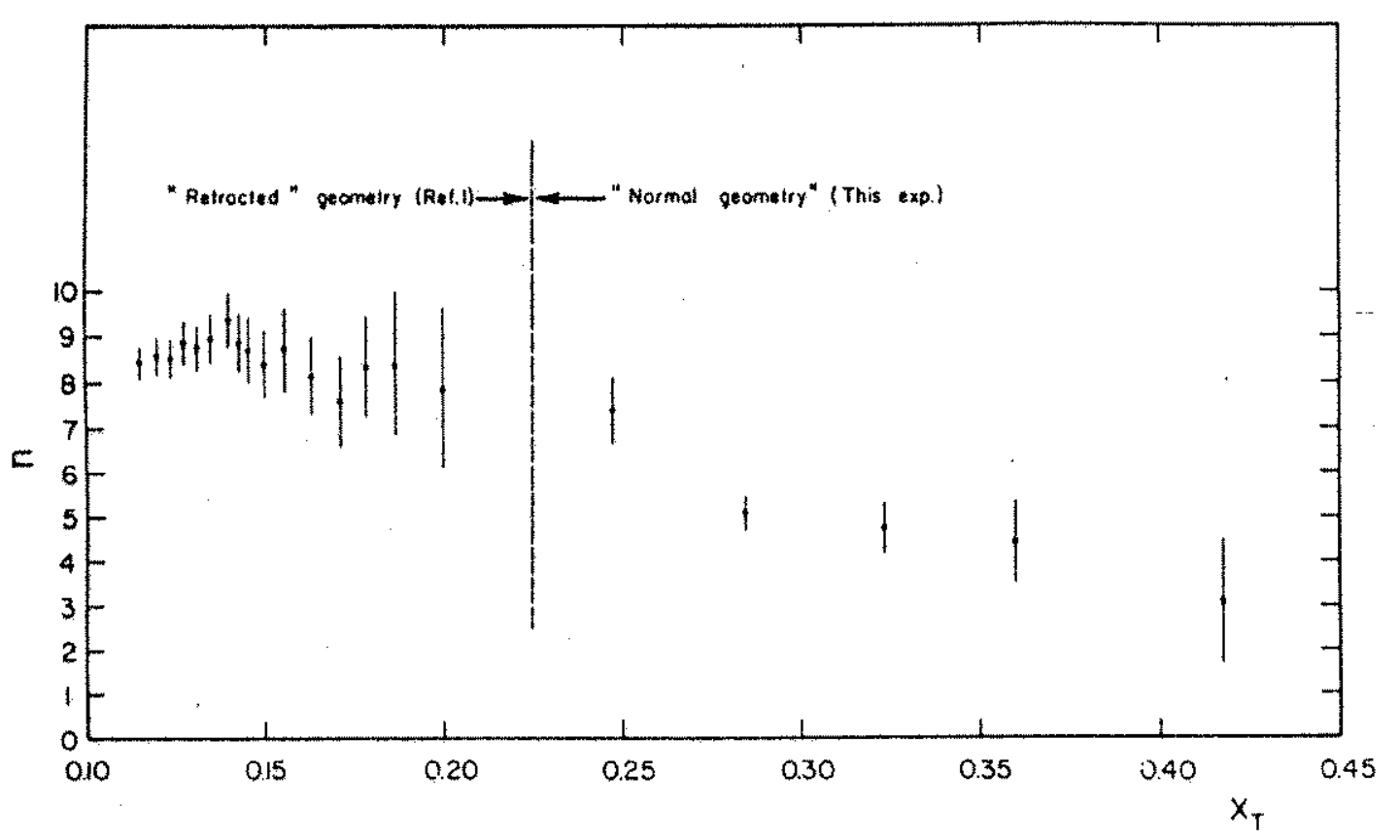}
\end{minipage}}
\caption[] {\cite{CCORPLB79} a)log-log plot of CCOR $\pi^0$ invariant cross section $Ed^3\sigma/dp^3$ vs $x_T$ for 3 values of $\sqrt{s}$; (b) $n_{\rm eff}(x_T, \sqrt{s_1},\sqrt{s_2} )$ from (a). \cite{WillisPLB84} c) $\pi^0$ cross section vs $p_T$ from ABCS~ at $\sqrt{s}=62.4$ GeV with some other $\pi^0$ measurements from the ISR; d) $n_{\rm eff}(x_T, 52.7, 62.4)$ from ABCS data. Only statistical errors are shown. There are  additional common systematic uncertainties in $n_{\rm eff}$ of $\pm 0.33$ in (b) and $\pm 0.7$ in (d).   
\label{fig:goodnxt}}\vspace*{-1.0pc}
\end{figure}

\subsubsection{$\mathbf{x_T}$ scaling with $\mathbf{n_{\rm eff}(x_T, \sqrt{s})}$}
$x_T$ scaling fits to the CCOR data in Fig.~\ref{fig:ccorpi0} for $\sqrt{s}=53.1$ and 62.4 GeV~\cite{CCORPLB79} confirmed the validity of the idea that $n_{\rm eff}$ is a function of $x_T$ and $\sqrt{s}$ with results $n_{\rm eff}=8.0\pm0.4$ for the $p_T$ interval $3.7<p_T< 7.0$ GeV/c and $n_{\rm eff}=5.13\pm0.14$ for $7.5<p_T<14.0$ GeV/c. The variation of $n_{\rm eff}$ with $x_T$ and $\sqrt{s}$ is easier to observe with 
a log-log plot of the invariant cross section $\sigma^{\rm inv}=Ed^3\sigma/dp^3$ vs $x_T$ (Fig.~\ref{fig:goodnxt}a) where $n_{\rm eff}$ can be read off by inspection:
\begin{equation}
n_{\rm eff}(x_T, \sqrt{s_1},\sqrt{s_2} )=\frac{\ln\left[\sigma^{\rm inv}(x_T,\sqrt{s_1})
/\sigma^{\rm inv}(x_T,\sqrt{s_2})\right]}{\ln\left[\sqrt{s_2}/\sqrt{s_1}\right]}\quad . \label{eq:neff}
\end{equation}
The fact that the data for $\sqrt{s}=53.1$ and 62.4 GeV are essentially parallel for $x_T\geq 0.3$ in Fig.~\ref{fig:goodnxt}a indicates that $n_{\rm eff}\approx 5$ is roughly constant in this range as shown in Fig.~\ref{fig:goodnxt}b, where all the point by point calculated $n_{\rm eff}(x_T, \sqrt{s_1},\sqrt{s_2} )$ are plotted. Also the increase in the spacing in Fig.~\ref{fig:goodnxt}a for $x_T\leq 0.2$ compared to $x_T\geq 0.3$ indicates the region where $n_{\rm eff}(x_T, 53.1, 62.4)\approx 8$. An interesting observation about the fact that $n_{\rm eff}\approx 5$ for $x_T\geq 0.3$ is that this value is now close to the value $n_{\rm eff}=4$, originally expected by BBG~\cite{BBG1972} for QED or vector gluon exchange (Section \ref{sec:BBG}), but modified by the scale breaking from \QCD\ evolution as suggested by Ref.~\cite{CahalanPRD11}.

Another nice feature of $x_T$ scaling for experimentalists is that in the calculation of $n_{\rm eff}(x_T, \sqrt{s_1},\sqrt{s_2})$ from Eq.~\ref{eq:neff} the systematic uncertainty in the absolute value of the $p_T$ scale cancels, which eliminates the dominant source of experimental systematic uncertainty. This is nicely illustrated by the $\pi^0$ measurements from another `second generation' ISR experiment~\cite{WillisPLB84} based on liquid-argon--Pb plate calorimeters of high spatial resolution and large solid angle acceptance (Fig.~\ref{fig:goodnxt}c). Notably, the $\pi^0$ cross-sections from `this experiment' and `Ref.4' (CCOR)~\cite{CCORPLB79} disagree by a large factor $\approx 3$ for $p_T\geq 7$ GeV/c but the values of $n_{\rm eff}(x_T, 53.1, 62.4)$ (Fig.~\ref{fig:goodnxt}d) are in excellent agreement with the CCOR values (Fig.~\ref{fig:goodnxt}b). \vspace*{-1.0pc}
\subsection{First theoretical calculations of pion $\bm{p_T}$ spectra using \QCD: 1978}  
\begin{figure}[!h]
    \centering
\raisebox{0.0pc}{\includegraphics[angle=+0.0,width=0.35\linewidth]{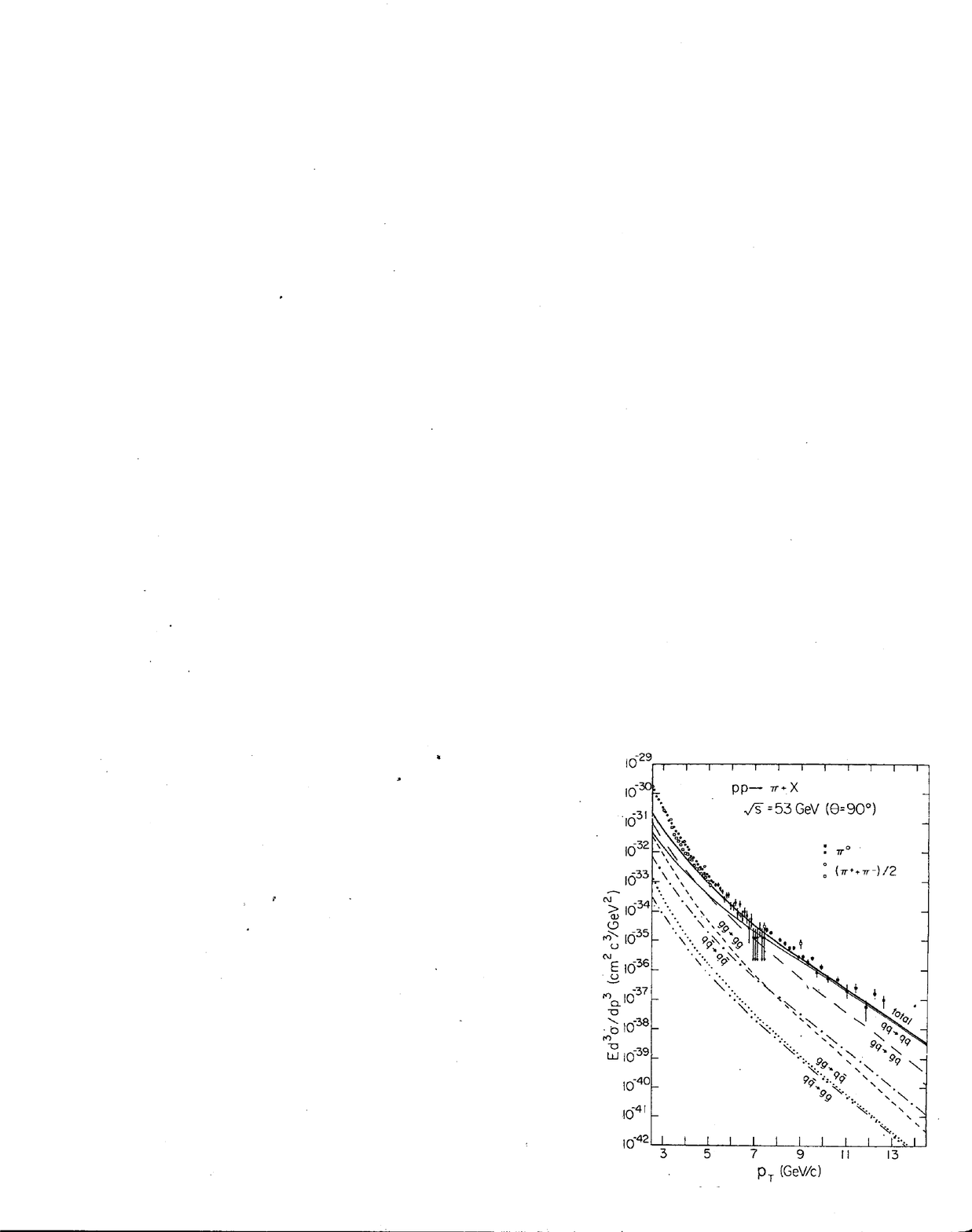}} 
\raisebox{-0.0pc}{\includegraphics[width=0.34\linewidth]{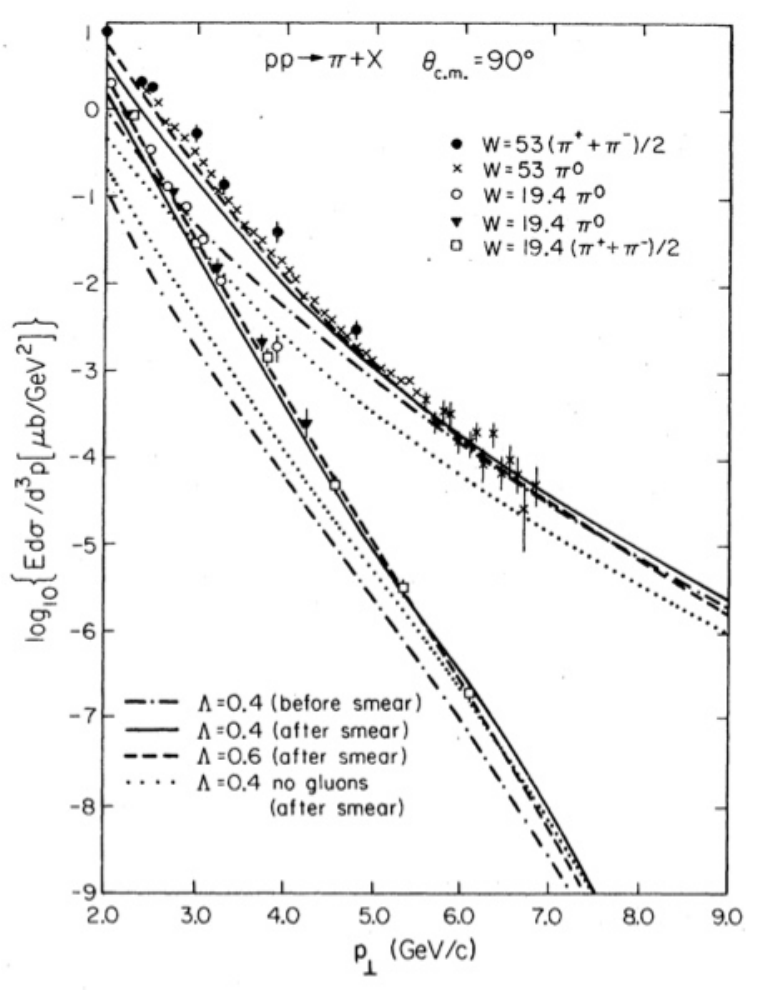}}
\raisebox{-0.0pc}{\includegraphics[angle=0.0,width=0.30\linewidth]{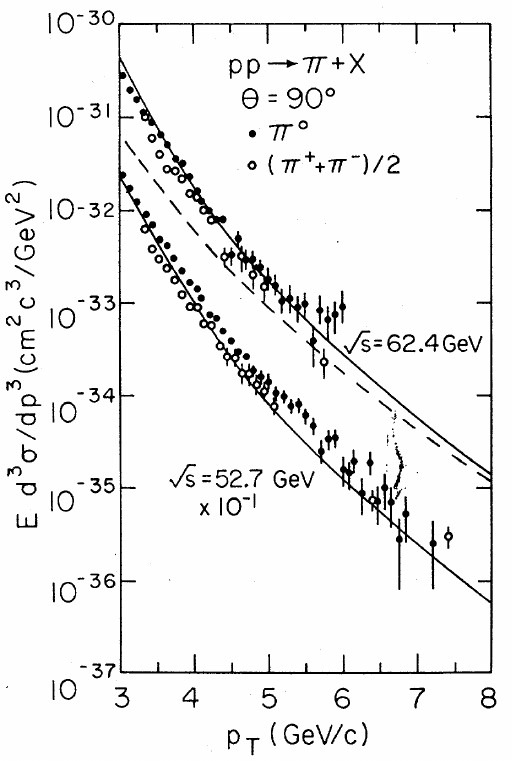}}
\caption[]{a) (left) \QCD\ calculation~\cite{OwensPRD18}, with subprocess contributions shown, compared to preliminary CCOR $\pi^0$ measurements~\cite{PopeCCORPrelim}. \QCD\ calculations including the $k_T$ effect: (b) (center) Feynman Field and Fox~\cite{FFFPRD18} and (c) (right) Owens and Kimel~\cite{OwensKimel} compared to CERN-ISR and Fermilab pion measurements indicated. 
\label{fig:QCDtodata}} 
\end{figure}

The first calculations of inclusive high $p_T$  pion cross sections using \QCD\  were performed by Owens, Reya and Gl\"uck~\cite{OwensPRD18} (Fig.~\ref{fig:QCDtodata}a) and Feynman, Field and Fox (FFF)~\cite{FFFPRD18} (Fig.~\ref{fig:QCDtodata}b). 
Figure~\ref{fig:QCDtodata}a shows all the \QCD\ subprocesses that go into the calculation; but the total, which fits the data well for $p_T>5$ GeV/c,  falls below the data for $p_T<5$ GeV/c. Note that the data are preliminary results of CCOR~\cite{PopeCCORPrelim} which agree with the final publication~\cite{CCORPLB79} shown in Fig.~\ref{fig:ccorpi0}. The FFF calculations in Fig.~\ref{fig:QCDtodata}b follow the BS~\cite{AlperNPB100} $(\pi^+ +\pi^-)/2$ and CCRS $\pi^0$~\cite{CCRSNPB106}  measurements at $\sqrt{s}=53$ GeV/c, as well as $\sqrt{s}=19.4$ GeV/c measurements from Fermilab, from $p_T \approx 7$ GeV/c down to $p_T\approx 2$ GeV/c because the effect of $k_T$ smearing was included. $k_T$, the transverse momentum of a parton in a nucleon, had been introduced by FFF the previous year~\cite{FFF1} to explain ISR measurements of two-particle correlations which will be discussed in detail in a later section. A second calculation by Owens~\cite{OwensKimel} which included the $k_T$ effect also fit   the CCRS~\cite{CCRSNPB106} data down to $p_T=3$ GeV/c.

These calculations all assume that high $p_T$ particles are produced from states with two roughly back-to-back jets which are the result of scattering of a parton from each nucleon. The calculations require the purely theoretical elementary \QCD\  subprocesses plus other quantities which must come from measurements.
It is also important to understand the kinematics of parton-parton scattering and the definition of the various kinematic quantites. These are given in the Appendix (section~\ref{sec:appendix}). 

\section{Elements of \QCD\ for Experimentalists} 
\QCD\ calculations at the parton level are relatively straightforward. For example consider the scattering:
$$a+b\rightarrow c+d$$
where $a$ and $b$ are incident partons with c.m. energy $\sqrt{\hat{s}}$ that scatter elastically through angle $\theta^*$, with 4-momentum-transfer-square: 
\begin{equation}
{Q}^2=-\hat{t}=\hat{s} \frac{(1-\cos\theta^*)}{2} \label{eq:Qsqparton} \qquad.
\end{equation}
The \QCD\ cross section for this subprocess is simply: 
\begin{equation}
\left.\frac{d\sigma}{d\hat{t}}\right|_{\hat{s}}=\frac{\pi \alpha_s^2(Q^2)} {\hat{s}^2} \Sigma^{ab}(\cos\theta^*)  \qquad ;
\label{eq:dsigdthat}
\end{equation} 
where the elementary \QCD\ subprocesses involving quarks ($q$) and gluons ($g$) in hadron-hadron collisions and their characteristic angular distributions~\cite{CombridgePLB70} \cite{CutlerPRD17} 
{ $\Sigma^{ab}(\cos\theta^*)$} are enumerated in Fig.~\ref{fig:someQCD}a-h. 

\subsection{The Elementary \QCD\ Subprocesses}
\begin{figure}[!h]
\begin{center}
\begin{minipage}[b]{0.49\textwidth}
\includegraphics[width=0.96\textwidth]{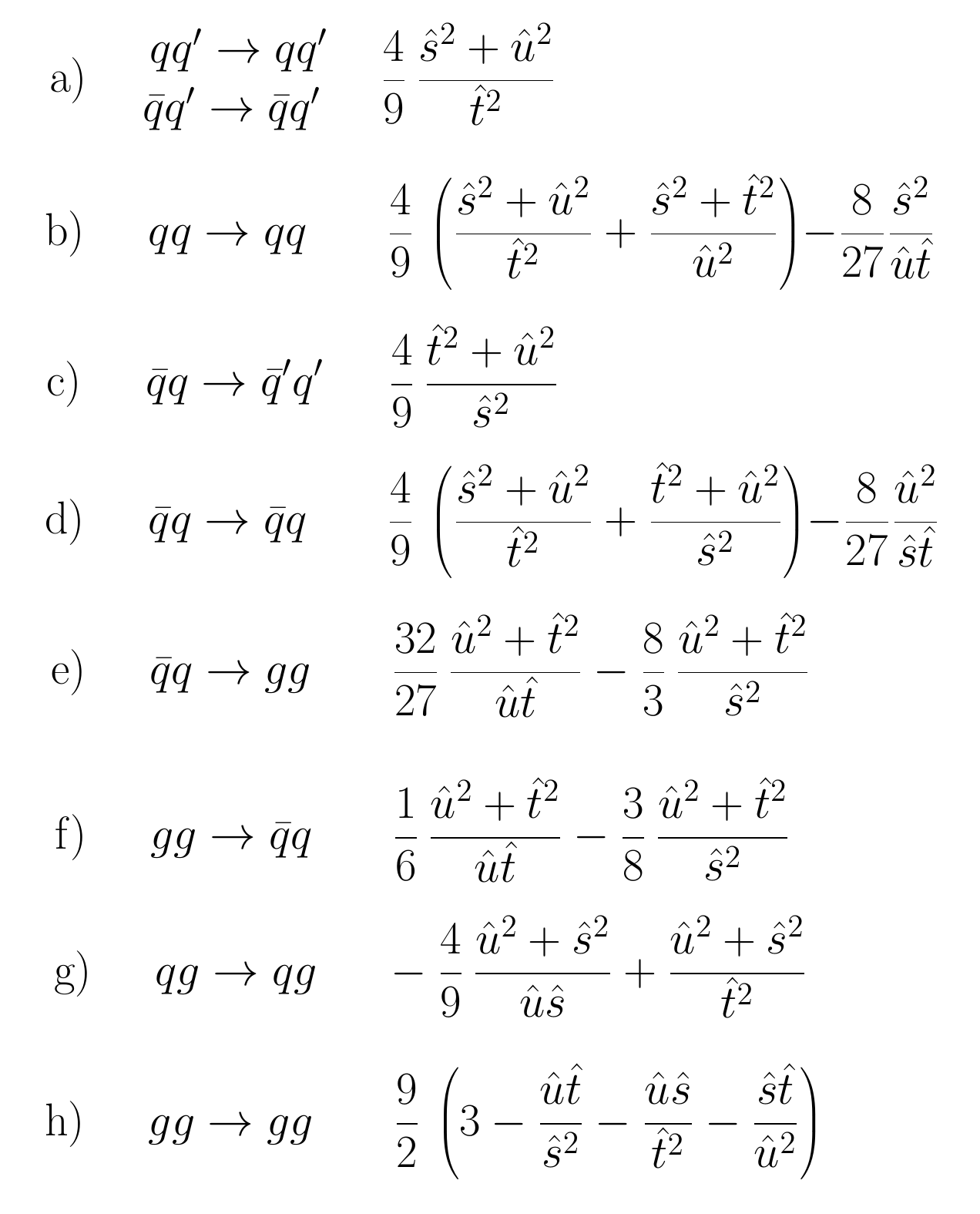}
\end{minipage}
\hspace*{0.01\textwidth}
\begin{minipage}[b]{0.48\textwidth}
\includegraphics[width=0.96\textwidth]{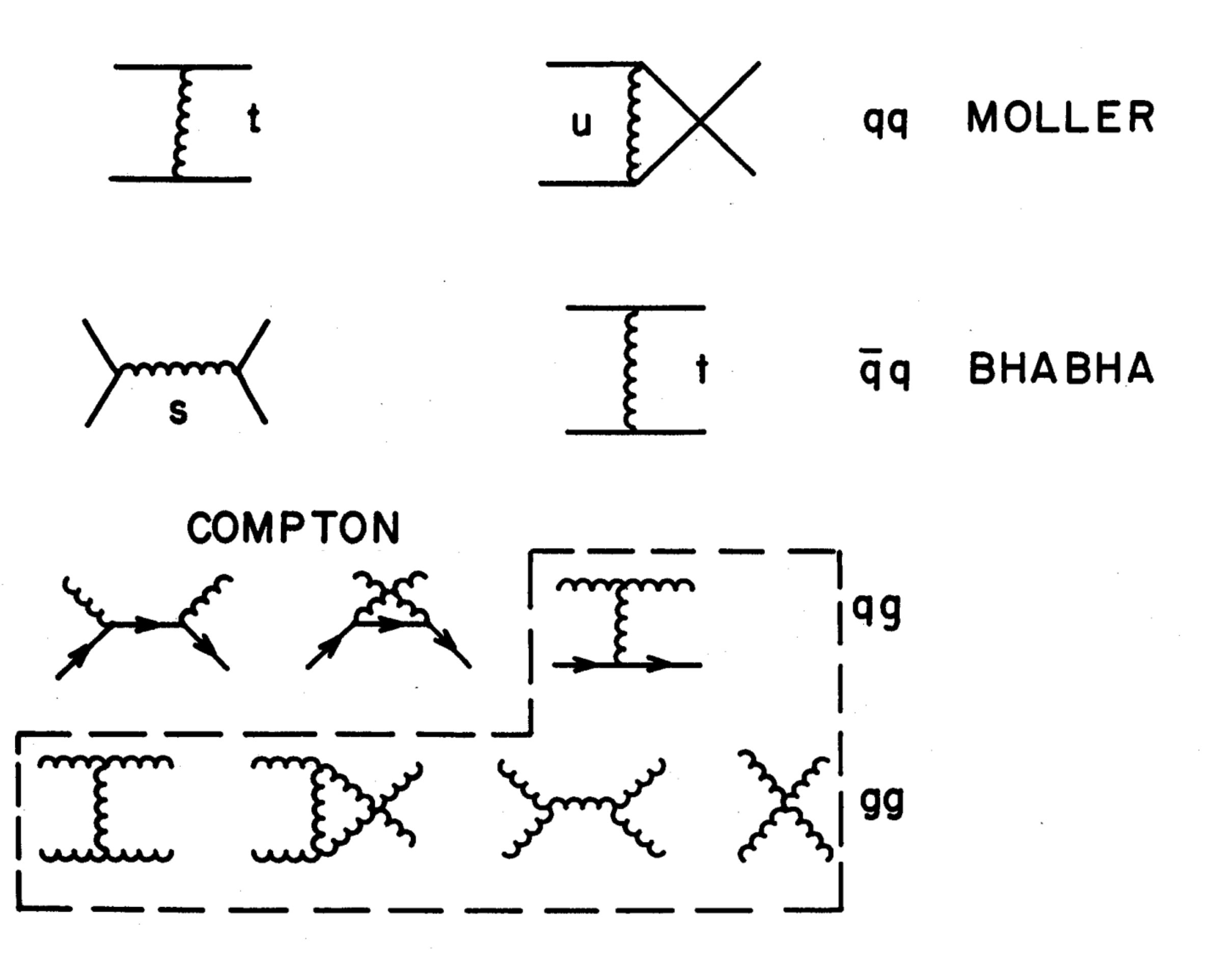}
\includegraphics[width=0.76\textwidth]{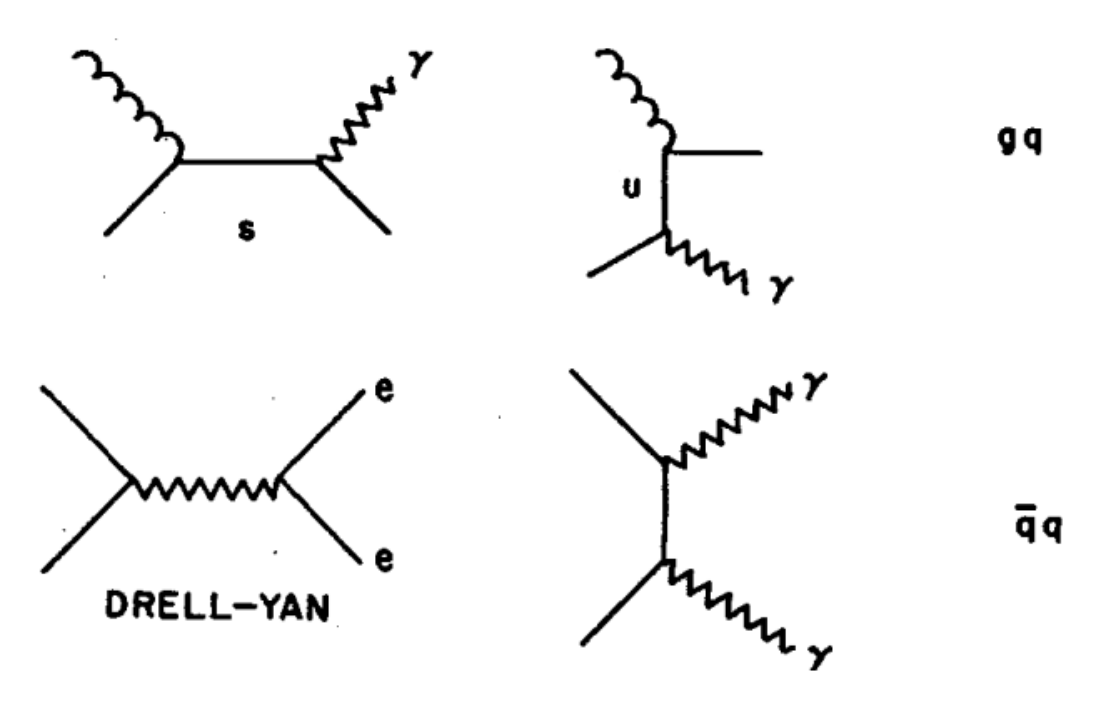}
\vspace*{-0.12in}
\end{minipage}
\end{center}
\caption[]{a)-h) (left) Hard scattering subprocesses in QCD~\cite{CombridgePLB70} \cite{CutlerPRD17} for quarks $q$ and gluons $g$ where the $q'$ indicates a different flavor quark from $q$, with their lowest order cross section angular distribution $\Sigma^{ab}(\cos\theta^*)$ as defined in Eq.~\ref{eq:dsigdthat}. (right) Lowest order diagrams involving initial state $q$ and $g$ scattering. }
\label{fig:someQCD} 
\end{figure}
At the elementary level, the lowest order \QCD\ Born diagrams, shown in Fig.~\ref{fig:someQCD}-(right), 
are analogous to the QED processes of M{\o}ller, Bhabha and Compton scattering indicated; and tests of their validity would be quite analogous to the studies of the validity of Quantum Electrodynamics (QED) which occupied the 1950's and 1960's~\cite{BrodskyDrellARNPS}. The $\hat{t}$ and $\hat{u}$ channels and  $\hat{s}$ and $\hat{t}$ channels are indicated for the $qq$ and $\bar{q}q$ diagrams respectively. It is important to note that the diagrams in the dashed box are unique to QCD since the gluons carry color charge and interact with each other while the photons in QED do not carry electric charge and so do not self-interact. 

Although the different combinations of $\hat{s}$, $\hat{t}$ and $\hat{u}$ for the cross sections in Fig.~\ref{fig:someQCD}a-h may at first seem to be formidable, it can be seen by substituting $\hat{t}=-\hat{s}(1-\cos\theta^*)/2$, $\hat{u}=-\hat{s}(1+\cos\theta^*)/2$, that indeed the $\Sigma^{ab}(\cos\theta^*)$ are nothing other than angular distributions. For example, for $qq'\rightarrow qq'$ Fig.~\ref{fig:someQCD}a:
\begin{equation}
\Sigma^{qq'}(\cos\theta^*)=\frac{4}{9} \,\frac{\hat{s}^2 +\hat{u}^2}{\hat{t}^2}= \frac{4}{9} \left[ \left(\frac{2}{1-\cos\theta^*}\right)^2
+ \left(\frac{1-\cos\theta^*}{1+\cos\theta^*}\right)^2 \right]
\label{eq:qq'scat}
\end{equation}

However, bringing Eqs.~\ref{eq:dsigdthat} and \ref{eq:qq'scat} to the p$+$p collision level leads to  complications.

\subsection{p$+$p hard-scattering in \QCD\ }
\begin{figure}[!h]
\centering
\includegraphics[width=0.69\linewidth]{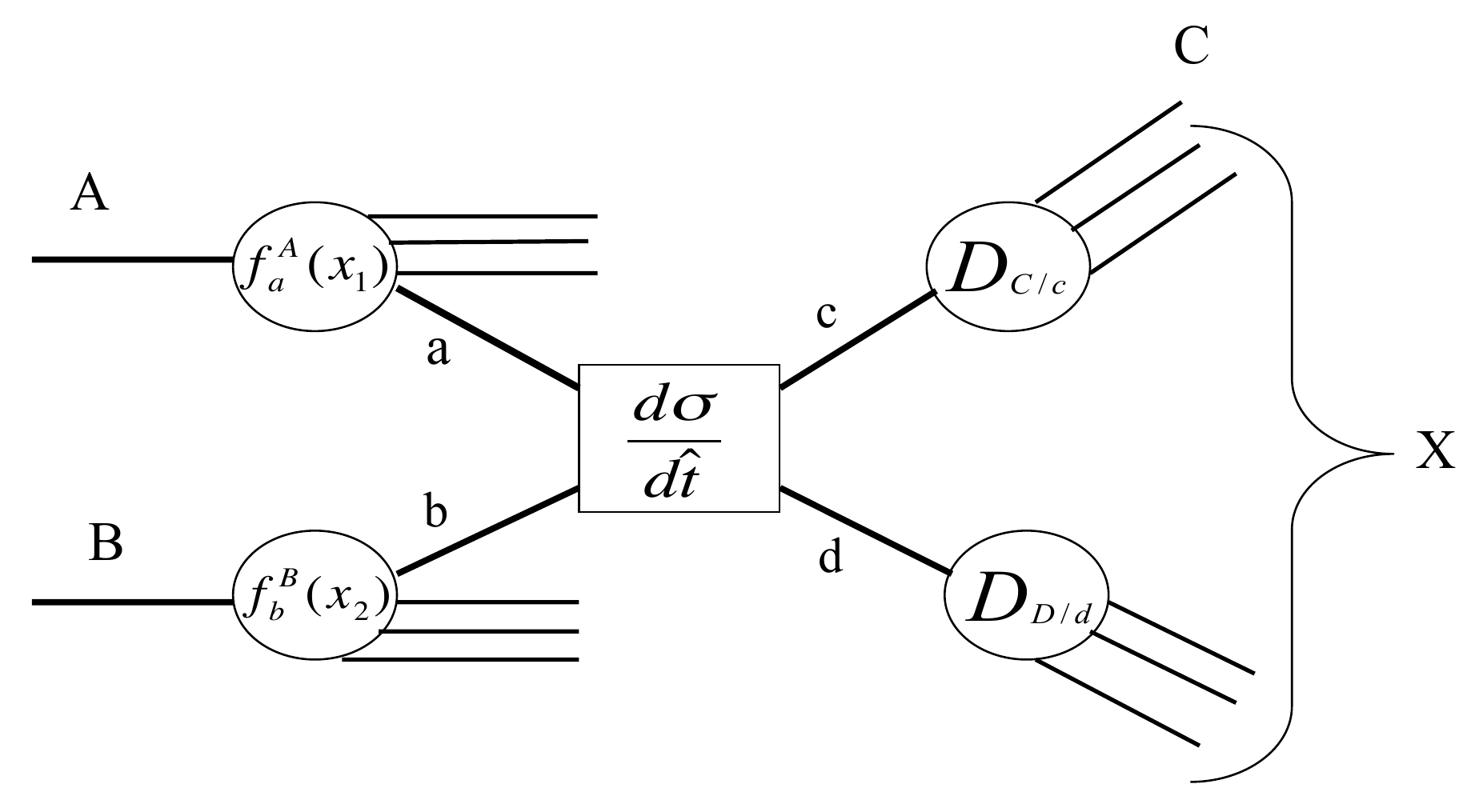}
\caption[]{Sketch (inspired by Fig.~3 of \cite{FFFPRD18}) of the reaction of initial colliding particles A+B producing a particle C in the final state where X represents all other particles.~\label{fig:ppQCDscat}}
\end{figure}

	The overall hard-scattering cross section $A+B\rightarrow C + X$ (Fig.~\ref{fig:ppQCDscat})
in ``leading logarithm'' p\QCD\    is the sum over parton reactions $a+b\rightarrow c +d$ 
(e.g. $g+q\rightarrow g+q$) at parton-parton center-of-mass (c.m.) energy $\sqrt{\hat{s}}$ as given in Eq.~\ref{eq:QCDabscat}~\cite{OwensRMP59}  
 
\begin{equation}
\frac{d^3\sigma}{dx_1 dx_2 d\cos\theta^*}=
\frac{s d^3\sigma}{d\hat{s} d\hat{y} d\cos\theta^*}=
\frac{1}{s}\sum_{ab} f^A_a(x_1) f^B_b(x_2) 
\frac{\pi\alpha_s^2(Q^2)}{2x_1 x_2} \Sigma^{ab}(\cos\theta^*)
\label{eq:QCDabscat}
\end{equation} 
where $f^A_a(x_1)$, $f^B_b(x_2)$, are parton distribution functions (PDF), 
the differential probabilities for partons
$a$ and $b$ to carry momentum fractions $x_1$ and $x_2$ of their respective 
protons (e.g. $u(x_2)$), and where $\theta^*$ is the scattering angle in the parton-parton c.m. system. 
The parton-parton c.m. energy squared is $\hat{s}=x_1 x_2 s$,
where $\sqrt{s}$ is the c.m. energy of the p$+$p  collision. The parton-parton 
c.m. system moves with rapidity $\hat{y}=(1/2) \ln (x_1/x_2)$ in the p$+$p  c.m. system. The quantities $f_a(x_1)$ and $f_b(x_2)$, the ``number'' 
distributions of the constituents, are related 
(for the electrically charged quarks) to the structure functions measured in $e+A$ lepton-hadron 
Deeply Inelastic Scattering (DIS), where $A$ is p or n, e.g. 
\begin{equation}
F^A_{1}(x,Q^2)={1\over 2} \sum_a e_a^2\; f^A_a(x,Q^{2}) \;\;\;\mbox{and}\;\;\;
F^A_{2}(x,Q^2)=x\sum_a e_a^2\; f^A_a(x,Q^{2})
\label{eq:F2}
\end{equation}
where $e_a$ is the electric charge of the quark $a$. 

The transverse momentum of a scattered constituent is:
\begin{equation}
p_T = p_T^* = { \sqrt{\hat{s}} \over 2 } \; \sin\theta^* \: ,
\label{eq:cpT}
\end{equation}
and the scattered constituents $c$ and $d$ in the outgoing parton-pair have equal and opposite momenta in the parton-parton (constituent) c.m. system. 
A naive experimentalist would think of $Q^2 = -\hat{t}$ for a scattering 
subprocess and $Q^2 = \hat{s}$ for a Compton or annihilation subprocess.

Equation~\ref{eq:QCDabscat} gives the $p_T$ spectrum of outgoing parton $c$, which then
fragments into a jet of hadrons, including e.g. $\pi^0$.  To go to the particle level the fragmentation function $D^{\pi^0}_{c}(z)$ which is the probability for a $\pi^0$ to carry a fraction
$z=p^{\pi^0}/p^{c}$ of the momentum of outgoing parton $c$ must be multiplied in along with its differential $dz$. Equation~\ref{eq:QCDabscat}
must be summed over all subprocesses leading to a $\pi^0$ in the final state weighted by their respective fragmentation functions. In this formulation, $f^A_a(x_1)$, $f^B_b(x_2)$ and $D^{\pi^0}_c (z)$ 
represent the ``long-distance phenomena" to be determined by experiment;
while the characteristic subprocess angular distributions,
{\bf $\Sigma^{ab}(\cos\theta^*)$} (Fig.~\ref{fig:someQCD}) 
and the coupling constant,
$\alpha_s(Q^2)=\frac{12\pi}{25\, \ln(Q^2/\Lambda^2)}$,
are fundamental predictions of \QCD~\cite{CombridgePLB70} \cite{CutlerPRD17}
for the short-distance, large-$Q^2$, phenomena. 
When higher order effects are taken into account, it is necessary to specify  factorization scales $\mu$ for the distribution and fragmentation functions in addition to renormalization scale $\Lambda$ which governs the running of 
$\alpha_s(Q^2)$. 
As noted above, the momentum scale $Q^2\approx p_T^2$ for the scattering
subprocess, while $Q^2\approx\hat{s}$ for a Compton or annihilation subprocess,
but the exact meanings of $Q^2$ and $\mu^2$ tend  to be treated as parameters  rather than as dynamical
quantities. 
\section{Direct single photon production, the most elegant \QCD\ reaction}
Direct single photon ($\gamma$) production with a ratio $\gamma/\pi^0\sim 10-20\%$~\cite{Glennys76} was one of the first proposed explanations of the direct single $e^\pm$ discovery at the ISR but was excluded with  95\% confidence level to $\gamma/\pi^0< 5\%$ for $p_T>1.3$ GeV/c ~\cite{CCRSNPB106}~\footnotemark[1]. 

\subsection{Direct-$\gamma$ \QCD\ elementary subprocesses: 1977}
The first \QCD\ calculation of direct-$\gamma$ production, ``the inverse \QCD\ Compton Effect'', via the constituent reaction $g+q\rightarrow \gamma+q$ was presented by Fritzsch and Minkowski in 1977~\cite{FritzschPLB69}.  This reaction has many beautiful aspects as a hadronic probe. 
\begin{enumerate}
\item[1.] The $\gamma$-ray participates directly in the hard-scattering and then emerges freely and unbiased from the reaction, isolated, with no accompanying particles. 
\item[2.] The energy of this outgoing parton (the $\gamma$-ray) can be measured precisely. 
\item[3.] No fragmentation function is required for Eq.~\ref{eq:QCDabscat}.
\item[4.] Since there are many fewer $\bar{q}$ in a nucleon than quarks or gluons, the $g+q$ reaction dominates so that the parton opposite the $\gamma$-ray is most likely a quark. 
\item[5.] The scattered quark has equal and opposite transverse momentum to the direct-$\gamma$, so the transverse momentum of the jet from the outgoing quark is also precisely known (modulo $k_T$). \end{enumerate} 

The cross-section for the elementary \QCD\ subprocess, $g+q\rightarrow \gamma+q$ is simply:
\begin{equation}
\left.\frac{d\sigma}{d\hat{t}}\right|_{\hat{s}}=\frac{\pi\alpha_s\,\alpha\, e^2_q}{3\,\hat{s}^2} \left(\frac{\hat{s}+\hat{t}}{\hat{s}}+\frac{\hat{s}}{\hat{s}+\hat{t}}\right ) \qquad, \label{eq:QCDCompton-dsigdt}
\end{equation}
Using the same argument as item 4 above, the \QCD\ cross-section for the reaction $A+B\rightarrow \gamma +q$ with the direct-$\gamma$ produced at rapidity $y_c$ with $p_T$, and the quark-jet at $y_d$ with $p_T$ is analytical: 
\begin{eqnarray}
\frac{d^3\sigma}{dp_T^2\, d{y_c}\, d{y_d}}
&=&
 x_1 G_{A}(x_1)\, F_{\rm 2B}(x_2) 
\frac{\pi\alpha\alpha_s(Q^2)}{3\hat{s}^2} \left(\frac{1+\cos\theta^*}{2}+\frac{2}{1+\cos\theta^*}\right ) \nonumber \\ 
&+& F_{\rm 2A}(x_1)\, x_2 G_{B}(x_2) 
\frac{\pi\alpha\alpha_s(Q^2)}{3\hat{s}^2} \left(\frac{1-\cos\theta^*}{2}+\frac{2}{1-\cos\theta^*}\right ) \nonumber\\ \label{eq:QCDComptonfull} 
\end{eqnarray}
where 
\begin{equation} 
\cos\theta^*=\tanh\frac{(y_c -y_d)}{2} \qquad x_{1,2}=x_T\frac{e^{\pm y_c} + e^{\pm y_d}}{2} \quad
\sqrt{\hat{s}}=2 p_T \cosh\frac{(y_c -y_d)}{2} .
\end{equation} 
Here $F_2(x,Q^2)$ is the sum of the PDF's over all the quarks and anti-quarks (predominantly $u$ and $d$) in the nucleons or nuclei A and B and  $G(x,Q^2)$ is the gluon PDF which is the principal theoretical problem in the calculation since it is the least well known. However, the main difficulty with the measurement of direct photon production is experimenatal: a huge background from the decays $\pi^0\rightarrow \gamma +\gamma$ and $\eta\rightarrow \gamma+\gamma$. 
\subsubsection{$R_{\gamma}$, `the double ratio' for signal/background}
For measurements at mid-rapidity, the $\pi^0$ cross section is typically a power law with $d\sigma/p_T dp_T\propto p_T^{-n}$. In this case the $p_T$ spectrum of the background $\gamma$ rays from \mbox{$\pi^0\rightarrow\gamma+\gamma$} relative to the $\pi^0$ spectrum  is given by the simple expression~\cite{FerbelMolzon}\footnotemark[1]:
\begin{equation} 
\left.\frac{\gamma}{\pi^0}\right|_{\pi^0}=\frac{2}{n-1} \Longrightarrow  \left.\frac{\gamma}{\pi^0}\right|_{\rm bkg}\approx1.19\times\frac{2}{n-1} \label{eq:gambkg}
\end{equation}
where $(\gamma/\pi^0)_{\rm bkg}$ includes the $\eta\rightarrow\gamma+\gamma$ background ($\eta/\pi^0 \approx 0.5$ for $p_T>3$ GeV/c with $\eta\rightarrow \gamma+\gamma$ branching ratio 0.38~\cite{CCRSPLB55eta}).

Generally, the background is calculated using Monte Carlo calculations including efficiency, acceptance, etc. However the signal/background ratio is best presented by the double ratio:
\begin{equation}
R_{\gamma}=\frac{(\gamma/\pi^0)_{\rm Measured}}{(\gamma/\pi^0)_{\rm Background}}\approx \frac{\gamma_{\rm Measured}}{\gamma_{\rm Background}} \label{eq:Rgamma}
\end{equation} 
because the calculated background in the form $(\gamma/\pi^0)_{\rm Background}$ can be checked by comparison to Eq.~\ref{eq:gambkg}. $R_{\gamma}>1.0$ indicates a direct-$\gamma$ signal.
\subsection{Experimental results}
As in many reactions studied for the first time, the experimental results at the ISR, where direct-$\gamma$ were eventually discovered, started with an incorrect measurement~\cite{DarriulatNPB110} while  other early experiments set limits using low mass $e^+ e^-$ pairs\footnotemark[1]. The first correct results from an experiment specifically designed to detect real single photons at the ISR~\cite{YuanAmaldi}  set a 95\% confidence upper limit of $\gamma/\pi^0<4\%$ for $2.3\leq p_T\leq 3.4$ GeV/c. 

\begin{figure}[!h]
    \centering
\raisebox{0.3pc}{\includegraphics[angle=+0.0,width=0.29\linewidth]{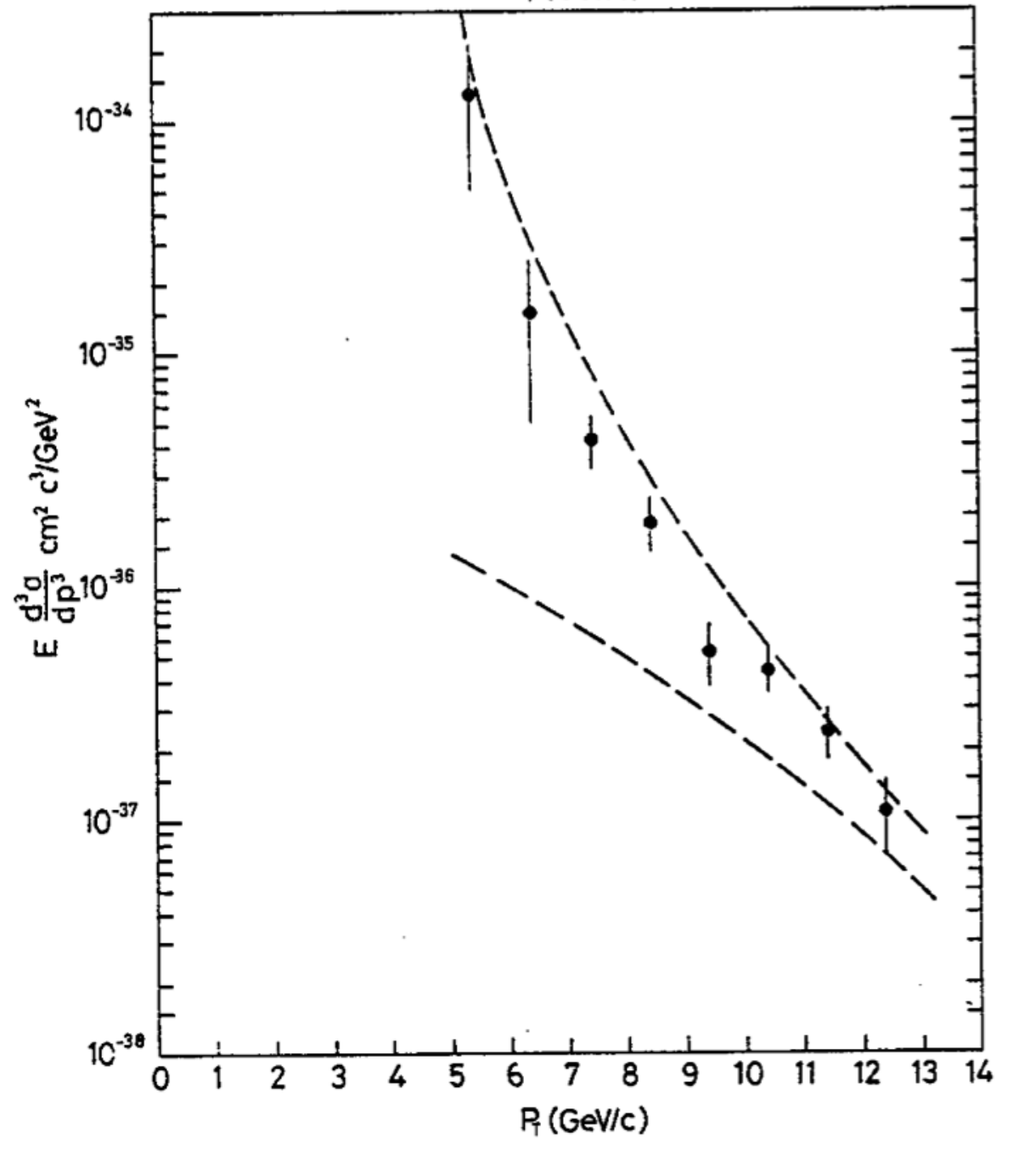}} 
\raisebox{-0.0pc}{\includegraphics[width=0.34\linewidth]{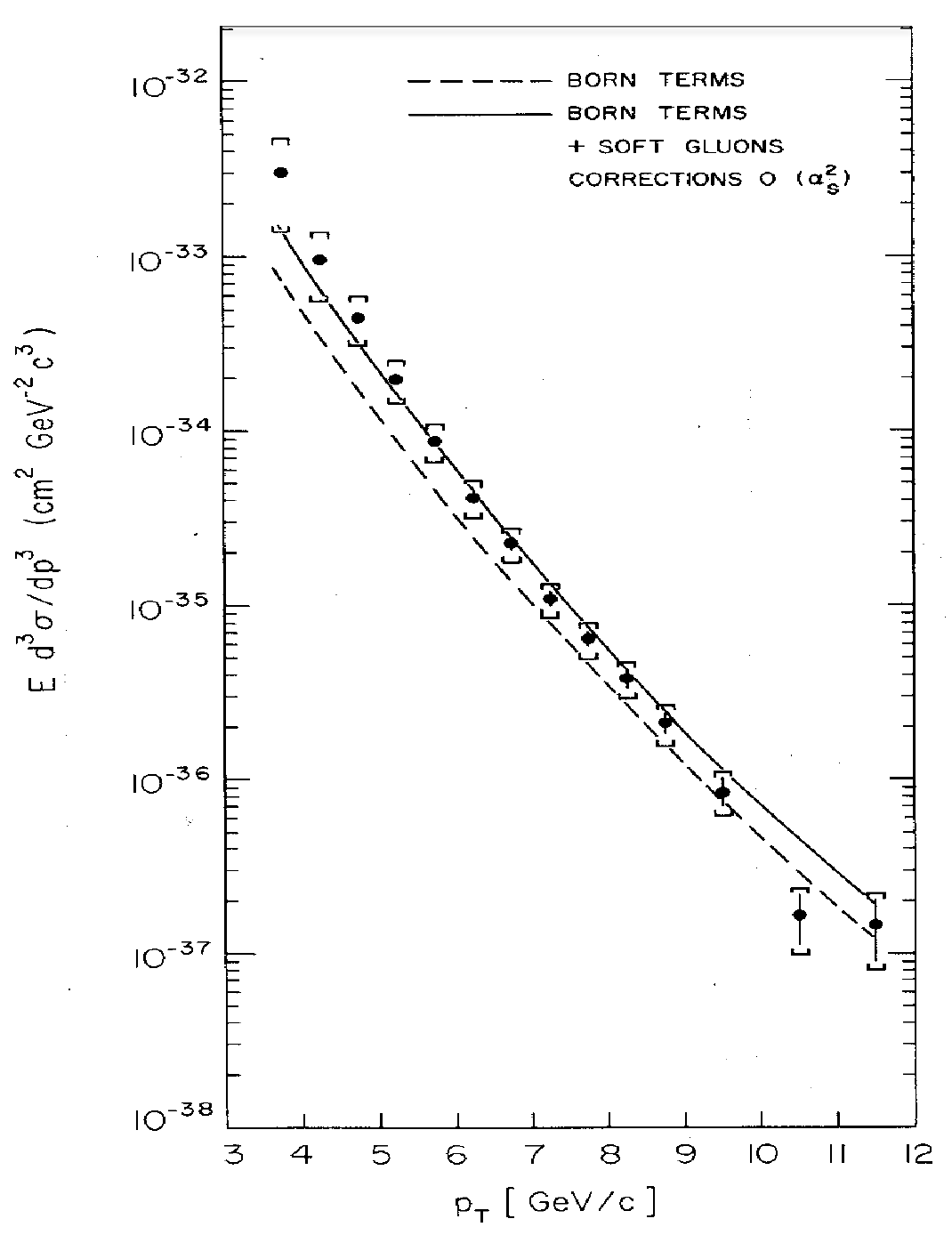}}
\raisebox{-0.0pc}{\includegraphics[angle=0.0,width=0.35\linewidth]{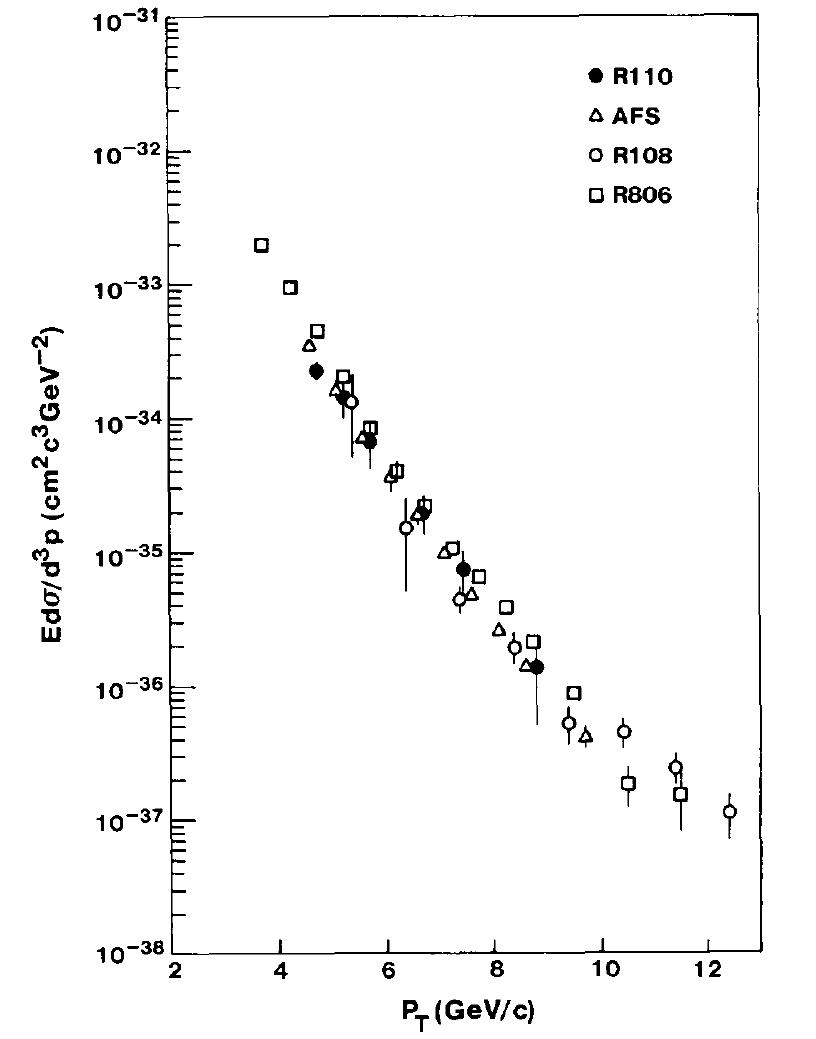}}
\caption[]{Inclusive direct-$\gamma$ cross sections at $\sqrt{s}=63$ GeV: a)(left) CCOR~\cite{CCORPLB94} (dashed lines are systematic uncertainty), b)(center) R806~\cite{WillisZPC13} (lines are \QCD\ calculations~\cite{ContogourisPLB104}), c) (right) CMOR~\cite{CMORNPB327}. 
\label{fig:Dirgamma}} 
\end{figure}

The first experiment to actually correctly claim the observation of a signal for direct-$\gamma/\pi^0 \approx 0.2$ for $p_T\geq 4.5$ GeV/c and generally given credit for the discovery of direct-$\gamma$ production was the measurement by the AABC experiment ~\cite{WillisPLB87} a modification of the ABCS experiment using the high resolution calorimeters which could resolve the two $\gamma$'s from $\pi^0$ decay. The first measurement of the direct-$\gamma$ cross section was by CCOR~\cite{CCORPLB94} who could not resolve the two $\gamma$s in a cluster but instead measured the fraction of clusters of a given $p_T$ that would pass through the 1.0 radiation length thin-wall of the superconducting solenoid without making a conversion(47\% for a single $\gamma$ and $\approx20-24$\% for 2 or more $\gamma$ rays in the cluster). This led to a large systematic uncertainty which was not too bad for $9\leq p_T\leq 13$ GeV/c (Fig~\ref{fig:Dirgamma}a). The cross-sections in this region were in surprisingly good agreement with the final AABC (R806) results (Fig~\ref{fig:Dirgamma}b)~\cite{WillisZPC13} as discussed in detail by ~\cite{FerbelMolzon}. The final ISR direct-$\gamma$ measurement ~\cite{CMORNPB327} also includes a summary of the previous measurements, which are all in impressive agreement over the range $4\leq p_T\leq 13$ GeV/c as shown in Fig~\ref{fig:Dirgamma}c. 

\vspace*{-2.0pc} 

\subsubsection{Theoretical predictions of direct-$\gamma$ data show that \QCD\ really works.}
\begin{figure}[!h]
    \centering
\raisebox{1.0pc}{\includegraphics[width=0.40\linewidth]{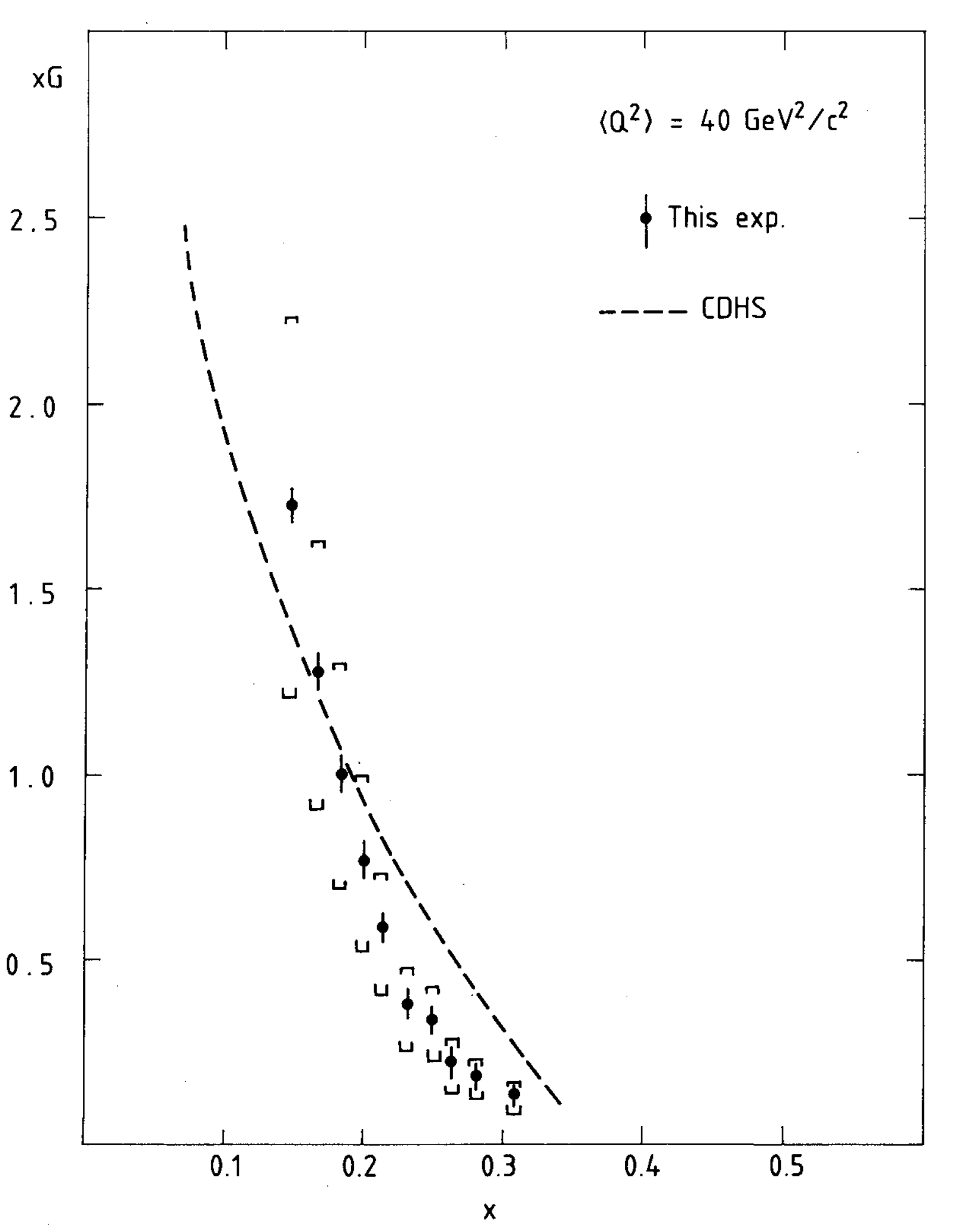}}\hspace*{1.0pc}
\raisebox{0.0pc}{\includegraphics[angle=+0.0,width=0.58\linewidth]{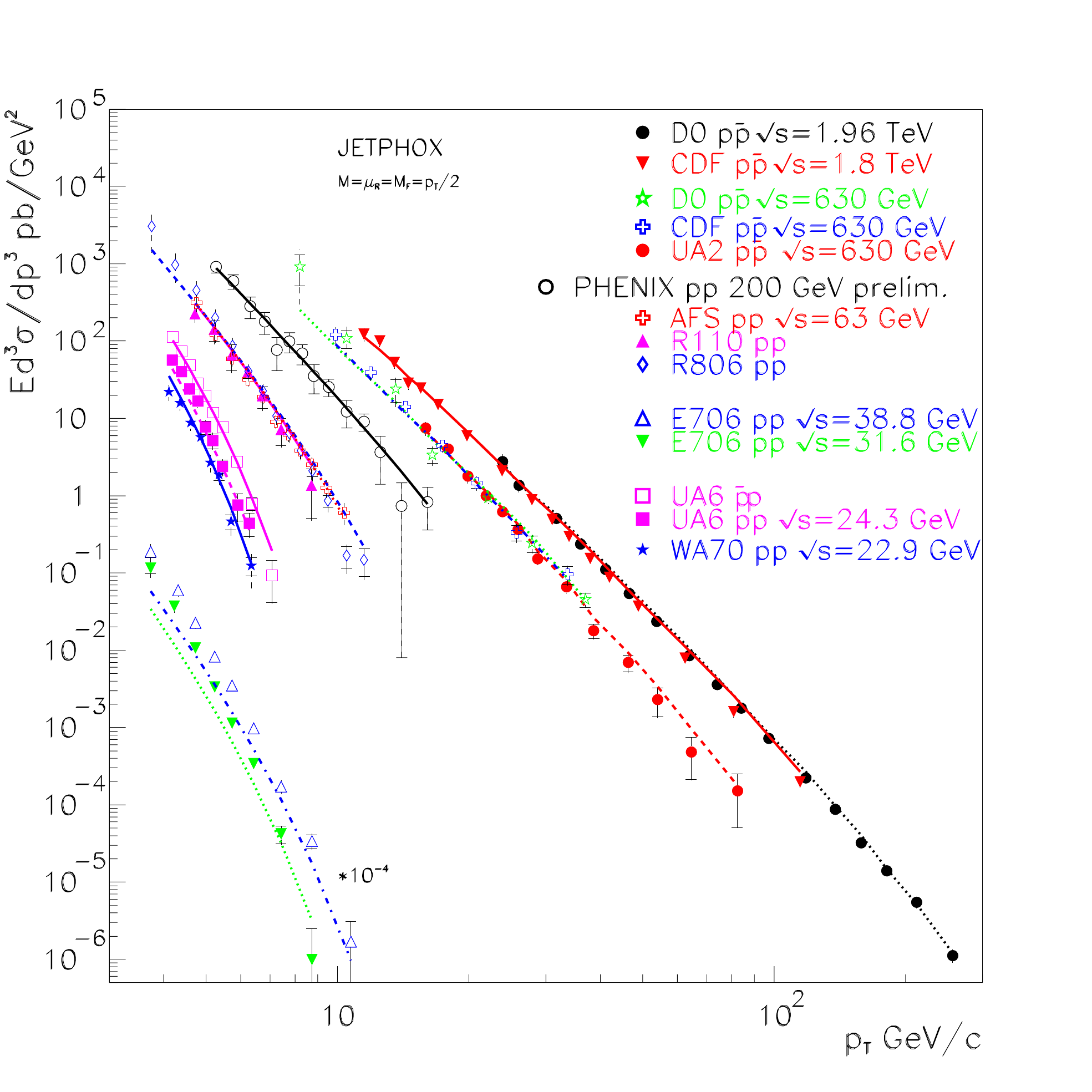}}\vspace*{-1.0pc}
\caption[]{a)(left) $x G(x)$ \cite{AFSZPC34} compared to \cite{CDHSZPC12} dashes. b)(right) Plot of all direct-$\gamma$ $Ed^3\sigma/dp^3$ measurements in p$+$p and $\bar{\rm p}+$p collisions circa 2006 compared to ``JETPHOX'' NLO \QCD\ predictions~\cite{AurenchePRD73}. 
\label{fig:Aurenchedoesit}} 
\end{figure}

The theoretical predictions of direct-$\gamma$ cross-sections lagged behind the experimental measurements because the gluon structure functions (PDF), $G(x,Q^2)$, in deeply inelastic e$+$p or neutrino scattering are not measured directly as are the quark PDF's, $F_2(x,Q^2)$, but are measured via the scaling violations, the $Q^2$ evolution of $F_2(x,Q^2)$~\cite{CDHSZPC12}. In fact, the first attempted direct measurement of $G(x,Q^2)$ was made at the ISR by the AFS experiment~\cite{AFSZPC34} with measurements of direct-$\gamma$+jet, both at mid-rapidity, solving Eq.~\ref{eq:QCDComptonfull} for $x G(x)$ (Fig.~\ref{fig:Aurenchedoesit}a).

It is interesting and informative to skip ahead to the present, where next-to-leading-order \QCD\  calculations using the latest PDF's as well as ``joint resummation of both threshold and recoil effects due to soft multigluon emission'' are in excellent agreement with all existing direct-$\gamma$ measurements at the time of publication~\cite{AurenchePRD73} (Fig.~\ref{fig:Aurenchedoesit}b.)\vspace*{-2.5pc}
\vspace*{-2.0pc} 

         \begin{figure}[!h]
   \begin{center}
\raisebox{0.2pc}{\includegraphics[width=0.452\textwidth]{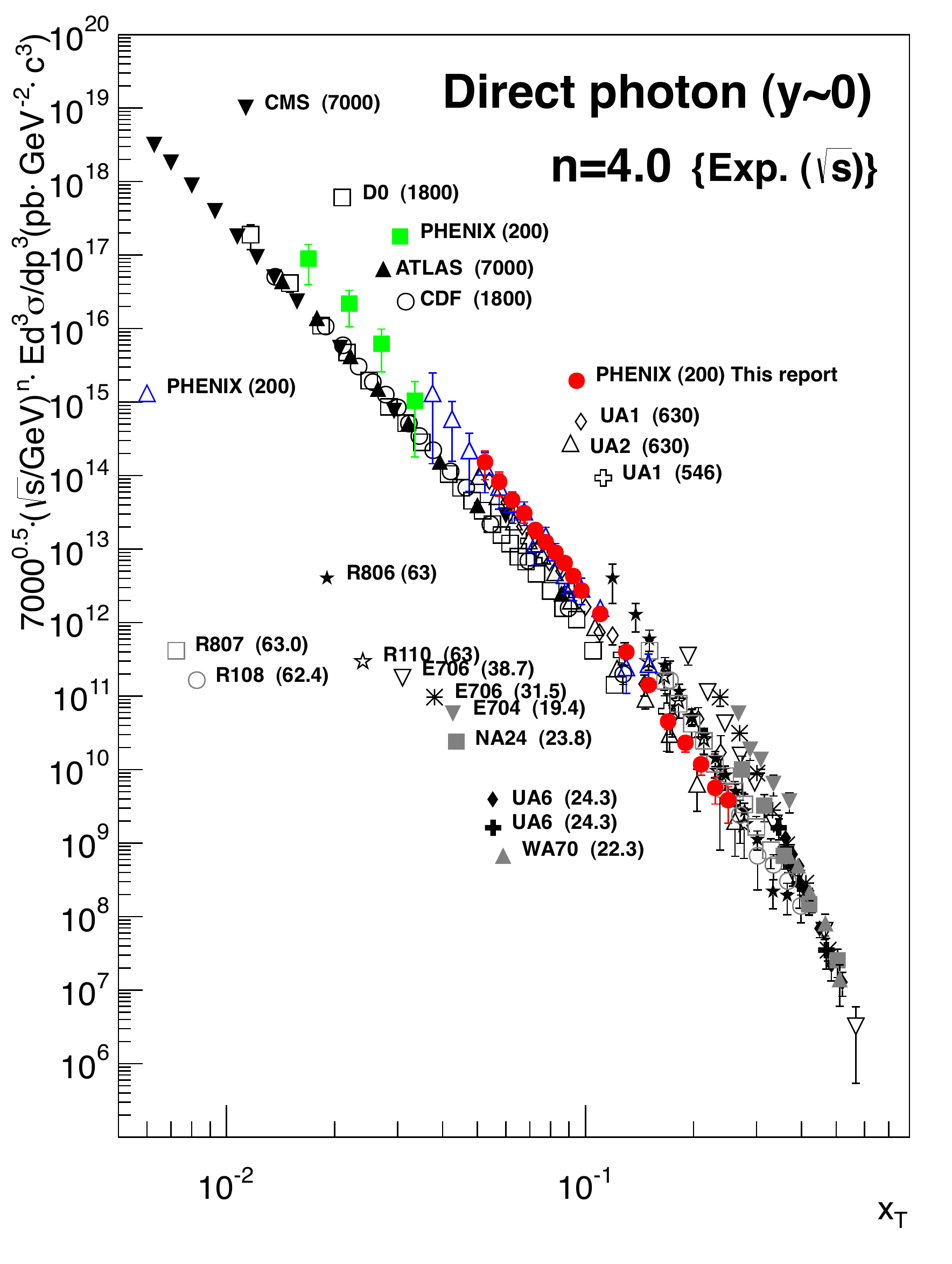}}
\includegraphics[width=0.45\textwidth]{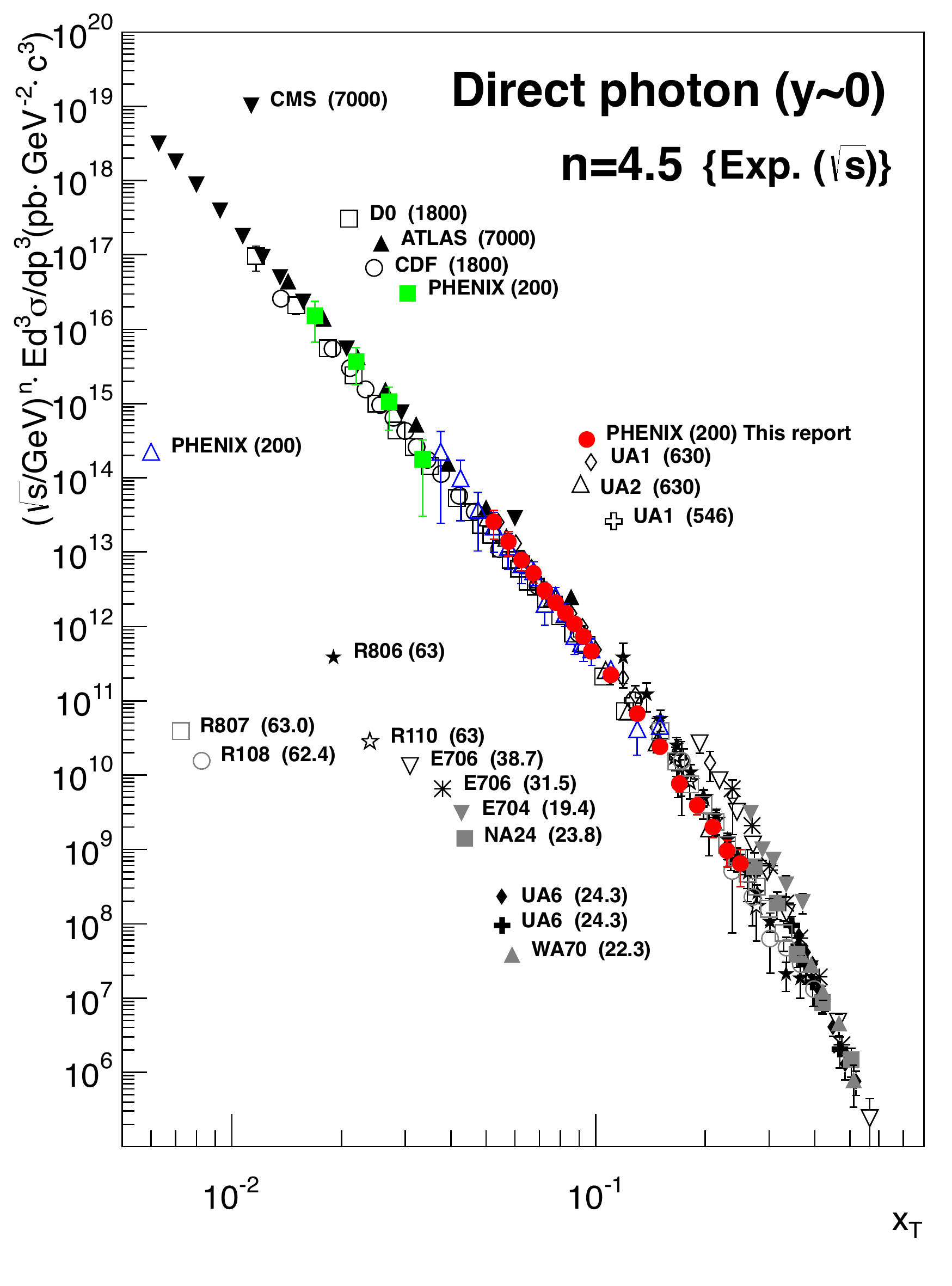}
\end{center}\vspace*{-1.5pc}
\caption[]{a)(left) $\sqrt{s}^{\,n_{\rm eff}} \times E d^3\sigma/dp^3$, as a function of $x_T=2p_T/\sqrt{s}$, with $n_{\rm eff}=4.0$,  for direct-$\gamma$ measurements in p$+$p  and $\bar{\rm p}$-p experiments at the ($\sqrt{s}$ GeV) indicated.~\cite{ppg136}. b)(right) same as (a) with $n_{\rm eff}=4.5$  
\label{fig:ggg}}\vspace*{-1pc}
\end{figure}

From the experimental point of view, a better illustration of \QCD\ in action is given by the $x_T$ scaling of all the existing direct-$\gamma$ measurements (Fig.~\ref{fig:ggg}). For $x_T$ scaling with $n_{\rm eff}$=4.0 (the parton model) \QCD\ non-scaling is visible; but $x_T$ scaling with $n_{\rm eff}$=4.5 accounts for the non-scaling evolution which is a key element of \QCD\ . 

\section{Two-particle correlations.}
\begin{figure}[!h]
\centering 
\raisebox{0.0pc}{\includegraphics[width=0.6\textwidth]{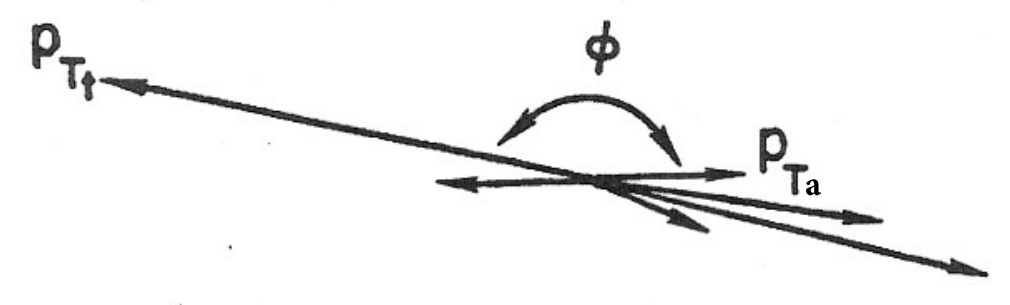}}
\caption[]{Sketch of view down the beam axis of dijet event trigged by particle with $p_{Tt}$ (circa 1973)    \label{fig:Bjazimuth}}
\end{figure}

It was natural for the experiments that had searched for single leptons and lepton-pairs at mid-rapidity but were overwhelmed by pions at large $p_T$ to look for what was balancing the $p_T$ of the pions: an outgoing parton with opposite $p_T$, as described by Bjorken~\cite{BjCornell} \cite{BjPRD8}, that fragmented into a collimated group of particles, ``a jet'' with a configuration like ordinary particle production e.g. $e^{-6p'_T}$ where $p'_T$ is perpendicular to the outgoing parton. In this case, the jets should be coplanar with the beam direction and balance transverse momenta. Viewed down the beam axis, in azimuthal projection (Fig.~\ref{fig:Bjazimuth}), the events should show strong azimuthal correlation.
Another possibility was balance as required by kinematics~\cite{DiLellaLondon}. 
 
 In Fig.~\ref{fig:Bjazimuth} the different distribution of fragments in the jet $\hat{p}_{Tt}$, triggered by a particle with $p_{Tt}$, and the away jet is no accident. Bjorken made a very important point in his parton scattering prediction, that the single high $p_T$ particle used as a trigger to search for the opposite jet, would carry ``a major fraction (60-80\%) of the total trigger parton (or jet) momentum. 
This is known as ``trigger bias''~\cite{JacobNPB113} and is related to the  Bjorken parent-child relation~\cite{JacobPLC48}. 

The cross-section for a pion with $p_{Tt}$ which is a fragment with momentum fraction $z=p_{Tt}/\hat{p}_T$ in a jet from a parton $q$ with $\hat{p}_{T}$ where $D^q_{\pi} (z)$ is the fragmentation function (e.g $\sim e^{-bz}$) is:
   \begin{equation}
   {{d^2\sigma_{\pi} (\hat{p}_T,z) }\over {\hat{p}_T d\hat{p}_T dz }}={{d\sigma_q}\over {\hat{p_T}d\hat{p}_{T}}}\times D^q_{\pi} (z)={A \over {\hat{p}_{T}^{n}}}  \times D^q_{\pi} (z) 
 \qquad .  \label{eq:mjt-zgivenq}
   \end{equation}
The change of variables, $\hat{p}_{T}=p_{Tt}/z$, ${d\hat{p}_{T}}/{dp_{Tt}}|_{z}=1/z$, 
then gives the joint probability of the pion with transverse momentum $p_{Tt}$ and fragmentation fraction $z$: 
\begin{equation}
{{d^2\sigma_{\pi} (p_{Tt},z)} \over {p_{Tt} dp_{Tt} dz}} 
={A \over {p_{Tt}^{n}}} \times z^{n-2} D^q_{\pi}(z) \qquad . 
\label{eq:mjt-zgivenpi}
\end{equation}
Thus, the effective fragmentation function, for a trigger particle with $p_{Tt}$ is weighted upward in $z$ by a factor $z^{n-2}$, where $n$ is the simple power fall-off of the jet invariant cross section (Eq.~\ref{eq:mjt-zgivenq}). This is the `trigger bias'~\cite{JacobNPB113}. 

The pion $p_{Tt}$ distribution is the integral of Eq.~\ref{eq:mjt-zgivenpi} over all values of the parton $\hat{p}_{T}$ from $\hat{p}_{T}=p_{Tt}$, $z=1$, to $\hat{p}_{T}=\sqrt{s}/2$, $z=x_{Tt}$, which has the same power $n$ as the parton $\hat{p}_{T}$ distribution: 
\begin{equation}
{{d\sigma_{\pi} (p_{Tt})} \over {p_{Tt} dp_{Tt}}} 
={1 \over {p_{Tt}^{n}}} \int^1_{x_{Tt}} A\ z^{n-2}\ D^q_{\pi}(z)\ dz\approx {{\rm constant}\over{p_{Tt}^{n}}},  
\label{eq:mjt-pisig}
\end{equation}
since typically $x_{Tt}\ll 1$, so the integral depends only weakly on $p_{Tt}$. Thus the invariant $p_{Tt}$ spectrum of the $\pi^0$ fragment is a power law with the same power $n$ as the original parton $\hat{p}_T$ spectrum, for $p_{Tt}\ll \sqrt{s}/2$. This is the Bjorken parent-child relation.   
\subsection{Two particle correlation measurements 1975-1977.}
\label{sec:xEFFISR}
\begin{figure}[!h]
\centerline{\includegraphics[width=0.95\linewidth]{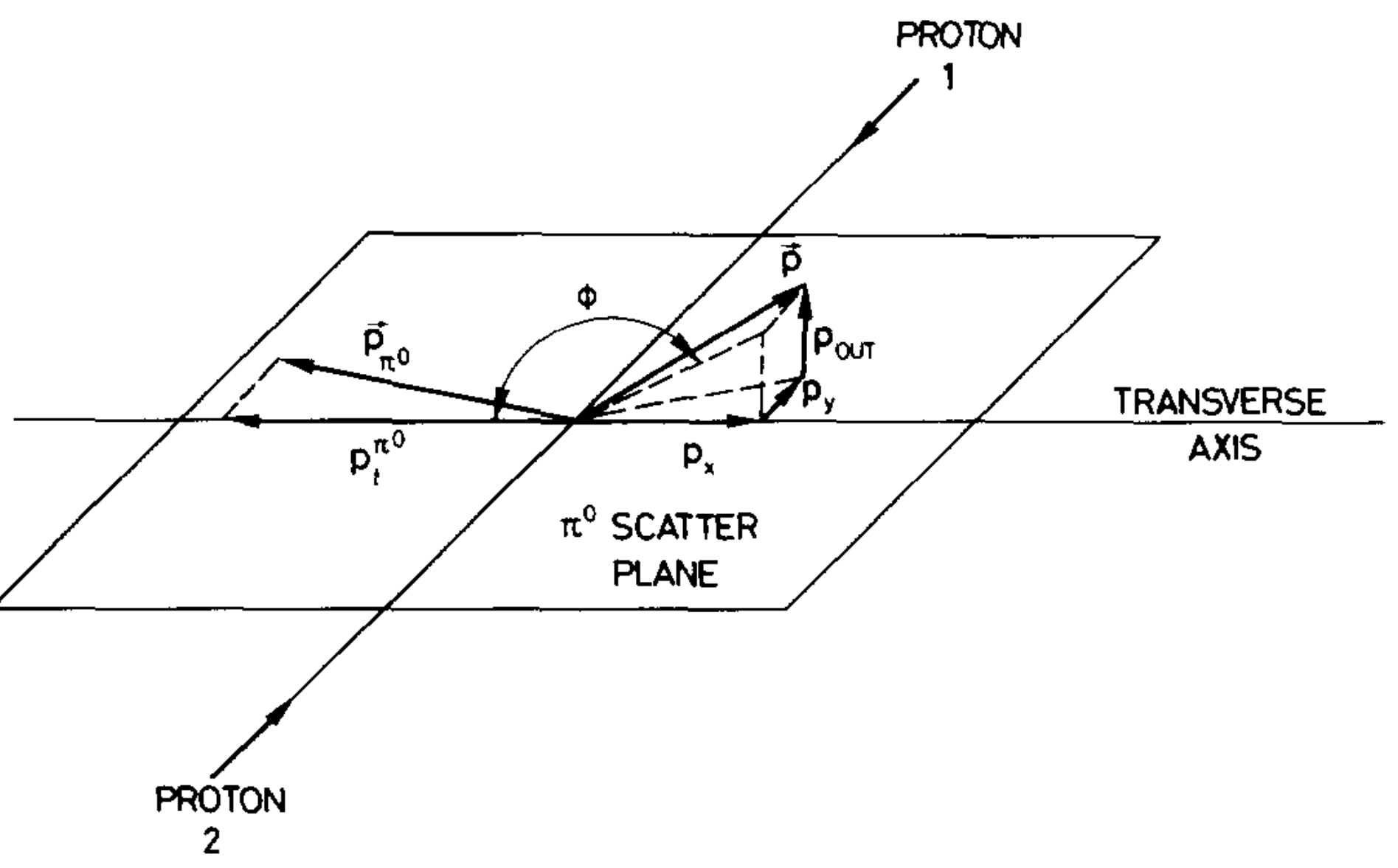}}	  
\caption[]{Diagram~\cite{Darriulatpoutxe} of kinematical quantities for $\pi^0$-hadron correlations. The trigger $\pi^0$ has momentum 
${\bm{p}}_{\pi^0}$ and transverse momentum $p_{T_t}^{\pi^0}$. The away hadron momentum ${\bm p}$ is broken into four components: i) $p_{\rm out}$, perpendicular to the scattering plane formed by the colliding protons and ${\bm{p}}_{\pi^0}$; ii) $p_{y}$ parallel to the p$+$p collision axis in the scattering plane; iii)  ${{p}}_{T}$, not labelled, the component of ${\bm p}$ transverse to the p$+$p collision axis; and iv) $p_x=p_T\cos\phi$, opposite to the direction of $p_{T_t}^{\pi^0}$ in the scattering plane, where $\phi$ (often called $\Delta\phi$) is the azimuthal angle between $p_{T_t}^{\pi^0}$ and $p_T$.    
}
\label{fig:poutxedef}
\end{figure}

Historically, at the ISR~\cite{Darriulatpoutxe,CCHK}, since the trigger bias implied that the $p_{Tt}$ of the trigger particle was a reasonable approximation to the $\hat{p}_{T}$ of the triggered jet, and the away jet would be approximately equal and opposite to the trigger jet, the transverse momentum of the away particle $p_{Ta}$ was decomposed into components (Fig.~\ref{fig:poutxedef}) of which two are commonly used: one perpendicular to the trigger plane $p_{\rm out}=p_{Ta} \sin{\Delta\phi}$,  and one in the trigger plane:
 \begin{equation}
x_E=\frac{-p_{x}}{p_{Tt}}=\frac{-p_{Ta} \cos(\Delta\phi)}{p_{Tt}}\simeq\frac{p_{Ta}/\hat{p}_{Ta}}{p_{Tt}/\hat{p}_{Tt}} \approx \frac {z}{z_{\rm trig}} \qquad. 
\label{eq:mjt-xE}
\end{equation}

\begin{figure}[!h]
    \centering
\raisebox{-1.0pc}{\includegraphics[angle=+0.5,width=0.46\linewidth]{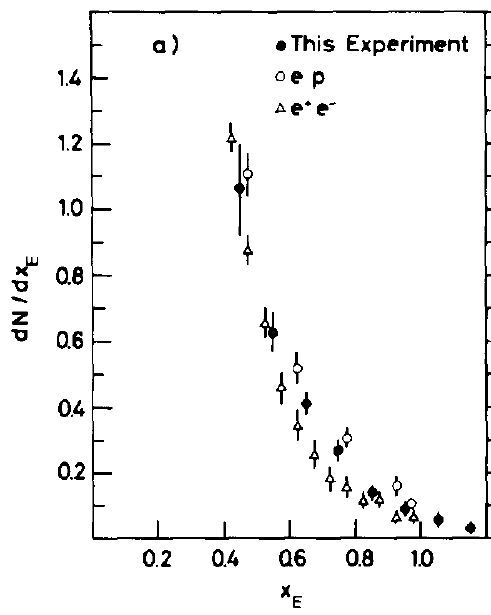}} 
\raisebox{-0.0pc}{\includegraphics[angle=0.0,width=0.46\linewidth]{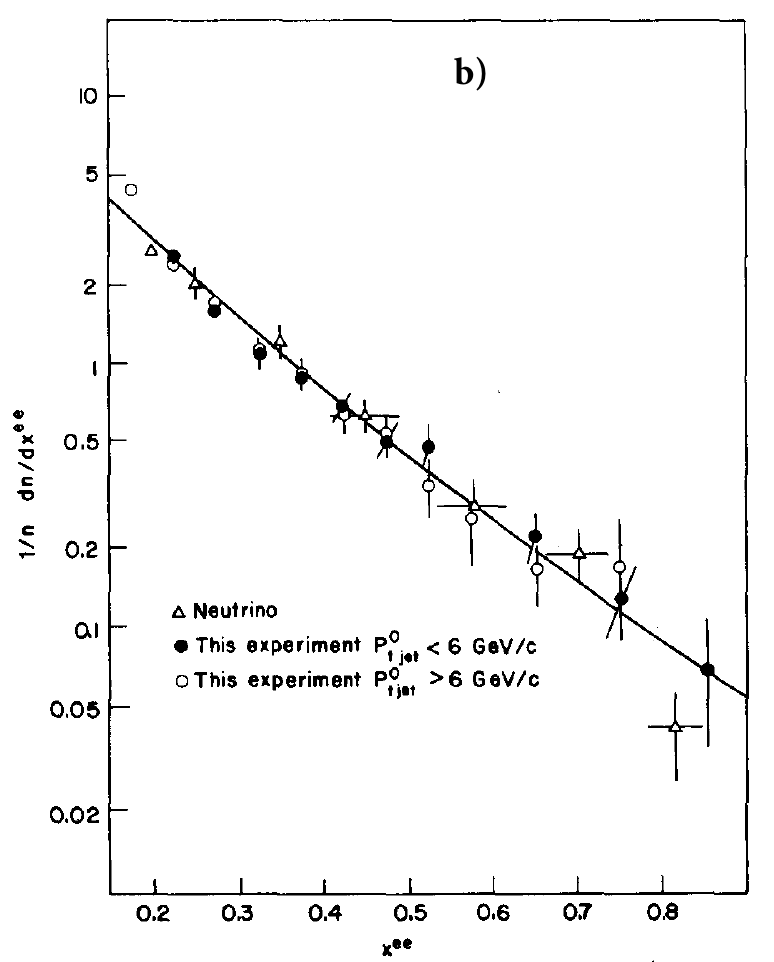}}
\caption[]{a) $x_E$ distribution for $\pi^0$--$h$ correlations with $2.0<p_{Tt}<4.1$ GeV/c and $1.2<p_x<3.2$ GeV/c together with e$+$p DIS and $e^+ e^-$ measurements~\cite{Darriulatpoutxe}. b) jet fragmentation functions from $\nu+$p (triangles), $e^+ e^-$ (full line) compared to measured p$+$p $x_E$ distributions~\cite{ClarkNPB160}.
\label{fig:xEfrag1}} 
\end{figure}

With the assumption that the trigger and away jets balance transverse momenta, $\hat{p}_{Ta}=-\hat{p}_{Tt}$, as assumed in the last step of Eq.~\ref{eq:mjt-xE}, the variable $x_E$ was thought to measure the fragmentation fraction $z$ of the away jet from the highly biased trigger jet with $z_{\rm trig}\rightarrow 1$. It was generally assumed that the $p_{Ta}$ distribution of away side hadron fragments from an away-side parton opposite a single particle trigger with $p_{Tt}$, would be the same as that from a jet-trigger and follow the same fragmentation function of partons as observed in $e^+ e^-$  or DIS~\cite{Darriulatpoutxe}
(Fig.~\ref{fig:xEfrag1}). Because of the relatively large trigger bias at ISR energies and small range of $\mean{z_{\rm trig}}$, $0.8\lsim \mean{z_{\rm trig}}\lsim 0.9$, it was also assumed that $x_E$ scaling~\cite{JacobNPB113} would hold, i.e. all $x_E$ distributions measured at different $p_{Tt}$ would be the same. 
    \vspace*{-2pc} 

\begin{figure}[!h]
\begin{center}
\raisebox{0.0pc}{\includegraphics[width=0.49\linewidth]{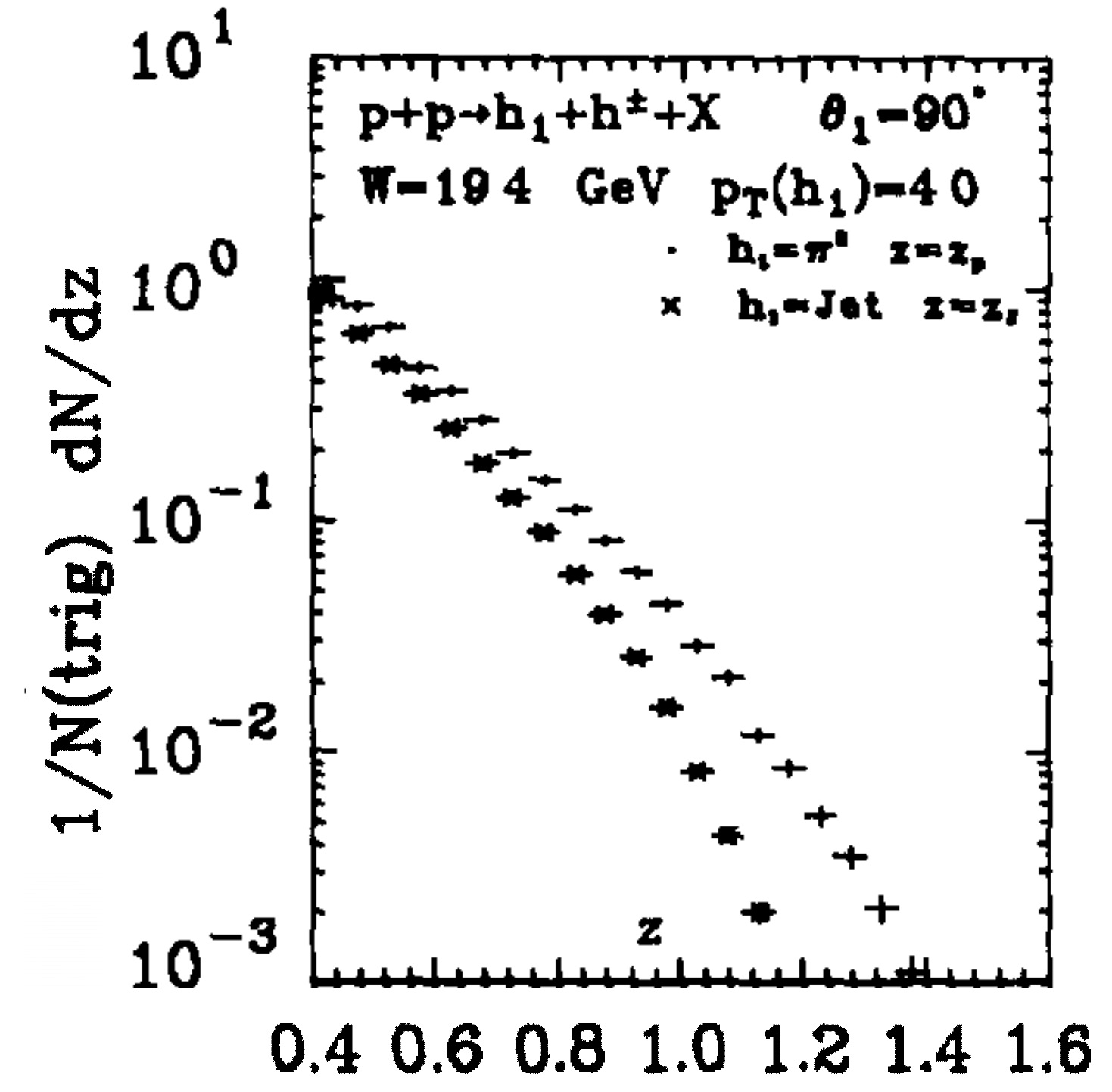}}
\raisebox{0.3pc}{\includegraphics[width=0.49\linewidth]{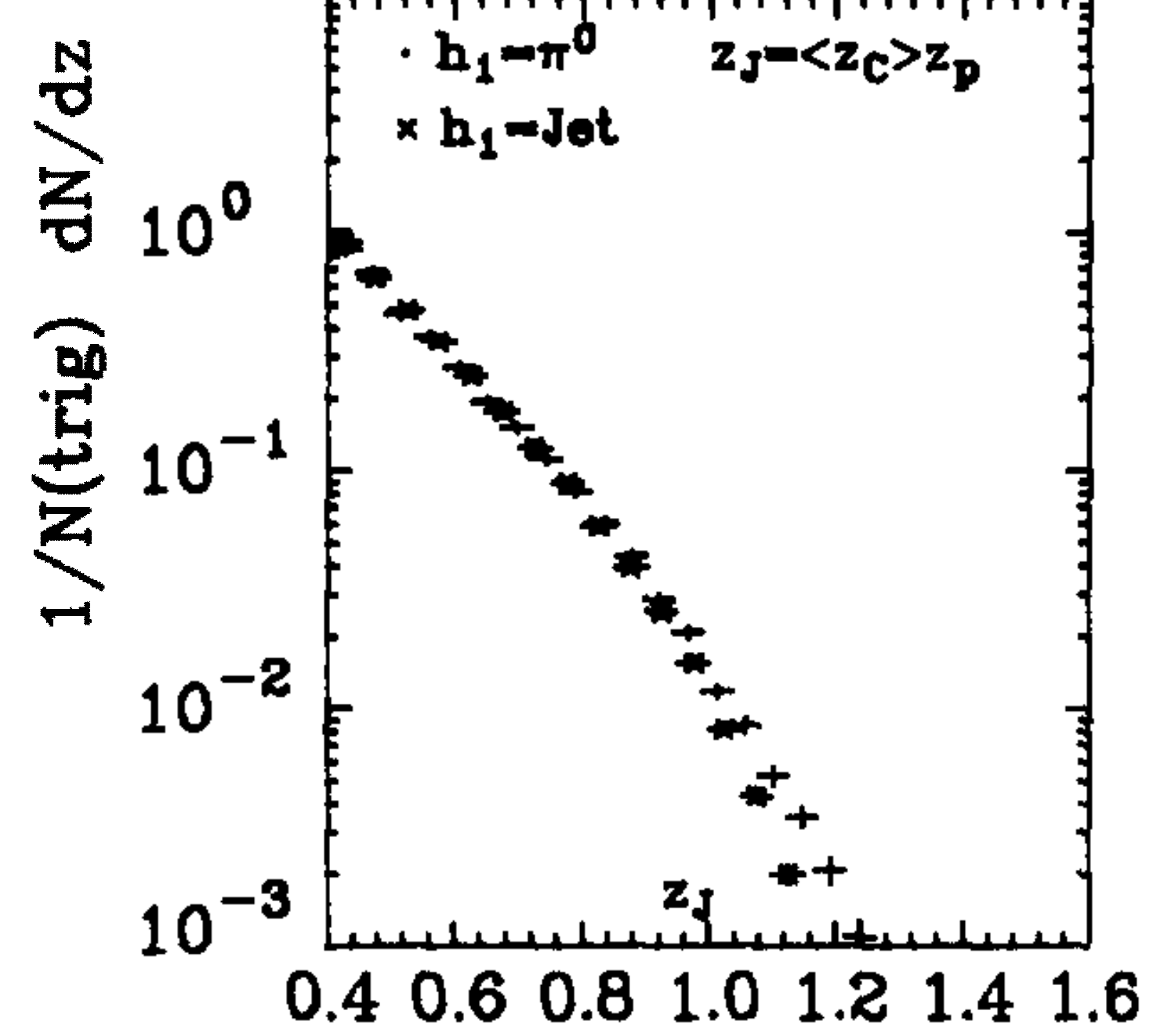}}
\end{center}\vspace*{-1pc}
\caption[]{\cite{FFF1}: a)(left) $dn/dz|_{trig}$ plotted vs. $z=x_E$ for a single particle trigger with $p_{Tt}$ (dots) and $z=z_J=p_x/p_{T{\rm jet}}$ for a trigger  by the jet (crosses); b)(right) same plot except now $z_J=\mean{z_{trig}}x_E$ is plotted for the single particle trigger instead of $z=x_E$. \label{fig:FFFfig23}}
\end{figure}

These ideas were further clarified by Feynman, Field and Fox~\cite{FFF1}, who showed that the away side correlations for a single-particle trigger with $p_{Tt}$ would be roughly the same as the away-side correlations for a jet trigger with $\hat{p}_{Tt}=p_{Tt}/\mean{z_{\rm trig}}$ due to the fact that ``the `quark' from which the single hadron trigger came had a higher $\hat{p}_{Tt}$ than  
did the quark producing the jet trigger (by a factor of $1/\mean{z_{\rm trig}}$)''. (Fig~\ref{fig:FFFfig23}).
The ideas of $x_E$ scaling according to a fragmentation function were supported by later ISR measurements but first a contemporary experiment (CCHK)~\cite{CCHK} did not find $x_E$ scaling for values of $p_{Tt}<4$ GeV/c which led to an important discovery.\vspace*{-1.0pc}

\subsubsection{ {${k_T}$}, the transverse momentum of a parton in a proton}

\begin{figure}[ht]
\begin{center}
\begin{tabular}{cc}
\includegraphics[width=0.47\linewidth]{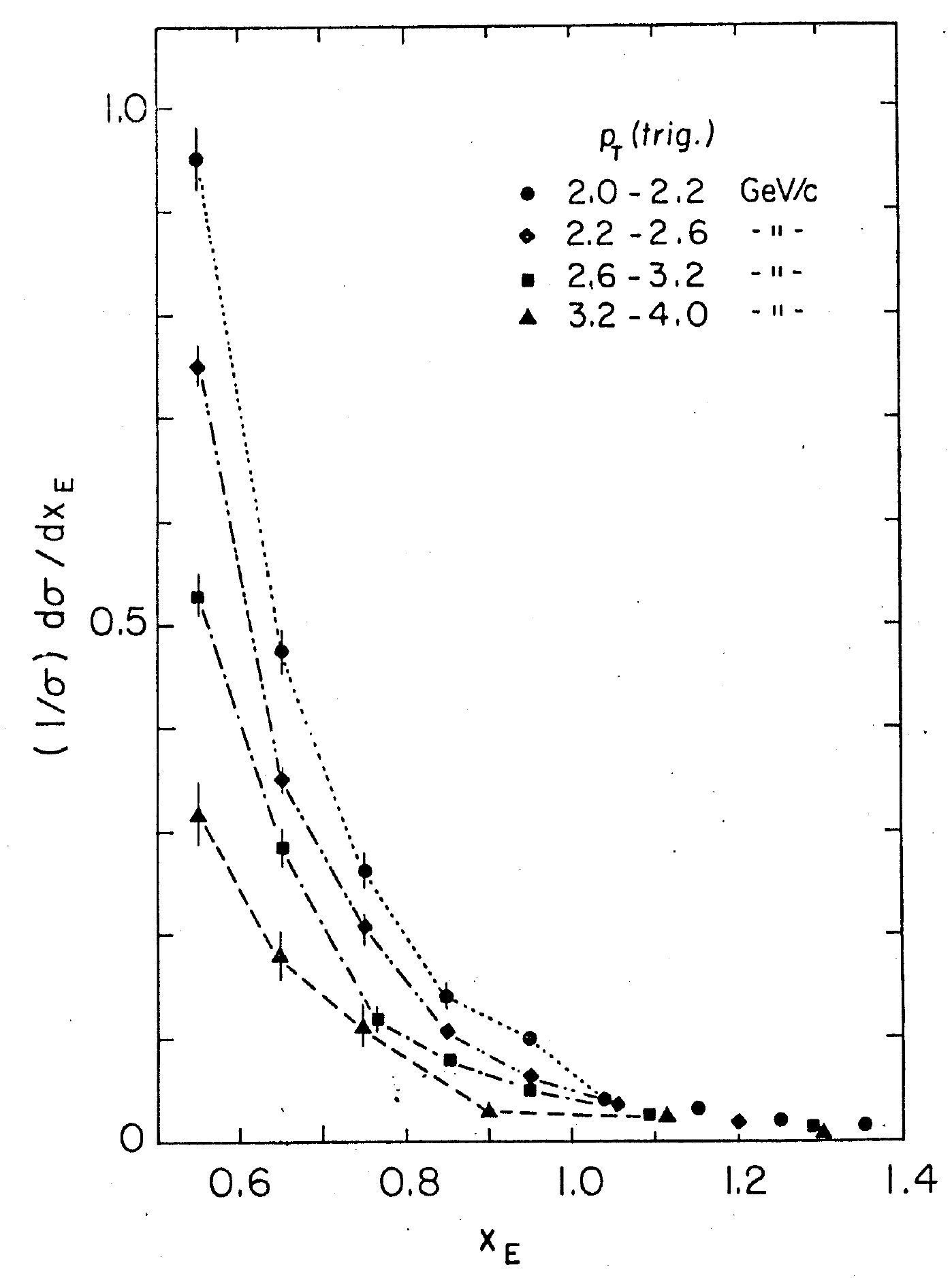} &
\includegraphics[width=0.50\linewidth]{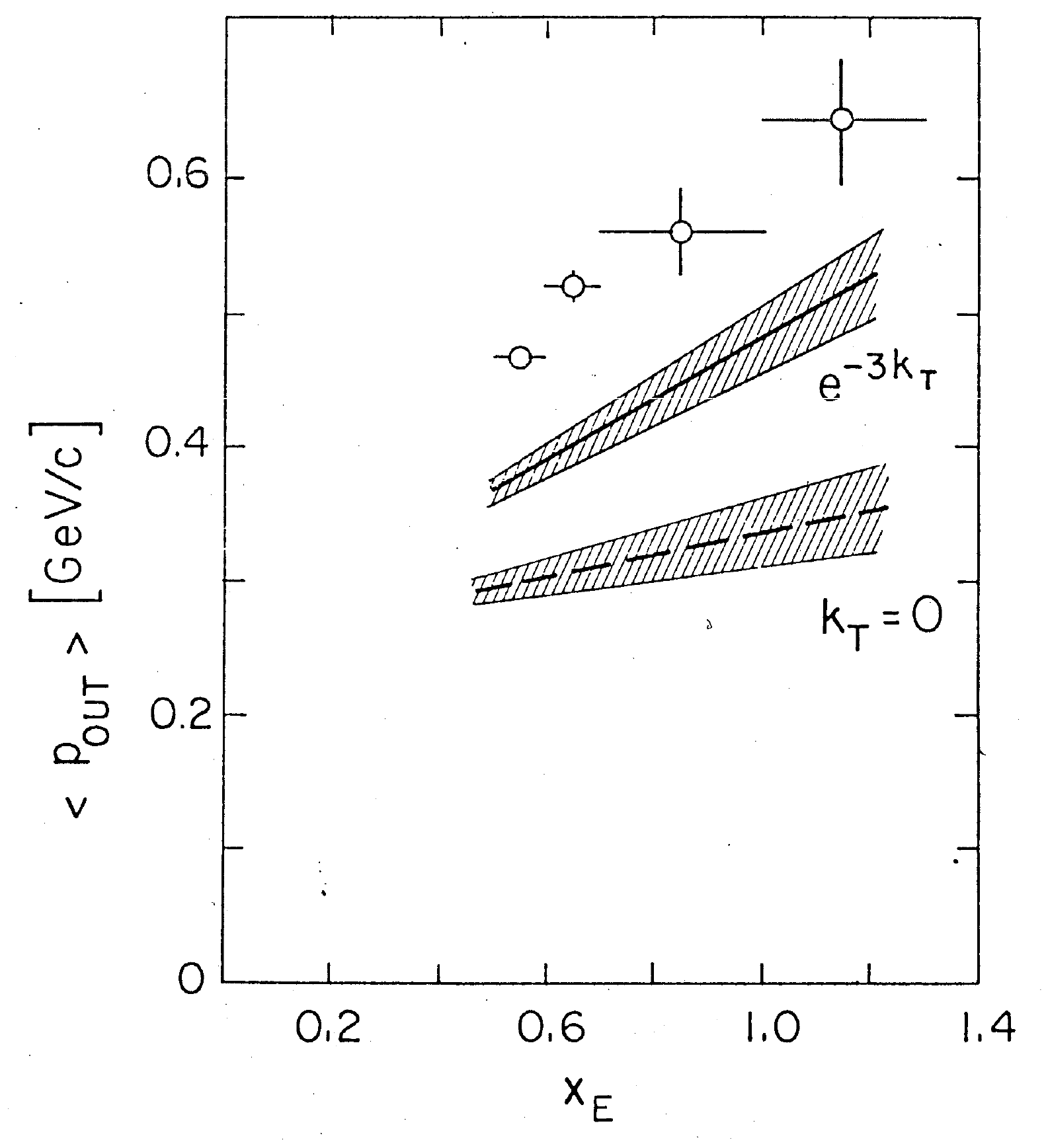}
\end{tabular}
\end{center}
\vspace*{-6mm}
\caption[]
{a)(left) CCHK~\cite{CCHK} measurement of $x_E$ distributions,  $dn/dx_E|_{p_{T_t}}$, for intervals of $p_{Tt}$ in the range 2.0--4.0 GeV/c.  b) (right) CCHK~\cite{CCHK} measurement of $\mean{p_{\rm out}}$ versus $x_E$ for triggers in the range $2.0\leq p_{T_t}\leq 3.2$ GeV/c. Also shown are the predictions of their parton scattering Monte Carlo model with $k_T=0$ and $dN/k_T dk_T\propto e^{-3k_T}$ (with the restriction $k_T<5/3$ GeV/c).  
\label{fig:CCHKxE} }
\end{figure}

The CCHK experiment at the CERN-ISR~\cite{CCHK}  observed that $x_E$ scaling~\cite{JacobNPB113} didn't work in the range $2.0\leq p_{T_t}\leq 3.2$ GeV/c; i.e. different values of trigger $p_{T_t}$ did not produce a universal $x_E$ distribution (Fig.~\ref{fig:CCHKxE}a). 
CCHK also looked at the $p_{\rm out}$ variable and plotted $\mean{p_{\rm out}}$ versus $x_E$ for triggers in the range $2.0\leq p_{T_t}\leq 3.2$ GeV/c. They found that the $\mean{p_{\rm out}}$  increased with increasing $x_E$ up to a maximum value of $\mean{p_{\rm out}}\sim 0.65$ GeV/c (Fig.~\ref{fig:CCHKxE}b). 
The original parton model did not assign transverse motion to quarks in the proton, only longitudinal $x$,  but it had been  proposed by Levin and Ryskin~\cite{LevinRyskin75} that quarks in the proton also carry   transverse momentum. This idea, coupled with the lack of $x_E$ scaling for $p_{T_t} < 3$ GeV/c, was taken by CCHK~\cite{CCHK} as evidence for the transverse momentum of quarks inside the proton. Calculations from their parton scattering model with $k_T=0$ and $\mean{k_T}=610$ MeV/c are shown in Fig.~\ref{fig:CCHKxE}b. This result led Feynman, Field and Fox (FFF)~\cite{FFF1} to formally introduce $\overrightarrow{k_T}$, the transverse momentum of a parton in a nucleon into their model of parton-parton scattering.  

\subsection{The second round of two-particle correlation measurements 1978-1979}
The two experiments on two particle correlations at the CERN-ISR discussed above both used the Split Field Magnet Facility (SFM)~\cite{SFM} which could measure the momenta of charged particles in nearly the full polar angular range but was optimized for measurements at forward and backward angles. One experiment~\cite{Darriulatpoutxe} moved their small E.M. calorimeter to mid-rapidity of the SFM  transverse to the beam axis while the CCHK measurement~\cite{CCHK} used only the SFM in a self-triggering mode at polar angles of $20^\circ$ and $45^\circ$. 

\begin{figure}[ht]
\begin{center}
\includegraphics[width=0.55\linewidth]{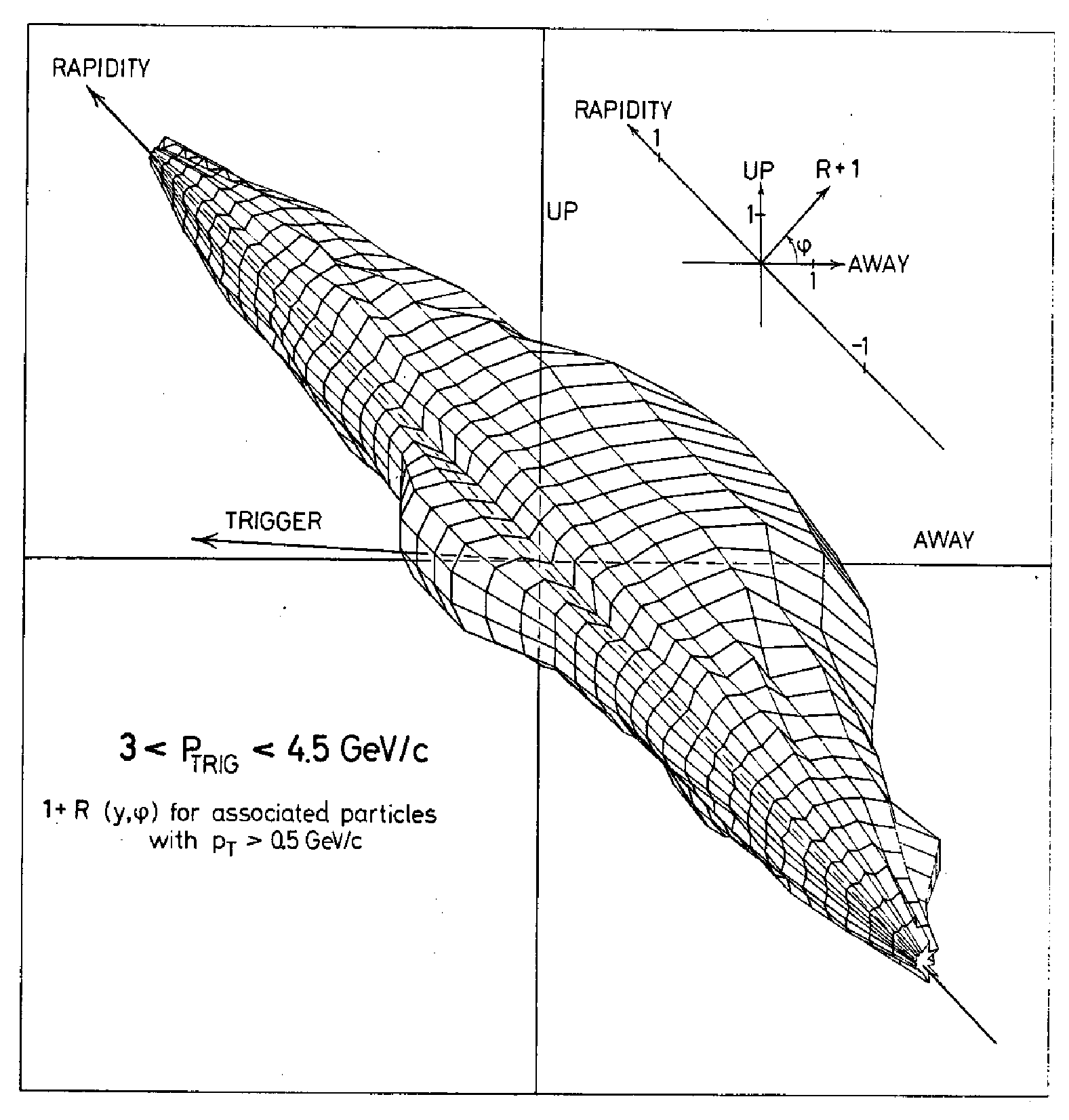}
\end{center}
\vspace*{-1pc}
\caption[]
{BFS~\cite{BFSNPB145} measurement of $R+1$, ratio of charged particle track distribution from a high $p_T$ trigger to that from minimum bias events\label{fig:BFS} }
\end{figure}

The second round of two particle correlations with improved detectors started to present data in 1978. The BFS collaboration~\cite{BFSNPB145} had moved the British Scandinavian wide angle spectrometer to the SFM at an angle of  90$^\circ$ at mid rapidity. They could trigger on identified charged particles with detection of the associated charged particles in the Split Field Magnet. The acceptance for associated charged particles, which was the same as CCHK, covered a rapidity range $-4 \leq y \leq +4$, with an azimuthal aperture of $\pm 40^\circ$ on the trigger side and $\pm 25^\circ$ on the away side. This experiment presented one of the most interesting results, a dramatic map of the two particle correlation function over nearly the full region in rapidity. Figure~\ref{fig:BFS} shows the ratio of the density of tracks with $p_T>0.5$ GeV/c, as a function of rapidity $y$ and azimuthal angle $\phi$, from a high $p_{T_t}$ trigger with $3< p_{T_t}<4$, GeV/c located at $y=0$ $\phi=180^\circ$, relative to minimum bias events ($p_{Tt}\geq 0.5$ GeV/c with elastic scattering events removed).  They called this ratio, which cancels the effects of acceptance variation, $R+1$. 
The main features of Fig.~\ref{fig:BFS} are an increase in the value of the correlation function in a small region in $y$ and $\phi$ near the trigger particle and a much larger increase on the away side, mainly within $|\phi|<45^\circ$ (the limit of their acceptance), but extending in rapidity to $|y|\simeq 3$. This illustrates the di-jet coplanar structure of high $p_{T_t}$ triggered events. One can also see a smaller same side correlation extending out to $|y|\simeq 2$.

\begin{figure}[ht]
\begin{center}
\includegraphics[width=0.50\linewidth]{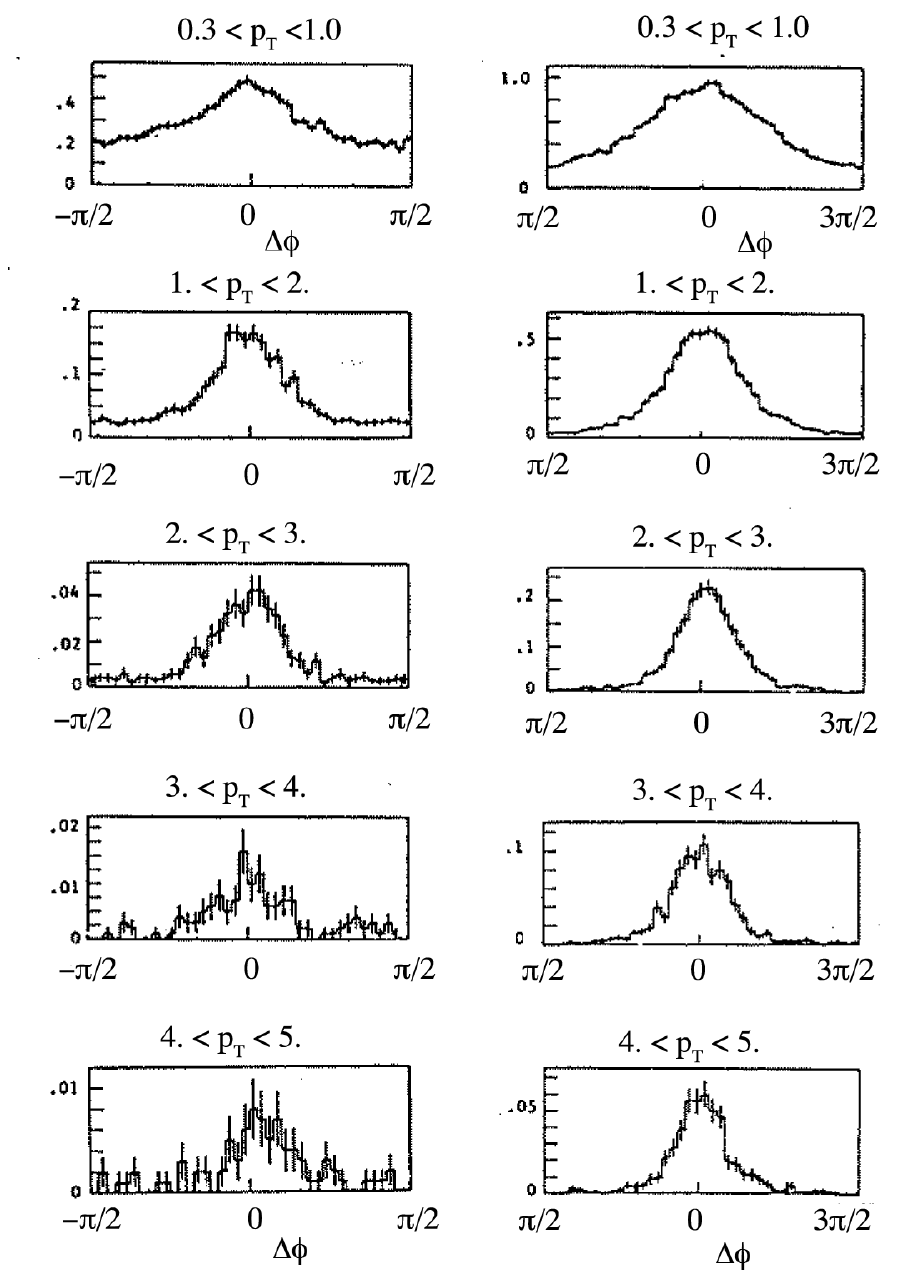} 
\includegraphics[width=0.48\linewidth]{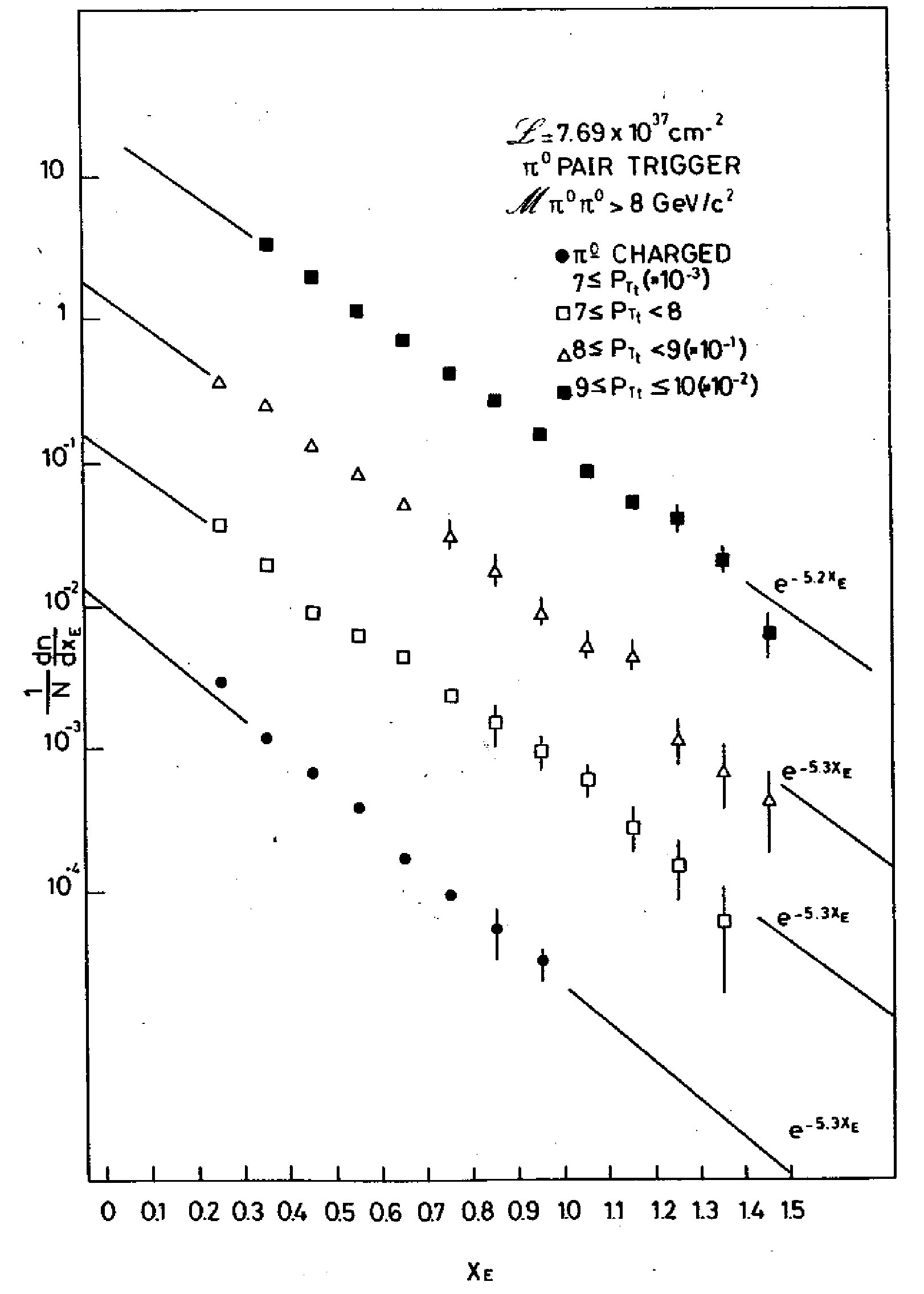}
\end{center}
\vspace*{-0.12in}
\caption[]
{a,b) Azimuthal distributions of charged particles of transverse momentum $p_{T}$, with respect to a trigger $\pi^0$ with $p_{Tt}\geq 7$ GeV/c, for 5 intervals of $p_{T}$~\cite{Angelis79}: a) (left-most panel) for $\Delta\phi=\pm \pi/2$ rad about the trigger particle, and b) (middle panel) for $\Delta\phi=\pm \pi/2$ about $\pi$ radians (i.e. directly opposite in azimuth) to the trigger. The trigger particle is restricted to $|\eta|<0.4$, while the associated charged particles are in the range $|\eta|\leq 0.7$. c) (right panel) $x_E$ distributions corresponding to the data of the center panel.   
\label{fig:mjt-ccorazi} }
\end{figure}

 The CCOR experiment, which was proposed in May 1973~\cite{CCOR73} and approved in March 1974, was installed in the ISR as R108 (8th experiment at Intersection 1) at the end of 1976 along with a low $\beta$ insertion for higher luminosity. It was debugged, tuned-up and started taking data in July 1977. This experiment~\cite{Angelis79} with its thin-coil superconducting solenoid and cylindrical drift chambers was the first at the CERN-ISR to provide charged particle measurement with full and uniform acceptance over the entire azimuth, with pseudorapidity coverage $-0.7\leq \eta\leq 0.7$, so that the jet structure of high $p_T$ scattering could be easily seen and measured. However, from August to October the liquid He refrigerator was broken, so the solenoid was turned off and only $\pi^0$ measurements in the EMcalorimeter could continue~\cite{CCORPLB79} (Section~\ref{sec:QCD1}). The measurements of charged-particles associated to a $\pi^0$ trigger came a few months later (Fig.~\ref{fig:mjt-ccorazi})~\cite{Angelis79}. 
 
In  Fig.~\ref{fig:mjt-ccorazi}a,b, the azimuthal distributions of associated charged particles 
relative to a $\pi^0$ trigger with transverse momentum $p_{Tt} > 7$ GeV/c are shown for five intervals of associated particle transverse momentum $p_T$. In all cases, strong correlation peaks on flat backgrounds are clearly visible, indicating the di-jet structure which is contained in an interval $\Delta\phi=\pm 60^\circ$ about a direction towards and opposite the to trigger for all values of associated $p_T\, (>0.3$ GeV/c) shown. The width of the peaks about the trigger direction (Fig.~\ref{fig:mjt-ccorazi}a), or opposite to the trigger (Fig.~\ref{fig:mjt-ccorazi}b) indicates out-of-plane activity from the individual fragments of jets. If the width of the away distributions (Fig.~\ref{fig:mjt-ccorazi}b) corresponding to the out of plane activity were due entirely to jet fragmentation, then  
$\langle |\sin(\Delta\phi)|\rangle=\langle |j_{T_{\phi}}|/p_T \rangle$ would decrease in direct proportion to $1/p_T$, since $j_{T_{\phi}}$, the component of  the jet fragmentation transverse momentum, $\overrightarrow{j_T}$, in the azimuthal plane, should be independent of $p_T$. These data were also further analyzed to measure $\overrightarrow{j_T}$ as well as $\overrightarrow{k_T}$, the transverse momentum of a parton in a nucleon, as originally shown by the CCHK collaboration~\cite{CCHK}, and elaborated by Feynman, Field and Fox (FFF)~\cite{FFF1}.
 
The $x_E$ distributions~\cite{Angelis79}, \cite{JacobEPS79} from the data of Fig.~\ref{fig:mjt-ccorazi}b are shown in Fig.~\ref{fig:mjt-ccorazi}c.  The $x_E$ scaling is evident for all values of $p_{T_t}$, with the expected fragmentation behavior, $e^{-6z}\sim e^{-6 x_E \langle z_{\rm trig}\rangle}$. This figure also showed that there were no di-jets, each of a single particle, as claimed by another ISR experiment of that period~\cite{WillisPLB86}, and by Jacob and Landshoff~\cite{JacobNPB113}, since there is no peak at \mbox{$x_E=1$}. There is a small anecdote concerning this measurement and Maurice Jacob's talk at the EPS1979 HEP Conference~\cite{JacobEPS79}. Maurice was originally only going to show a plot from the ABCS experiment~\cite{WillisPLB86} which appeared to show ``a systematic wiggle departure from an exponential...The wiggle could bear witness to a specific process where the two jets would each only consist of one high $p_T$ $\pi^0$'' as he had predicted~\cite{JacobNPB113}. I insisted that Maurice also show Fig.~\ref{fig:mjt-ccorazi}c, a plot that I had made with three higher $p_T$ points than the CCOR plot in Ref.~\cite{Angelis79}, ``which challenged the one high $p_T$ $\pi^0$ only jet idea'' after I nearly broke the telephone in the ISR counting room when I heard that Maurice was only going to show the plot with the ``wiggle''~\cite{WillisPLB86}. 

Even though this period ended with the strong belief~\cite{Darriulat1980} that jet fragmentation functions from $\nu+$p and $e^+ e^-$ reactions are the same as p$+$p $x_E$ distributions, with the same dependence of the exponential slope $b$ on $p_{T_t}$ or $\sqrt{s}/2$ for $e^+ e^-$ (Fig.~\ref{fig:xEfrag1}b), it turned out that nature had the last laugh. This belief was one of the very few results from the ``high $p_T$ discovery period'' that did not stand the test of time. It was discovered at RHIC~\cite{PXppg029PRD74}, a quarter of a century later, that the shape of the $x_E$ distribution triggered by the fragment of a jet, such as a $\pi^0$ had nothing to do with fragmentation functions but instead measured the ratio of the away jet to the trigger jet ($\hat{x}_h\equiv \hat{p}_{Ta}/\hat{p}_{Tt}$) and depended only on the power $n$ of the invariant single particle high $p_T$ cross section:
    \begin{equation}
\left.{d{\cal{P}} \over dx_E}\right|_{p_{T_t}}  = {N\,(n-1)}{1\over\hat{x}_h} {1\over {(1+ {x_E \over{\hat{x}_h}})^{n}}} \,  
\qquad .  
\label{eq:condxe2N}
\end{equation}\vspace*{-2.0pc}   
\subsection{Measurement of the jet fragmentation transverse momentum, $\vec{j_T}$, and $\vec{k_T}$, the transverse momentum of a parton in a nucleon} 
Following the idea of Levin and Ryskin~\cite{LevinRyskin75} and CCHK~\cite{CCHK}, Feynman, Field and Fox~\cite{FFF1} established the formalism for $\overrightarrow{k_T}$, the transverse momentum of a parton in a nucleon. 
\begin{figure}[!h]
\centering \vspace*{-1.0pc}
\includegraphics[width=0.7\linewidth]{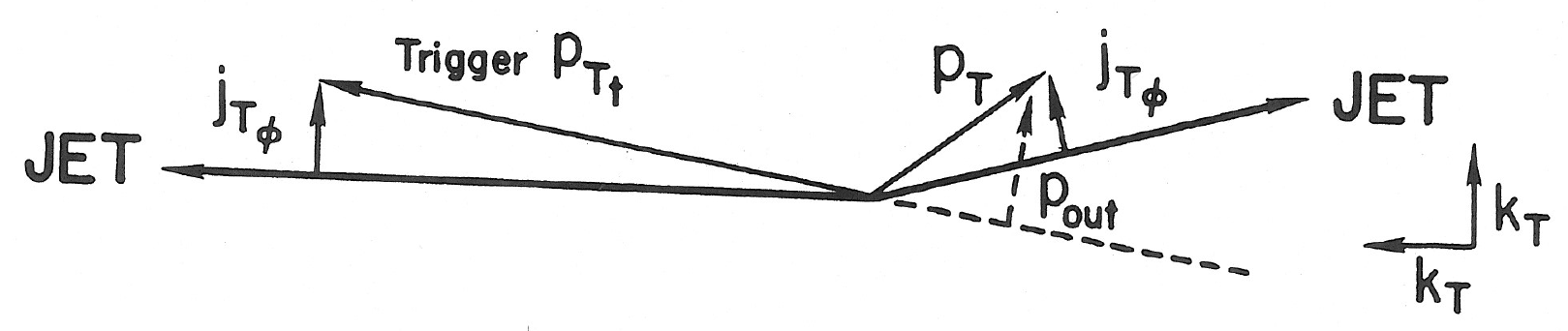}
\caption[]{Azimuthal projection of di-jet with trigger particle $p_{Tt}$ and associated away-side particle $p_{Ta}$, and the azimuthal components $j_{T\phi}$ of the fragmentation transverse momentum. The initial state $\overrightarrow{k_T}$ of a parton in each nucleon is  shown schematically: one vertical which gives an azimuthal decorrelation of the jets and one horizontal which changes the transverse momentum of the trigger jet.} 
\vspace*{-0.5pc}
\label{fig:kTdwg}
\end{figure}
In this formulation (Fig.~\ref{fig:kTdwg}), the net transverse momentum of an outgoing parton pair, where the two $\overrightarrow{k_T}$ add randomly, is $\sqrt{2} k_T$, which is composed of two orthogonal components, $\sqrt{2} k_{T_{\phi}}=k_T$, out of the scattering plane, which makes the jets acoplanar, i.e. not back-to-back in azimuth, and $\sqrt{2} k_{T_x}=k_T$, along the axis of the trigger jet, which makes the jets unequal in energy. 

	FFF~\cite{FFF1} gave the approximate formula to derive $k_T$ from the measurement of $p_{\rm out}$ as a function of $x_E$:
\begin{equation}
\langle |p_{\rm out}|\rangle^2=x_E^2 [2\langle |k_{T_{\phi}}|\rangle^2 +  \langle |j_{T_{\phi}}|\rangle^2 ] + \langle |j_{T_{\phi}}|\rangle^2 \qquad .
\label{eq:mjt-FFFpoutkT}
\end{equation}
This formula assumed that $\mean{z_{\rm trig}}=1$ and that the jet energies are equal.~\vspace*{-1.0pc}
\begin{figure}[!hb]
\begin{minipage}[b]{0.50\linewidth}
\includegraphics[width=\linewidth]{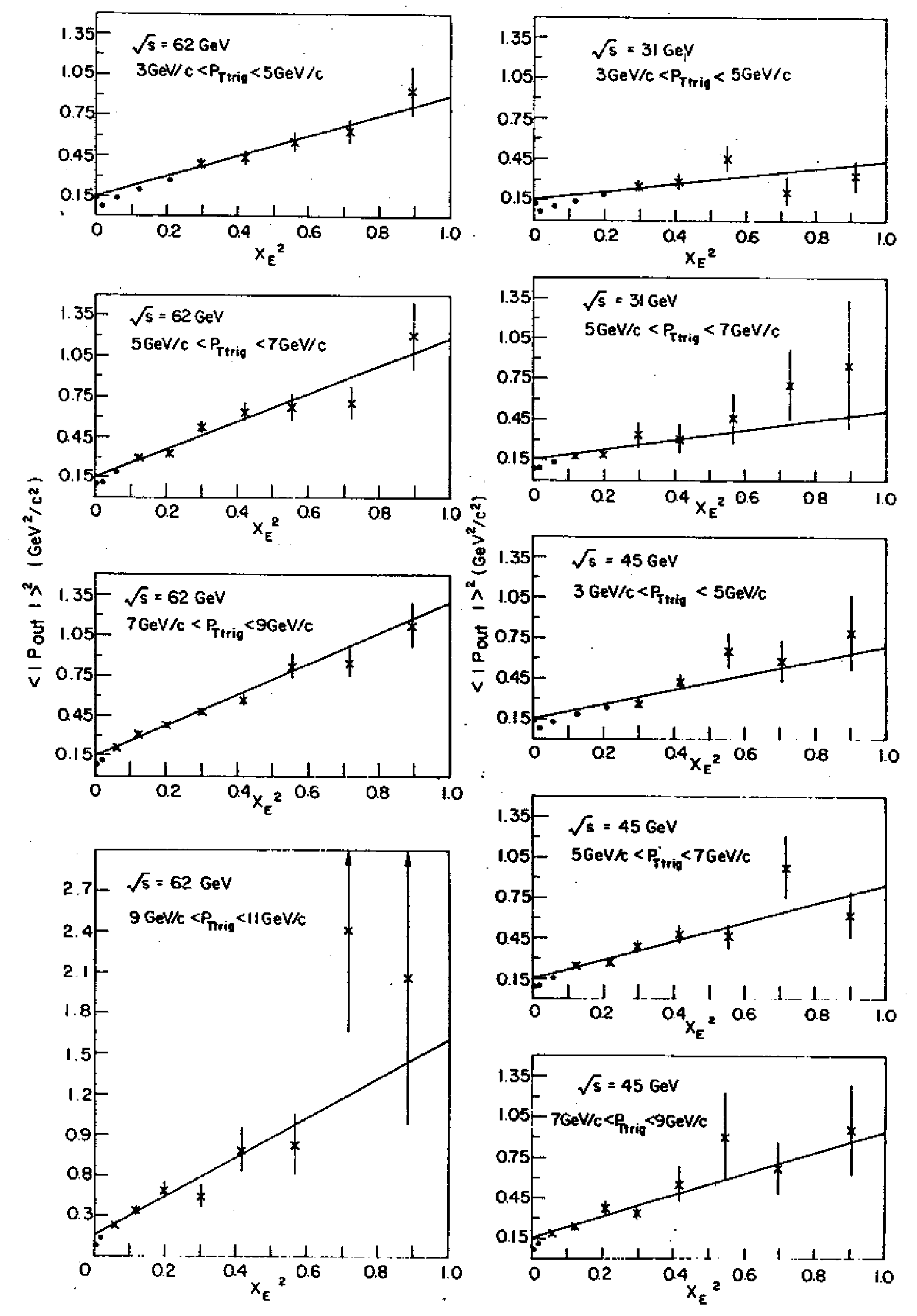} 
\end{minipage}\hspace*{0.2pc}
\begin{minipage}[b]{0.49\linewidth}
\includegraphics[width=\linewidth]{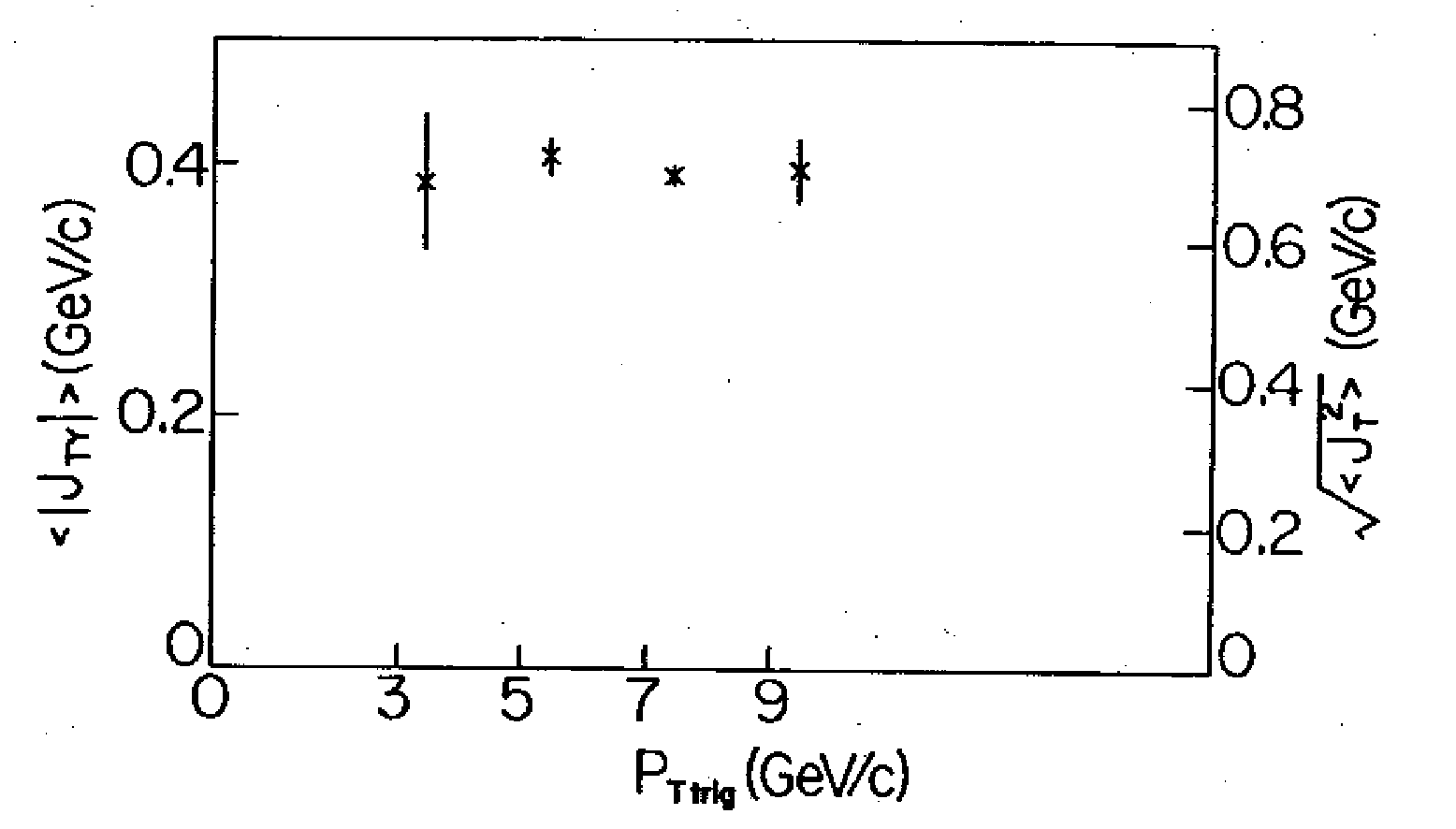}
\includegraphics[width=0.96\linewidth,angle=-0.8]{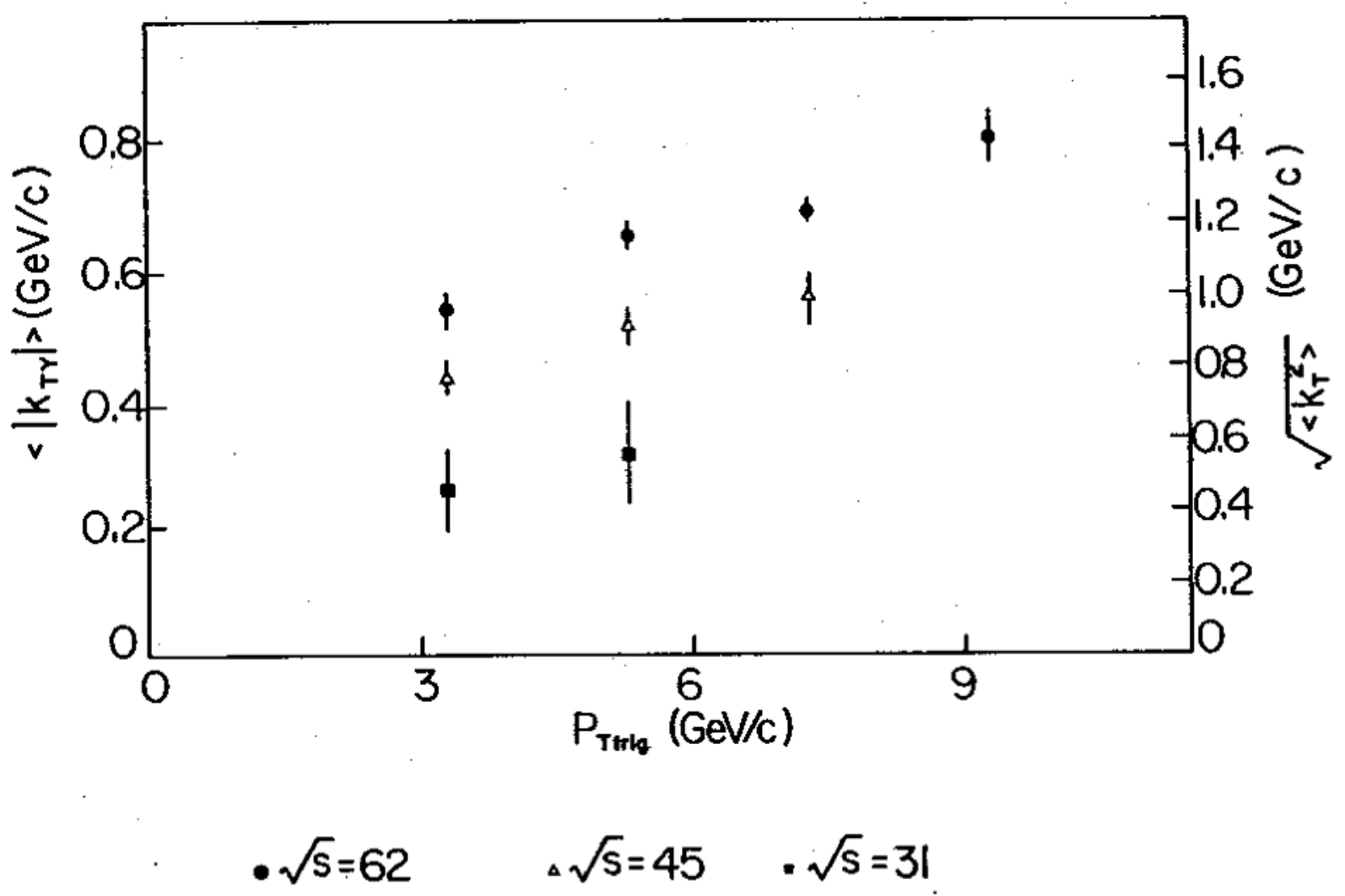}
\end{minipage}
\caption[]{~\cite{CCORjTkT} a)(left) CCOR $\mean{|p_{\rm out}|}^2$ versus $x_E^2$ for nine different data samples. The crosses are those points used in the fit to Eq.~\ref{eq:mjt-FFFpoutkT}. The straight lines shown are obtained keeping the intercept the same for all nine data samples. b)(right): (top) Fitted values of $\langle |j_{T_{\phi}}|\rangle$ and $\sqrt{\mean{j_T^2}}$ as a function of $p_{T_t}$. These values were all equal as a function of $\sqrt{s}$ and constrained to be equal in the fit shown; (bottom) $\langle |k_{T_{\phi}}|\rangle$ and $\sqrt{\mean{k_{T}^2 }}$ as a function of $p_{T_t}$ for the 3 values of $\sqrt{s}$ indicated.}
\label{fig:CCORjTkT}
\end{figure}\vspace*{-0.5pc}

CCOR~\cite{CCORjTkT} used a fit to this formula (Fig.~\ref{fig:CCORjTkT}a) to derive $\langle |k_{T_{\phi}}|\rangle$ and $\langle |j_{T_{\phi}}|\rangle$ as a function of $p_{Tt}$ and $\sqrt{s}$ (Fig.~\ref{fig:CCORjTkT}b) from the data of Fig.~\ref{fig:mjt-ccorazi}b.  This important result  showed that $\langle |j_{T_{\phi}}|\rangle$ is constant, independent of $p_{T_t}$ and $\sqrt{s}$, as expected for fragmentation, but that $\langle |k_{T_{\phi}}|\rangle$ varies with both $p_{T_t}$ and $\sqrt{s}$, suggestive of a radiative, rather than an intrinsic origin for $k_T$. Large values of $\sqrt{\mean{k_{T}^2 }}\approx 1$ GeV/c were also reported in other CERN-ISR measurements~\cite{ClarkNPB160},\cite{WillisNPB158}  but it took the $e^+ e^-$ people several more years to get $j_T$ correct~\cite{TASSOZPC22} because they hadn't understood the ``seagull effect''~\cite{seagull}: with increased momentum of fragments in a jet, their momentum transverse to the jet axis also increases until it reaches its true $j_T$. 
\section{Multiparticle correlations--the search for jets 1977--1980}
The jet searches like many other new measurements, but perhaps the worst in this regard, started off with a major incorrect claim of the observation of the jets of hard scattering at Fermilab. Several fixed target experiments searched for jets using the energy detected in calorimeters with limited aperture ($\Delta\phi=\pm45^\circ$, $\Delta y\approx \pm 0.35$ to 0.55) at mid-rapidity in the p$+$p c.m. system. The first claim for discovering the jets from the ``quark-quark scattering model'' was by Fermilab E260~\cite{FoxNPB134}  who found that for $\sqrt{s}=19.4$ GeV the cross section for the jet, which was defined by summing all the energy in their calorimeter, was ``similar in shape to the single particle cross section but two orders of magnitude larger.'"

As calorimeters in experiments got larger, the ``jet'' to single particle ratio kept getting larger until the jet fiasco ended dramatically when the NA5 fixed target experiment at CERN, with a hadron calorimeter which covered the full azimuth with a c.m. polar angular interval $54^\circ<\theta^*<135^\circ$ ($\Delta y\approx \pm 0.9$) in the p$+$p c.m. system at $\sqrt{s}=23.8$ GeV, presented their results~\cite{PretzlICHEP80}. The coup de grace was their conclusion:``The events selected by the full calorimeter trigger show no dominant jet structure. They appear to originate from processes other than two constituent scattering.'' The actual publication of the NA5 result~\cite{DeMarzoPLB112} was perhaps even clearer:``The large transverse energy observed in the calorimeter is the result of a large number of particles with a rather small transverse momentum''. 

The problem with the mistaken claim of `jets' was that the difference between single particle and multiparticle measurements was not understood. The principal multiparticle variables are the charge multiplicity distribution $dN_{ch}/d\eta$ and the transverse energy ($E_T$) distribution~\cite{LandshoffPRD18} :
\begin{equation}
     E_T\equiv\sum_i E_i \sin\theta_i  \quad,  \label{eq:ETdef}
\end{equation}
where the sum is over all particles emitted on an event into a fixed but large solid angle. The NA5 result~\cite{DeMarzoPLB112} was the first $E_T$ measurement in the form that is still in use at present~\cite{MJTIJMPA4}\cite{AdlerPRC89}.

\begin{figure}[!b]
\centering
\raisebox{-0.0pc}{\includegraphics[width=0.49\textwidth]{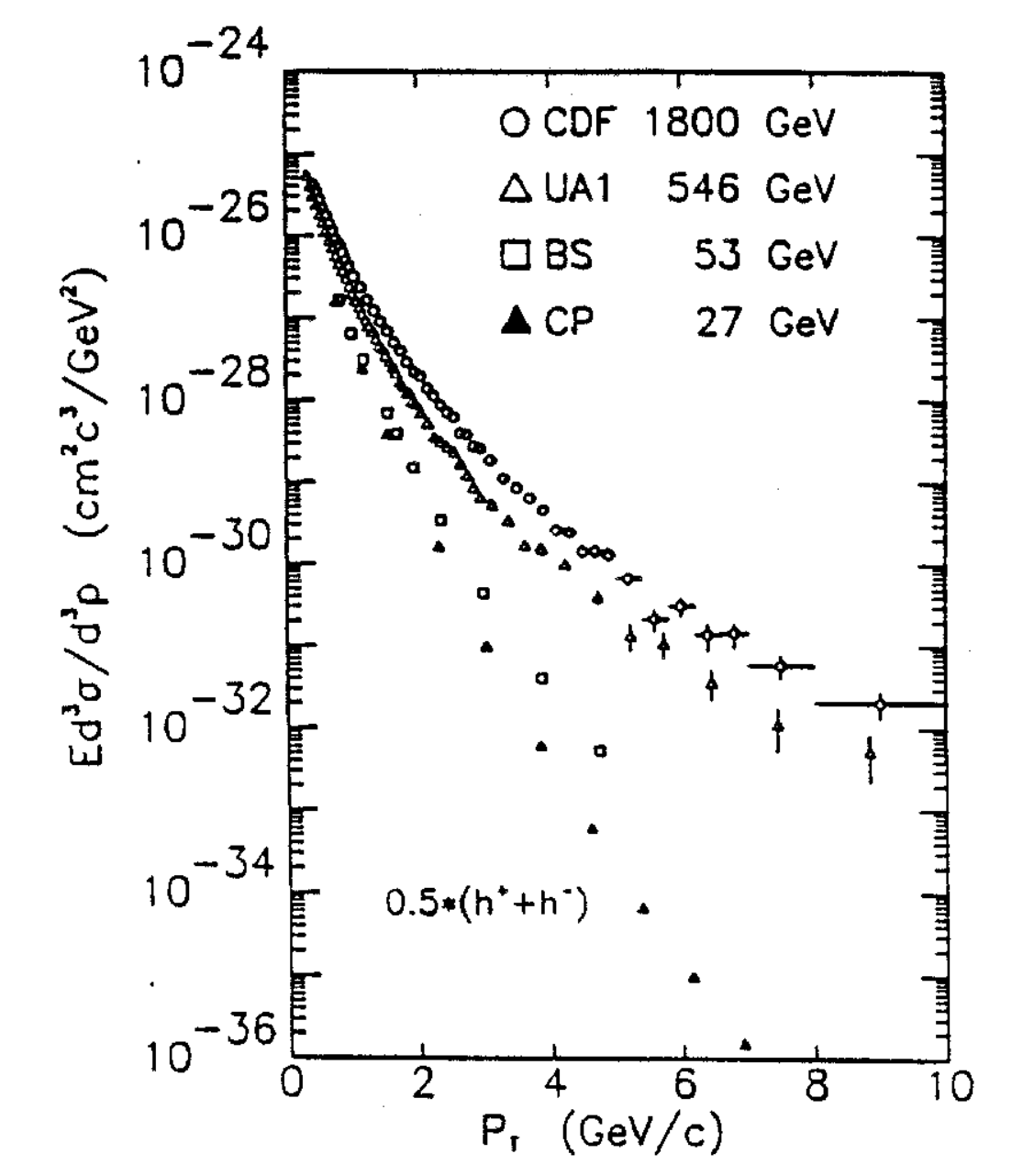}}
\raisebox{-0.0pc}{\includegraphics[angle=+1.0,width=0.45\textwidth]{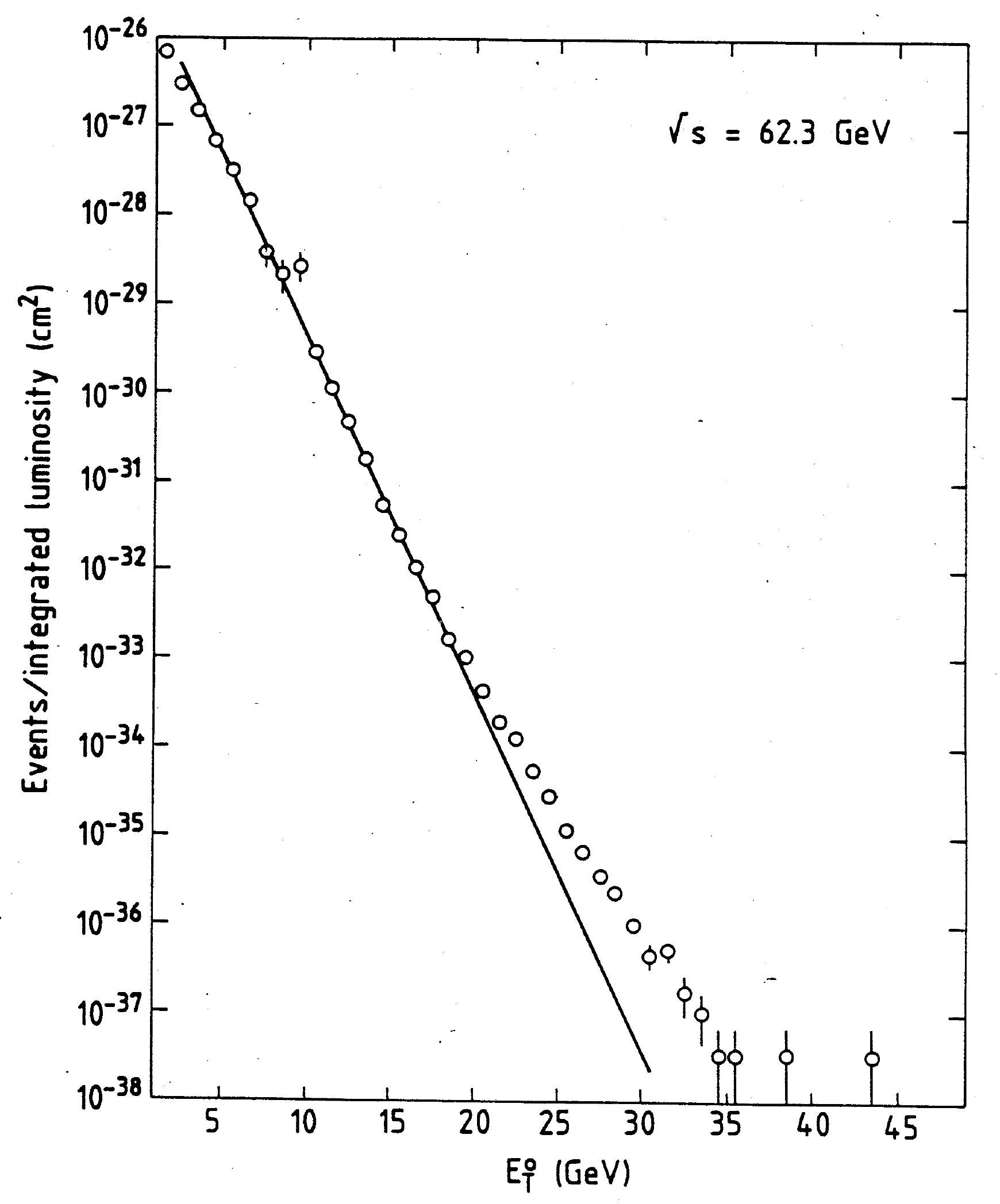}}
\caption[]{a) (left) $E {d^3\sigma}(p_T)/{d^3p}$ for single hadrons at mid-rapidity as a function of $\sqrt{s}$ in p$+$p and $\bar{p}+p$ collisions ~\cite{CDF88}. b) (right) $E_T^0$ spectrum at $\sqrt{s}=62$ GeV~\cite{ET83}. }
\label{fig:singlemultiple}\vspace*{-1.0pc}
\end{figure}

The difference between single and multi-particle distributions is shown in Fig.~\ref{fig:singlemultiple}. Single particle $p_T$ distributions (left) follow the   
$e^{-6p_T}$~\cite{Cocconi} soft physics particle spectrum until $p_T\approx 1-2$ GeV/c where the hard-scattering power law begins to dominate, roughly 2--3 orders of magnitude down in cross section.  On the other hand, the spectrum of the neutral transverse energy $E_T^0$ at 
$\sqrt{s}=62.4$ GeV (right) falls exponentially for $\geq 6$ orders of magnitude until the hard scattering component of the $E_T^0$ distribution becomes evident by the break from the exponential spectrum. In the exponential region there was a uniform azimuthal disribution of  $E_T^0$, with a dominant 2 jet structure for $E_T^0>24$ GeV~\cite{ET83}.

\begin{figure}[!h]
\centering
\includegraphics[width=0.85\linewidth]{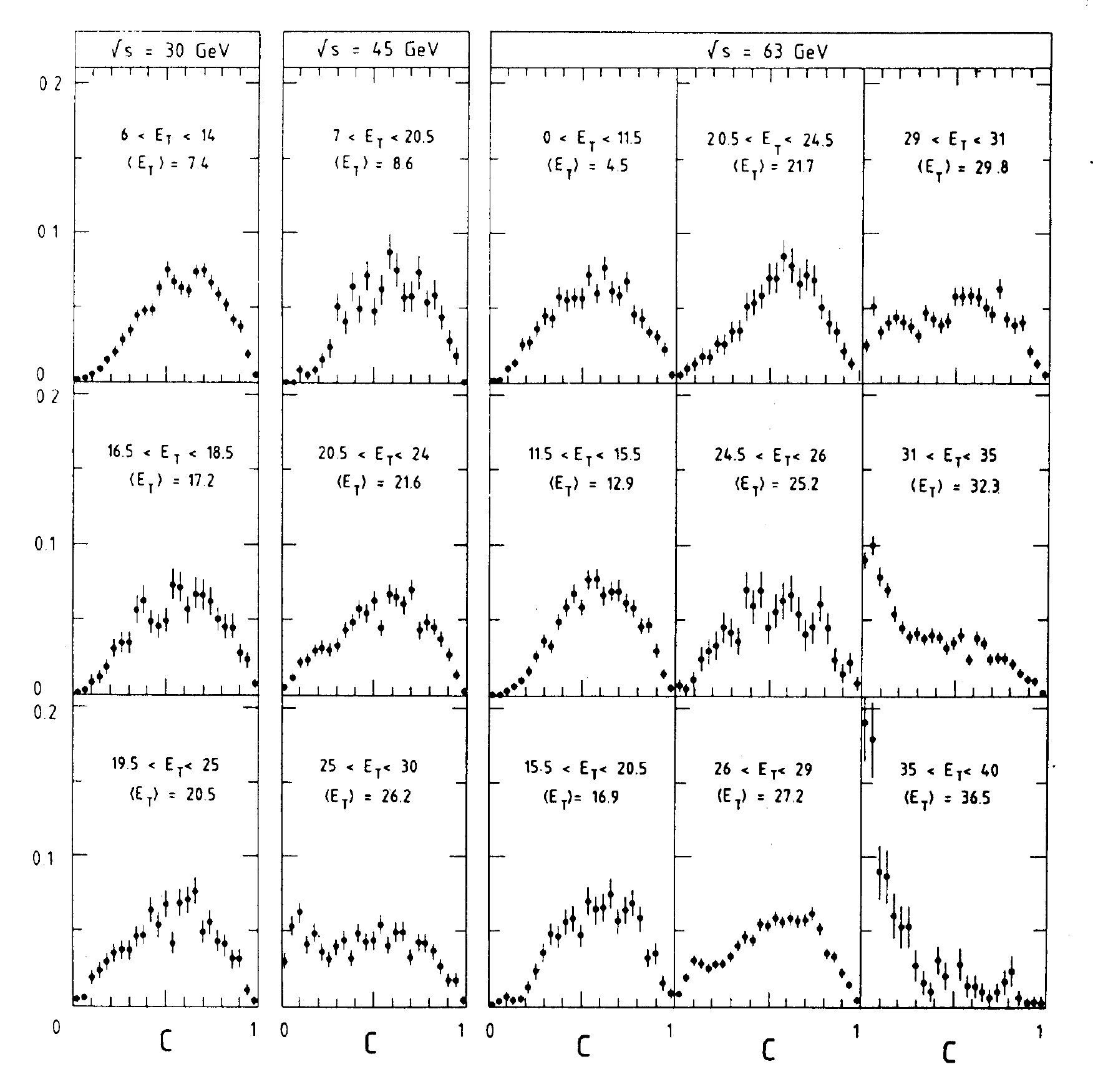}\vspace*{-1.0pc}
\caption[] {\label{fig:ET29}  Circularity distributions for bins in $E_T$ and $\sqrt{s}$. 
The error bars are statistical only \cite{WillisPLB126}.}
\end{figure}

A much clearer separation between ``soft" and ``hard" physics in such  
spectra was determined at the ISR by the AFS collaboration~\cite{WillisPLB126} with a full azimuth EM and hadron calorimeter covering the rapidity range $|y|<0.9$. A study of the 
event shape as a function of $E_T$, was performed using a principal axis 
analysis. A quantity, ``circularity", was defined which 
would be 1 for a totally uniform azimuthal distribution of the components of 
$E_T$ and zero if all the $E_T$ were in two narrow jets back to back in 
azimuth. The distributions (Fig.\ref{fig:ET29}) show no evidence of jets for $E_T< 25$ GeV at 
any c.m. energy. However for $E_T>25$ GeV, there is an increase in 
low circularity events, leading up to a dominance of low-circularity events from a two-jet structure for 
$E_T\geq 31$ GeV at $\sqrt{s}=63$ GeV. \vspace*{-1.0pc} 
\vspace*{-0.5pc} 
 
     These beautiful measurements from the year 1983 made it clear that the jets of hard 
scattering could indeed be observed using $E_T$ distributions; but that hard 
scattering effects have negligible influence on the shape of $E_T$ spectra in 
proton-(anti)proton collisions for the first 4, or even 6, orders of magnitude 
of cross section, depending on the c.m. energy. However the first convincing evidence of jets and the validity of \QCD\ as the mechanism for hard-scattering came at the ICHEP1982 conference in Paris.
\section{The final proof of jets and \QCD\ by measurements 1980--1982}
The rejection of the Fermilab E260~\cite{FoxNPB134} jet claim by the observation of no jets in a better measurement at CERN~\cite{PretzlICHEP80}, presented at the ICHEP1980, led to confusion in the High Energy Physics community in the U.~S. during the period from the ICHEP1980 to the ICHEP1982. There was no clear understanding of why the jets of \QCD\  were not observed in the full azimuth calorimeter which led to doubts by many of the utility and validity of \QCD\  tests in hadron-hadron collisions. To get an idea of the thinking during this period, I had to give a seminar at my laboratory, BNL, in March 1982 with the title:``Why I believe in jets in spite of calorimeter experiments. For \QCD\ tests try to minimize the effect of jets.''

In March 1980, I had moved from the Rockefeller University to BNL to help save ISABELLE, the $\sqrt{s}=800$ GeV superconducting p$+$p collider under construction. My assignment was to help sort out the problems they were having with the superconducting magnets that quenched at a much lower magnetic field than expected. I described my work in this period in a previous EPJH article~\cite{MJTW}, so I'll skip to March 1982 when I transferred to the BNL Physics Department and resumed work on the CCOR experiment. One of the problems with the CCOR full azimuth superconducting solenoid detector was that it did not have any specific electron identification device like the Cherenkov counters in CCRS, so that in trying to measure $e^+ e^-$ pairs we had to trigger on clusters with energy $> 2.5$ GeV in each arm of our EMcalorimeters (as described in Section~\ref{sec:ISRproposals}). This left us with $\approx 750,000$ $\pi^0$ pairs which 
we decided to analyze in the same way we would analyze $e^+ e^-$ pairs in the detector, namely the invariant mass $M_{\pi\pi}$ of the pair, its net $P_t$ and rapidity $Y$ and $\cos\theta^*$ in the c.m. system of the pair.  

\begin{figure}[!ht]
\includegraphics[width=0.74\linewidth]{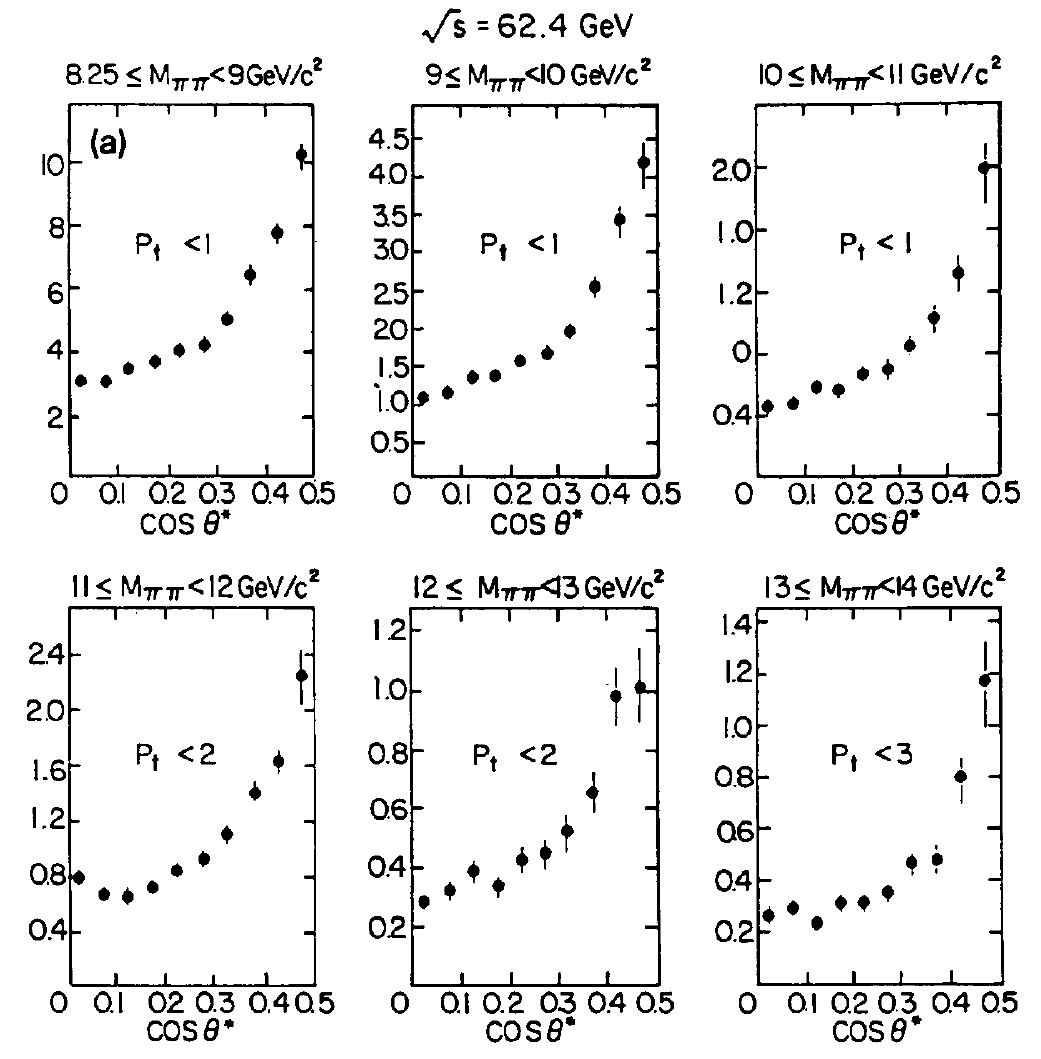} 
\begin{minipage}[b]{0.25\linewidth}
\raisebox{+0.2pc}{\includegraphics[width=\linewidth]{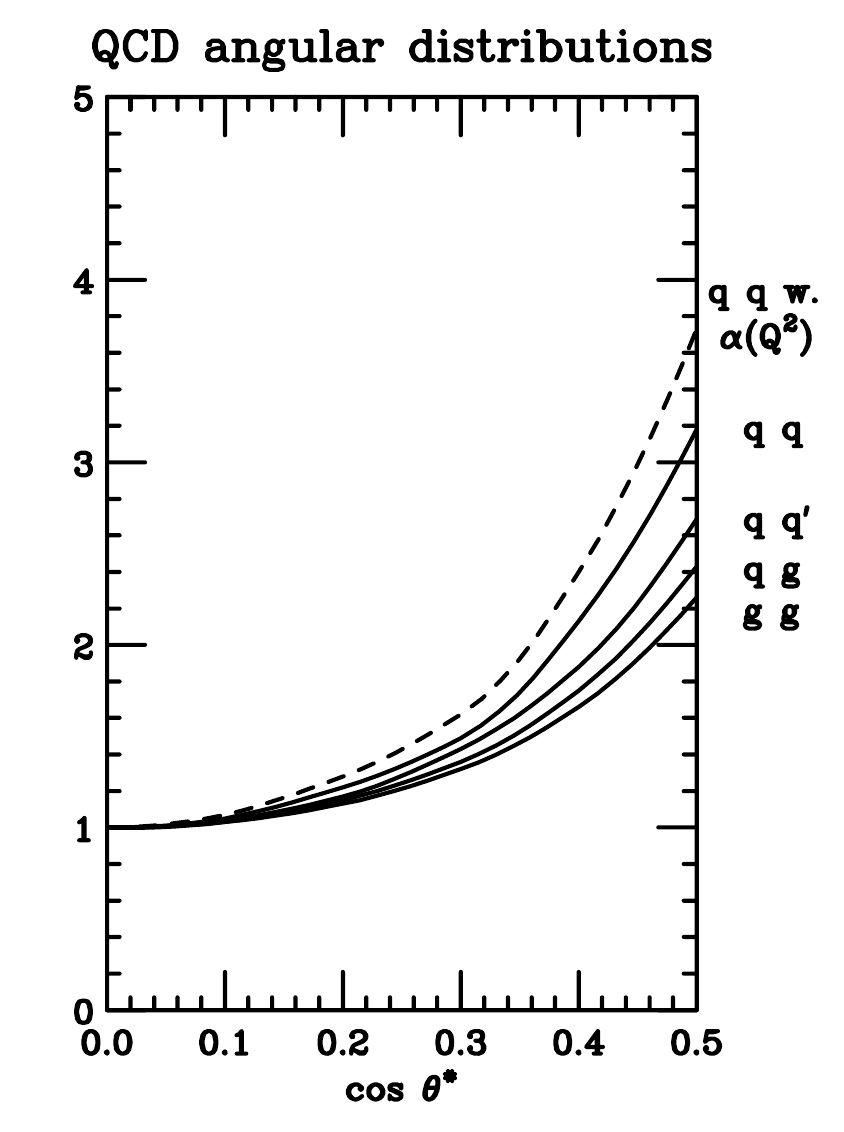}}
\includegraphics[width=\linewidth]{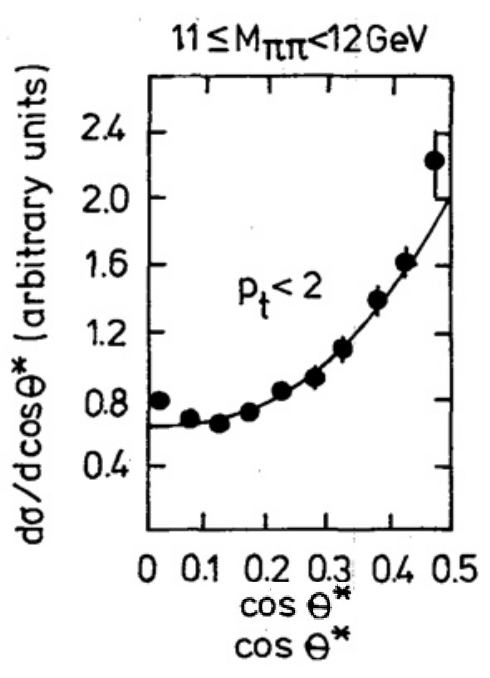}
\end{minipage}
\caption[]{a) (left 6 panels) CCOR measurement~\cite{CCORNPB209} of polar angular distributions of $\pi^0$ pairs, with $M_{\pi\pi}$ and net $P_t$ indicated, in p$+$p collisions at $\sqrt{s}=62.4$ GeV. b) (right 2 panels) (top) \QCD\ elementary subprocess angular distributions at fixed $\hat{s}$, normalized at 90$^\circ$, $\cos\theta^*=0$, which I showed as an overlay on the $9\leq M_{\pi\pi}\leq10$ Gev/c$^2$ plot~\cite{TannenbaumParis}. (bottom) G\"unter Wolf's calculation of $\Sigma^{ab}(\cos\theta^*)$ for $qq$ overlaid on the CCOR $11\leq M_{\pi\pi}\leq10$Gev/c$^2$ measurement~\cite{WolfParis}. }
\label{fig:CCORcostheta}
\end{figure}

There were enough events so that we could select $\pi^0$ pairs with net $P_t<1$ GeV/c (or $P_t<2$ GeV/c for $M_{\pi\pi}>11$ GeV/c$^2$) and $|Y|<0.35$. Then we moved to the c.m. system of the pair, in which the two $\pi^0$ are back to back at an angle $\cos\theta^*$, by the simple rapidity shift $y*=y-Y$ (Sec.~\ref{sec:rapidity}) for each $\pi^0$. We knew from FFF~\cite{FFF1} that the trigger $\pi^0$ had almost the same $p_{Tt}$ as the parton $\hat{p}_{Tt}$ $\mean{z_{\rm trig}}\approx 1.0$. Thanks to the full azimuth charged particle tracking we were able to measure $\mean{z_{\rm trig}}$ from our own data by summing over all charged particles within an azimuthal angle $\pm 60^\circ$ of the trigger in the rapidity range $|y|<0.7$, multiplying by 1.5 to account for the missing neutrals~\cite{CCORNPB209}.
We noted that the large measured values of $\mean{z_{\rm trig}}>0.9$ for $M_{\pi\pi}>8$ GeV/c$^2$ and measured value of $\mean{|j_{Ty}|}=0.44$ GeV/c justify the assumption that the axis of the di-pion system in the pair c.m. system follows closely that of the original parton-parton scattering with $\sqrt{\hat{s}}\approx M_{\pi\pi}$. These measured $\cos\theta^*$ distributions are shown in Fig.~\ref{fig:CCORcostheta} and should be able to be compared directly to the $\cos\theta^*$ distributions of the elementary \QCD\ subprocesses in terms of $d\sigma/d\cos\theta^*$ at fixed $\hat{s}$, Eq.~\ref{eq:dsigdcosth}, easily derived from Eq.~\ref{eq:dsigdthat}:
\begin{equation}
\left.\frac{d\sigma}{d\cos\theta^*}\right|_{\hat{s}}=\frac{\pi \alpha_s^2(Q^2)} {2\hat{s}} \Sigma^{ab}(\cos\theta^*)  \qquad .
\label{eq:dsigdcosth}
\end{equation} 
The effect of $\mean{|j_{Ty}|}$ is negligible, well within the bins in $\cos\theta^*$.  
\begin{figure}[!h]
\centering
\raisebox{+0.0pc}{\includegraphics[width=0.81\linewidth]{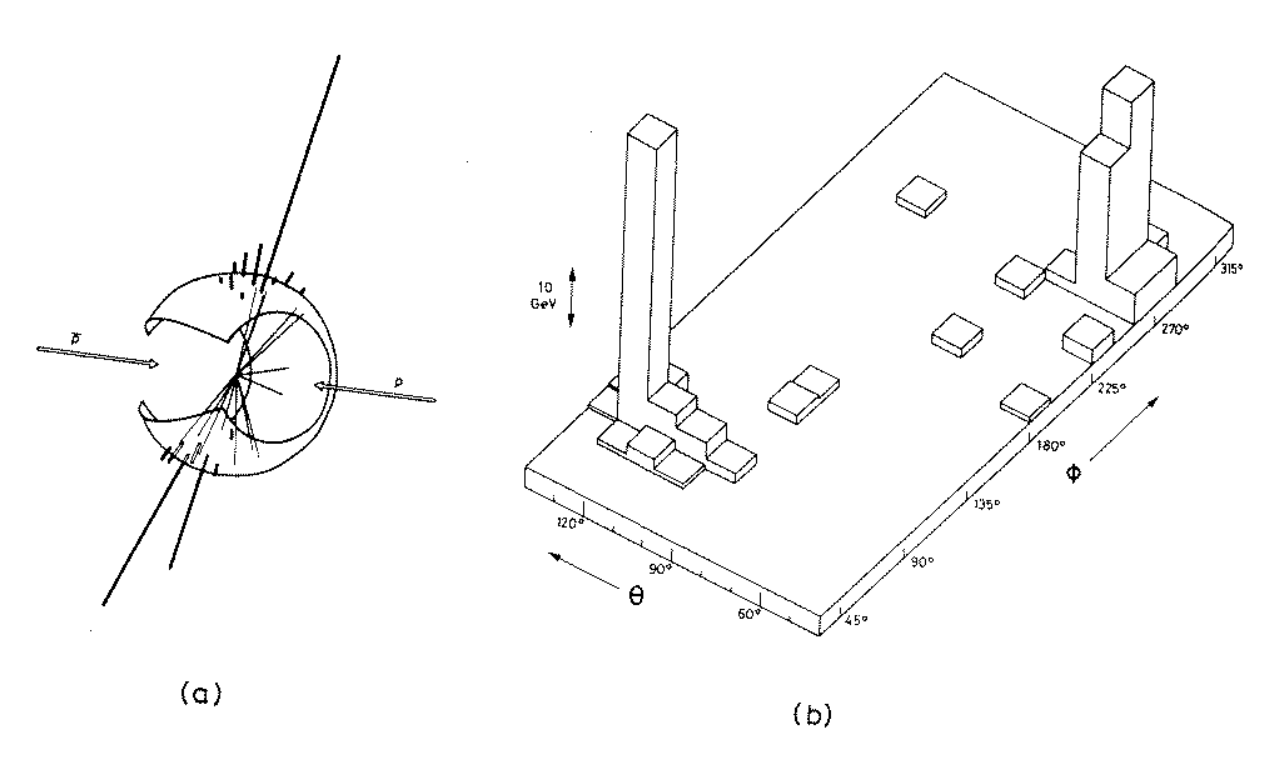}}\vspace*{-1.0pc}
\caption[]{Configuration of the UA2 event~\cite{RepellinParis} with the largest value of $\Sigma E_T$, 127 GeV (M = 140 GeV): (a) charged tracks pointing to the inner face of the central calorimeter are shown together with cell energies (indicated by heavy lines with lengths proportional to cell energies). (b) the cell energy distribution as a function of polar angle $\theta$ and azimuth $\phi$.   \label{fig:UA2jet}}
\end{figure}

In fact the measured distributions for the various $M_{\pi\pi}\approx\hat{s}$ all have the same shape which we compared directly to the \QCD\ angular distributions from Eq.~\ref{eq:dsigdcosth} for the subprocesses, $gg$, $qg$, $qq'$ and $qq$ elastic scattering~\cite{TannenbaumParis} also shown in Fig.~\ref{fig:CCORcostheta}. The measured distributions are steeper than all the \QCD\ constituent scattering subprocess distributions with constant $\alpha{_s}(Q^2)$. However once the increase of the \QCD\ coupling constant $\alpha{_s}(Q^2)$, with decreasing $Q^2=-\hat{t}$ as $\cos\theta^*$ increases, is taken into account as shown as the dashed curve for $qq$ with $\alpha{_s}(Q^2)$, the measurement and \QCD\  predictions agree very well. We had this data for two years until Dave Levinthal and Steve Pordes realized that we had to include the \QCD\ $\alpha{_s}(Q^2)$ evolution.

I was lucky enough to present this result~\cite{TannenbaumParis} at the ICHEP82 meeting in Paris after the session where UA2~\cite{RepellinParis} revealed their di-jet LEGO plot measured in a highly segmented total absorption calorimeter covering the azimuthal range $\Delta\phi=300^\circ$ with $|y|<1.0$. This now famous plot (Fig.~\ref{fig:UA2jet}) clearly  indicated di-jets in collisions at $\sqrt{s}=540$ GeV, one of the first results at the CERN $S\bar{p}pS$ $\bar{p}$+$p$ collider, which immediately convinced everybody that jets existed.

These two measurements and G\"unter Wolf's outstanding rapporteur talk, in which he also verified the $\cos\theta^*$ calculation with his own overlay plot (lowest right plot in Fig.~\ref{fig:CCORcostheta}), convinced all observers, and as the news spread, everybody else, that \QCD\ and jets were the cornerstone  of the Standard Model. Quotes from G\"unter's proceedings~\cite{WolfParis} are worth repeating:``Amongst the most exciting results are the direct measurement of the parton-parton scattering angular distribution and the observation of very energetic jets at the SPS collider." ``\QCD\ provides a consistent description for the underlying constituent scattering processes." Since that time \QCD\ and jets have become the standard tools of high energy particle physics. \vspace*{-0.0pc}
\subsection{A few more recent collider results on \QCD\ and jets}\vspace*{-1.0pc}
\begin{figure}[!h]
\centering
\raisebox{-0.0pc}{\includegraphics[width=0.51\linewidth]{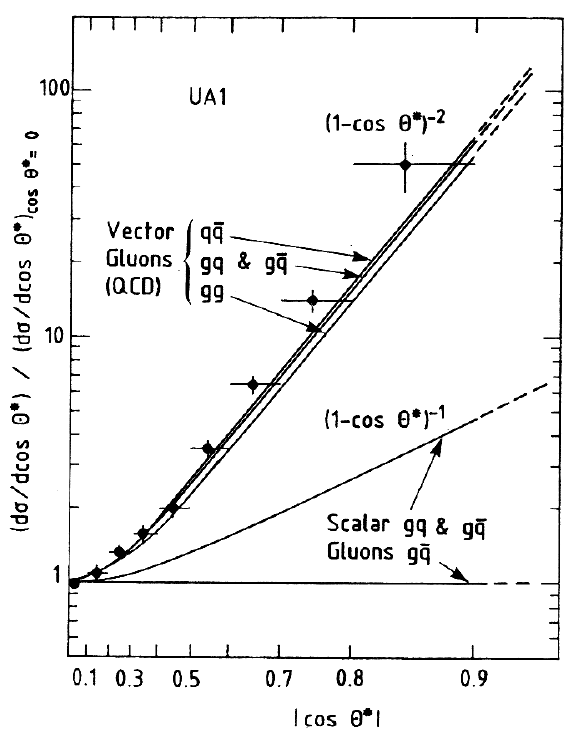}}\hspace{1.0pc}
\raisebox{+1.2pc}{\includegraphics[width=0.44\linewidth]{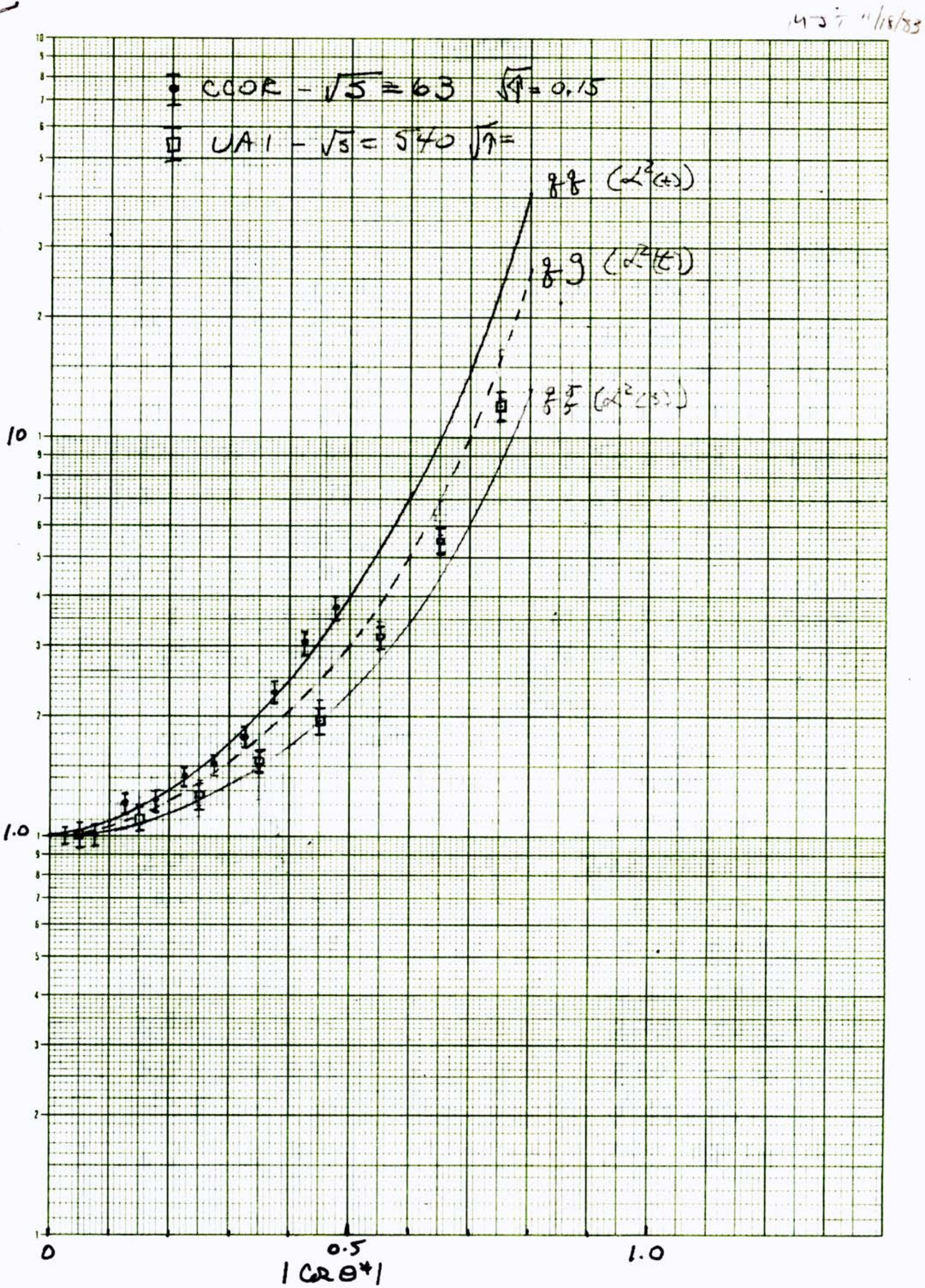}}
\caption[]{a) (left) UA1 $\cos\theta^*$ distribution for dijets in $\bar{p}+p$ collisions at $\sqrt{s}=540$ GeV with \QCD\ $\Sigma^{ab}(\cos\theta^*)$ calculations~\cite{UA1PLB136}. b) (right) My comparison of UA1 $\bar{p}+p$ and CCOR $p+p$ $\cos\theta^*$ distributions and \QCD\ subprocesses, November 1983.}
\label{fig:latercostheta}\vspace*{-1.0pc}
\end{figure}
When the CERN $\bar{p}+p$ collider experiments started to publish di-jet angular distribution measurements  compared to \QCD\ constituent scattering subprocess distributions, e.g. UA1~\cite{UA1PLB136} (Fig.~\ref{fig:latercostheta}a), I made a plot (Fig.~\ref{fig:latercostheta}b) of their data for $\sqrt{s}=540$ GeV and the CCOR measurement at 
 $\sqrt{s}=62.4$ GeV with my calculations of the \QCD\ $qq$, $gq$ and $\bar{q}q$ subprocess angular distributions . The $p+p$ data aligned best with the $qq$ calculation while the $\bar{p}+p$ data aligned beautifully with the $\bar{q}q$ calculation for $|\cos\theta^*|\leq 0.5$ which I thought was pretty neat in 1983 but I never published it.
 
\begin{figure}[!t]
\begin{center}
\raisebox{0.9pc}{\includegraphics[width=0.47\textwidth]{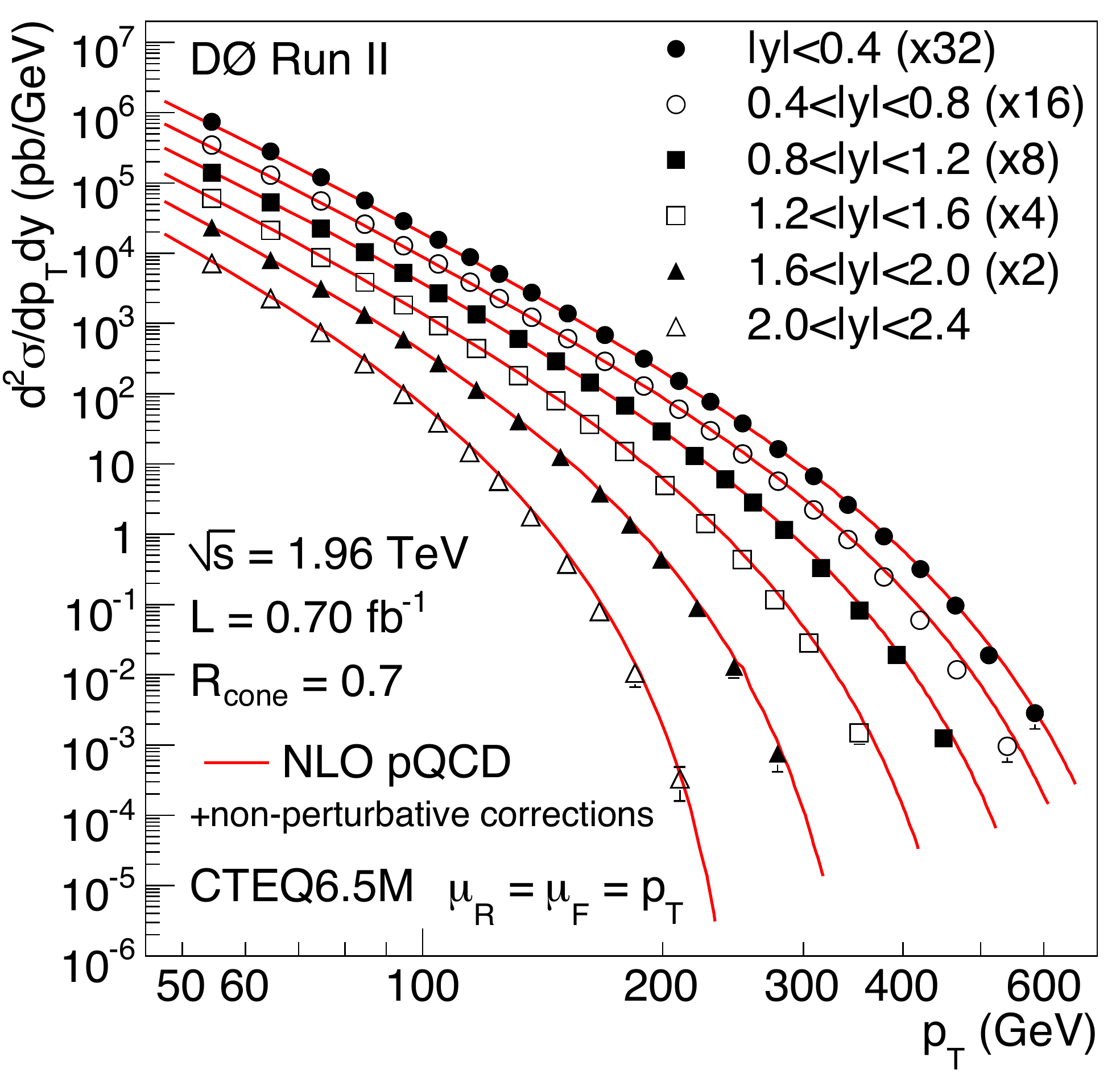}}
\includegraphics[width=0.51\textwidth]{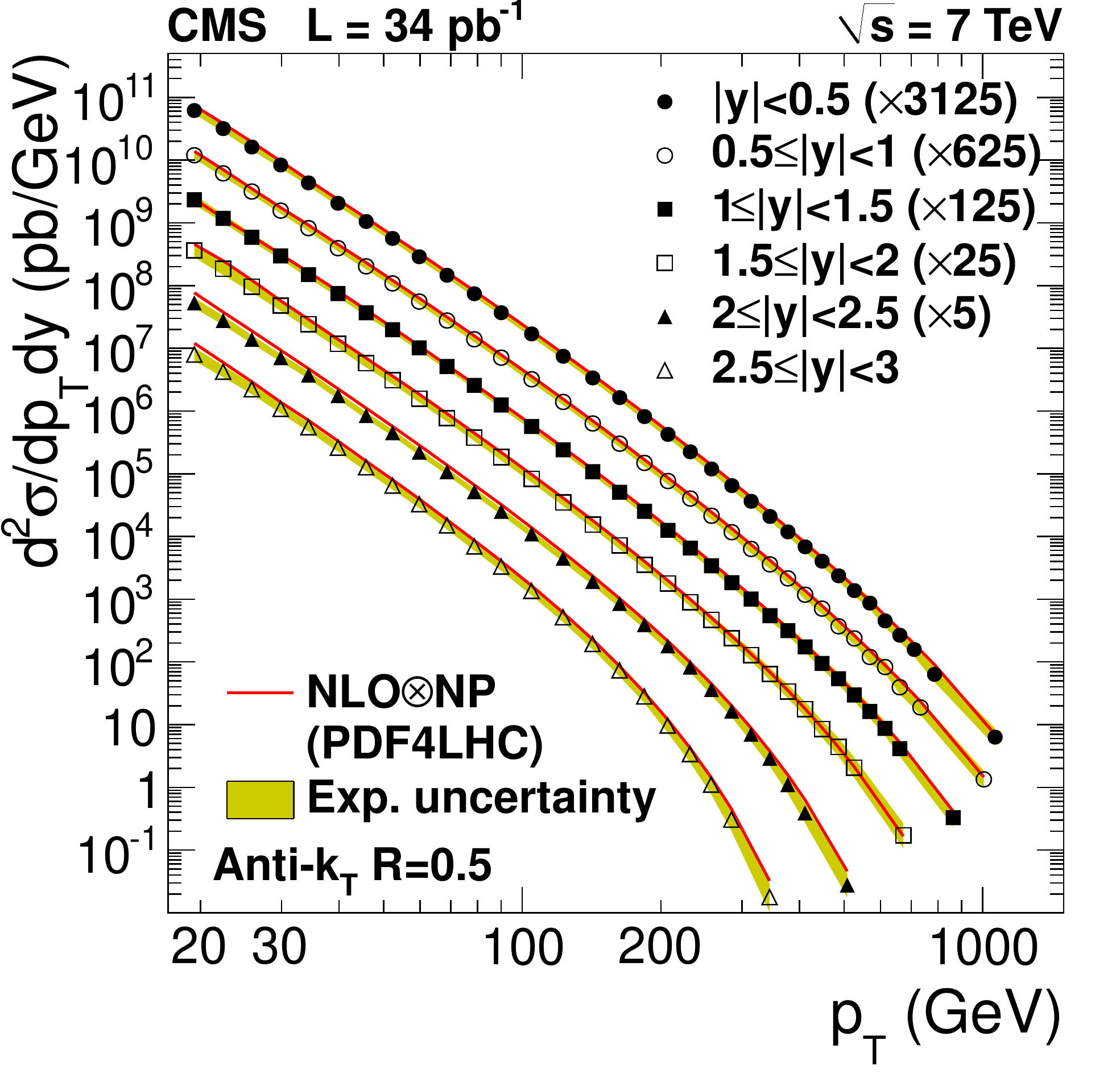}
\end{center}
\caption[]{a) (left)  D0~\cite{D0jetPRL101} inclusive jet cross sections at $\sqrt{s}=1.96$ TeV as a function of jet $p_T$ in bins of jet rapidity $y$ with NLO pQCD predictions.  b) CMS~\cite{CMSjetPRL107} measurements at $\sqrt{s}=7$ TeV (data points) with NLO theoretical predictions. }
\label{fig:2011jets} 
\end{figure}
The relatively recent jet cross section measurements (Fig.~\ref{fig:2011jets}) as a function of $p_T$ and rapidity at the Fermilab Tevatron in $\bar{p}+p$ collisions at $\sqrt{s}=1.96$ TeV and the CERN LHC $p+p$ collider at $\sqrt{s}=7$ TeV agree incredibly well with Next to Leading Order (NLO) \QCD\ . Note that at mid-rapidity the D0 data follow the typical hard-scattering power law but drop sharply at large $p_T$ and $y$, due to conservation of energy. The drop is much weaker at large $p_T$ and $y$ for the CMS data because of the 3.5 times larger c.m. energy. 

\begin{figure}[h]
\centering
\includegraphics[width=0.48\textwidth]{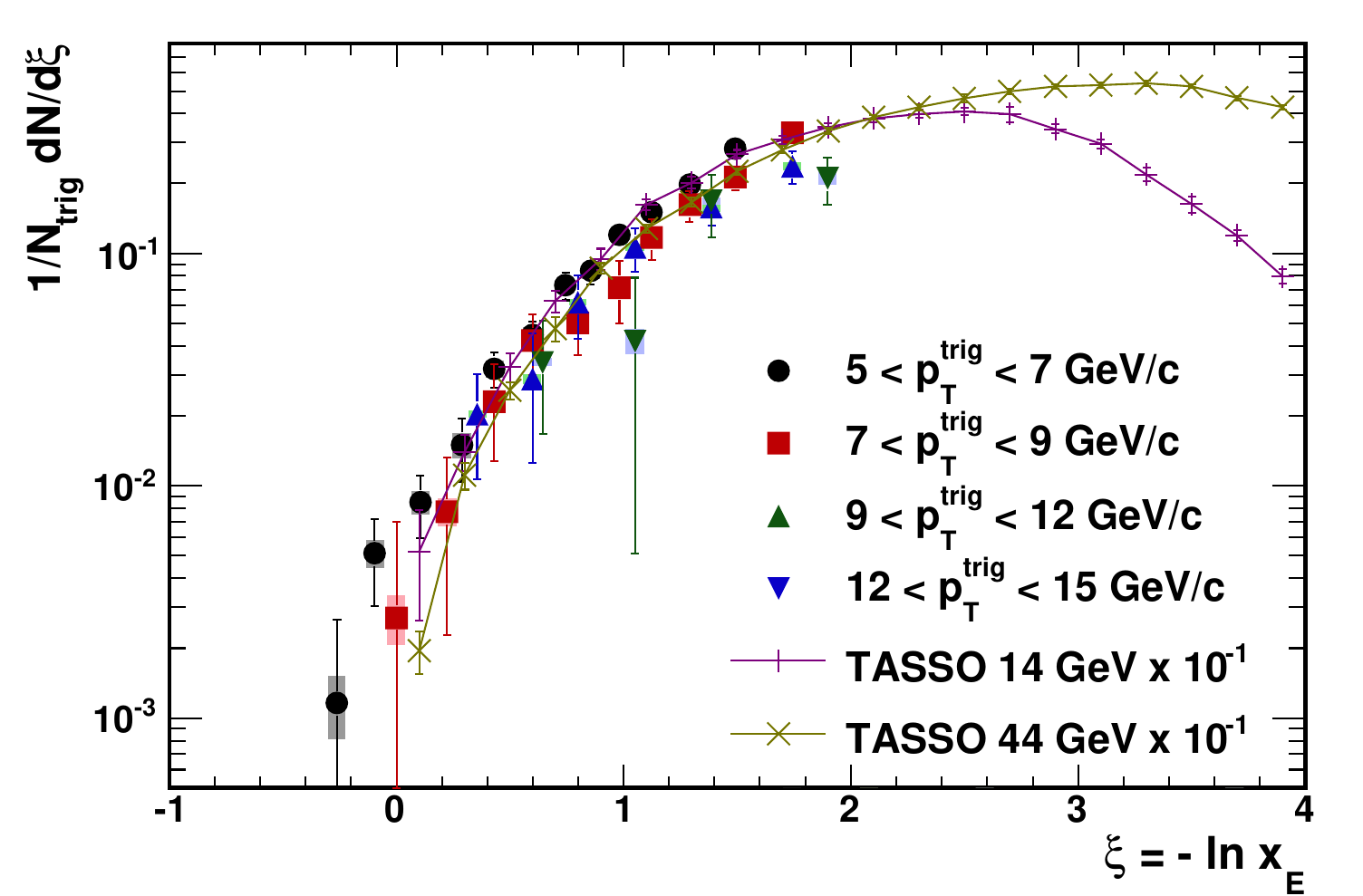}
\includegraphics[width=0.50\textwidth]{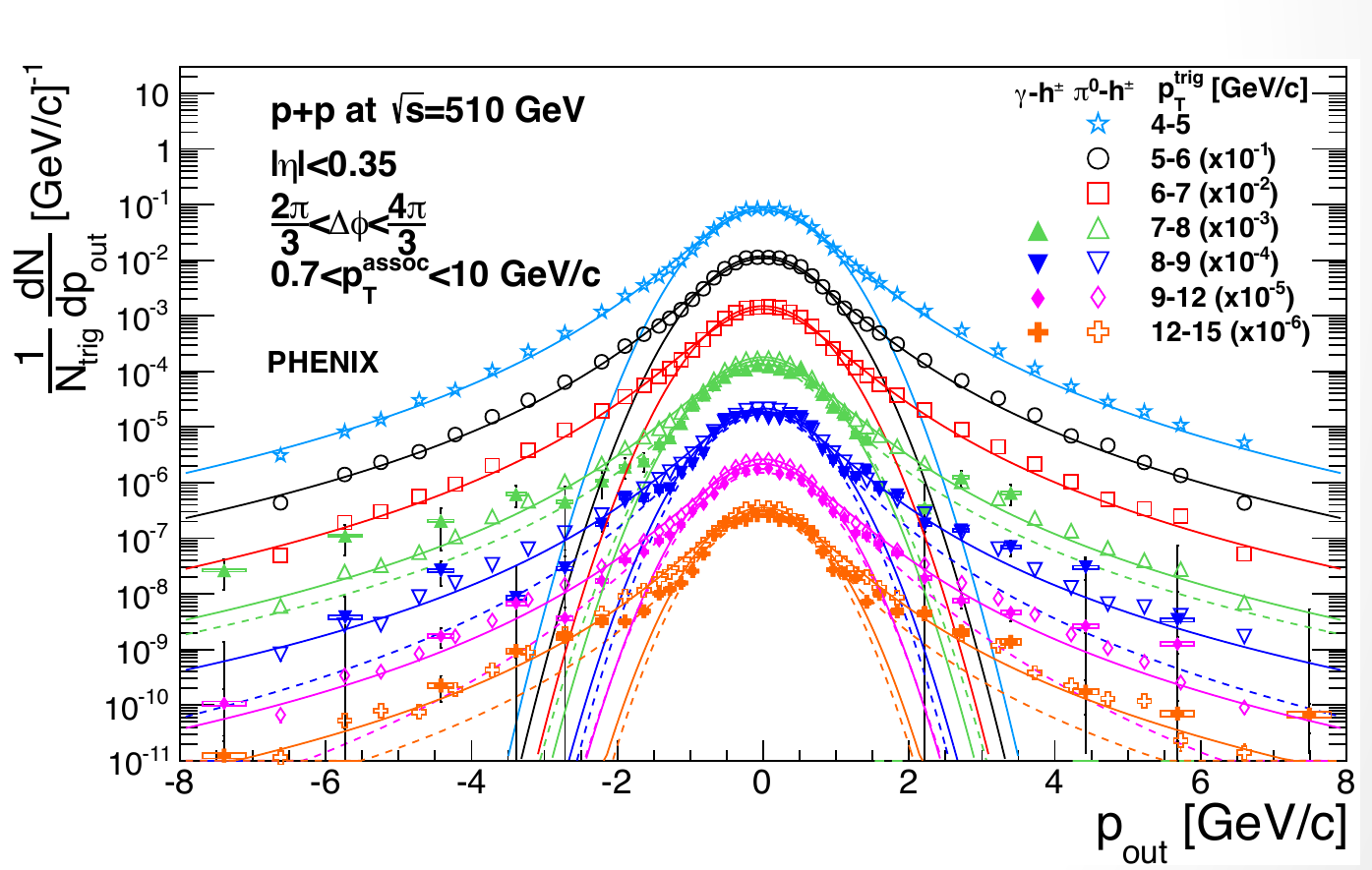}
\caption[]{a) (left) Direct-$\gamma$-h correlations in $\sqrt{s}=200$ GeV p$+$p collisions as a function of $\xi=-\ln x_E$~\cite{PXPRD82}   b) (right) $p_{\rm out}$ distributions of charged hadrons in $\pi^0+h$ and $\gamma+h$ correlations with $0.7<p_{Ta}<10$ GeV/c for 7 values of $p_{Tt}$ in $\sqrt{s}=510$ GeV p$+$p collisions~\cite{PXPRD95}.} 
\label{fig:PHENIXppQCD} \vspace*{-1.0pc}
\end{figure}
Recent two-particle correlation measurements at RHIC (Fig.~\ref{fig:PHENIXppQCD}) nicely show that the $x_E$ distribution of direct-$\gamma-h$ correlations plotted as a function of $\xi=-\ln x_E$ really does measure the fragmention function as measured in $e^+ e^-$ collisions at $\sqrt{s}=14$ and 44 GeV~\cite{TASSOZPC47} and that the $p_{\rm out}$ distribution has two components, a Gaussian likely to be from the intrinsic $k_T$ of partons in the nucleon, and a power-law tail from NLO \QCD\ gluon emission as suggested in~\cite{PXPRD95}.  
\pagebreak
\section{Hadron Collider discoveries, not exactly \QCD\ 1983--2012}
\label{sec:WZTH}
\vspace*{-2.0pc}\begin{figure}[h]
\centering
{\footnotesize a)}\raisebox{+0.5pc}{\includegraphics[width=0.30\textwidth]{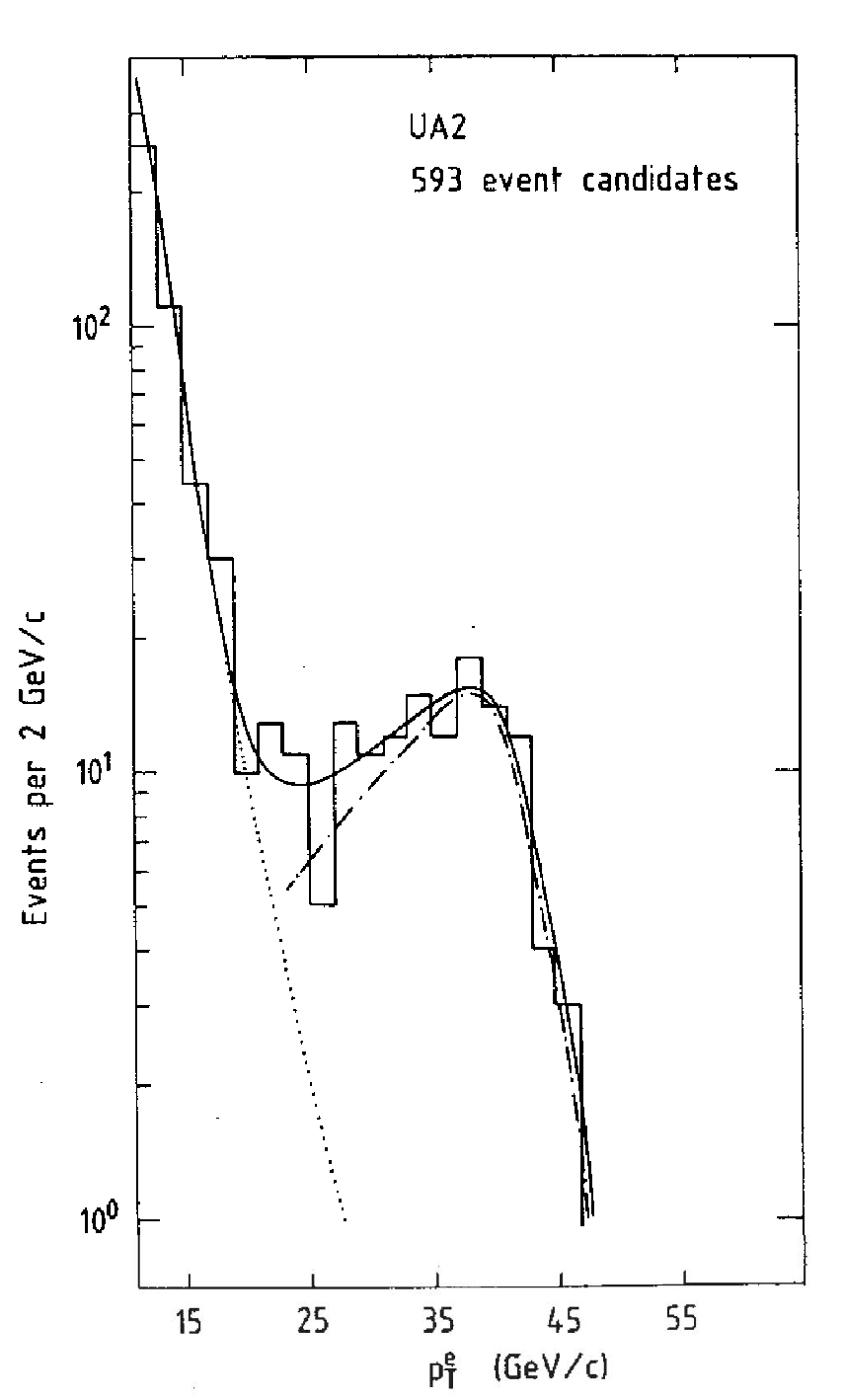}}\hspace*{2.0pc}
{\footnotesize b)}\raisebox{+0.4pc}{\includegraphics[width=0.45\textwidth]{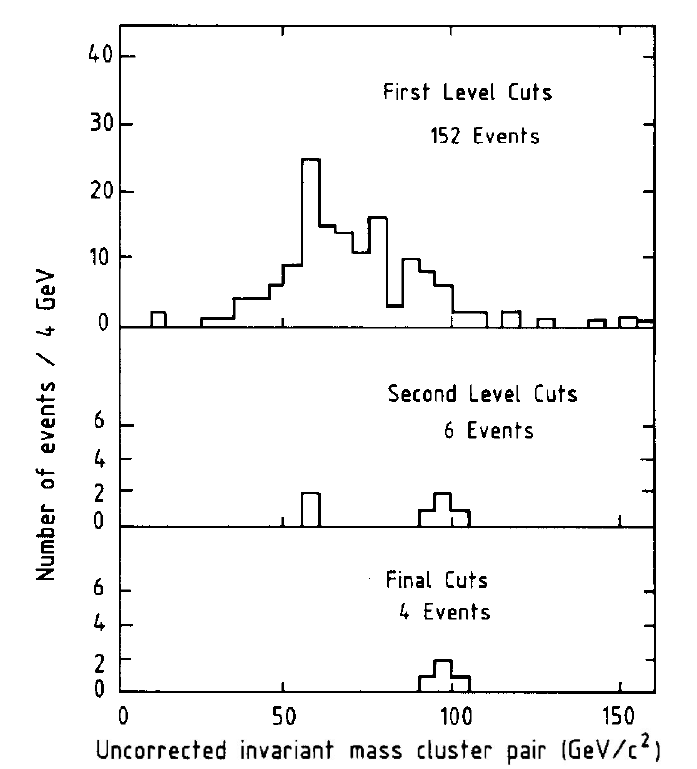}}\vspace*{0.5pc}
{\footnotesize c)}\raisebox{+0.4pc}{\includegraphics[width=0.40\textwidth]{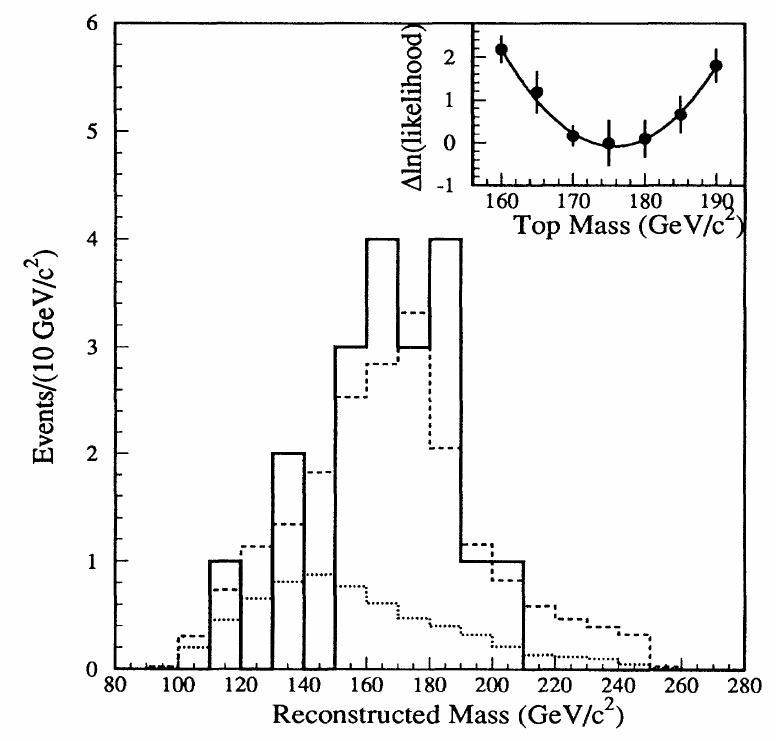}}\hspace*{0.5pc}
{\footnotesize d)}\includegraphics[width=0.45\textwidth]{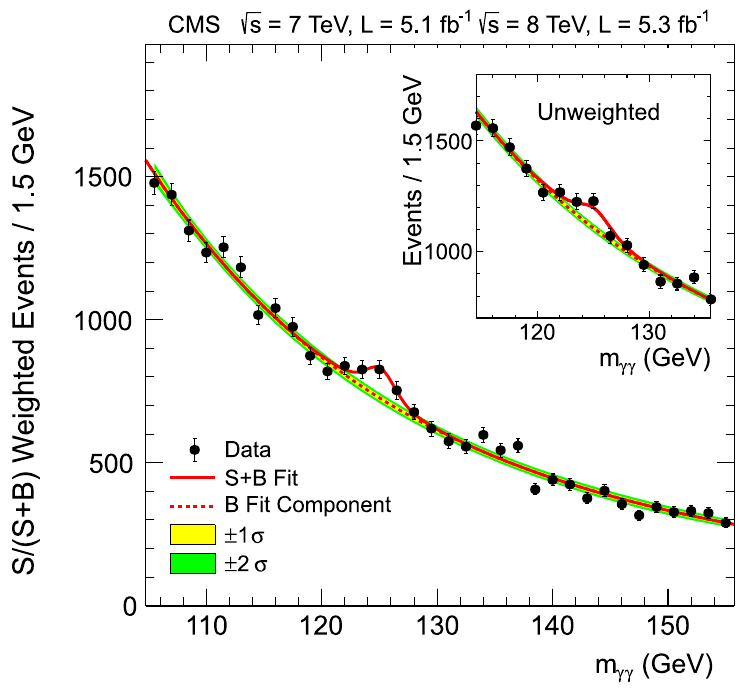}
\caption[]{a) UA2 $p^e_T$ spectrum~\cite{UA2ZPC30}. b) UA1 Z$^0\rightarrow e^+ e^-$ discovery~\cite{UA1Z}. c) CDF $b$-tagged $W+\geq4$ jet top quark mass plot~\cite{CDFtop}. d) CMS $m_{\gamma\gamma}$ plot with 125 GeV Higgs $\rightarrow \gamma\gamma$~\cite{CMSHiggs}. }  
\label{fig:discoveries1} 
\end{figure}
To keep the record straight, I think that it is important to note several major discoveries at hadron colliders that are not exactly \QCD\ but nevertheless are key elements of the Standard Model.
The $W$ and $Z$ bosons of the Weak interactions were discovered at the CERN $\bar{p}+p$ collider in 1983 by experiments UA1 (W)~\cite{UA1W}, (Z)~\cite{UA1Z};  and UA2 (W)~\cite{UA2W} (Z)~\cite{UA2Z}.   
Figure \ref{fig:discoveries1}a shows a UA2 measurement of $W^{\pm} \rightarrow e^{\pm} + X $ with a nice Zichichi signature and Fig.~\ref{fig:discoveries1}b the actual UA1 discovery plot of the $Z^0\rightarrow e^+ + e^-$.\footnote {The $Z^0$ plot reminds me of Fig.~\ref{fig:Tingetal}b; but for Fig.~\ref{fig:discoveries1}b it was Carlo Rubbia who had the last laugh, a well deserved Nobel Prize.} The top quark was discovered at the Fermilab Tevatron $\bar{p}p$ collider by D0~\cite{D0top} and CDF\cite{CDFtop} (Fig.~\ref{fig:discoveries1}c) and the Higgs Boson was discovered at the CERN-LHC by ATLAS~\cite{ATLASHiggs} and CMS~\cite{CMSHiggs}(Fig.~\ref{fig:discoveries1}d).  
\pagebreak
\section{\QCD\ at Relativistic Heavy Ion Colliders 2000--2017 }
    \subsection{From Bjorken Scaling to \QCD\ to the \QGP}  
    
      Bjorken scaling not only led to the parton model and \QCD\ as we have already discussed, but also led to the conclusion~\cite{CollinsPerry} that ``superdense matter (found in neutron-star cores, exploding black holes, and the early big-bang universe) consists of quarks rather than of hadrons", because the hadrons overlap and their individuality is confused. This is different from earlier models which take hadrons as the basic entities~\cite{Hagedorn1994}. Collins and Perry~\cite{CollinsPerry} called this state ``quark soup" but used the equation of state of a gas of free massless quarks from which the interacting gluons acquire an effective mass which provides long-range screening. They also pointed  out that for the theory of strong interactions (\QCD), ``high density matter is the second situation where one expects to be able to make reliable calculations---the first is Bjorken scaling". In the Bjorken scaling region, the theory is asymptotically free at large momentum transfers while in high-density nuclear matter long range interactions are screened by many-body effects, so they can be ignored and short distance behavior can be calculated with the asymptotically-free \QCD\  and relativistic many-body theory. Shuryak~\cite{Shuryak80} codified and elaborated on these ideas and provided the name ``QCD (or quark-gluon) plasma" \QGP\  for ``this phase of matter", a plasma being an ionized gas.   
   
        It didn't take long for others to realize that relativistic heavy ion (RHI) collisions could provide the means of obtaining superdense nuclear matter in the laboratory~\cite{BearMountain}~\cite{BjorkenPRD27}. 
 The kinetic energy of the incident projectiles would be dissipated in the large 
volume of nuclear matter involved in the reaction.  The system is expected 
to come to equilibrium, thus heating and compressing the nuclear matter so that it undergoes a phase transition from a state of nucleons containing bound quarks and gluons to a state of deconfined quarks and gluons, the Quark Gluon Plasma (\QGP), in chemical and thermal equilibrium, covering the entire
volume of the colliding nuclei or a volume that corresponds to many units of the characteristic length scale. 

In the terminology of high energy physics, this is called a ``soft'' (low $Q^2$) process, related to the \QCD\  confinement scale
\begin{equation}
\Lambda^{-1}_{\rm QCD} \simeq {\rm (0.2\ GeV)}^{-1} \simeq 1 \, 
\mbox{fm}\qquad .
\label{eq:LambdaQCD}
\end{equation}
   With increasing temperature, $T$, in analogy to increasing $Q^2$, the strong coupling constant $\alpha_{s}(T)$ becomes smaller, reducing the binding energy,  and the string tension, $\sigma(T)$, becomes smaller, increasing the confining radius, effectively screening the potential\cite{SatzRPP63}: 
  \begin{equation}
  V(r)=-{4\over 3}{\alpha_{s}\over r}+\sigma\,r \rightarrow 
-{4\over 3}{\alpha_{s}\over r} e^{-\mu_D\,r}+\sigma\,{{(1-e^{-\mu_D\,r})}\over \mu_D}
\label{eq:VrT}
\end{equation} 
where $\mu_D=\mu_D(T)=1/r_D$ is the Debye screening mass~\cite{SatzRPP63}. For $r< 1/\mu_D$ a quark feels the full color charge, but for $r>1/\mu_D$, the quark is free of the potential and the string tension, effectively deconfined. The properties of the \QGP\ can not be calculated in \QCD\  perturbation theory but only in Lattice \QCD\ Calculations~\cite{SoltzARNPS65}.

\subsection{From ISABELLE to CBA to RHIC 1983}
Following the discovery of the $W$ and $Z$ bosons at CERN, the $\sqrt{s}=800$ GeV p$+$p collider at BNL, ISABELLE, which had been renamed Colliding Beam Accelerator (CBA), was cancelled by HEPAP (the U.S.  High Energy Physics Advisory Panel) on July 11, 1983, but was miraculously immediately resuscitated by NSAC (the U.S. Nuclear Science Advisory Committee) to eventually become the $\sqrt{s_{NN}}=200$ GeV A$+$A Relativistic Heavy Ion Collider (RHIC)~\cite{Crease2008} whose purpose was to discover the properties of nuclear matter under extreme conditions with possible discovery of new states of matter, e.g. the \QGP. In Ref.~\cite{MJTW}, I discussed several lucky breaks that I had in being able to attend the  ICHEP82 and then to present the lecture with the results shown in Fig.~\ref{fig:CCORcostheta}. Well, it turned out that I had another important lucky break in August 1983 just after RHIC became a gleam in the eye of NSAC. The chair of the BNL physics department, Arthur Schwartzschild, a Nuclear Physicist,  offered me five BNL Nuclear Physicists to participate in the set-up and data taking of an $\alpha+\alpha$ run which had been scheduled for 16--30 August 1983 at the CERN ISR,  as well as to help analyze the data. The purpose was to get collider experience for the RHIC proposal and to help understand an ``exciting result'' from the previous $\alpha+\alpha$ run at the ISR. 
\subsubsection{``Exciting result''? How I became a nuclear physicist!}
It started in 1979, when Martin Faessler~\cite{ISRC79-10} and collaborators proposed to measure $p+\alpha$ and $\alpha+\alpha$ collisions in the CERN-ISR using the SFM. The proposal was approved which  led to runs with $\alpha+\alpha$ at $\sqrt{s_{NN}}=31$ GeV and $p+\alpha$ at $\sqrt{s_{NN}}=44$ GeV in 1980, with a subsequent run in 1983 with $\alpha+\alpha$, $p+\alpha$, $d+d$ and p$+$p interactions all at the same $\sqrt{s_{NN}}=31$ GeV. The high energy physicists at the ISR had the option of turning off their detectors and resting for a few weeks or continuing to operate their detectors for the nuclear collisions, with the possibility of exciting new physics results. They all opted to continue. Exactly same option and, predictably, exactly the same outcome occurred at the LHC 30 years later.

Once again, the publications from measurements in a new field started out with an incorrect result from the 1980 run, this time  where I was a co-author~\cite{CORPLB116}. Incredibly, it eventually turned out that this result was helpful. The 1980 $\alpha+\alpha$ run was at the full ISR energy, $\sqrt{s}=62.4$ GeV for p$+$p collisions which was  only $\sqrt{s_{NN}}=31.2$ GeV for $\alpha+\alpha$ where $Z/A$=1/2.  There was no comparison p$+$p data at $\sqrt{s}=31.2$ GeV in the 1980 run, only $\sqrt{s}=62.4$ GeV data. However, there was comparison p$+$p data at $\sqrt{s}=31.2$ GeV from the 1979 run which was not used because of uncertainty of a change in the absolute $p_T$ scale for the EM calorimeter by $\approx 5\%$, a huge effect when trying to measure a cross section that drops like $1/p_T^n$ with $n\approx 10$. The $\sqrt{s}=62.4$ GeV p$+$p data were used for comparison, but extrapolated to $\sqrt{s}=31.2$ GeV by a method that I kept saying was wrong whenever I was able to communicate with my collaborators adequately (no internet!!!) because I was still making magnets at BNL $\approx 6000$ km away. I told them to use $x_T$ scaling but they ignored my advice. 

For high $p_T$ hard-scattering, which is the result of scattering of pointlike partons, the ratio of the cross sections in p$+$A or B$+$A collisions to the p$+$p cross section should be simply equal to the product of the number of nucleons in the projectile and target, a factor $A$ times larger for p$+$A and $B\times A$ for B$+$A collisions.  However an `anomalous nuclear enhancement' was found in p$+$A collisions by Jim Cronin and collaborators at Fermilab~\cite{CroninPRD19}. The pion cross section ratio increased as $A^{\alpha(p_T)}$ where $\alpha(p_T)$ peaked at $\sim 1.15$ for $p_T=4-5$ GeV/c for $19.4\leq \sqrt{s_{NN}}\leq 27.4$ GeV. By contrast, the COR measurement in $\alpha+\alpha$ collisions~\cite{CORPLB116} for $p_T\geq 5$ GeV/c was equivalent to $\alpha(p_T)=1.3$ a factor of $\sim 1.6$ larger for the $\alpha\alpha$/pp cross section ratio than the extrapolation of Cronin's measurement. To quote a review by~\cite{FaisslerPLC115} of the results from the 1980 $\alpha+\alpha$ run:``If this trend is confirmed, it could eventually signify that something very interesting is going on in nucleus nucleus collisions.'' 

The interesting results from the $\alpha+\alpha$ collisions spread to the CERN management and through the nuclear physics grapevine to the chairman of the BNL physics department, Arthur Schwartzschild, who offered me the five BNL Nuclear Physicists. My CERN collaborators were happy about this because they also wanted Nuclear Physicists to help them understand the ``exciting result''.
  
Well, of course, we found out that the ``exciting result'' was wrong because our new result from the 1983 $\alpha+\alpha$ and p$+$p runs at $\sqrt{s_{NN}}=31$ GeV~\cite{BCMORPLB185} was $\alpha(p_T)=1.14\pm 0.01$ which now agreed with Cronin's result. We also had a preliminary result which Sanki Tanaka~\cite{ETQM83} was able to complete in time to present at the Quark Matter 1983 conference, the last week in September 1983, which had been moved from Helsinki to BNL. This result showed that the ratio of the cross sections for $\alpha\alpha$/pp as a function of $E_T$ varied by 2 to 6 orders of magnitude, so that the $A^{\alpha(p_T)}$ Cronin formalism was completely inadequate. We soon understood that this was because $E_T$ was a multiparticle distribution~\cite{MJTQM1984} which turned out to be very useful in Relativistic Heavy Ion collisions. Based on these ISR $\alpha+\alpha$ results, I joined with Chellis Chasman, Ole Hansen, Andy Sunyar of BNL, Lee Grodzins of MIT~\footnote{Sunyar and Grodzins along with Maurice Goldhaber had made the famous measurement of the helicity of the neutrino at BNL~\cite{GGSPR109}.}, Shoji Nagamiya of Columbia and others in an experiment (E802) to explore all aspects of relativistic heavy ion collisions at the new heavy ion beam at the BNL-AGS built to prepare for RHIC.
\subsection{RHIC and its Experiments: Design and Construction-1991--2000} 
The initial proposal by BNL to the U.S. Department of Energy to build RHIC was in 1984; the first funds for construction of RHIC were in the U.S. budget for Fiscal Year 1991; construction was completed and the first Au$+$Au collisions at $\sqrt{s_{NN}}=130$ GeV were in 2000 and with the standard c.m. energy $\sqrt{s_{NN}}=200$ GeV in 2001~\cite{NIMA499}. The first call for Letters of Intent for experiments was in April 1990 which were evaluated in November 1990 with updated LOI submitted in July 1991. The first Program Advisory Committee (PAC) to evaluate these updated LOI's approved the STAR proposal to join with a BNL TPC proposal to build a large Time Projection Chamber (TPC) concentrating on hadrons.  Three other proposals: di-muon, OASIS and TALES/SPARHC were rejected but told by the Associate Lab Director Mel Schwartz to merge into an experiment (most like TALES/SPARHC, my affiliation) to study electrons and photons emerging from the \QGP\ , which became the PHENIX experiment. Two smaller experiments were also approved~\cite{NIMA499}. 

     The STAR experiment is similar to a conventional solenoid collider detector of the late 1980's except that its solenoid with 2.6m radius and magnetic field B=0.5 T is not superconducting. The TPC covers the full azimuth at mid-rapidity, $|\eta|\leq1.0$. Particle identification is done with $dE/dx$ in the TPC and Time of Flight counters. An EMcalorimter outside the solenoid has segmentation $\Delta\eta \times \Delta\phi=0.05\times 0.05$ with a shower-maximum pre-converter/detector 5 $X_o$ deep for improved $\gamma/\pi^0$ separation. 
     
     The PHENIX experiment is a two-arm spectrometer with a fine grain EMcalorimeter $\Delta\eta \times \Delta\phi\approx0.01\times0.01$ to separate single-$\gamma$ and resolve $\pi^0\rightarrow\gamma+\gamma$ with $p_T$ up to $\approx 20$ GeV/c. Each arm has a Ring Imaging Cerenkov counter, time-of-flight (TOF) and drift-chamber tracking for $e^{\pm}$, $\gamma$ and identified hadron measurements at mid-rapidity; with full azimuth muon spectrometers at forward and backward rapidity $1.1<|y|<2.2-2.4$. Full azimuth Beam Beam Counters at $3.0\leq|\eta|\leq 3.9$ and Zero Degree hadron Calorimeters that measure forward going neutrons within a 2 mrad ($|\eta|>6$) cone are used for triggering and luminosity measurement.
     
\begin{figure}[!h]
\begin{center}
\includegraphics[width=0.49\textwidth]{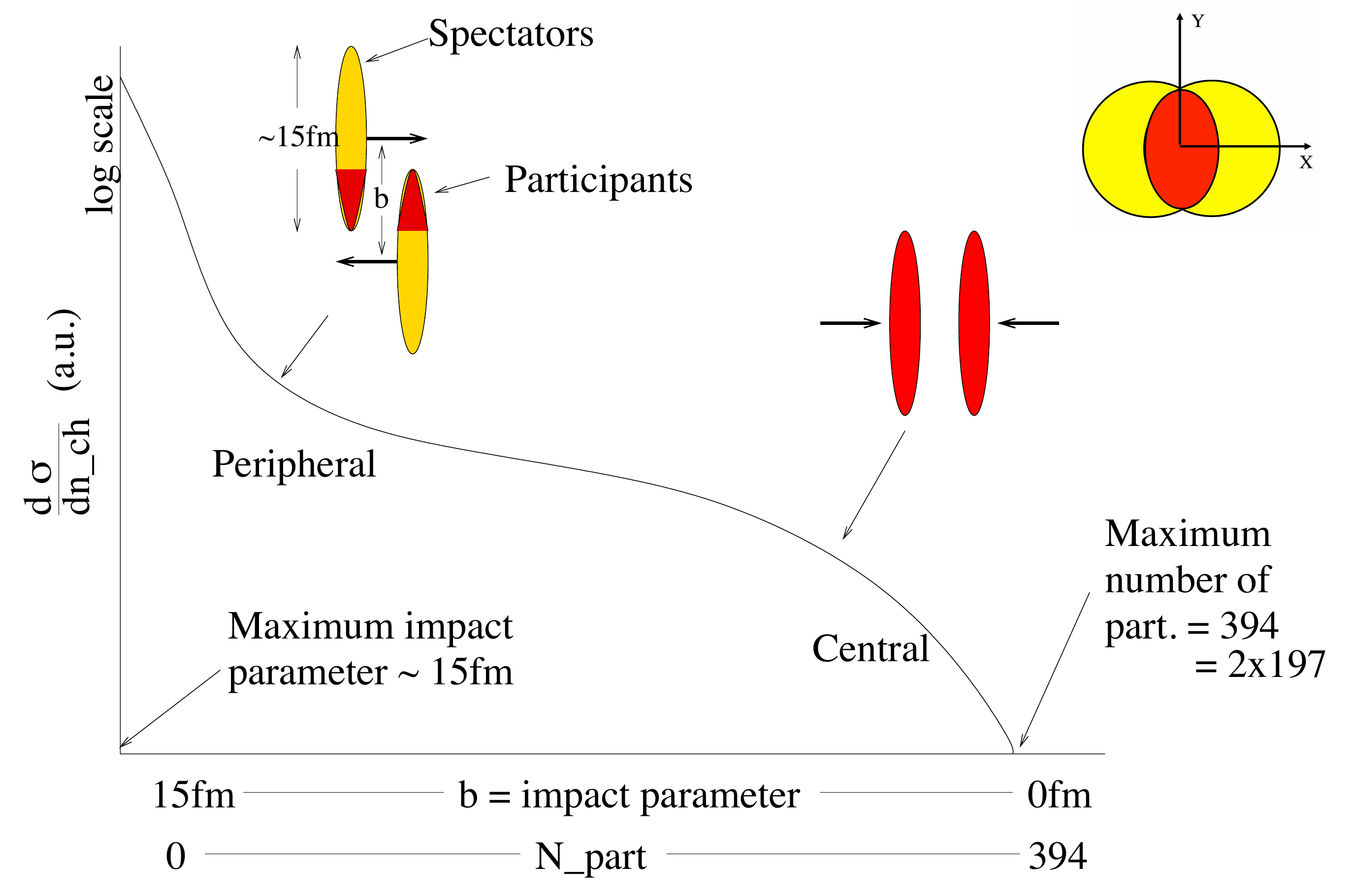}
\includegraphics[width=0.49\textwidth]{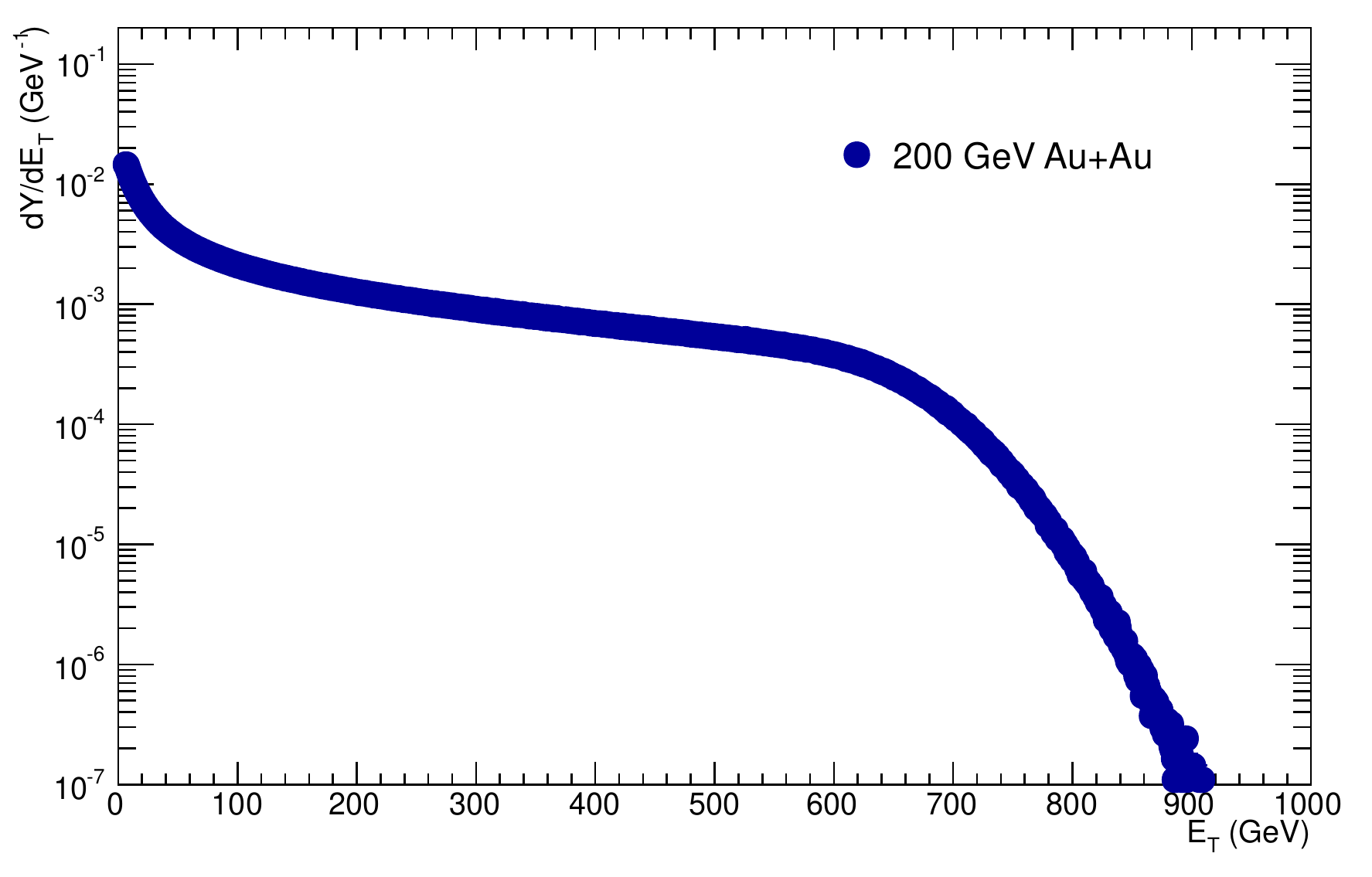}
\end{center}\vspace*{-0.5pc}
\caption[]{\footnotesize a) (left) \cite{MJTBuffalo} Schematic of collision in the c.m. system of two Lorentz contracted Au$+$Au nuclei with radius $R$ and impact parameter $b$. $N_{\rm part}$ is the number of nucleons struck in the collision.  The curve with the ordinate labeled $d\sigma/d n_{\rm ch}$ represents the relative probability of charged particle  multiplicity $n_{\rm ch}$. The upper right corner shows the almond shape overlap of the nuclei in peripheral collisions. It should be rotated so that the X axis is perpendicular to the page. More particles will be emitted along the X axis than the Y axis because of the stronger pressure gradient. This is called elliptical flow. b)(right) $E_T$ distribution in Au+Au at  $\sqrt{s_{NN}}=200$ GeV from PHENIX~\cite{AdlerPRC89}. The $E_T$ is corrected to the region $\Delta\eta=1.0$, $\Delta\phi=2\pi$. The charged multiplicity, $dN_{ch}/d\eta\approx 1.14\times E_T$. \label{fig:nuclcoll}}\vspace*{-0.5pc}
\end{figure}

     It was no accident that PHENIX had all the features of CCRS (Section:\ref{sec:CCRS1}) plus precision TOF for particle identification: I was one of the principal proponents of TALES/SPARHC. This gave PHENIX the possibility of measuring single $e^\pm$ for $p_T>0.3$ GeV/c which was crucial for $J/\Psi$ and charm measurements. Measurements of identified hadrons (including charm), although not part of Mel Schwartz's directive, were obviously necessary for understanding the background from their decay. Also, it was always my intention to use single $e^\pm$ to measure charm~\cite{MJT96}, because there is no combinatoric background. This is a serious problem in A$+$A collisions where charged multiplicities rise to $\approx$ A times that in p$+$p for the most central collisions where the two nuclei fully overlap~\cite{ppg019} (Fig.~\ref{fig:nuclcoll}). 

\subsection{How to find the \QGP\ 1986--2000}
At the time of the proposals for experiments at RHIC (also for the LHC~\cite{ALICEJINST}) in 1990-91, there were two proposed signatures of the \QGP: strangeness enhancement~\cite{PLC142} and the ``gold-plated''signature for  deconfinement in the \QGP, $J/\Psi$ suppression~\cite{MatsuiSatz86}. 
Matsui and Satz predicted  that $J/\Psi$ production in A$+$A
collisions will be suppressed by Debye screening of the quark
color charge in the \QGP. The $J/\Psi$ is produced when two gluons
interact to produce a $c \bar c$ pair which then resonates to form the
$J/\Psi$. In the \QGP, the $c \bar c$ interaction is screened so that the 
$c \bar c$ go their separate ways and eventually pick up other quarks at
the periphery to become {\it open charm}.  However, as pointed out a year later~\cite{MatsuiLBL24604}, enhanced production of $c$ and $\bar{c}$ quarks in A$+$A collisions, so that many of the ``other quarks'' picked up are $c$ or $\bar{c}$, could lead to recombination of $c\bar{c}$ into $J/\Psi$ which might ``hinder'' $J/\Psi$ suppression as evidence for the \QGP. Another problem was that the $J/\Psi$ is suppressed in p$+$A collisions~\cite{PrinoISMD2000}. These issues were worked out in further detail by analysis of $J/\Psi$ measurements at the CERN fixed-target heavy ion program~\cite{PBMPLB490} with the prediction of enhancement of $J/\Psi$ at LHC energies. 

\subsubsection{A hard-scattering \QGP\ signature based on \QCD\ 1997--1998}
\begin{figure}[!h]
\centering
\raisebox{+0.0pc}{\includegraphics[width=0.40\textwidth]{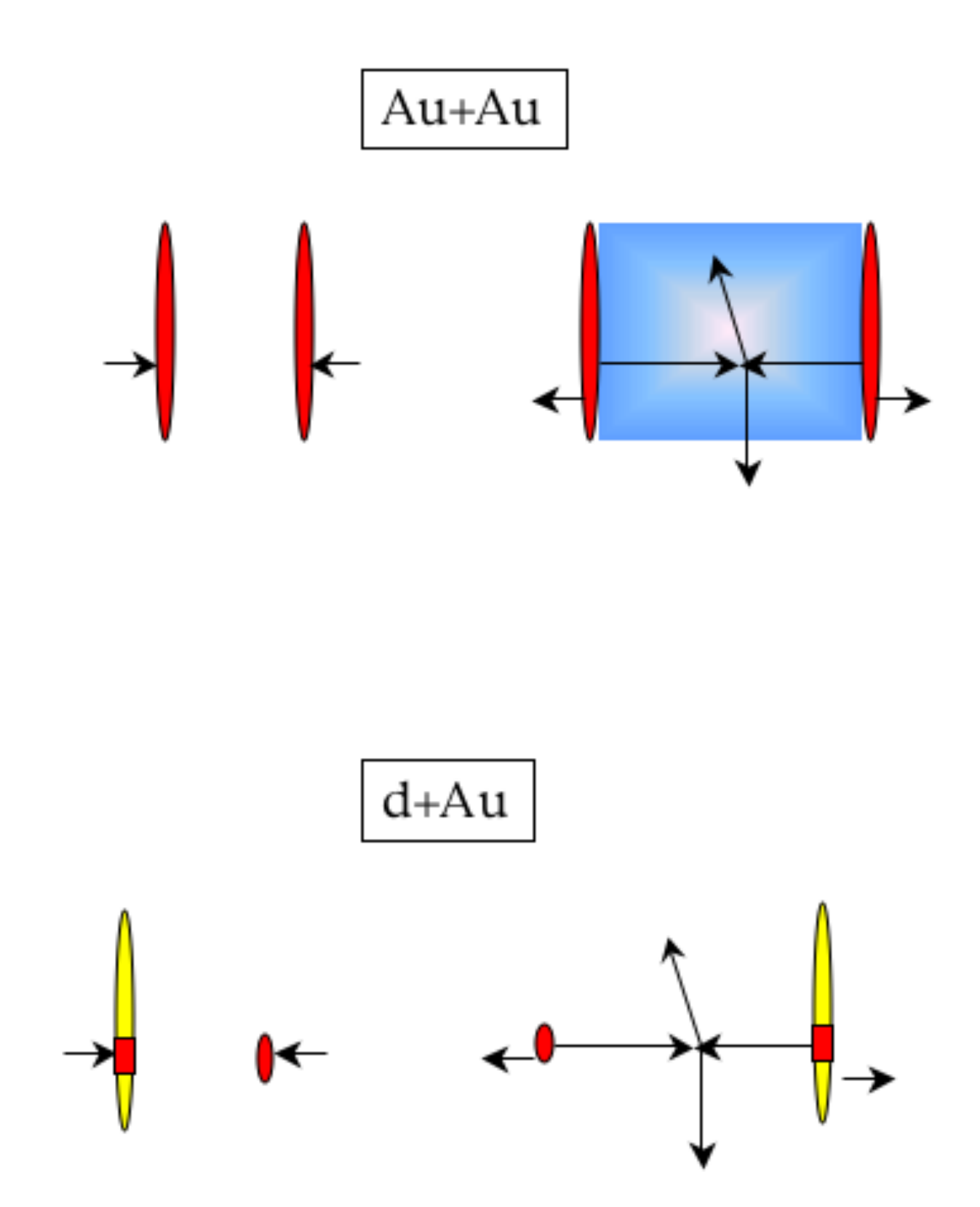}}\hspace*{1.0pc}
\raisebox{+0.0pc}{\includegraphics[width=0.54\textwidth]{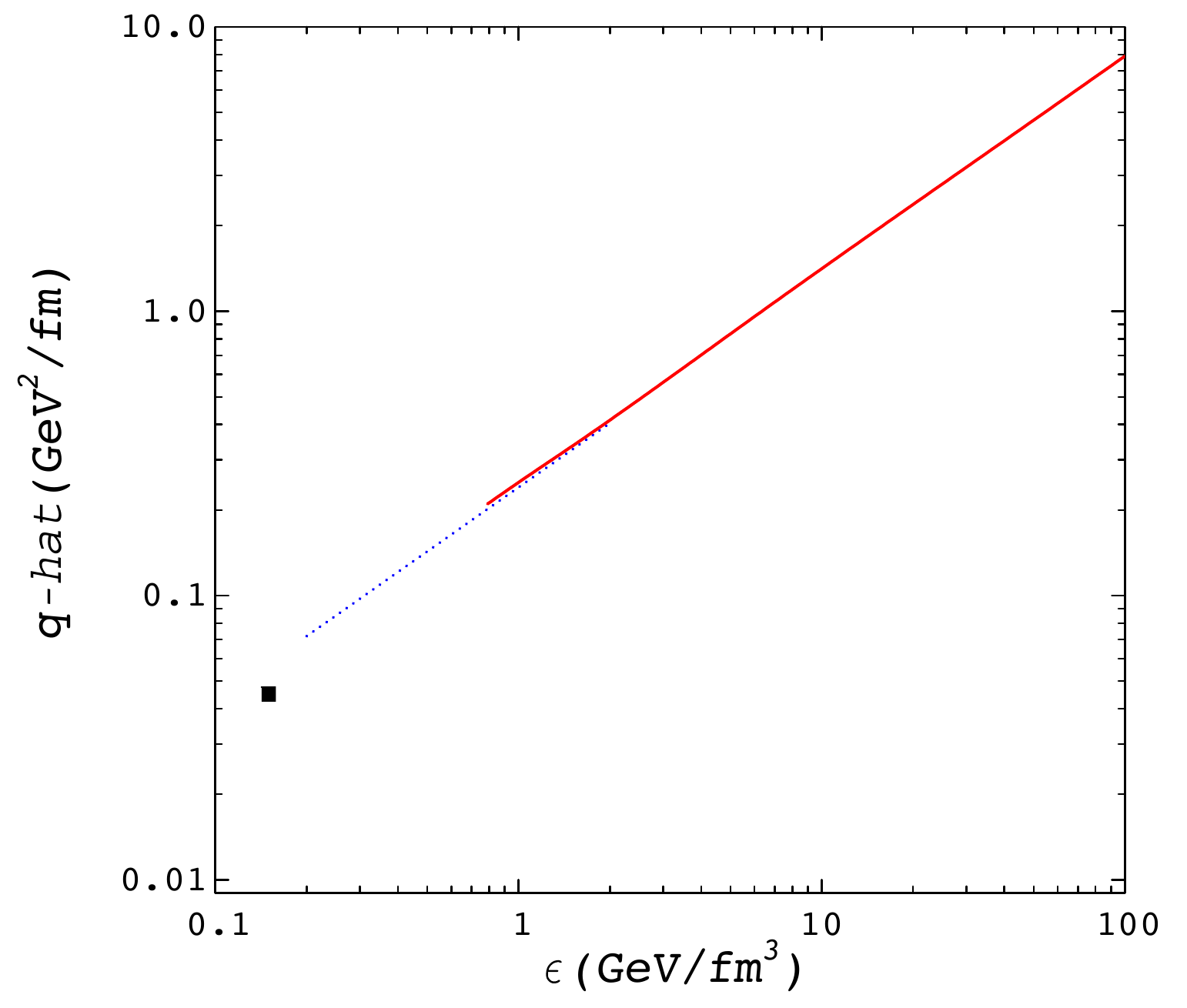}}
\caption[]{ a) (left)~\cite{MJTRPP69} Schematic diagram of hard scattering in Au$+$Au and d(p)$+$Au collisions.  Parton scattering occurs when the nuclei overlap, and for Au$+$Au the scattered high $p_T$ partons emerge sideways through the medium formed. For d+Au no medium is formed and the outgoing partons travel in vacuum until they fragment. b) (right) Transport coefficient $\hat{q}$ as a function of energy density $\epsilon$ for different media~\cite{BaierQM02}: cold nuclear matter (filled square), massless hot pion gas (dotted) and `ideal' \QGP\ (solid curve).  \label{fig:dAuAuAu} }
\end{figure}

A new tool for probing the color response function of the \QGP\ with a firm basis in \QCD\  was developed shortly before RHIC turned on. I found out about this in 1998 at the \QCD\ workshop in Paris~\cite{4thQCDWks}, when Rolf Baier asked me whether jets could be measured in Au$+$Au collisions because he had made studies in p\QCD~\cite{BDMPS2} of the energy loss, by `coherent' (LPM) gluon bremsstrahlung, of hard-scattered partons ``with their color charge fully exposed'' traversing a medium, ``with a large density of similarly exposed color charges''. This leads to a reduction of the $p_T$ of both the outgoing partons and their fragments, hence a reduction in the number of partons or fragments at a given $p_T$, which is called jet quenching. The effect is absent in p$+$A or d$+$A collidons beause no medium is produced (Fig.~\ref{fig:dAuAuAu}a)

The energy loss of an outgoing parton, $-dE/dx$, per unit length $(x)$ of a medium with total length $L$ due to coherent gluon bremsstrahlung is proportional to the 4-momentum-square, $q^2(L)$, transferred to the medium and takes the form:
\begin{equation}
{-dE \over dx }\simeq \alpha_s \mean{q^2(L)}=\alpha_s\, \hat{q}\, L \qquad . \label{eq:dEdx}
\end {equation}
$\hat{q}$, the transport coefficient of a gluon in the medium, is defined as the mean 4-momentum transfer-square, $q^2$, to the medium by a radiated gluon, {\it per gluon mean free path}, which can be calculated with \QCD~\cite{BDMPSZ}. Figure~\ref{fig:dAuAuAu}b shows an early calculation of $\hat{q}$~\cite{BaierQM02} . 

I told Rolf (and put in the proceedings) that because the expected energy in a typical jet cone $R=\sqrt{(\Delta\eta)^2+ (\Delta\phi)^2}$ in central Au$+$Au collisions at $\sqrt{s_{NN}}=200$ GeV would be $\pi R^2\times1/2\pi \times dE_T/d\eta=R^2/2 \times dE_T/d\eta~\sim 300$ GeV for $R=1$, where the kinematic limit is 100 GeV, jets can not be reconstructed in Au$+$Au \underbar{central} collisions at RHIC. This is still correct at present (19 years later) where the solution is to make smaller jet cones, which may (or may not) be a problem. 

I also told Rolf the good news that the jet suppression could be measured by single particle inclusive and two-particle correlations at RHIC as we had done at the CERN-ISR and that the PHENIX detector had actually been designed to make such measurements. 

\subsection{Measurements relevant to \QCD\  in A+A collisions at RHIC and LHC}
Measuring and understanding the properties of nuclear matter under extreme conditions and possibly 
the \QGP\ requires knowledge of Statistical Physics, Thermo- and Hydro- dynamics without much \QCD\ input. However, there were several important discoveries and measurements which did involve \QCD.
\subsubsection{Baryon chemical potential measured without particle identification}
In an equilibrated thermal medium, particles should follow a Boltzmann distribution in the local rest frame~\cite{CooperFrye}
\begin{equation}
{{d^2\sigma} \over {dp_L p_T dp_T}}={{d^2\sigma} \over {dp_L m_T dm_T}} \propto {1\over {e^{(E-\mu)/T} \pm 1}}\sim e^{-(E-\mu)/T} \qquad ,
\label{eq:boltz}
\end{equation}
where $m_T=\sqrt{p_T^2+m^2}$ and $\mu$ is a chemical potential. In fact,   the ratios of particle abundances (which are dominated by low $p_T$ particles) for central Au+Au collisions at RHIC, even for strange and multi-strange particles, 
are well described~\cite{STWP} by fits to a thermal distribution, 
          \begin{equation}
{{d^2\sigma} \over {dp_L p_T dp_T}}\sim e^{-(E-\mu)/T} \rightarrow {\bar{p} \over p}=\frac{e^{-(E+\mu_B)/T}}{e^{-(E-\mu_B)/T}}=e^{-(2\mu_B)/T} \qquad ,
\label{eq:boltz2}
\end{equation}
 with similar expressions for strange particles. $\mu_B$ (and $\mu_S$) are chemical potentials associated with each conserved quantity: baryon number, $\mu_B$,  (and strangeness, $\mu_S$). 

However for this problem there is also Lattice \QCD\ thermodynamics which can calculate $\mu_B$ from the net electric charge distributions of non-identified charged particles ($n^+ -n^-$)~\cite{BazavovPRL109}.
The theoretical analyses are made by a Taylor expansion of the free energy \hbox{$F=-T\ln Z$} around the  freezeout temperature $T_f$ where $Z$ is the partition function, or sum over states, which is of the form \vspace*{-0.5pc}
\begin{equation}\large
Z\propto e^{-(E-\sum_i \mu_i Q_i)/kT} \label{eq:partitionfn}  \end{equation} \normalsize
 and $\mu_i$ are chemical potentials associated with conserved charges $Q_i$.  The terms of the Taylor expansion, which are obtained by differentiation, are called susceptibilities, denoted $\chi$. 

The only connection of this method to mathematical statistics is that the Cumulant generating function in mathematical statistics for a random variable $x$ is also a Taylor expansion of the $\ln$ of an exponential:
\begin{equation}
g_x (t)=\ln\mean{e^{tx}}=\sum_{n=1}^\infty \kappa_n \frac{t^n}{n!} \qquad \kappa_m =\left.\frac{d^m g_x (t)}{dt^m}\right|_{t=0} \qquad .
\label{eq:cumgenfn}
\end{equation}
Thus, the susceptibilities are Cumulants in mathematical statistics terms.  
The first four Cumulants are $\kappa_1=\mu\equiv\mean{x}$,  
$\kappa_2=\mean{(x-\mu)^2}\equiv \sigma^2$, $\kappa_3=\mean{(x-\mu)^3}$, \hfill\break  \mbox{$\kappa_4=\mean{(x-\mu)^4}-3\kappa_2^2$}. Two so-called normalized or standardized Cumulants are common in this field, the skewness, $S\equiv\kappa_3/\sigma^3$ and the kurtosis, $\kappa\equiv\kappa_4/\sigma^4$. The theoretical results are presented as ratios of Cumulants so that the volume dependences of $\mu$, $\sigma$ $S$, $\kappa$ cancel. 

\begin{figure}[h]
\centering
\raisebox{+0.0pc}{\includegraphics[width=0.51\textwidth]{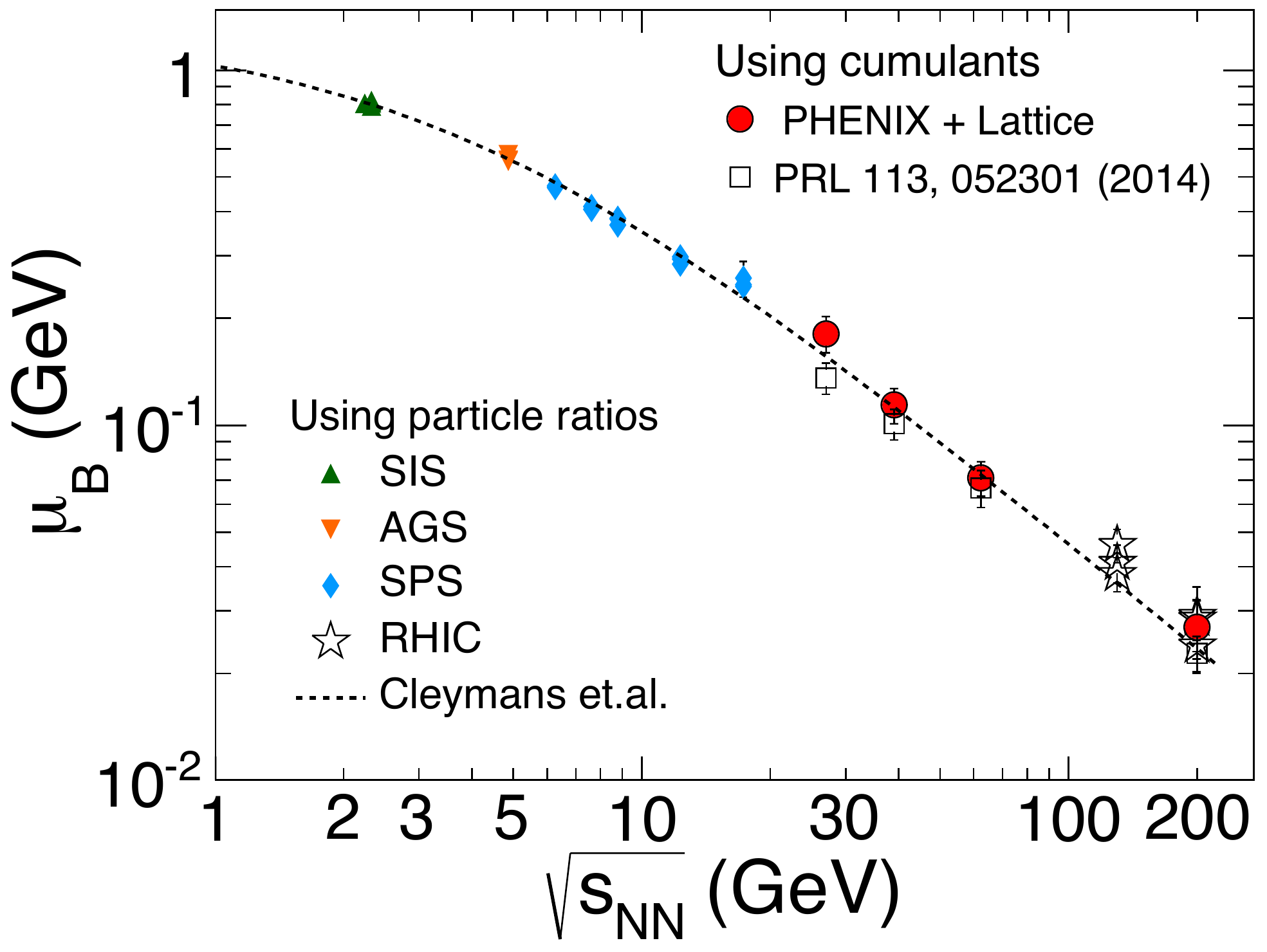}}
\raisebox{+0.0pc}{\includegraphics[height=0.38\textwidth,width=0.48\textwidth]{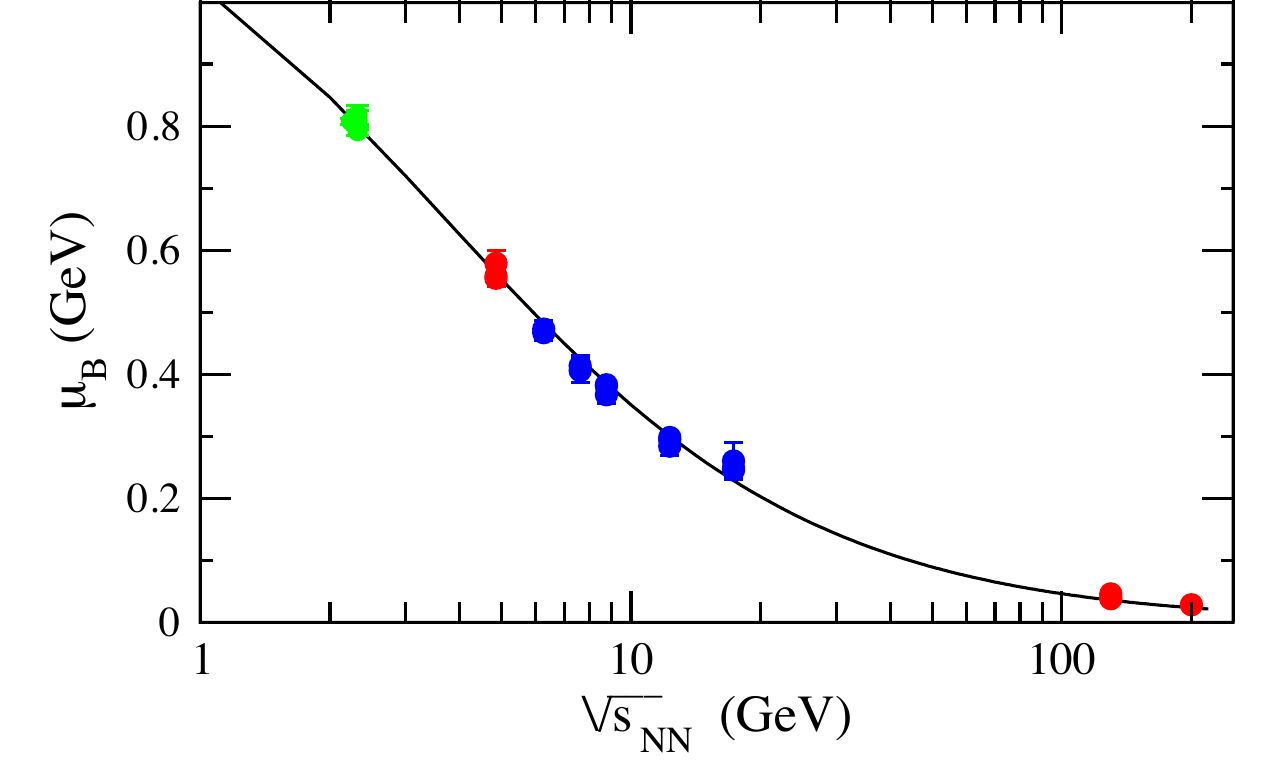}}
\caption[]{a)(left) $\sqrt{s_{NN}}$ dependence of $\mu_B$ from PHENIX+Lattice~\cite{ppg179} net-charge results. Open squares are from STAR net-charge~\cite{STARPRL113} together with net-protons~\cite{BorsanyiPRL113}. Dashed line and other data points are from b)(right), the best accepted analysis of $\mu_B$ vs $\sqrt{s_{NN}}$ from baryon/anti-baryon ratios~\cite{CleymansOeschler}.} \vspace*{-1.0pc}
  
\label{fig:QCDThermo} 
\end{figure}

The measured values of the temperature $T_f$ and $\mu_B$ are obtained~\cite{ppg179} by comparing the measured values of $\kappa_1/\kappa_2$ and $\kappa_3/\kappa_1$ to the Lattice \QCD\ calculations~\cite{BazavovPRL109}, which are given as functions of $T_f$ and $\mu_B$. Figure~\ref{fig:QCDThermo} shows that the PHENIX + Lattice results for $\mu_B$ from net-charge fluctuations, {\it with no particle identification}, are in excellent agreement with the best accepted analysis of $\mu_B$ from baryon/anti-baryon ratios~\cite{CleymansOeschler}. Both the $\mu_B$ and $T_f$ (not shown) results~\cite{ppg179} also agree with the more conventional best accepted values, which I believe was a first for measurements plus Lattice \QCD\  calculations in A$+$A collisions!

\subsubsection{Discovery of the \QGP\ at RHIC by suppression of high $p_T$ particles---2002}
The discovery at RHIC ~\cite{ppg003} that high $p_T$ $\pi^0$ produced by hard parton-parton scattering in the colliding Au$+$Au nuclei are suppressed in central Au+Au collisions by a factor of $\sim5$ compared to pointlike scaling from $p$$+$$p$ collisions is arguably {\em the}  major discovery in Relativistic Heavy Ion Physics. 
 
 	\begin{figure}[!htb]
\begin{center}
\includegraphics[width=0.8\linewidth]{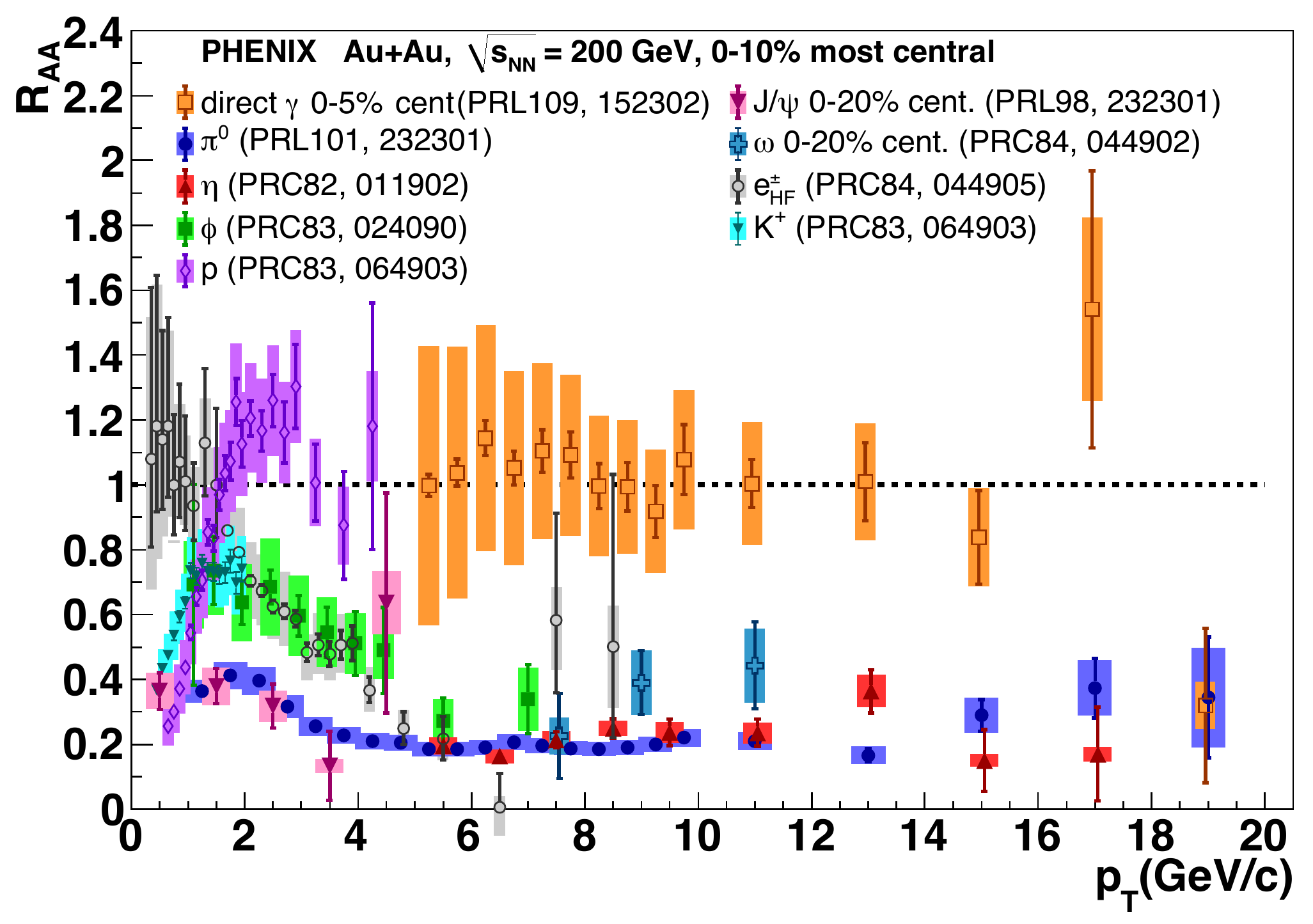}
\end{center}\vspace*{-0.12in}
\caption[]{\cite{MJTErice52} PHENIX measurements of the suppression $R_{AA}$ of identified particles with references to publication as a function of transverse momentum $p_T$. }
\label{fig:Tshirt}\vspace*{+0.5pc}
\end{figure}
In Fig.~\ref{fig:Tshirt}, the suppression of the many identified particles measured by PHENIX at RHIC is presented as the Nuclear Modification Factor, 
$R_{AA}(p_T)$, the ratio of the yield of e.g. $\pi$ per central Au+Au collision (upper 10\%-ile of observed multiplicity)  to the pointlike-scaled $p$$+$$p$ cross section at the same $p_T$, where $\mean{T_{AA}}$ is the average overlap integral of the nuclear thickness functions: 
   \begin{equation}
  R_{AA}(p_T)=\frac{(1/N_{AA})\;{d^2N^{\pi}_{AA}/dp_T dy}} { \mean{T_{AA}}\;\, d^2\sigma^{\pi}_{pp}/dp_T dy} \quad . 
  \label{eq:RAA}
  \end{equation}
  
The striking differences of $R_{AA}(p_T)$ in central Au+Au collisions for the many particles measured by PHENIX  (Fig.~\ref{fig:Tshirt}) illustrates the importance of particle identification for understanding the physics of the medium produced at RHIC. The most notable observations are: 
\begin{enumerate}
\item[1.] the equal suppression of $\pi^0$ and $\eta$ mesons by a constant factor of 5 ($R_{AA}=0.2$) for $4\leq p_T \leq 15$ GeV/c, with suggestion of an increase in $R_{AA}$ for $p_T > 15$ GeV/c;
\item[2.] the equality of suppression of direct-single $e^{\pm}_{\rm HF}$ from heavy flavor ($c$, $b$ quark) decay, and $\pi^0$ at $p_T\gsim 5$ GeV/c; 
\item[3.] the non-suppression of direct-$\gamma$ for $p_T\geq 4$ GeV/c. 
\end{enumerate}
For $p_T\gsim 4$ GeV/c, the hard-scattering region,  the fact that all hadrons are suppressed, but direct-$\gamma$ are not suppressed, indicates that suppression is a medium effect on outgoing color-charged partons, likely due to energy loss by coherent Landau-Pomeranchuk-Migdal radiation of gluons, predicted in p\QCD~\cite{BDMPSZ}. 
\subsubsection{\QGP, the perfect liquid, from charm suppression and flow--2007}
\begin{figure}[h]  
\centering
\raisebox{+0.0pc}{\includegraphics[width=0.48\textwidth]{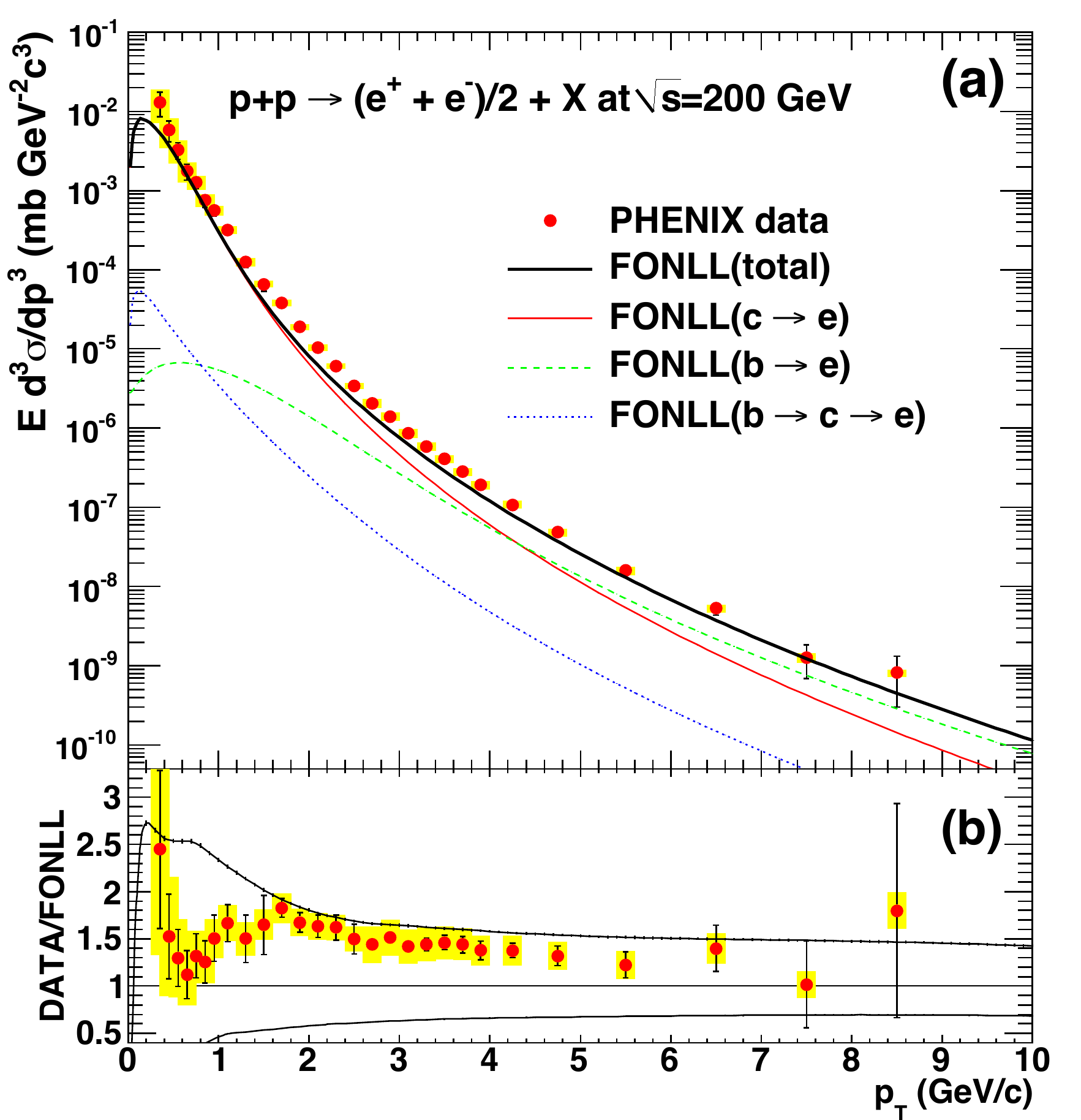}}
\raisebox{-0.2pc}{\includegraphics[width=0.51\textwidth]{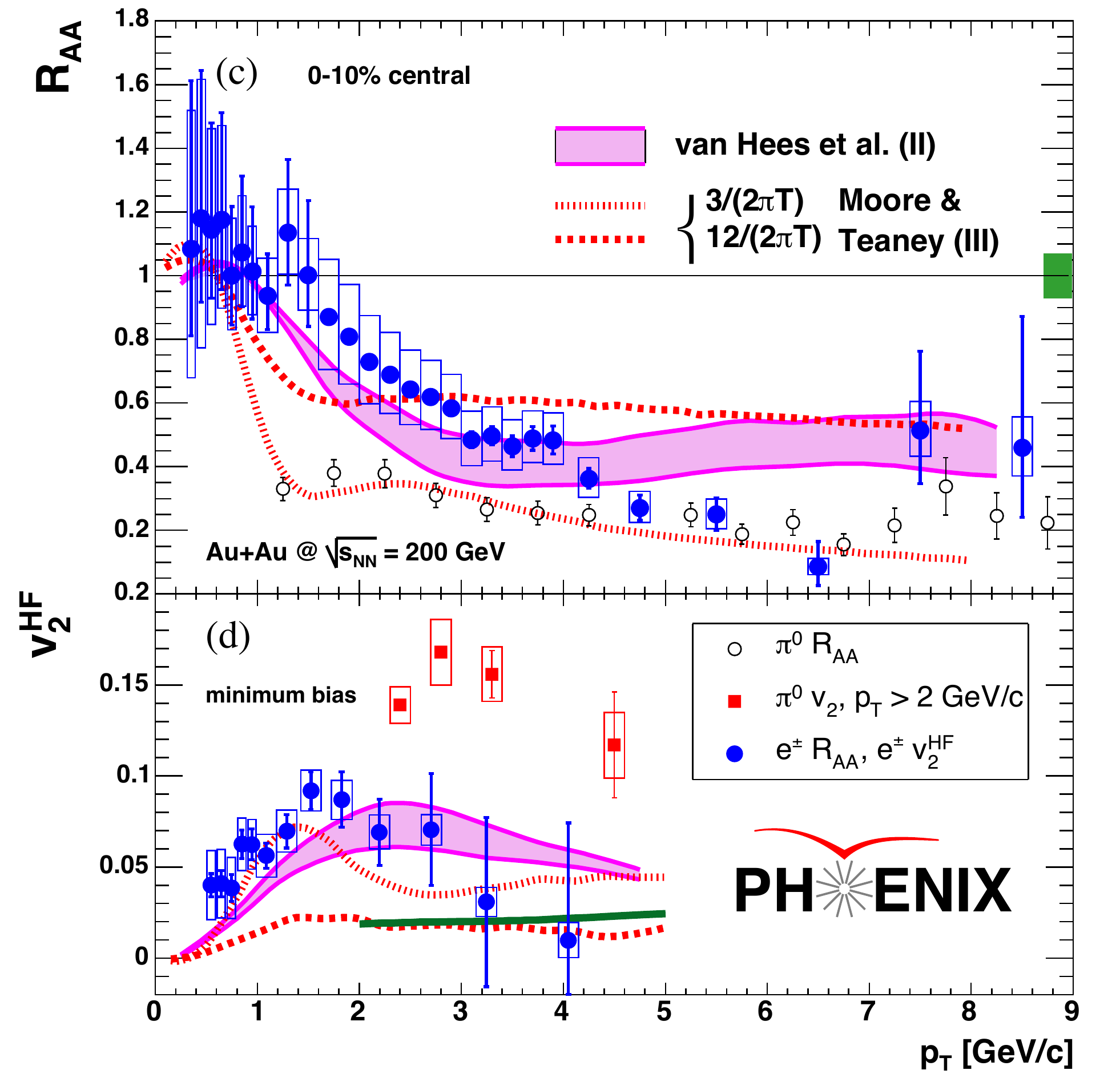}}
\caption[]{(left) a) \cite{ppg065} Invariant differential cross sections of electrons from heavy flavor quark decays. The curves are FONNL calculations~\cite{FONNL} b) The ratio of the measurement to the FONNL calculation. c) ~\cite{ppg066} $R_{AA}$ from direct-single $e^{\pm}$ and $\pi^0$, d) elliptical flow $v_2$ for Heavy Flavor quarks and $\pi^0$. Dashed and filled lines are theoretical predictions of the diffusion coefficient D.}
\label{fig:PerfectLiquid}
\end{figure}

The measurement of direct single $e^{\pm}$ (also called heavy flavor $e^{\pm}$ at RHIC) at mid-rapidity by PHENIX~\cite{ppg065} in p$+$p collisions at $\sqrt{s}=200$ GeV is shown in Fig.~\ref{fig:PerfectLiquid}a,b. The background from internal and external $\gamma$ conversions was determined by the converter method as in section~\ref{sec:convertermethod}. The data are compared to a fixed-order-plus-next-to-leading-log (FONLL) p\QCD\ calculation~\cite{FONNL} which is in excellent agreement with the measurement and gives the relative contributions of $c$ and $b$ quark decay.    

Figure~\ref{fig:PerfectLiquid}c shows the $R_{AA}$ of direct single heavy flavor $e^{\pm}$ and $\pi^0$ in $\sqrt{s_{NN}}=200$ GeV Au$+$Au collisions~\cite{ppg066} which become equal in the range $4\leq p_T\leq7$ GeV/c.  Initially this result was a surprise because heavy flavor quarks were predicted to lose less energy than light quarks due to the ``dead cone effect'' which happens in QED bremsstrahlung~\cite{YuriDima2001}. Figure~\ref{fig:PerfectLiquid}d shows the anisotropic eliptical flow measurement $v_2^{\rm HF}$ for the heavy flavor quarks and for $\pi^0$~\cite{ppg066}. The heavy quarks at $p_T\approx 2$ GeV/c are actually barely relativistic, because $\gamma\beta=p_T/m=2.0/1.3\approx 1.5$, but have a significant $v_2^{\rm HF}$. 

The $v_2^{\rm HF}$ in this region suggested~\cite{MooreTeaneyPRC71} that the charm quarks thermalize in the medium  which responds as a thermalized fluid with a small transport mean free path. Thus they treat the heavy quark in the medium as a thermal diffusion problem with diffusion coefficient $D=6\eta/(\epsilon+p)$ where $\eta$ is the shear viscosity, $\epsilon$ is the energy density, and $p$ the pressure. The enthalpy $(\epsilon +p)=Ts$ for zero baryon chemical potential, $\mu_B$ (a reasonable assumption at $\sqrt{s_{NN}}=200$ GeV, Fig.~\ref{fig:QCDThermo}), where $T$ is the temperature and $s$ is the entropy density. Obviously this is a thermodynamic/hydrodynamic problem with some \QCD\ in the Monte Carlo to get $R_{AA}$ but the results, shown as the dashes for two different values of $D=3/(2\pi T)$ and $D=12/(2\pi T)$ on Fig.~\ref{fig:PerfectLiquid}c,d, lead to a spectacular conclusion. Taking $D=6\eta/Ts \approx (6\ \rm{to}\ 4)/(2\pi T)$ as the most reasonable range that fits both $R_{AA}$ and $v_2^{\rm HF}$ in Fig.~~\ref{fig:PerfectLiquid}c,d~\cite{van Hees}, gives the result:
\begin{equation} 
\eta/s=(2\ \rm{to}\ 4/3)/4\pi \label{eq:PerfectLiquid}
\end{equation}           
which is intriguingly close to the conjectured~\cite{QuantumBound} quantum lower bound, $\eta/s=1/4\pi$. 

This is why we claimed the discovery of the \QGP, the perfect liquid, at RHIC ~\cite{NPA740}, \cite{NPA757B}, \cite{NPA757P},\cite{STWP}, \cite{NPA757PX}. \vspace*{-1.0pc}
\subsubsection{$J/\Psi$ enhancement not suppression proves the existence of the \QGP\ at the LHC}
\begin{figure}[h]
\centering
\raisebox{+0.5pc}{\includegraphics[width=0.45\textwidth]{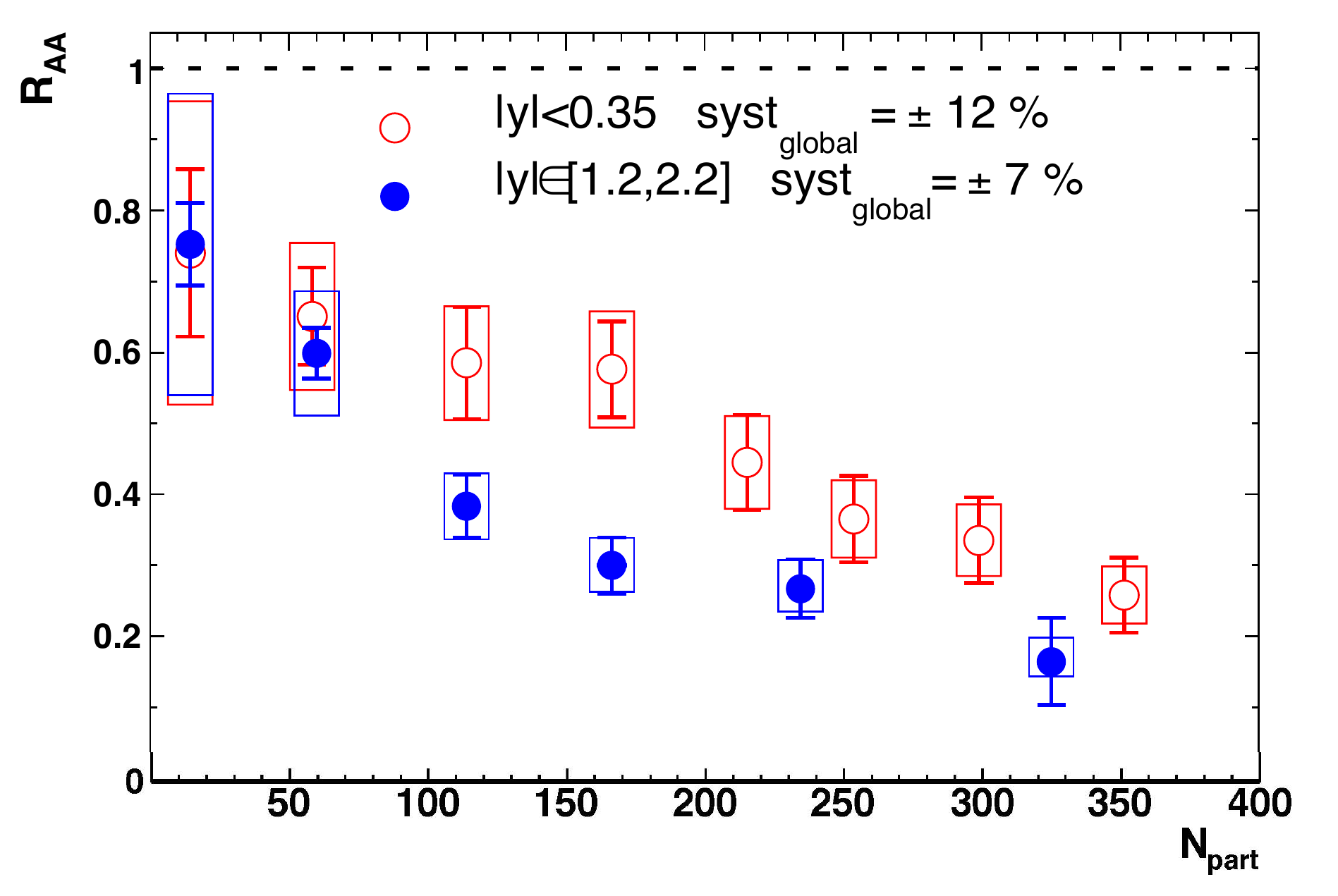}}
\raisebox{+0.0pc}{\includegraphics[width=0.52\textwidth]{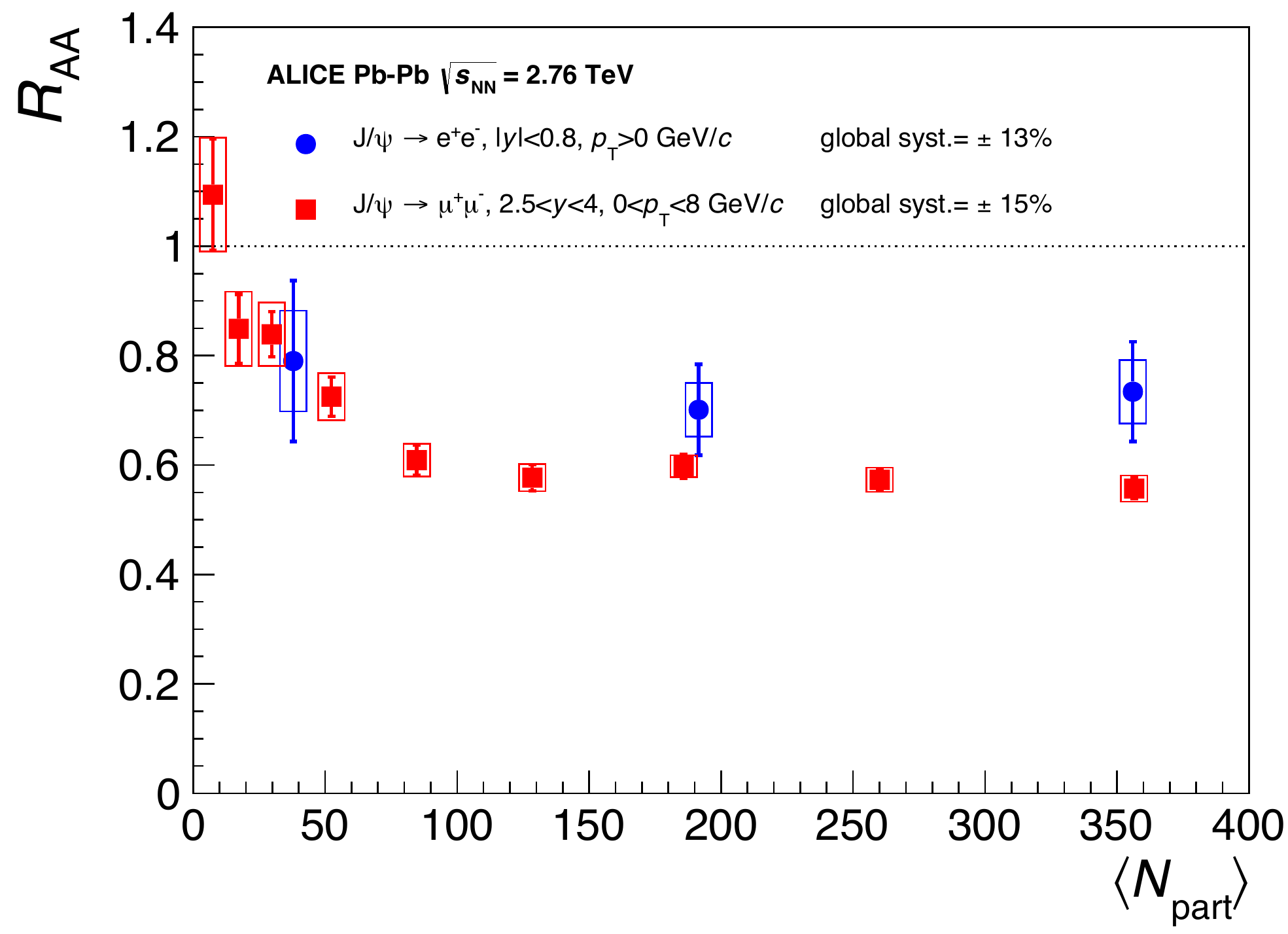}}
\caption[]{a) (left) PHENIX~\cite{PXPRL98J} $R_{AA}$ of $J/\Psi$ in  Au$+$Au at $\sqrt{s_{NN}}=200$ GeV for $e^+ e^-$ ($|y|<0.8$), and $\mu^+ \mu^-$ ($1.2\leq y\leq 2.2$) decay. b)(right) ALICE~\cite{ALICEPLB734} $R_{AA}$ of $J/\Psi$ in  Au$+$Au at $\sqrt{s_{NN}}=2.76$ TeV for $e^+ e^-$ ($|y|<0.35$), and $\mu^+ \mu^-$ ($2.5\leq y\leq 4.0$) decay.   } 
\label{fig:alljpsi} 
\end{figure}

PHENIX measurements of $J/\Psi$ suppression ($R_{AA}$) in Au$+$Au collisions at $\sqrt{s_{NN}}=200$ GeV  relative to point-like scaling of p$+$p collisions at mid-rapidity~\cite{PXPRL98J} Fig.~\ref{fig:alljpsi}a turned out to be nearly identical to the NA50 fixed-target measurements in $\sqrt{s_{NN}}=17.2$ Pb$+$Pb collisions at CERN~\cite{NA50EPJC39}.
The equality of $J/\Psi$ suppression at $\sqrt{s_{NN}}=17.2$ and 200 GeV was was elegantly explained as recombination or coalescence of $c$ and $\bar{c}$ quarks in the \QGP\ to regenerate $J/\Psi$~\cite{RappPLB664}. 

Miraculously this made the observed $R_{AA}$ equal at SpS and RHIC c.m. energies. I called this my ``Nightmare Scenario'' because I thought that nobody would believe it. The good news was that such models are testable because they predicted the reduction of $J/\Psi$ suppression or even an enhancement ($R_{AA}> 1$) at LHC energies~\cite{PBMPLB490}, \cite{ThewsPRC63},\break \cite{AndronicNPA789}, which would be spectacular, if observed. 

As shown in Fig.~\ref{fig:alljpsi}b, the most recent ALICE~\cite{ALICEPLB734} measurement of $R_{AA}$ for $J/\psi$ at $\sqrt{{s}_{NN}}=2.76$ TeV exhibits considerably less suppression for both $J/\psi\rightarrow e^+ e^-$ at mid rapidity and $J/\psi\rightarrow \mu^+ \mu^-$ at forward rapadity than the PHENIX measurements at $\sqrt{s_{NN}}=200$ GeV in  Fig.~\ref{fig:alljpsi}a. The reduction of $J/\Psi$ suppression at the LHC compared to RHIC is a clear observation of regeneration at LHC which directly proves the existence of the \QGP, since it is evidence that the large number of $c$ and $\bar{c}$ quarks produced (with their color charge hidden by Debye screening) freely traversed the medium (with a large density of similarly screened color charges) until they met another quark close enough within the screening radii to form $J/\Psi$'s.   

However~\cite{Satz2013}, these beautiful results do not prove that $J/\Psi$ are deconfined in the \QGP. According to Satz, the crucial issue is whether the medium modifies the fraction of produced $c\bar{c}$ pairs which form $J/\Psi$. Dissociation of $J/\Psi$ in the medium would reduce the observed $J/\Psi/c\bar{c}$ ratio in A$+$A compared to p$+$p collisions, i.e $R_{AA}^{J/\Psi}/R_{AA}^{c\bar{c}}\ll 1$. Personally, I think that Debye screening which is the basis of suppression of $J/\Psi$ in the \QGP\ is proved by Fig.~\ref{fig:alljpsi} but Satz's requirement may be important to find how low in $\sqrt{s_{NN}}$ the \QGP\ is formed.
\subsubsection{jet quenching at the LHC}
\begin{figure}  
\centering
\raisebox{+0.0pc}{\includegraphics[width=0.81\textwidth]{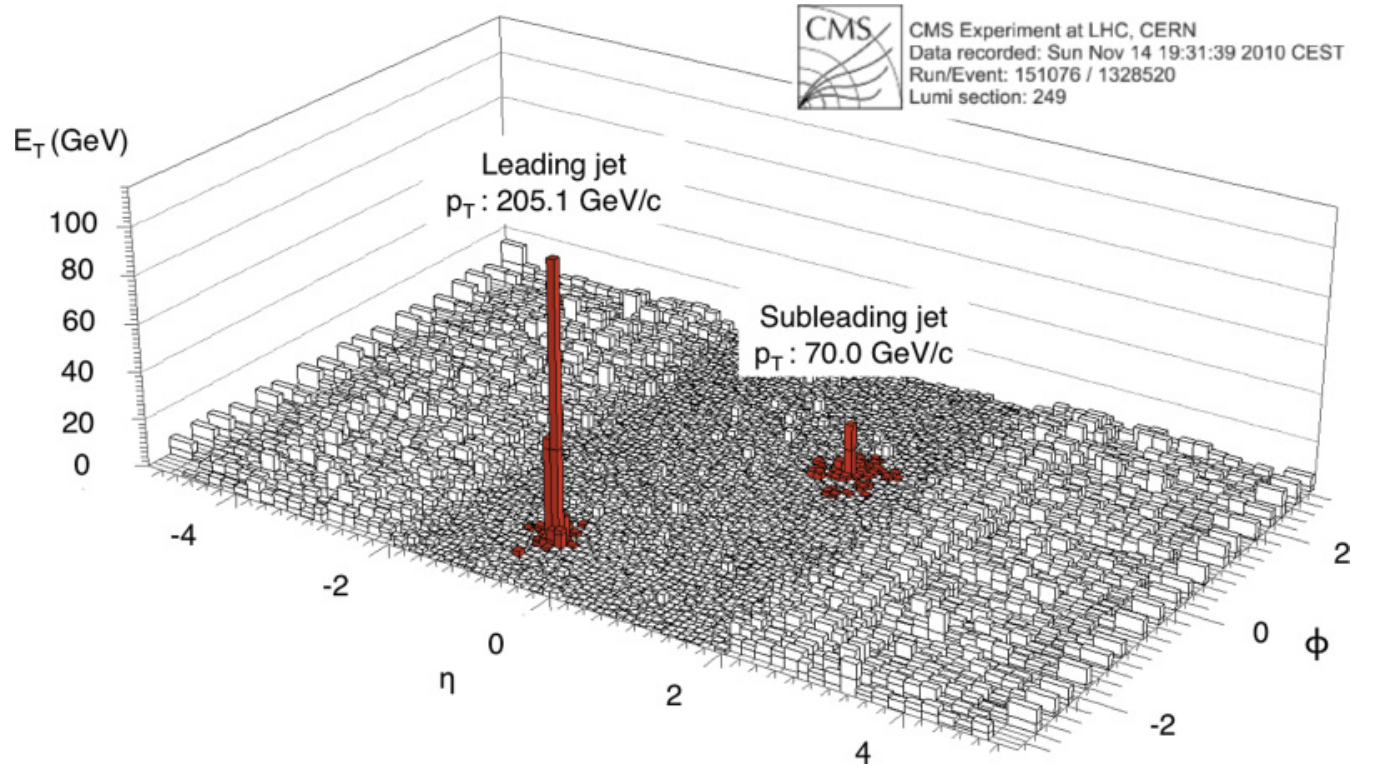}}
\caption[]{Example~\cite{CMSPRC84} of an unbalanced dijet in a PbPb collision event at $\sqrt{s_{NN}}=2.76$ TeV. The summed transverse energy in the electromagnetic and hadron calorimeters is plotted vs. $\eta$ and $\phi$, with the identified jets highlighted, and labeled with the corrected jet transverse momentum. } 
\label{fig:CMSnojet}
\end{figure}

Although nobody at RHIC had claimed to measure the quenching of fully reconstructed jets in $\sqrt{s_{NN}}=200$ GeV Au$+$Au collisions, because the multiparticle background is too large, as soon as the LHC started its 2760 GeV run, unbalanced jets were visibile on the event displays for both ATLAS~\cite{ATLASPRL105} and CMS (Fig.~\ref{fig:CMSnojet})~\cite{CMSPRC84}

\section{Summary and Conclusions}
In the years 1972-1982 experiments at Hadron Colliders made major contributions to the theory of \QCD\ which was developed in this same period. More recently, Heavy Ion Colliders have opened up a new field of the study of superdense nuclear matter in the laboratory, leading to the discovery of the Quark Gluon Plasma which can be analyzed with lattice \QCD.  Highlights of these achievements can be summarized as follows.

\subsection{Hadron Colliders and \QCD\ }
\noindent {\bf The CERN Intersecting Storage Rings (ISR), Geneva, SZ, 1971--1983.}\\ 
This was the first hadron collider, with p$+$p collisions at c.m. energy $\sqrt{s}=23.5-62.4$ GeV. Also d$+$d $\alpha+\alpha$ p$+\alpha$ and p$+\bar{\rm p}$ experiments were performed. The largest contributions for the development of \QCD\ were from the ISR.
\begin{enumerate}
\item{Discovery of particle production at high $p_T\geq 3$ GeV/c, 1972.}
\item{Discovery of charm particles by direct single $e^{\pm}$ production in 1974, but not understood until 1976.}
\item{First application of \QCD\ theory to ISR $\pi^{\pm}$ and $\pi^0$ spectra with $3\leq p_T\leq 14$ GeV/c, 1978.}
\item{Discovery of direct-$\gamma$ production 1979-1980 as predicted by \QCD\ in 1977.} 
\item{Observation of the di-jet structure of hard-scattering via two-particle correlations including discovery and measurements of $k_T$, the transverse momentum of a parton in a nucleon, and measurement of $\mean{j_T}$ the average momentum of jet fragments transverse to the jet axis, 1977-1979.}
\item{First measurement of the $\cos\theta^*$ distribution of elementary \QCD\ subprocesses including the increase of the \QCD\ coupling constant $\alpha_s(Q^2)$ with decreasing $Q^2$ at more forward angles, 1982.}
\item{First direct measurement of the gluon structure function $G(x,Q^2)$, 1987.} 
\end{enumerate}\vspace*{0.5pc} 

\noindent {\bf The CERN S$\bar{p}p$S Collider 1981--1991.} Nobel Prize winning hadron collider with $\bar{\rm p}+$p collisions at $\sqrt{s}=546-630$ GeV.
\begin{enumerate}
\item{First definitive observation of jets in hadron collisions, 1982.}
\item{Discovery of the $W^{\pm}$ `intermediate boson' of weak interactions, 1983.}
\item{Discovery of the $Z^0$ boson of weak interactions, 1983.}
\end{enumerate}\vspace*{0.5pc} 

\noindent{\bf ISABELLE at BNL, Upton, NY, USA, 1978--1983 (cancelled)}\\
ISABELLE was designed as p$+$p collider with $\sqrt{s}=800$ GeV. This was the first collider with superconducting magnets, which led to problems during construction. Eventually greatly improved superconducting accelerator magnets were developed and tested, known as the Palmer Magnet~\cite{PalmerMagnet}. This design has been used by all following colliders except for the Fermilab Tevatron Collider. \vspace*{0.5pc}

\noindent{\bf Tevatron Collider at Fermilab, Batavia, Illinois, USA, 1986--2011}\\
This was a $\sqrt{s}=1.8$ TeV $\bar{\rm p}+$p collider with a different superconducting magnet design than ISABELLE. However, the Palmer Magnet developed at ISABELLE was based on the superconducting cable developed for the Tevatron.
\begin{enumerate} 
\item{Discovery of the top quark, 1995}
\item{Measurement of jet cross sections over a wide range of $p_T$ and rapidity in agreement with \QCD\ }.
\end{enumerate}\vspace*{0.5pc} 

\noindent{\bf Superconducting Super Collider (SSC), Waxahachie, Texas, USA, 1990--1993 (cancelled)} \\
The SSC was designed as a $\sqrt{s}=40$ TeV p$+$p collider. Its demise hurt High Energy Physics in the U.S.A; but on the positive side, in 1994, the U.S. High Energy Physics Advisory Panel recommended that the U.S. siginificantly participate in the CERN Large Hadron Collider to help maintain U.S. involvement in High Energy Physics. In 1997 the DOE recommended spending a total of more than \$531 million for U.S. contributions to both the accelerator and the experiments, the first agreement between CERN and the U.S. government. This allowed CERN to build the LHC in a single stage of $\sqrt{s}=14$ GeV rather than the originally approved two stage construction with missing magnets~\cite{LHC2008}.\vspace*{0.5pc}

\noindent{\bf The CERN Large Hadron Collider, LHC, 2008--present.}\\
The largest and highest energy p$+$p collider with $\sqrt{s}=7-13$ TeV. Also Pb$+$Pb and p$+$Pb collisions.
\begin{enumerate} 
\item{Measurement of jet cross sections over a wide range of $p_T$ and rapidity in agreement with \QCD\ }.
\item{Discovery of the Higgs Boson, 2012}.
\item{Discovery of suppression of jets in Pb$+$Pb collisions, 2010.}
\item{Discovery of $J/\Psi$ enhancement compared to measurements at RHIC, evidence for the Quark Gluon Plasma, 2014.}
\end{enumerate}\vspace*{0.5pc} 

\noindent{\bf RHIC, the Relativistic Heavy Ion Collider at BNL, 2000--present.}\\
RHIC is a Heavy Ion Collider with c.m. energy per nucleon pair, $\sqrt{s_{NN}}=19.6-200$ GeV, and the first polarized p$+$p collider with $\sqrt{s}=500$ GeV. RHIC has provided collisions of p$+$p, p$+$Au, p$+$Al, d$+$Au, $^3$He$+$Au, Cu$+$Cu, Cu$+$Au, Au$+$Au and U$+$U.
\begin{enumerate}
\item{Discovery of jet quenching predicted for the \QGP, 2002.}
\item{Discovery of the \QGP\ as a perfect liquid with shear viscosity/entropy density near the quantum limit, 2005--2007.}
\end{enumerate}\vspace{-1.0pc}
 \subsection{Future Possibilities for \QCD}
A remaining problem with \QCD\ is the need to utilize structure and fragmentation functions, which come from experimental measurements, in the calculations. An excellent possibility for the future is to be able to also calculate the structure and fragmentation functions in \QCD.~\footnote{Thanks to Norman Christ for this suggestion.}

\section{Appendix}
\label{sec:appendix}
\subsection{The relativistic longitudinal variable, $\bm y$, rapidity}
\label{sec:rapidity}
Any particle with momentum $P$ and energy $E$ can be represented by its longitudinal momentum $P_L$, which is subject to a Lorentz transformation, and its transverse momentum $P_T$ which is not affected, so 
$E^2=p^2+m^2=p_T^2+p_L^2+m^2$ in the nomenclature where the speed of light $c\equiv 1$. The `transverse mass', $m_T\equiv\sqrt{p_T^2+m^2}=\sqrt{E^2-P_L^2}$, is invariant under a Lorentz transformation.  
The definition of the rapidity of this particle is:
\begin{equation} 
\cosh y={E}/{m_T} \qquad \sinh y={p_L}/{m_T} \qquad dy=dp_L/E \label{eq:ydef1}
\end{equation}
\begin{equation}
y=\frac{1}{2} \ln\left(\frac{E+P_L}{E-P_L}\right)=\ln \left(\frac{E+P_L}{m_T}\right) \qquad .
\label{eq:ydef2}
\end{equation} 

The advantage of rapidity is that if the rapidity of a particle is $y^*$ in a frame moving with velocity  $\beta=v/c$ with respect to our system, then the rapidity $y$ of the particle in our system is related to $y^*$ by simple addition, $y=Y+y^*$, where $Y$ is the rapidity of the moving frame   
$$Y=\frac{1}{2}\ln\left(\frac{1+\beta}{1-\beta}\right) \qquad .$$

Using the rapidity variable, the invariant differential single particle inclusive cross section for a scattered particle with longitudinal momentum $p_L$ along the collision axis, and transverse momentum $p_T$ at azimuthal angle $\phi$ in cylindrical coordinates, can be written in the Lorentz invariant form:
\begin{equation}
\frac{E d^3\sigma}{dp^3}=\frac{Ed^3\sigma}{p_T dp_T dp_L d\phi}=\frac{d^3\sigma}{p_T dp_T dy d\phi} \qquad .
\label{eq:sigmainv}
\end{equation}
\subsubsection{Pseudorapidity, $\bm{\eta}$}
In the limit when ($P\gg m$) for a particle, $E\rightarrow P$, $m_T\rightarrow p_T\rightarrow E \sin\theta$, $p_L\rightarrow E \cos\theta$, where $\theta$ is the polar angle, the rapidity $y$ (Eq.~\ref{eq:ydef1}) reduces to the pseudorapidity, $\eta$ :
\begin{equation}
\cosh\eta=\csc\theta \qquad \sinh\eta=\cot\theta \qquad \tanh\eta=\cos\theta
\label{eq:5}
\end{equation}
\begin{equation}
\eta=-\ln\tan\theta/2 \qquad .
\label{eq:4} \end{equation}

\subsection{Parton-parton scattering}
\label{app:pscat}

In hadron colliders, collisions usually take place with protons of equal and opposite vector momenta ${\bf P}$ and $-{\bf P}$ so that the p$+$p center of mass (c.m.) system is at rest in the laboratory. 
\begin{figure}[!h]
\centering
\includegraphics[width=0.7\linewidth]{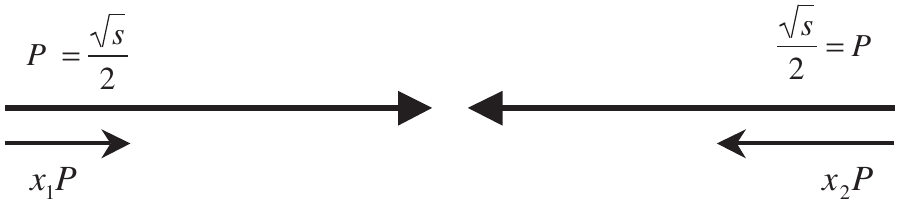}
\caption[] {Proton-proton collision with colliding parton momenta illustrated
\label{fig:ppscat}}
\end{figure}
However, the partons in each proton have fractional momenta $x_1$ and $x_2$ which are not generally equal so that the parton-parton c.m. system moves longitudinally in the p$+$p c.m. system (Fig.~\ref{fig:ppscat}). 
For kinematic calculations, the gluon and quark partons are massless; and for simplicity the proton mass $M$ can be ignored because $P\gg M$ for the colliders discussed, so that $E^2=P^2+M^2\rightarrow P^2$ and the particles all have relativistic velocities so that Lorentz Transformations are required.

 The c.m. energy of the p$+$p collision is given by the Lorentz invariant Mandelstam variable $\sqrt{s}=2P$, where 
\begin{equation} 
s=-(p_1+p_2)^2=-[(0,0,P,iP)+(0,0,-P,iP)]^2=4P^2 \qquad .
\label{eq:sdef}
\end{equation} 
The partons (with notation $\hat{}$\ ) are assumed to travel along the longitudinal directions of their protons so that the c.m. energy of the parton-parton collision $\sqrt{\hat{s}}$ is given by:
\begin{eqnarray} 
\hat{s}=-(\hat{p}_1+\hat{p}_2)^2&=&-[(0,0,x_1 P,ix_1 P)+(0,0,-x_2 P,ix_2 P)]^2=4P^2x_1x_2 \nonumber \\
                                &=& s x_1 x_2 \qquad .
\label{eq:shat}
\end{eqnarray} 
The parton-parton c.m. system has longitudinal momentum $P_L=(x_1 - x_2)P$ and total energy $E=(x_1 + x_2)P$ in the p$+$p  c.m. system, transverse mass \mbox{$m_T=\sqrt{E^2-P_L^2}$}=$\sqrt{\hat{s}}$ and thus has rapidity (Eq.~\ref{eq:ydef2}) $\hat{y}$:
\begin{equation}
\hat{y}=\ln \left(\frac{E + P_L}{\sqrt{\hat{s}}}\right)=\ln\sqrt{\frac{x_1}{x_2}}
\qquad. \label{eq:haty}
\end{equation}
Also, Eqs.~\ref{eq:haty},~\ref{eq:shat} can be solved for $x_1$ and $x_2$:
\begin{equation}
x_1=\sqrt{\frac{\hat{s}}{s}}\,e^{\hat{y}} \qquad x_2=\sqrt{\frac{\hat{s}}{s}}\,e^{-\hat{y}}
\qquad . \label{eq:x1x2hats}
\end{equation}
\subsubsection{Kinematics of parton-parton scattering}
\begin{figure}[!h]
\begin{center}
\includegraphics[width=0.55\linewidth]{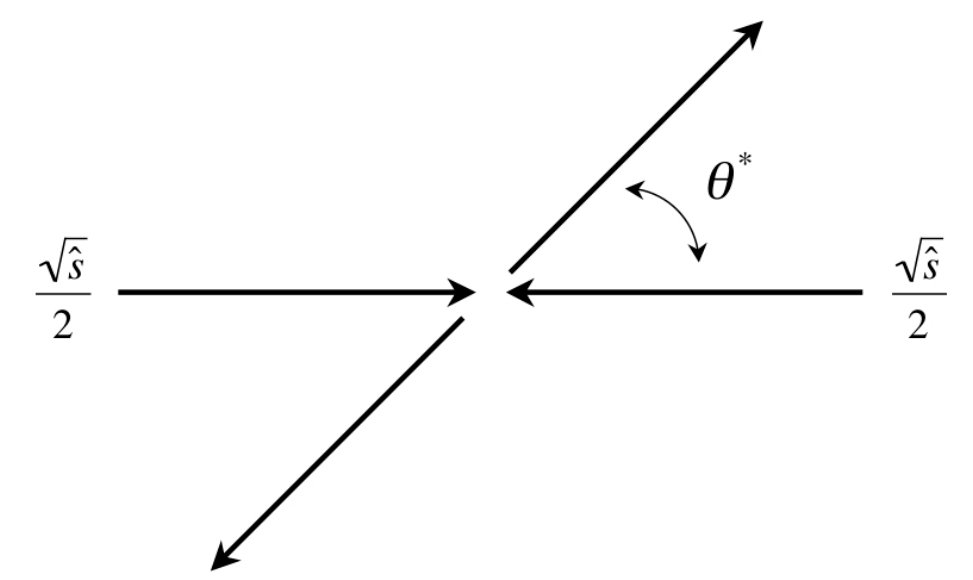}
\end{center}\vspace*{-1.0pc}
\caption[]{Elastic scattering in parton-parton c.m. system}
\label{fig:pscat-theta}
\end{figure}

Figure~\ref{fig:pscat-theta} shows the elastic scattering of the two initial partons through angle $\theta^*$ in the parton-parton c.m. system, where the two colliding partons have equal and opposite momenta, $\sqrt{\hat{s}}/2$. 
The scattered parton 3-momenta are equal and opposite, ${\bf P_3^*=-P_4^*}$, and their energies are equal, and equal to the magnitude of their 3-momenta,  $E_3^*=E_4^*=P_3^*=P_4^*=\sqrt{\hat{s}}/2$, because both outgoing partons are assumed to be massless. For the $z$ axis along the direction of the initial partons and the $y$ axis perpendicular to the $z$ axis in the scattering plane,  
the parton 4-momenta in the parton-parton c.m. system can be written as:
\begin{eqnarray} p_1^*=\frac{\sqrt{\hat{s}}}{2} (0,0,1, i) & & p_2^*=\frac{\sqrt{\hat{s}}}{2} (0,0,-1, i) 
\label{eq:pscat2init}\\
p_3^*=\frac{\sqrt{\hat{s}}}{2} (0,\sin\theta^*,\cos\theta^*, i) & & p_4^*=\frac{\sqrt{\hat{s}}}{2} (0,-\sin\theta^*,-\cos\theta^*, i)  \qquad , \label{eq:pscat2all}
 \end{eqnarray}
so that 
\begin{equation}
 P_{L_3}^*=-P_{L_4}^*=\frac{\sqrt{\hat{s}}}{2}\cos\theta^*  \qquad P_{T_3}^*=-P_{T_4}^*\equiv p_T=m_T=\frac{\sqrt{\hat{s}}}{2}\sin\theta^* \quad . \label{eq:PL*PTdefs} 
 \end{equation} 
 For the outgoing partons, the rapidities are equal and opposite in the parton-parton c.m. system:
\begin{equation}
y_3^*=\sinh^{-1} P_{L_3}^*/m_T=-y_4^*
\end{equation}

The other two Mandelstam invariants of the parton-parton scattering, $\hat{t}$, the 4-momentum-transfer-squared, and $\hat{u}$ can be easily computed from Eqs.~\ref{eq:pscat2init} and \ref{eq:pscat2all}. 
The invariants are related by the c.m. scattering angle:
\begin{equation}
 \hat{Q}^2=-\hat{t}=(p_1^* - p_3^*)^2 =\hat{s} \frac{(1-\cos\theta^*)}{2} \qquad;\label{eq:hatt}
 \end{equation}  
and likewise
\begin{equation} 
-\hat{u}=(p_1^* - p_4^* )^2=\hat{s} \frac{(1+\cos\theta^*)}{2} \qquad.\label{eq:hatu}
\end{equation}

\section*{Acknowledgements}
 Research supported by U.S. Department of Energy, Contract No. DE-SC0012704.   

\end{document}